\documentclass{aa}   

\usepackage{graphicx}
\usepackage{txfonts}
\usepackage[breaklinks=true,colorlinks=true,linkcolor=blue,citecolor=gray,urlcolor=magenta]{hyperref}
\usepackage[]{listings} 
\usepackage{color}
\definecolor{light-gray}{gray}{0.95}
\newcommand{\simgt}{\,\rlap{\lower 3.5 pt \hbox{$\mathchar \sim$}} \raise
1pt \hbox {$>$}\,}
\newcommand{\simlt}{\,\rlap{\lower 3.5 pt \hbox{$\mathchar \sim$}} \raise
1pt \hbox {$<$}\,}
\newcommand{\lya}{Ly$\alpha$}

\newcommand{\civ}{\ion{C}{iv}}
\newcommand{\civone}{\ion{C}{iv}$_1$}
\newcommand{\civtwo}{\ion{C}{iv}$_2$}

\newcommand{\nv}{\ion{N}{v}}
\newcommand{\nvone}{\ion{N}{v}$_1$}
\newcommand{\nvtwo}{\ion{N}{v}$_2$}

\newcommand{\heii}{\ion{He}{ii}}

\newcommand{\oiii}{\ion{O}{iii}}
\newcommand{\oiiione}{\ion{O}{iii}]$_1$}
\newcommand{\oiiitwo}{\ion{O}{iii}]$_2$}

\newcommand{\siiii}{\ion{Si}{iii}}
\newcommand{\siiiione}{[\ion{Si}{iii}]}
\newcommand{\siiiitwo}{\ion{Si}{iii}]}

\newcommand{\ciii}{\ion{C}{iii}}
\newcommand{\ciiione}{[\ion{C}{iii}]}
\newcommand{\ciiitwo}{\ion{C}{iii}]}

\newcommand{\mgii}{\ion{Mg}{ii}}

\newcommand{\ciiifull}{[\ciii]~$\lambda$1907~+~\ciii]~$\lambda$1909}
\newcommand{\siiiifull}{[\siiii]~$\lambda$1883~+~\siiii]~$\lambda$1892} 
\newcommand{\oiiifull}{[\oiii]~$\lambda$1661~+~\oiii]~$\lambda$1666} 
\newcommand{\civfull}{\civ~$\lambda$1548~+~\civ~$\lambda$1551} 

\newcommand{\oiifull}{[\ion{O}{ii}]~$\lambda$3726~+~[\ion{O}{ii}]~$\lambda$3729}
\newcommand{\siifull}{[\ion{S}{ii}]~$\lambda$6716~+~[\ion{S}{ii}]~$\lambda$6730}

\usepackage{multirow}

\usepackage{xcolor}

\newcommand{\orcid}[1]{\href{https://orcid.org/#1}{\includegraphics[scale=0.3]{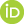}\,}}

\usepackage{natbib,twoopt}
\bibpunct{(}{)}{;}{a}{}{,}             
\makeatletter
  \newcommandtwoopt{\citeads}[3][][]{\href{http://adsabs.harvard.edu/abs/#3}%
    {\def\hyper@linkstart##1##2{}%
     \let\hyper@linkend\@empty\citealp[#1][#2]{#3}}}
  \newcommandtwoopt{\citepads}[3][][]{\href{http://adsabs.harvard.edu/abs/#3}%
    {\def\hyper@linkstart##1##2{}%
     \let\hyper@linkend\@empty\citep[#1][#2]{#3}}}
  \newcommandtwoopt{\citetads}[3][][]{\href{http://adsabs.harvard.edu/abs/#3}%
    {\def\hyper@linkstart##1##2{}%
     \let\hyper@linkend\@empty\citet[#1][#2]{#3}}}
  \newcommandtwoopt{\citeyearads}[3][][]%
    {\href{http://adsabs.harvard.edu/abs/#3}
    {\def\hyper@linkstart##1##2{}%
     \let\hyper@linkend\@empty\citeyear[#1][#2]{#3}}}
\makeatother
\begin{document} 

\title{Recovery and analysis of rest-frame UV emission lines in 2052 galaxies observed with MUSE at $1.5<z<6.4$\thanks{The catalogs described in Appendix~\ref{sec:mastercat} (Table \ref{tab:mastercatcol}) and \ref{sec:litcol} (Table~\ref{tab:litcatcol}) are available in electronic form at the CDS via anonymous ftp to cdsarc.u-strasbg.fr (130.79.128.5)
or via \href{http://cdsweb.u-strasbg.fr/cgi-bin/qcat?J/A+A/}{http://cdsweb.u-strasbg.fr/cgi-bin/qcat?J/A+A/}.}}
\titlerunning{UV emission at $1.5<z<6.4$}
\authorrunning{Schmidt et al. (2021)}

\author{K.~B.~Schmidt\inst{\orcid{0000-0002-3418-7251} 1}\thanks{E-mail: kbschmidt@aip.de},
J.~Kerutt\inst{\orcid{0000-0002-1273-2300} 1,2}, 
L.~Wisotzki\inst{\orcid{0000-0003-2977-423X} 1}, 
T.~Urrutia\inst{\orcid{0000-0001-6746-9936} 1}, 
A.~Feltre\inst{\orcid{0000-0001-6865-2871} 3},
M.~V.~Maseda\inst{\orcid{0000-0003-0695-4414} 4},
T.~Nanayakkara\inst{\orcid{0000-0003-2804-0648} 5}, \newline
R.~Bacon\inst{6},
L.~A.~Boogaard\inst{\orcid{0000-0002-3952-8588} 4,7},
S.~Conseil\inst{\orcid{0000-0002-3657-4191} 8},
T.~Contini\inst{\orcid{0000-0003-0275-938X} 9},
E.~C.~Herenz\inst{\orcid{0000-0002-8505-4678} 10},
W.~Kollatschny\inst{\orcid{0000-0002-0417-1494} 11},
M.~Krumpe\inst{1}, 
F.~Leclercq\inst{\orcid{0000-0002-6085-5073} 2}
G.~Mahler\inst{\orcid{0000-0003-3266-2001} 12,13},
J.~Matthee\inst{\orcid{0000-0003-2871-127X} 14},
V.~Mauerhofer\inst{\orcid{0000-0003-0595-9483} 2,6},
J.~Richard\inst{\orcid{0000-0001-5492-1049} 6}, and
J.~Schaye\inst{\orcid{0000-0002-0668-5560} 4}.
}

\institute{
$^1$ Leibniz-Institut f\"{u}r Astrophysik Potsdam (AIP), An der Sternwarte 16, 14482, Potsdam, Germany\\
$^2$ Observatoire de Gen\`eve, Universit\'e de Gen\`eve, Chemin Pegasi 51, 1290 Versoix, Switzerland\\
$^3$ INAF -- Osservatorio di Astrofisica e Scienza dello Spazio di Bologna, Via P. Gobetti 93/3, 40129 Bologna, Italy\\
$^4$ Leiden Observatory, Leiden University, P.O. Box 9513, 2300 RA, Leiden, The Netherlands\\
$^5$ Centre for Astrophysics \& Supercomputing, Swinburne University of Technology, PO Box 218, Hawthorn, VIC 3112, Australia\\
$^6$ Univ. Lyon, Univ. Lyon1, ENS de Lyon, CNRS, Centre de Recherche Astrophysique de Lyon UMR5574, 69230 Saint-Genis-Laval, France\\
$^7$ Max Planck Institute for Astronomy, K\"{o}nigstuhl 17, 69117 Heidelberg, Germany\\
$^8$ Gemini Observatory/NSF's NOIRLab, Casilla 603, La Serena, Chile\\
$^9$ Institut de Recherche en Astrophysique et Plan\'etologie (IRAP), Universit\'e de Toulouse, CNRS, UPS, CNES, Toulouse, France\\
$^{10}$ European Southern Observatory, Av. Alonso de C\'ordova 3107, 763 0355 Vitacura, Santiago, Chile\\
$^{11}$ Institut f\"ur Astrophysik, Universit\"at G\"ottingen, Friedrich-Hund Platz 1, D-37077 G\"ottingen, Germany\\
$^{12}$ Institute for Computational Cosmology, Durham University, South Road, Durham DH1 3LE, UK\\
$^{13}$ Centre for Extragalactic Astronomy, Durham University, South Road, Durham DH1 3LE, UK\\
$^{14}$ Department of Physics, ETH Z\"urich, Wolfgang-Pauli-Strasse 27, 8093 Z\"urich, Switzerland\\
}
\date{Received March 25, 2021 / Accepted July 8, 2021}

\abstract{
Rest-frame ultraviolet (UV) emission lines probe electron densities, gas-phase abundances, metallicities, and ionization parameters of the emitting star-forming galaxies and their environments. 
The strongest main UV emission line, \lya, has been instrumental in advancing the general knowledge of galaxy formation in the early universe. 
However, observing \lya{} emission becomes increasingly challenging at $z\gtrsim6$ when the neutral hydrogen fraction of the circumgalactic and intergalactic media increases. 
Secondary weaker UV emission lines provide important alternative methods for studying galaxy properties at high redshift. 
We present a large sample of rest-frame UV emission line sources at intermediate redshift for calibrating and exploring the connection between secondary UV lines and the emitting galaxies' physical properties and their \lya{} emission.
The sample of 2052 emission line sources with $1.5<z<6.4$ was collected from integral field data from the MUSE-Wide and MUSE-Deep surveys taken as part of Guaranteed Time Observations.
The objects were selected through untargeted source detection (i.e., no preselection of sources as in dedicated spectroscopic campaigns) in the three-dimensional MUSE data cubes.
We searched optimally extracted one-dimensional spectra of the full sample for UV emission features via emission line template matching, resulting in a sample of more than 100 rest-frame UV emission line detections. 
We show that the detection efficiency of (non-\lya) UV emission lines increases with survey depth, and that the emission line strength of \heii~$\lambda$1640\AA, \oiiifull, and \siiiifull{} correlate with the strength of \ciiifull.
The rest-frame equivalent width (EW$_0$) of \ciiifull{} is found to be roughly $0.22\pm0.18$ of EW$_0$(\lya).
We measured the velocity offsets of resonant emission lines with respect to systemic tracers. 
For \civfull{} we find that $\Delta v_\textrm{\civ}\lesssim250$~km~s$^{-1}$, whereas $\Delta v_\textrm{\lya}$ falls in the range of 250-500~km~s$^{-1}$ which is in agreement with previous results from the literature.
The electron density $n_e$ measured from \siiiifull{} and \ciiifull{} line flux ratios is generally $<10^{5}\textrm{cm}^{-3}$ and the gas-phase abundance is below solar at $12+\log_{10}(\textrm{O/H})\approx8$.
Lastly, we used ``PhotoIonization Model Probability Density Functions'' to infer physical parameters of the full sample and individual systems based on photoionization model parameter grids and observational constraints from our UV emission line searches. 
This reveals that the UV line emitters generally have ionization parameter $\textrm{log$_{10}$(U)}\approx-2.5$ and  metal mass fractions that scatter around $Z\approx10^{-2}$, that is $Z\approx0.66\,Z_\odot$.
Value-added catalogs of the full sample of MUSE objects studied in this work and a collection of UV line emitters from the literature are provided with this paper. 
}
 

\maketitle
%

\newpage
\section{Introduction}
\label{sec:intro}

Over the last decade there have been increased efforts to characterize and study rest-frame ultraviolet (UV) emission lines from star-forming galaxies at increasingly higher redshifts, with a recent record holder at $z=11.09$ presented by \cite{2020NatAs.tmp..246J}, as such studies provide extensive knowledge about the emitting galaxy population.
Also at lower redshift and $z\approx0$, UV lines have recently provided the means necessary to leverage the extensive information provided by rest-frame optical emission, which is often unavailable at higher redshift, and they have provided analogs of high-redshift systems, being a key diagnostic for analyzing and understanding galaxy evolution in broader terms.

Rest-frame UV emission lines probe the physical conditions of the ionized gas, provide constraints on the physical properties of the emitting galaxies (such as from photo-ionization models), and have proven to be valuable probes of the surrounding environment of their host galaxies. 
Hence, they provide insights into general conditions for the galaxy formation and assembly at the targeted redshifts. 

For instance, the \siiiifull{} and \ciiifull{} doublet components provide information about the electron density of the gas from which they are emitted \citep{2006agna.book.....O,2019ApJ...880...16K}.
The resonant \civfull{} doublet is produced by highly ionizing radiation mainly from young stars, shocks, and/or active galactic nuclei (AGN), 
and it can be used to trace the physical conditions in the ionized gas and the interstellar medium (ISM) in star-forming galaxies as well as winds from the emitting massive O and B stars \citep{2017MNRAS.470.3532V,2018ApJ...863...14B,2018ApJ...859..164B,2020A&A...641A.118F}. 
By combining the \oiiifull{} and \ciiifull{} doublet fluxes, the C/O ratio in the galaxy can be approximated \citep{2018ApJ...859..164B,2016ApJ...827..126B}, and if information on the optical [\oiii]~$\lambda$5007 flux can be obtained, the \oiiifull{} flux allows for a determination of the electron temperature \citep{2018ApJ...859..164B}, avoiding the need for detections of the faint auroral [\oiii]~$\lambda$4363 emission \citep[e.g.,][]{2016ApJ...825L..23S,2020MNRAS.491.1427S,2018ApJ...859..175B}.
Determining the emission flux ratios of different UV line species and comparing them to grids of photoionization models predicting UV line fluxes provides estimates of, for instance, the metallicity and gas-phase abundances, the ionization parameter, and hydrogen number density 
\citep[e.g.,][]{
2016MNRAS.462.1757G,
2016MNRAS.456.3354F,
2016ApJ...833..136J,
2017MNRAS.472.2468H,2019MNRAS.487..333H,
2017ApJ...840...44B,2018ApJ...863...14B,2020ApJ...893....1B,
2018MNRAS.477.2098N,
2019MNRAS.490..978P,
2019ApJ...880...16K}.
The \heii~$\lambda$1640 emission likely includes both nebular emission and emission from stellar winds \citep[][]{2018ApJ...863...14B,2020ApJ...893....1B} and is therefore a valuable probe and diagnostic of these parameters, even though it has been proven challenging to reproduce the total observed \heii{} emission from star-forming galaxies without invoking increased ionizing photon production from binary stars and/or X-ray binaries \citep[e.g.,][]{2016ApJ...826..159S,2018ApJ...869..123S,2019ApJ...878L...3B,2019A&A...624A..89N,2020arXiv200809780S}.

The intrinsically strongest UV emission line
\ion{H}{i}~$\lambda1216$ (\lya) is the most common spectroscopic
probe of high-$z$ galaxies.  
But, due to the high \lya{} absorption
cross-section in neutral hydrogen, \lya{} photons scatter resonantly in
the ISM and circumgalactic medium (CGM).  
This scattering results in \lya{} emission being more extended than the stellar continuum emission \citep{2011ApJ...736..160S,2014MNRAS.442..110M,2016A&A...587A..98W,2018Natur.562..229W,2017A&A...608A...8L,2020A&A...635A..82L}
and imprints kinematic and spatial properties of the ISM and
CGM gas distribution into the observable \lya{} signal from the \lya{} emitters (LAEs; used here for any galaxy with detected \lya{} flux irrespective of the rest-frame equivalent width, EW$_0$).
The scattering furthermore makes \lya{} photons more prone to dust extinction, 
which can lead to a complete suppression of \lya{} from galaxies
\citep[e.g.,][see also Section~\ref{sec:UVandLya}]{2011MNRAS.414.2139D,2011ApJ...728...52L,2019A&A...627A..84L,2017arXiv170403416D,2020A&A...638A..12K}.
By combining the \lya{} observables with inferences
from the other rest-frame UV lines we may constrain this interplay
between the galaxies gas and the \lya{} radiation field.  This is of
particular value for predicting \lya{} observables from other UV lines
during the Epoch of Reionization (EoR) at $z\gtrsim6$, where the neutral hydrogen in the
intergalactic medium (IGM) scatters \lya{} out of the observers line of sight
\citep[e.g.,][]{
2013ApJ...775L..29T,
2014ApJ...793..113P,
2014ApJ...794....5T,
2017A&A...608A.123D,
2018MNRAS.473...30C,
2020A&A...638A..12K}.
The differences between observed and expected \lya{} can therefore be
used to constrain the temporal evolution of the degree of ionization
in the universe during the EoR.  

Hence, in addition to providing detailed knowledge about the detected galaxies themselves, rest-frame UV emission lines other than \lya{} can serve as redshift
confirmation of galaxies in neutral environments in the EoR when \lya{} is completely absent and provide probes of the systemic redshift of LAEs when \lya{} is offset due to to the scattering in the ISM and CGM.
To assess and quantify the feasibility of such approaches, the characteristics of these secondary UV emission lines need to be explored and understood for LAEs at non-EoR redshifts, that is $z<6$. 
Such efforts to characterize and study UV emission lines at both intermediate and high redshift will support the analysis of data from upcoming space-based sensitive infrared observatories like the James Webb Space Telescope (JWST) and the Nancy Grace Roman Space Telescope (formerly known as WFIRST) which are expected to revolutionize studies of galaxy evolution and assembly deep into the epoch of reionization (EoR).

Previously, sources with UV emission lines have been collected through three main methods \cite[see also][]{2020ApJ...893....1B}. 
First, creating composite spectra from stacking large samples of individual galaxy spectra provided some of the first results \citep[e.g.,][]{2003ApJ...588...65S} and has since been expanded \citep{2016ApJ...826..159S,2020A&A...641A.118F}.
Secondly, objects with extreme emission line ratios and fluxes high enough for direct detection have also been found \citep[e.g.,][]{2010ApJ...719.1168E,2014MNRAS.445.3200S,2017NatAs...1E..52A}. 
Thirdly, targeting gravitationally lensed sources uses nature's own telescopes to improve the observation's sensitivity to be able to detect the often intrinsically faint UV emission at high redshift \citep[e.g.,][]{2014ApJ...790..144B, 2016MNRAS.456.4191P, 2017ApJ...839...17S,2018AJ....155..104R,2018ApJ...853...87R,2018ApJ...859..164B}. 
To gain insight in the physical properties governing galaxy formation from the EoR to the main epoch of star formation at $z\approx2$, in this paper we used a fourth method.
We obtain samples of rest-frame UV emitters, by exploring observations from the Multi Unit Spectroscopic Explorer \citep[MUSE;][]{2010SPIE.7735E..08B} at ESO's Very Large Telescope (VLT).
A similar approach, that is searching for UV emission in large parent samples with rest-frame UV spectroscopy has also been pursued by, for example, \cite{2017A&A...608A...4M} and \cite{2019A&A...624A..89N} using MUSE-Deep data, \cite{2020A&A...636A..47S} using the VANDELS data, and \cite{2018MNRAS.477.2098N} and \cite{2019A&A...625A..51L} using the VIMOS Ultra-Deep Survey.
Basing these studies on samples of spectroscopically confirmed objects ensures confident redshifts aiding the efficiency of the search for the intrinsically faint UV emission.
Additionally, studying objects from samples based on emission line objects as opposed to objects with redshift determined from continuum spectroscopy, allows probing populations of fainter galaxies and ensures the presence of recent star formation responsible for the nebular UV line emission. 

Here we explore data taken with MUSE of 2052 spectroscopically confirmed emission line galaxies at $1.5<z<6.4$ including all LAEs and potential \ciiifull{} emitters. 
The MUSE data are presented in Section~\ref{sec:data}.
Sections~\ref{sec:objsel}~and~\ref{sec:1Dspec} explain how we selected and determined the redshifts of our parent sample and how we extracted optimal one-dimensional (1D) spectra of each of these sources.
Section~\ref{sec:UVEmissionLineSearch} describes our emission line template fitting approach (detailed in Appendix~\ref{sec:felis}) used to search for UV emission red-wards of \lya{} in the 1D spectra. 
This results in a large sample of UV line emitters of both LAEs and non-LAEs (summarized in Table~\ref{tab:UVESdetections}) which we explore the physical parameters of in the remainder of the paper.
We infer EW$_0$ of the emission lines (Section~\ref{sec:EWcoor}) and investigate correlations between them and the emission line fluxes with properties of the LAEs in our sample (Section~\ref{sec:UVandLya}). 
We explore the \lya{} emission line velocity offsets with respect to the secondary UV emission lines which provide systemic redshifts (Section~\ref{sec:voffset}), and assess the range of electron densities for objects showing \siiiifull{} and \ciiifull{} (Section~\ref{sec:neTe}).
In Section~\ref{sec:logOH} we determine the gas-phase abundances of systems with multiple UV emission features detected. 
Finally, we explore how further insight can be gained on the physical parameters of the targeted galaxies from photoionization models by introducing ``PhotoIonization Model Probability Density Functions'' (PIM-PDFs) in Section~\ref{sec:pimodelinference}, before summarizing and concluding our study in Section~\ref{sec:conc}. 
With the paper, we provide a catalog of the complete set of UV emission line measurements from the 2052 galaxies studied (described in Appendix~\ref{sec:mastercat}). Throughout the paper we compare these measurements to a collection of measurements from the literature which we also provide a catalog of (Appendix~\ref{sec:litcol}).

In the remainder of this paper, we use the short notations for the rest-frame UV emission lines red-wards of \lya{} listed in Table~\ref{tab:UVlines}.
The minimum required energy of ionizing photons responsible for these UV lines are also quoted together with the expected size of the emission doublet line flux ratios.
We use AB magnitudes \citep{1983ApJ...266..713O} and assume flat cosmological parameters of H$_0 = 70$~km~s$^{-1}$~Mpc$^{-1}$, $\Omega_m = 0.3$, and $\Omega_\Lambda = 0.7$.
\begin{table*}
\caption{\label{tab:UVlines}UV emission line notations and ionization energies.}
\centering
\begin{tabular}{llllllccc}
\hline\hline
Atomic			& Wavelength 		&   \multicolumn{3}{c}{Short notation}		& Transition					&  \multicolumn{2}{c}{$E_\textrm{Ionization}$}		& Expected 			\\  %
\cline{3-5}
ion			& [\AA]			&  Main	& Blue comp.	& Red comp.		&							&  [eV]  					&    [Ry]					& flux ratio			\\  %
\hline   																																		
\ion{H}{i}		& 1216              	& \lya             	&			&			& H$^{0}$ recombination			&	13.60$^\star$			&		1.00$^\star$		&    					\\        			
\nv$^\dagger$	& 1239, 1243          	& \nv			&	\nvone{}	& \nvtwo{}		& N$^{3+}\rightarrow$ N$^{4+}$	&	77.47				&		5.69				&     $f(\textrm{\nvone}) / f(\textrm{\nvtwo}) =2^\textrm{a,b,c}$ 		\\        			
\civ$^\dagger$	& 1548,1551 		& \civ{} 		&	\civone{}	& \civtwo{}		& C$^{2+}\rightarrow$ C$^{3+}$	&	47.89				&		3.52				&     $f(\textrm{\civone}) / f(\textrm{\civtwo}) = 2^\textrm{b,c,d}$			\\
\heii{}		& 1640			& \heii{}           	&			&			& He$^{0}\rightarrow$ He$^{1+}$	&	24.59              			&		1.81				&     				\\        			
\heii{}		& 1640			& \heii{}           	&			&			& He$^{1+}$ recombination		&	54.42$^\star$	              	&		4.00$^\star$		&     				\\        
\oiii{}			& 1661,1666 		& \oiii{} 		&	\oiiione{}	& \oiiitwo{}		& O$^{1+}\rightarrow$ O$^{2+}$	&	35.12				&		2.58				&     $f(\textrm{\oiiione}) / f(\textrm{\oiiitwo}) \simeq 0.7^\textrm{c}$ 					\\  
\siiii{}			& 1883,1892  		& \siiii{} 		&	\siiiione{}	& \siiiitwo{}	& Si$^{1+}\rightarrow$ Si$^{2+}$	&	16.35				&		1.20				&    $f(\textrm{\siiiione}) / f(\textrm{\siiiitwo}) \lesssim1.7^\textrm{e}$ 					\\  
\ciii{}			& 1907,1909 		& \ciii{} 		&	\ciiione{}	& \ciiitwo{}		& C$^{1+}\rightarrow$ C$^{2+}$	&	24.38				&		  1.79			&    $f(\textrm{\ciiione}) / f(\textrm{\ciiitwo}) \lesssim1.6^\textrm{e}$ 	\\        			
\hline\hline   																																					
\end{tabular}
\tablefoot{
The atom and emission line rest-frame wavelength for the UV lines studied here are provided in the first two columns. 
The short notation columns list the names used throughout this paper for the main lines or emission line doublets and their blue and red doublet components.
The transition column provides that transitions for which the ionization energies are quoted in the column $E_\textrm{Ionization}$ \citep{NISTAtomicSpectra:uk,2011AJ....141...37L}.
These energies correspond to the minimum required ionization energy needed to generate the respective emission lines.
References for the expected emission line doublet component flux ratios presented in the last column are:\\
$^\textrm{a}$\cite{1984RMxAA...9..107T}, \\
$^\textrm{b}$\cite{1976JPCRD...5..537M},\\
$^\textrm{c}$\cite{1991ApJS...77..119M}, \\
$^\textrm{d}$\cite{1983A&A...122..335F}, and\\
$^\textrm{e}$\cite{2006agna.book.....O,2019ApJ...880...16K}.\\
$^\dagger$The line profiles of \nv{} and \civ{} are complicated by the fact that this emission is potentially a superposition of P-Cygni profiles from stellar winds, emission from the ISM, and nebular emission. \\
$^\star$These energies correspond to the energy required to ionize the atom before electrons recombine with the atom to emit the emission line. 
}
\end{table*}


\section{Data}\label{sec:data}
\begin{figure}
\begin{center}
\includegraphics[width=0.38\textwidth]{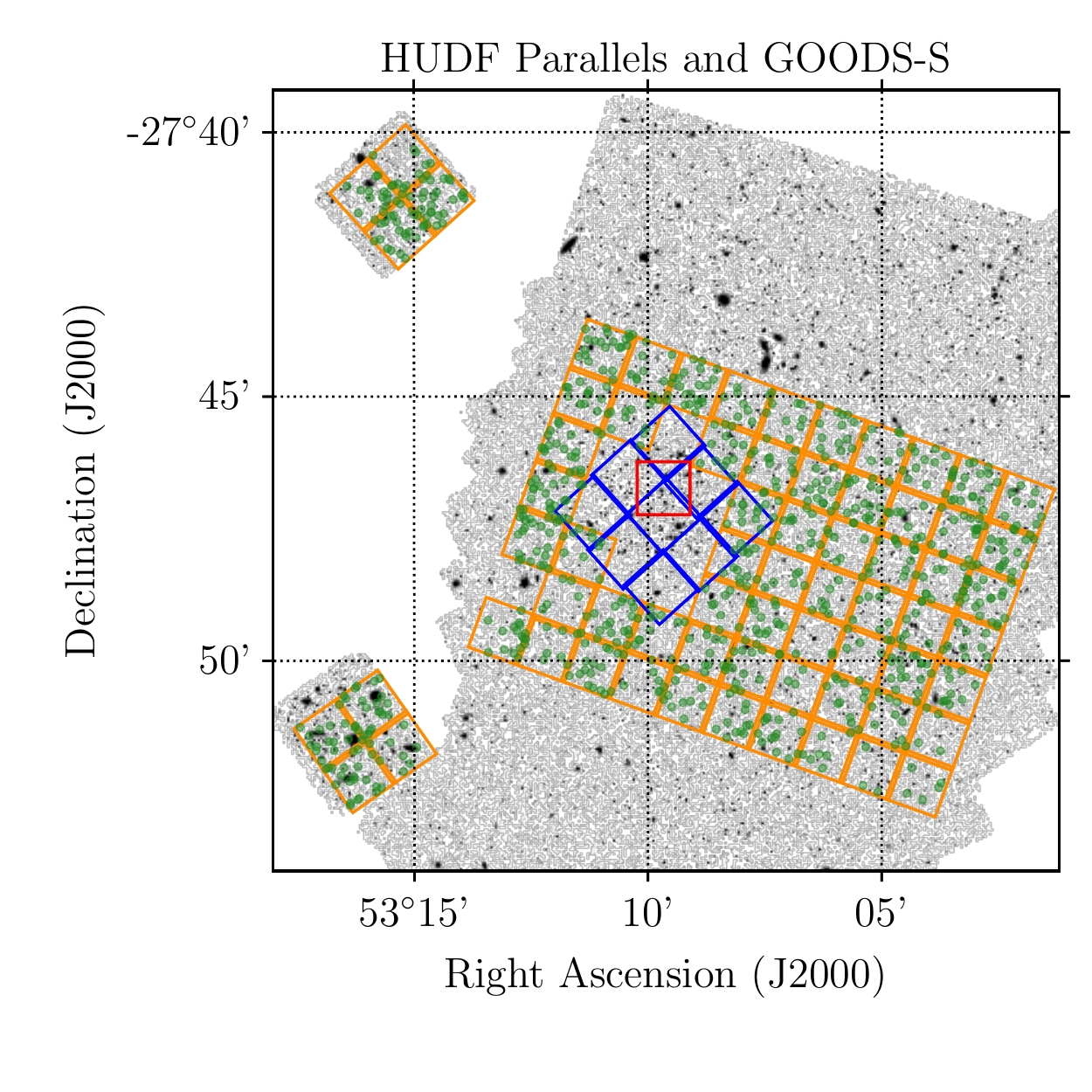}\\
\includegraphics[width=0.38\textwidth]{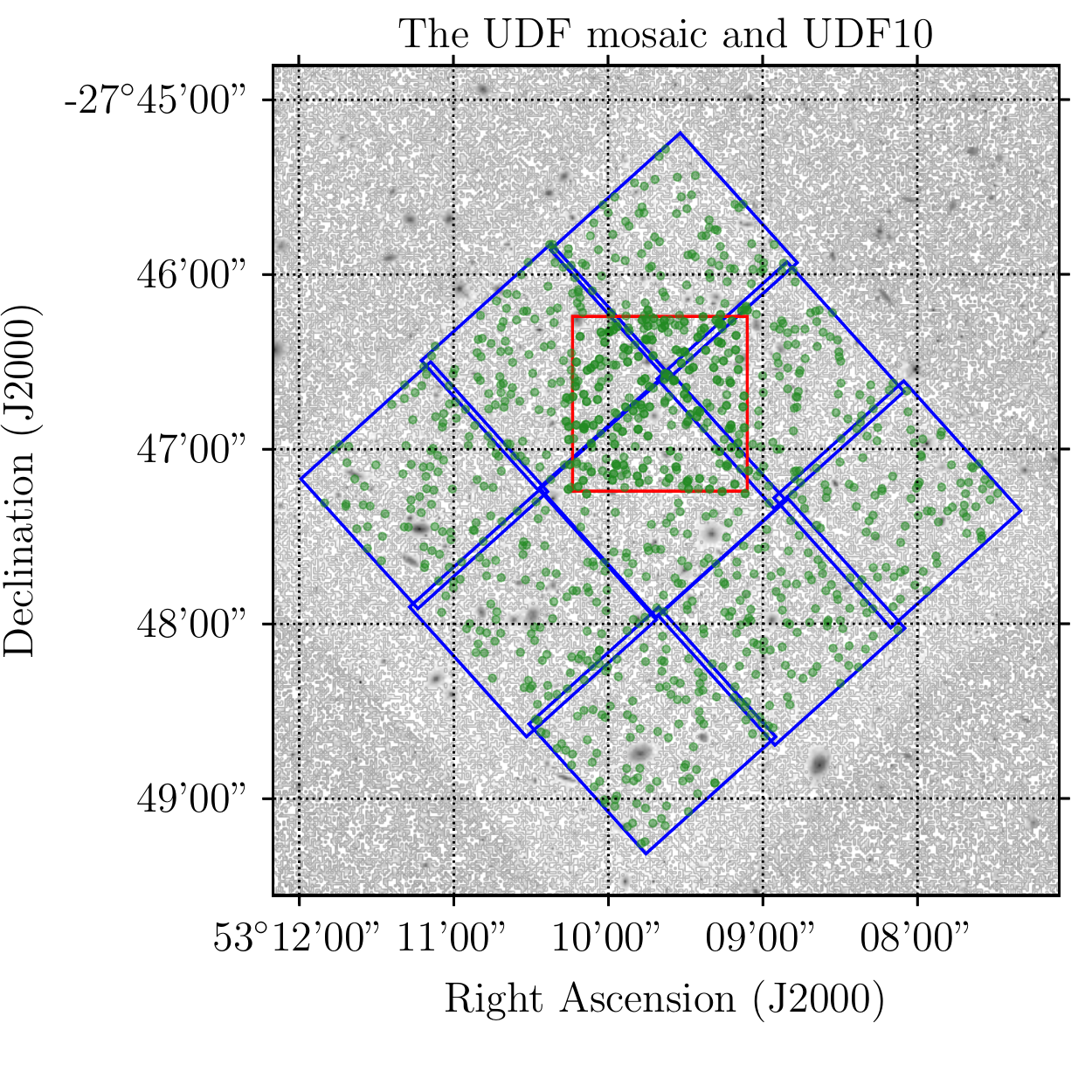}\\
\includegraphics[width=0.43\textwidth]{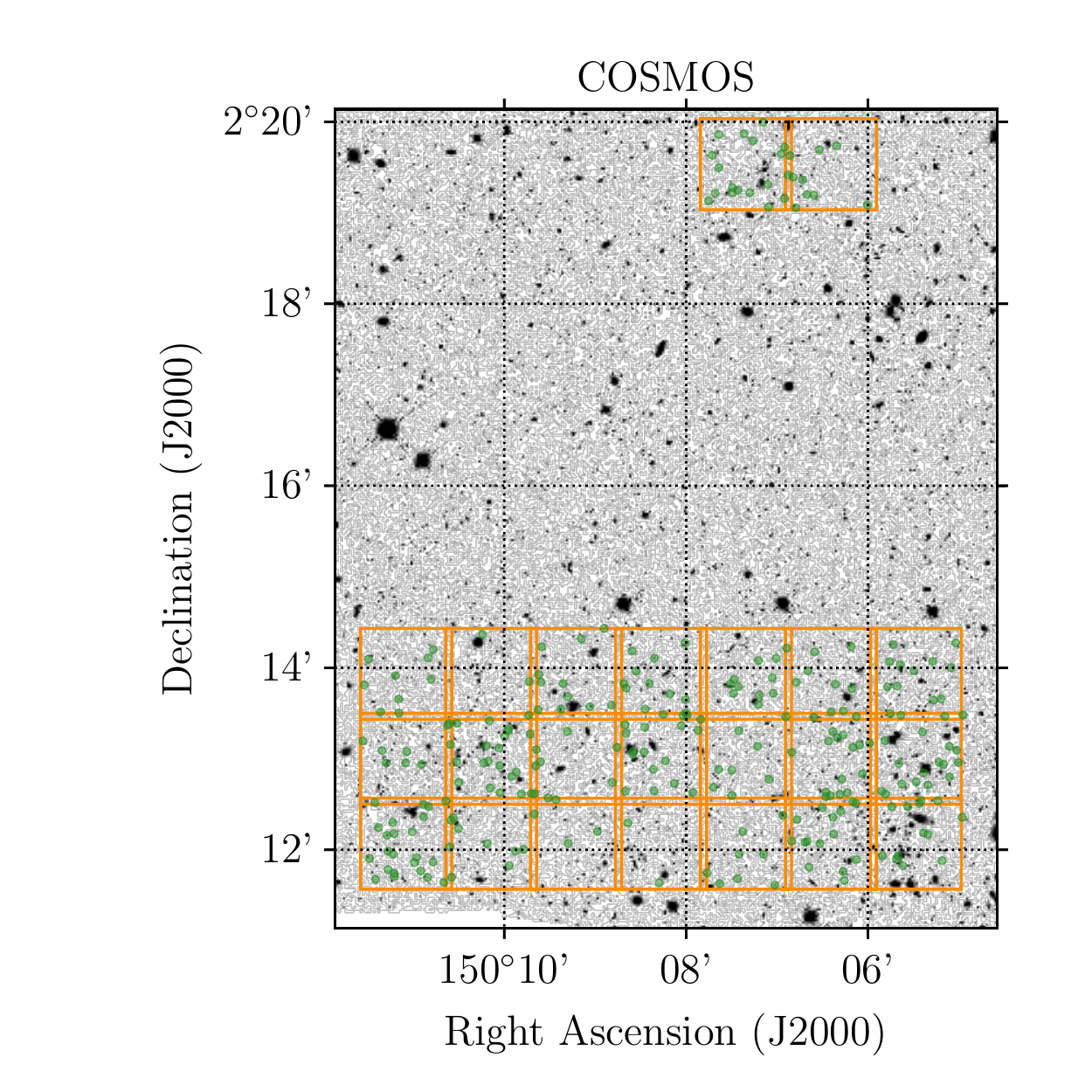}
\caption{
Location of the analyzed MUSE pointings in the GOODS-South (top), HUDF (center) and COSMOS (bottom) regions.
The MUSE-Wide pointings are shown in orange, the UDF mosaic in blue and the UDF10 pointing in red.
The location of the 2052 sources searched for rest-frame UV emission lines are marked with green circles. 
The gray scale images are the 3D-HST combined near-infrared detection images from \href{https://3dhst.research.yale.edu/}{https://3dhst.research.yale.edu/} (top and bottom) and the HLF F775W version 2.0 image from \href{https://archive.stsci.edu/prepds/hlf/}{https://archive.stsci.edu/prepds/hlf/} (center).
}
\label{fig:fields}
\end{center}
\end{figure}

In this paper we search for UV emission lines in a compilation of the MUSE consortium guaranteed time observations (GTO) data.
The MUSE integral field spectrograph provides spatial and spectral coverage of the 1~arcmin$^2$ field-of-view (FoV).
The spatial sampling in the reduced data cubes is $0\farcs2 \times 0\farcs2$. 
Each of these \emph{spaxels} contains a spectrum spanning the optical wavelength range from 4800~\AA{} to 9300~\AA{} at a wavelength sampling of 1.25~\AA{} per volume element, that is per \emph{voxel}.
The spectral resolution element of the MUSE data is roughly 2.5~\AA{} across the full wavelength range such that the spectral resolution $\textrm{R}=\lambda/\Delta\lambda \approx 3000$ \citep[$\textrm{R} \approx 1800$ in the blue up to 4000 in the red,][]{2020A&A...641A..28W}.
The data sets explored in this study come from the MUSE GTO team's ``wedding cake approach'' of a shallow, medium-deep and deep survey of multiple MUSE pointings. 
We describe each of these data sets in the following subsections.
The general outline and the location of each of the MUSE pointings are shown in Figure~\ref{fig:fields}.

\subsection{MUSE-Wide}\label{sec:mw}

The wide and shallow component of the GTO wedding cake approach is the MUSE-Wide survey.
It consists of 100 $1\times1$~arcmin$^2$ MUSE pointings distributed over the COSMOS \citep{2007ApJS..172....1S} and GOODS-South  \citep{2003mglh.conf..324D,2004ApJ...600L..93G} CANDELS footprints \citep{2011ApJS..197...35G,2011ApJS..197...36K}.

The first catalog and data release of the MUSE-Wide survey were presented by \cite{2017A&A...606A..12H} and \cite{2019A&A...624A.141U}, respectively.
These studies analyze 44
MUSE pointings of the MUSE-Wide data over the GOODS-South footprint.
Here we include data from these 44 fields adding the remaining 47 MUSE pointings of the full MUSE-Wide data set.
Formally, the MUSE-Wide survey consists of 100 MUSE pointings. 
But when referring to MUSE-Wide throughout this paper, we are excluding the nine pointings in the Hubble Ultra Deep Field \citep[HUDF;][]{2006AJ....132.1729B}, as these fields are included in the deeper MUSE UDF mosaic described in Section~\ref{sec:mosaic}.
The MUSE-Wide data were obtained through a series of MUSE GTO observing runs from September 2014 through September 2017.
Each of the 91 MUSE-Wide pointings have a depth of 1 hour and were observed on nights with a seeing of roughly $1\farcs0$.
Details of the seeing properties for the first data release from MUSE-Wide are given in Section~3.2.5 of \cite{2019A&A...624A.141U}. 
The remaining MUSE-Wide fields were observed under similar conditions.

The MUSE-Wide data were reduced using version 1.0 (or an early development version) of the data processing pipeline (DPP) for the MUSE instrument \citep{2014ASPC..485..451W,2020A&A...641A..28W} released in December 2014 to support the first observing runs with the seeing-limited wide-field mode of MUSE.
The DPP was complemented by enhanced so-called slice-based sky-subtraction, a self-calibration of the initial reductions using the self-calibration procedure from the MUSE Python Data Analysis Framework \citep[MPDAF;][]{2016arXiv161205308C,2017arXiv171003554P}, and accounting for the effect of varying shape of the LSF on the sky residuals with the Zurich Atmosphere Purge \citep[ZAP version 1.0,][]{2016MNRAS.458.3210S}.
Even though newer versions of the DPP have been released since 2014, all 91 MUSE-Wide pointings were reduced using the same versions of the DPP and the complementing software for consistency. 
For details on the MUSE-Wide data reduction see \cite{2019A&A...624A.141U}.
The reduced MUSE-Wide data on average reach a 1$\sigma$ emission line detection sensitivity of $1\times 10^{-19}$~erg~s$^{-1}$~cm$^{-2}$~\AA$^{-1}$ as estimated from point source and LAE insertion experiments \citep{2019A&A...621A.107H} assuming a 1 arcsec$^2$ aperture.

\subsection{MUSE-Deep UDF mosaic}\label{sec:mosaic}

Complementing the wide and relatively shallow data from MUSE-Wide, the GTO wedding cake program contains a medium depth survey of nine pointings at ten hours depth over the HUDF which we refer to as the UDF mosaic (blue squares in the top and central panels of Figure~\ref{fig:fields}).
The data over the UDF analyzed in the current study were first presented by \cite{2017A&A...608A...1B} and \cite{2017A&A...608A...2I}.
As described in Section~\ref{sec:objsel} below, we perform an independent and self-consistent source selection in all data analyzed in our study.
Hence, we are relying on these source lists instead of the source catalog presented by \cite{2017A&A...608A...2I}.
The UDF mosaic data were taken through a series of MUSE GTO observing runs from September 2014 through February 2016. 
Our source selection is based on an updated and improved reduction of these UDF data with the MUSE DPP which will be presented by Bacon et al. (in prep.).
Like the MUSE-Wide data reduction, the basic DPP reduction was complemented by the self-calibration and improved sky subtraction from MPDAF and ZAP for the reduction of the deeper data.
For details on the UDF mosaic (and UDF10) data reduction see Bacon et al. (in prep.).
As shown by \cite{2017A&A...608A...1B} the first version of the UDF mosaic data reach a 1$\sigma$ emission line flux sensitivity of $5.5 \times 10^{-20}$~erg~s$^{-1}$~cm$^{-2}$~\AA$^{-1}$ for a 1 arcsec$^2$ aperture averaged over the most sensitive part of the MUSE throughput curve at 7000--8500~\AA.

\subsection{MUSE-Deep UDF10}\label{sec:udf10}

The final layer of the MUSE GTO wedding cake program is the field referred to as UDF10.
UDF10 is a single $1\times1$~arcmin$^2$ MUSE pointing at 31 hours depth within the HUDF (red square in top and central panels of Figure~\ref{fig:fields}).
The UDF10 data are also part of the updated release of the MUSE-Deep data presented by Bacon et al. (in prep.) 
and were reduced using the same setup as the UDF mosaic data.
For the first version of the MUSE-Deep data presented by \cite{2017A&A...608A...1B}, the estimated 1$\sigma$ emission line flux for the UDF10 data is $2.8 \times 10^{-20}$~erg~s$^{-1}$~cm$^{-2}$~\AA$^{-1}$ .

\section{Object selection in the MUSE data}\label{sec:objsel}

To establish a sample of line emitters in the reduced MUSE GTO data cubes, all 101 individual data cubes were searched for emission lines using the dedicated Line Source Detection and Cataloguing (LSDCat) software described by \cite{2017A&A...602A.111H}.
Apart from the flux cube LSDCat requires a variance cube for the data too. 
As the variances obtained by formal error propagation in the MUSE DPP underestimate the true uncertainties of the data due to the voxel resampling in the cube construction, we estimated empirically calibrated \emph{effective variances} for all data cubes. 
In short, the effective variance cubes are constructed by 
first (i) measuring the typical variances between individual sky-voxels to capture the small-scale systematics like imperfect sky subtraction and flat fielding, 
then (ii) account for the covariances from the noise propagation in the DPP by rescaling the noise level by the average covariance level estimated by pulling a random noise cube (Gaussian with mean 0 and variance 1) through the DPP,
and lastly (iii) assuming that the effective noise is constant (modulo the number of exposures) across each individual wavelength slice of the data cubes, that is assuming background-limited noise. 
For further details on the effective variance procedure see Section~3.2.4 of \cite{2019A&A...624A.141U} and \cite{2020A&A...641A..28W}.
Furthermore, LSDCat requires continuum subtracted data cubes for line detection. 
These cubes were generated from the reduced data cubes by subtracting a 151 pixels ($\approx$189~\AA) wide running median from each of the spaxels.

The untargeted search for emission lines with LSDCat is based on these continuum subtracted versions of the reduced data cubes and the effective variance cubes.
LSDCat then performs a 3D template match over all voxels in the data cube. 
To optimize the recovery of Ly$\alpha$ emitters, we used a search template with a Gaussian spectral component of fixed full width at half maximum (FWHM) set to 250~km~s$^{-1}$ and a two-dimensional (2D) spatial Gaussian component with FWHM equal to the width of the point spread function (PSF) of the observations in the individual data cubes.
Hence, the detection significance is slightly biased towards narrow emission lines. 
However, this does not affect our independent search for rest-frame UV lines (Section~\ref{sec:UVEmissionLineSearch}) in the sources recovered by LSDCat.
The untargeted emission line detections from LSDCat with a signal-to-noise ratio (S/N) above five \citep[6.4 for the first 24 MUSE-Wide fields, see ][]{2019A&A...624A.141U} were grouped and then classified with the visual inspection tool QtClassify \citep{2017ascl.soft03011K,2017A&A...606A..12H}.
\cite{2017A&A...602A.111H} provide the analytic approximation for the minimum line flux recoverable by LSDCat at a given detection threshold\footnote{Due to an error in typesetting $\sigma_\textrm{G}$ in Equation~34 of \cite{2017A&A...602A.111H} was erroneously placed under the square root \cite{2021A&A...649.C5H}.} 
\begin{equation}
F_\textrm{line recover} \approx \textrm{S/N} \;\overline{\sigma}\;\sigma_\textrm{G}\;\Delta\lambda\;  \sqrt{8\pi^{3/2}\sigma_z}\;.
\end{equation}
Considering $\textrm{S/N}=5$ detections and inserting the $\Delta\lambda=1.25$~\AA{} wavelength sampling, $\sigma_\textrm{z} = 1.46$~pixels at 5000~\AA{} corresponding to the spectral detection template FWHM of 250~km~s$^{-1}$, and a spatial source model width of $\sigma_G=1.84$~pixels corresponding to a spatial Gaussian width of $0\farcs88$ the flux sensitivities $\overline{\sigma}$ quoted in Section~\ref{sec:data} translate into 5$\sigma$ limiting line fluxes of roughly 
$1.3\times10^{-17}$,  
$7.2\times10^{-18}$, and 
$3.7\times10^{-18}$~erg~s$^{-1}$~cm$^{-2}$,  
for the MUSE-Wide, UDF mosaic, and UDF-10 data, respectively. 

For the source classification, two (or sometimes three) people independently categorized all line-sets found by LSDCat by matching the shape and spectral location with rest-frame emission line lists.
The independent classifications were then consolidated in plenum with a third independent consolidator ensuring at least three independent assessments of all detected emission lines from the untargeted search.
The leading line of each line emitter, that is the associated line with the highest 3D LSDCat S/N was assigned a type (for instance, Ly$\alpha$, [\ion{O}{ii}] or H$\alpha$) and a confidence from C=0 to C=3. 
A confidence of 1 refers to line emitters with a single trustworthy line detection of uncertain nature, 
C=2 refers to a classification with high confidence, where there is support from line profiles (for example, a skewed asymmetric profile for Ly$\alpha$ emitters) or secondary S/N~$<5$ lines. For objects with C=3 there is little room for doubt about the line classification, either due to characteristic line profiles or multiple line detections at S/N~$>5$.
A confidence of 0 was given on rare occasions where it was hard to tell if the emission feature was real or not.

The QtClassify classification of the detected line emitters from LSDCat from all 101 MUSE data cubes leads to the self-consistent parent line emitter catalog from MUSE-Wide, the UDF mosaic and UDF10 used here.
This catalog therefore differs from both the line emitters presented by \cite{2017A&A...608A...2I} and the source catalog based on the improved rereduction of the MUSE-Deep data by Bacon et al. (in prep.).
The \cite{2017A&A...608A...2I} emission line sources were detected using the ORIGIN software \citep{2020A&A...635A.194M} in the data cubes from the first MUSE-Deep data release \citep{2017A&A...608A...1B}.
The updated release of the UDF mosaic and UDF10 catalogs by Bacon et al. (in prep.) merges the independent LSDCat and ORIGIN emission line source detections performed on the improved reductions of the MUSE-Deep data complemented by even deeper MUSE data in the HUDF from the MUSE eXtremely Deep Field \citep[MXDF; see][]{2021A&A...647A.107B}.
Bacon et al. (in prep.) also present continuum-selected non-emission line sources and sources at $z<1.5$. 
The self-consistent LSDCat-based catalog of line emitters at $z>1.5$ presented as part of our work is well suited for statistical studies. 
However, in terms of source completeness the MUSE-Deep catalog from the UDF mosaic and UDF10 by Bacon et al. (in prep.) will supersede it.

Each individual LSDCat detection in MUSE-Wide, the UDF mosaic or UDF10 was assigned a unique ID which we use throughout this paper when referring to and discussing individual objects.
For the sources from the MUSE-Wide pointings we follow the ID structure of the first data release of MUSE-Wide \citep{2019A&A...624A.141U} where each ID contains nine digits using the format ``ABBCCCDDD''.
Here A refers to the region in which the object was detected (1=GOODS-South, 2=COSMOS, 3=northern HUDF parallel, and 4=southern HUDF parallel), BB refers to the arbitrary numbering of the individual MUSE pointings, CCC refers to the LSDCat object identifier for that particular data cube, and DDD refers to a running number for the detected lead line of the source.   
The sources in the UDF mosaic have a nine digit ID using the format ``ACCCCDDDD''. 
Here A=6 refers to the UDF mosaic, 
CCCC is the four-digit LSDCat object identifier, and DDDD is the running number for the detected lead line of the source.
Similarly, the UDF10 objects were assigned IDs using the format ``A2CCCDDDD'' where A=7 refers to UDF10, and CCC and DDDD again indicate the LSDCat object identifier and the running number of the lead line.

To focus on potential rest-frame UV line emitters and for completeness, we selected all objects with a lead line at a redshift above 1.5 from the parent catalog irrespective of the confidence assignment. 
This includes all objects with Ly$\alpha$ as the lead line corresponding to $z\gtrsim2.9$ for the MUSE wavelength coverage and all objects in the so-called MUSE redshift desert at $z\approx1.5-2.9$.
At $z\approx1.5$ \ciii{} enters the MUSE wavelength range, whereas the prominent optical \oiifull{} doublet moves out of it.
Hence, neither \lya{} nor \oiifull{} is available for redshift determination in the redshift desert. 
In total there are 2197 unique IDs in the 101 MUSE pointings satisfying these selection criteria (1120 from MUSE-Wide, 842 from the UDF mosaic and 235 from UDF10).
The location of these emission line sources are marked by the green dots in Figure~\ref{fig:fields} 
and do not account for the overlaps between individual pointings. 

\section{Optimal spectral extraction}\label{sec:1Dspec}

Our search for faint UV emission line features in the parent sample of the selected line emitters from the MUSE GTO data is performed on 1D spectra extracted from the 3D data cubes.  
Here we use optimally extracted 1D spectra obtained with the dedicated tool for Three-Dimensional Optimal Spectral Extraction \citep[TDOSE\footnote{\href{https://github.com/kasperschmidt/TDOSE}{https://github.com/kasperschmidt/TDOSE}};][]{2019A&A...628A..91S}. 
In short, TDOSE bases the spectral extraction on a 2D spatial model of each object in the FoV.
This model is based on higher-resolution imaging from, for instance, the \emph{Hubble Space Telescope} (HST).
Hence, it is implicitly assumed that any emission line fluxes extracted from these spectra are well represented by the extent of the continuum emission in the modeled HST images, which is a poor assumption for extended emission like \lya{} but reasonable for UV emission lines.
Using the morphological HST models as object templates, TDOSE performs a simultaneous 3D template match of all sources in the FoV, scaling source fluxes at each wavelength layer in the data cube accounting for contaminating flux by neighboring sources.
Hence, TDOSE is based on an approach similar to PampelMuse \citep{2013A&A...549A..71K,2018ascl.soft05021K}, except that TDOSE focuses on extended sources (galaxies), whereas PampelMuse was developed for spectral extraction of point-sources (stars).  

All objects in our parent sample have spectra extracted using a single multivariate Gaussian source model for both the object of interest and each of the contaminants in the FoV.
For source extractions from the 91 MUSE-Wide data cubes the contamination models were based on the \cite{2014ApJS..214...24S} 3D-HST photometric catalog, as this catalog provides self-consistent source catalogs over GOODS-South, the HST UDF parallel fields and the COSMOS region mapped by the MUSE-Wide survey (see Figure~\ref{fig:fields}). 
In GOODS-South and COSMOS we modeled the source morphology in the available CANDELS HST F814W images corresponding to the \cite{2014ApJS..214...24S} photometry.
For the two HUDF parallel fields we based the models on the HST F160W images to ensure full coverage of the $2\times2$ MUSE pointings using the corresponding CANDELS and 3D-HST imaging and to ensure access to morphological priors from the \cite{2014ApJS..214...24S} catalog for the modeling of the HST images.
For extractions in the HUDF region we based the spatial models on the source catalog by \cite{2015AJ....150...31R} and the version 1.5 HST F775W images from the Hubble Legacy Field \citep[HLF\footnote{\href{https://archive.stsci.edu/prepds/hlf/}{https://archive.stsci.edu/prepds/hlf/}};][]{2016arXiv160600841I}.
No additional exposures were added to the F775W images for the recent HLF version 2.0 images presented by \cite{2019ApJS..244...16W}. 
Spectra for sources with no detections in the ancillary imaging, that is where either faint or no photometric counterparts could be identified, were extracted using a morphological model identical to a point-source convolved with the PSF of the observations.   

After an initial run of fully automatic spectral extractions, the resulting 1D spectra were visually inspected to identify suboptimal  source models or cases where the source flux scalings were unreliable.
Roughly 2\% (42/2197) of the spectra were selected for reextraction with more careful attention to source location, source numbers, FoV extent, etc.
Among the 2197 initial spectra only five were of LAEs with $z>6.44$. 
At these redshifts all of the considered secondary emission lines are redshifted beyond the red cutoff of the MUSE instrument at $\lambda\approx9300$~\AA.
These five spectra were therefore not searched for UV emission lines.    
For four objects (126042110, 602121764, 609223654, and 613534254) we rely on aperture spectra, as no satisfactory spatial morphological multivariate Gaussian or PSF-based extraction could be obtained when including the complete contamination model. 
They were also obtained with TDOSE, using an extraction aperture with a radius of $0\farcs6$, which corresponds roughly to the FWHM of the MUSE PSF in the deeper MUSE data \citep{2017A&A...608A...1B}.
Two objects (158002004, 601931670) were located directly behind bright foreground objects hampering a satisfactory spectral extraction. These two objects were also excluded in the further analysis.
Finally, two objects (208014258 and 600341002) were removed from the sample due to severe contamination and unreliable flux scalings caused by neighboring stars.

\begin{figure}
\begin{center}
\includegraphics[width=0.45\textwidth]{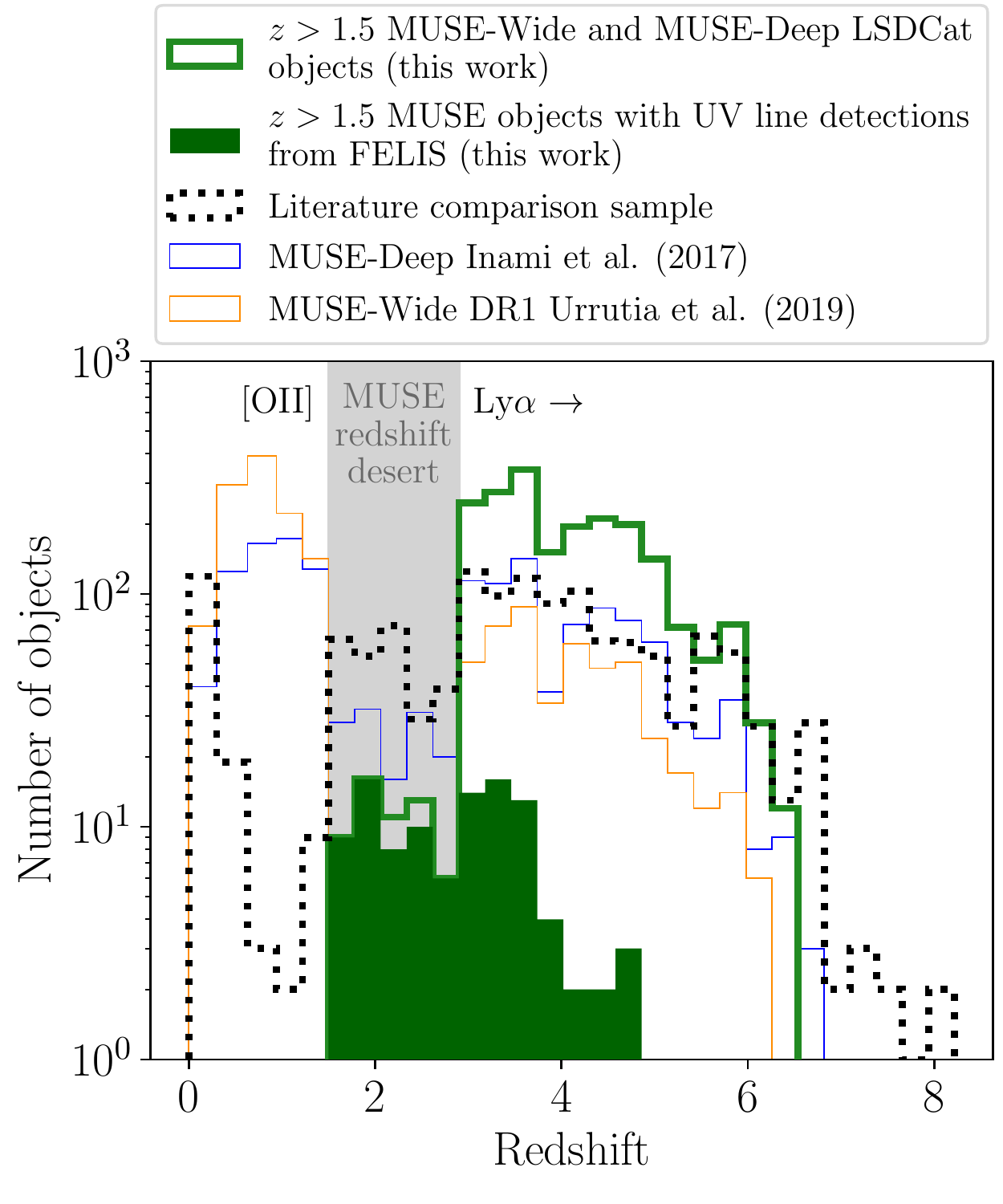}\\
\caption{Redshift distributions of the sources studied here (thick green histogram), the subsample of them with UV emission lines detected by FELIS (filled dark green histogram), the comparison sample of UV line emitters from the literature described in Appendix~\ref{sec:litcol} (dotted black histogram), the MUSE-Wide DR1 catalog by \citet[][thin orange histogram]{2019A&A...624A.141U}, and the independent MUSE-Deep UDF mosaic and UDF10 catalog presented by \citet[][thin blue histogram]{2017A&A...608A...2I}.
The MUSE redshift desert where neither \oiifull{} ($0.3<z<1.5$) nor \lya{} ($z>2.9$) emission is available for redshift identification in the MUSE wavelength range is marked by the gray band.
}
\label{fig:zdist}
\end{center}
\end{figure}

\begin{figure*}
\begin{center}
\includegraphics[width=0.95\textwidth]{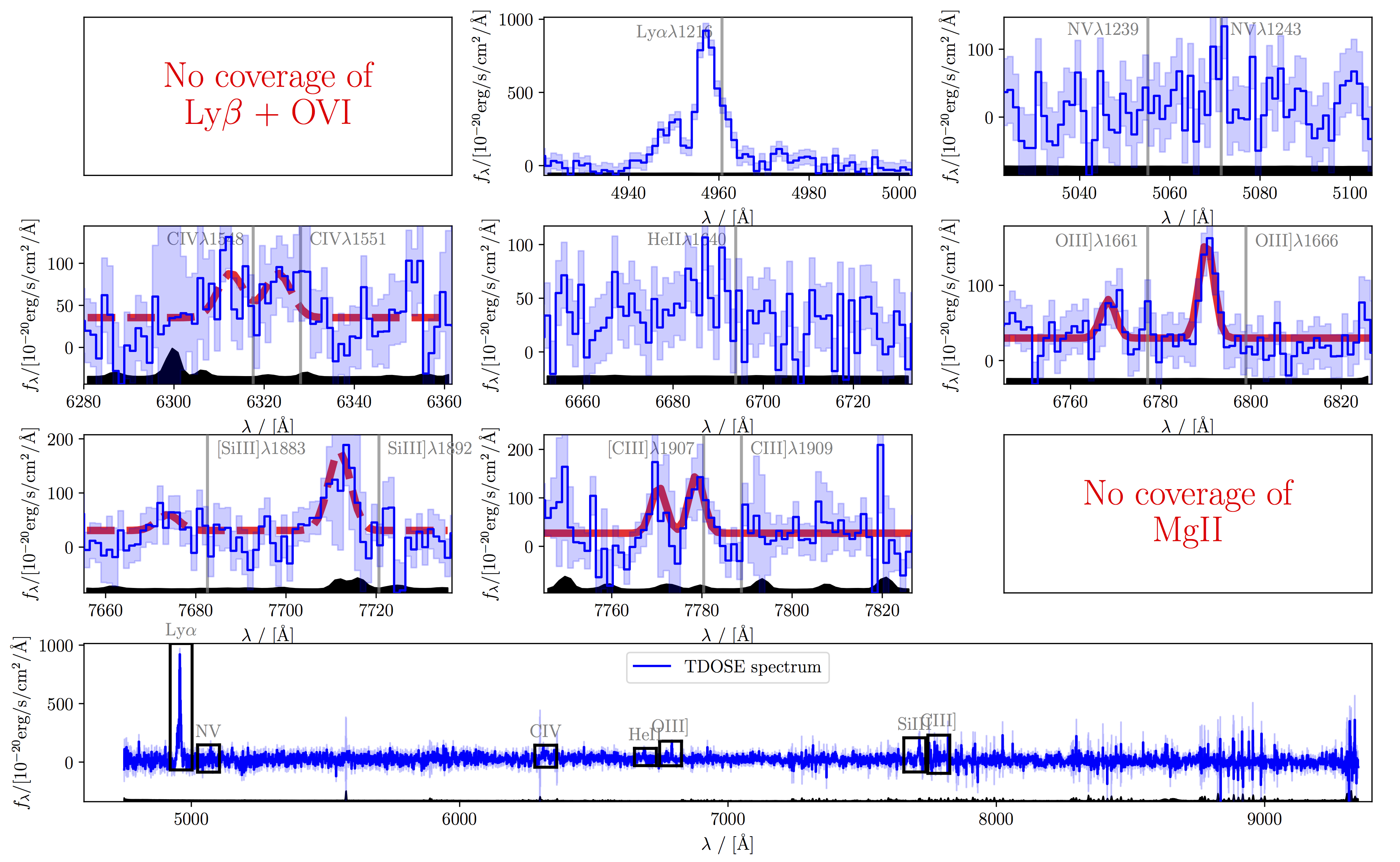}\\
\includegraphics[width=0.95\textwidth]{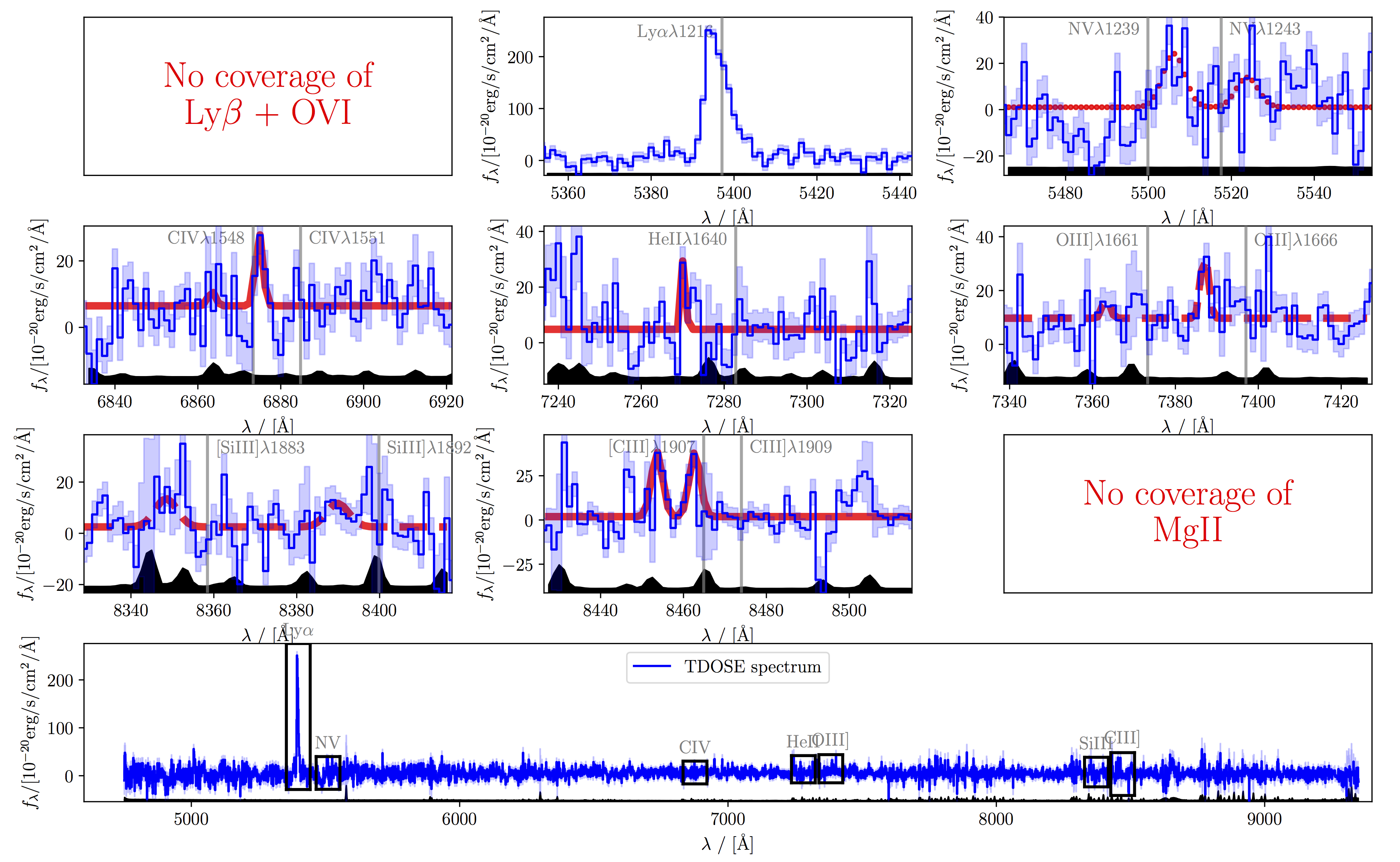}
\caption{Examples of TDOSE extracted 1D MUSE spectra (blue) of the two LAEs 201004081 ($z=3.08$; top) and 602922055 ($z=3.44$; bottom) with their 1$\sigma$ error spectrum indicated by the blue-shaded region.
Further examples can be found in Appendix~\ref{sec:examplespec}.
Each of the smaller panels shows a UV emission line region as marked by the black boxes in the bottom panels, which show the full MUSE wavelength range.
Vertical gray lines mark the approximate location of emission features based on the redshift of the LSDCat lead line.
No continuum subtraction has been performed on the spectra.
The UV emission lines detected by the software FELIS (Section~\ref{sec:UVEmissionLineSearch}) are marked by the red curves.
The solid red curves show FELIS detections with S/N~$>3$ and a velocity offset below 1000~km~s$^{-1}$ deemed trustworthy in the visual inspection process.
Dashed red curves show less secure detections, which were however deemed reliable.
Dotted red curves show S/N~$>3$ and $|\Delta v| < 1000$~km~s$^{-1}$ FELIS detections classified as unreliable.
In each panel 2$\times$ the median data cube noise in the respective fields (MUSE-Wide, MUSE mosaic and UDF10) are shown for reference as the black filled curves offset to the bottom of each panel for clarity.
These median noise spectra are also shown in Figure~\ref{fig:noisespec}.
}
\label{fig:ObjSpec}
\end{center}
\end{figure*}

As can be seen in Figure~\ref{fig:fields}, the UDF mosaic (blue pointings) overlaps with parts of the MUSE-Wide coverage (orange pointings) and the UDF10 field (red square) is fully within the UDF mosaic. Hence, the UDF mosaic and UDF10 contain duplicates of sources in MUSE-Wide and the UDF mosaic, respectively.
These duplicates were identified by searching for multiple objects within a search radius of $0\farcs25$ of all positions in the main catalog with lines identified at similar wavelengths. 
The $0\farcs25$ search radius corresponds to half the approximate coordinate precision between the MUSE detections and the HST reference images of $0\farcs5$.
We excluded the spectra of the UDF mosaic (MUSE-Wide) sources when there was a UDF10 (UDF mosaic) counterpart within this search radius at the same redshift, which corresponds to a total
 of 120 (16) sources.
Hence, the final sample of emission line sources with $1.5<z<6.4$ from MUSE-Wide, the UDF mosaic and UDF10 analyzed in the remainder of this paper 
amounts to 2052 unique objects of which 1997 are LAEs. 
The thick green histogram in Figure~\ref{fig:zdist} shows the redshift distribution of these objects compared to the 
independent MUSE-Deep UDF mosaic and UDF10 catalog presented by \citet[][thin blue histogram]{2017A&A...608A...2I} and the MUSE-Wide DR1 catalog by \citet[][thin orange histogram]{2019A&A...624A.141U}.
The set of objects at $z>1.5$ from the latter is a subsample of the parent sample studied here. 
The lack of \lya{} and \oiifull{} for redshift identification resulting in the MUSE redshift desert is clearly visible and marked by the gray shaded region.  
Figure~\ref{fig:ObjSpec} presents the TDOSE spectra of two example objects. 
Further examples of objects representing the breadth of data and supporting discussions of individual objects in the remainder of the paper are shown in Appendix~\ref{sec:examplespec}.

\section{Searching for UV emission lines in 1D spectra}\label{sec:UVEmissionLineSearch}

Knowing the (approximate) redshift for each of the 2052 unique objects, we searched the TDOSE spectra for signs of rest-frame UV emission lines through template matching using the publicly available Python software for Finding Emission Lines in Spectra (FELIS\footnote{\href{https://github.com/kasperschmidt/FELIS}{https://github.com/kasperschmidt/FELIS}, \cite{kasper_schmidt_2021_5131705}} described in Appendix~\ref{sec:felis}).
In short, FELIS matches a set of predefined emission line (doublet) templates independently to each spectrum via cross-correlation and $\chi^2$ minimization.
By cross-correlating each template ($\mathcal{T}$) with the spectrum around the expected location of the UV emission line, the template flux scaling that minimizes the disagreement with the data is estimated ($\alpha$ in Equation~\ref{eq:felisflux}).
Calculating the S/N of the best match of the cross-correlations provides an estimate for (S/N)$_{\mathcal{T}, \textrm{max}}$ for each template and its flux scaling.
The matched emission line (doublet) template with the highest (S/N)$_\textrm{max}$ then provides the best match to the considered part of the observed spectrum overall.
We refer to this best-match S/N as the ``FELIS S/N'' of the detected emission lines in the remainder of this work. 
Here we focus our search on the rest-frame UV emission lines red-wards of \lya{} listed in Table~\ref{tab:LinesAndTemplates}.

All emission line template matches were performed independently allowing for individual velocity offsets with respect to the LSDCat lead line of the spectra, which provides the selection redshift as described in Section~\ref{sec:objsel}.  
The lead line redshifts (and the targeted UV lines) are not guaranteed to be at systemic redshift, especially for objects where the lead line is the resonant \civ{} or \lya{} line, which can be offset from systemic by hundreds of km~s$^{-1}$, as indicated by the location of the vertical gray lines based on the LSDCat lead line redshift shown in Figure~\ref{fig:ObjSpec}.
Each spectrum was therefore searched around the expected location of the central UV line wavelengths (Table~\ref{tab:LinesAndTemplates}) in the range $[4790/(1+z)+10~\textrm{\AA},9310/(1+z)-10~\textrm{\AA}]$. 
Hence, if a central doublet wavelength is outside this range it will not be recovered by our search. 
This means that we search for the different rest-frame UV emission lines in the effective redshift ranges provided in Table~\ref{tab:LinesAndTemplates}.
\begin{table*}
\caption{\label{tab:LinesAndTemplates}Targeted rest-frame UV emission lines (left) and the FELIS template parameters (right) used for the search in the 1D TDOSE spectra.}
\centering
\begin{tabular}{lccc|cccc}
\hline\hline
Line 				&  $\lambda_\textrm{rest}$	&  $\lambda_\textrm{central}$	&    Effective $z$-range		&  Width			&    $\sigma_\textrm{Gauss}$		& Flux ratios						& N$_\textrm{Templates}$	\\  
			 	&  [\AA]					&  [\AA]					&    [\AA]				&  [\AA]			&    [\AA]						& F$_{\lambda1}$/F$_{\lambda2}$		&					\\  
\hline
\nv{} doublet          	& 1239, 1243				& 1240.8					&    2.8918 -- 6.4432		& 40				&    0.1--1.2; $\Delta=0.1	$		&  0.2--3.2; $\Delta=0.2$				&	192				\\  
\civ{} doublet		& 1548, 1551				& 1549.5					&    2.1114 -- 4.9699		& 30				&    0.1--1.2; $\Delta=0.1	$		&  0.2--3.2; $\Delta=0.2$				&	192				\\  
\heii               		& 1640           				& 1640.4           			&   1.9379 -- 4.6411 		& 30           		&    0.1--1.2; $\Delta=0.1	$		&  								&	12				\\  
\oiii{} doublet 		& 1661, 1666				& 1663.5					&    1.8969 -- 4.5632		& 60				&    0.1--1.2; $\Delta=0.1	$		&  0.1--1.3; $\Delta=0.1$				&	156				\\  
\siiii{} doublet		& 1883, 1892				& 1887.4					&    1.5514 -- 3.9067		& 60				&    0.1--1.2; $\Delta=0.1	$		&  0.1--1.8; $\Delta=0.1$				&	216				\\  
\ciii{} doublet		& 1907, 1909				& 1907.7					&   1.5241 -- 3.8548 		& 30				&    0.1--1.2; $\Delta=0.1	$		&  0.1--1.8; $\Delta=0.1$				&	216				\\  
\hline\hline
\end{tabular}
\tablefoot{
The columns provide a name for the line or doublet (Line), the rest-frame wavelength of the emission feature ($\lambda_\textrm{rest}$), the central wavelength of each line/doublet ($\lambda_\textrm{central}$) used as center for estimating the effective wavelength range the lines were searched for (Effective $z$-range), the width of the FELIS templates centered on the central wavelength (width), the Gaussian template line width ($\sigma_\textrm{Gauss}$), the sampled flux ratios of the emission doublets (Flux ratio), and finally the total number of templates used for each line or line doublet (N$_\textrm{Templates}$). 
The widths of the two components of emission line doublets were required to be the same.
The rest-frame wavelength sampling of each template was set to 0.05\AA{} per pixel. 
The sampling of the range of the Gaussian line widths and the doublet ratios are indicated with `$\Delta$'.
}
\end{table*}

%
%

The right part of Table~\ref{tab:LinesAndTemplates} lists the characteristics of the emission line templates used to search for each of the emission features in the 1D TDOSE spectra.
FELIS provides tools for generating search templates as described in Appendix~\ref{sec:FELIStemplates}. 
Here we generated Gaussian templates for all lines with a given width ($\sigma_\textrm{Gauss}$ in steps of $\Delta$ as given in Table~\ref{tab:LinesAndTemplates}).
For the emission line doublets we generated templates with two Gaussian components fixing the seperation between them. 
However, the doublet flux ratios were free to vary as listed in Table~\ref{tab:LinesAndTemplates}.
By using a fixed set of Gaussian templates we obtain robustness and efficiency when searching for undiscovered UV emission in the more than two thousand spectra compared to performing parametric fits of Gaussians at all possible locations in the spectra.

Of the 2052 objects searched for UV emission lines we obtained an initial candidate list of 705 
objects with at least one potential UV line detection with FELIS S/N $>3$ and a velocity offset with respect to the catalog redshift of less than 1000~km~s$^{-1}$.
Despite including the effective noise error information in the FELIS matches, a large fraction of these potential detections were spurious emission lines, 
where the cross-correlation latched on to sky-line residuals or minor variations in the overall spectrum matching template doublet spacings. 
Using FELIS S/N $>5$ reduces the number of objects with at least one candidate UV line detection to 167 objects removing the majority of the spurious detections but also discarding reliable detections.
To account for the spurious detections while still recovering reliable detections at $3<\textrm{S/N}<5$, we therefore visually vetted all 705 
$\textrm{S/N}>3$ emitter candidates to determine the amount and type of reliable FELIS template fits. 
Figure~\ref{fig:ObjSpec} shows examples of UV emission line detections deemed reliable (red solid curves over-plotted on the blue TDOSE spectra) and FELIS detections discarded in the vetting process (dashed and dotted red curves).
Hence, after visually vetting the 705 (167) objects with at least one potential $\textrm{S/N} > 3 (5)$ UV emission line detection we deemed 103 (57) of these objects to have reliable FELIS template matches to the MUSE TDOSE spectra.
A summary of the 103/705 objects with reliable UV emission line detections is presented in Table~\ref{tab:UVESdetections} and their redshift distribution is shown as the filled dark green histogram in Figure~\ref{fig:zdist}.
In Table~\ref{tab:UVESdetections} we list the number of detected lines, the number of available objects to search with FELIS and the corresponding fraction of detections for both the full sample (top), objects in the redshift desert (central), and the LAE sample (bottom) for all 101 fields, the MUSE-Wide pointings, the UDF mosaic and the deep UDF10 pointing. 
Of the listed 103 objects with UV line detections, 71 and 80 lines have $\textrm{S/N} > 5$ and $3 <\textrm{S/N} < 5$, respectively.
The objects with line detections in the redshift desert generally have high-confidence redshift determinations as 84\% of the sources (41/49) have at least one detection with $\textrm{S/N} > 5$. Furthermore, 53\% of the objects (26/49) have multiple lines detected, where at least one line has $\textrm{S/N} > 5$. 
Figure~\ref{fig:ObjSpec03} shows two example of such redshift desert objects with multiple UV lines detected at high confidence.
Appendix~\ref{sec:mastercat} and Table~\ref{tab:mastercatcol} describe the value-added catalog of the full sample of objects searched for UV emission lines including the corresponding fluxes, flux ratios, and EWs provided with this paper.
This catalog also includes the relevant measurements of the \lya{} properties determined by \cite{Kerutt:2021tr} described in Section~\ref{sec:UVandLya}.
\begin{table*}
\caption{\label{tab:UVESdetections}UV emission lines detected In MUSE-Wide and MUSE-Deep TDOSE spectra via FELIS template matches.}
\centering
\resizebox{\textwidth}{!}{ 
\begin{tabular}{lc|rrr|rrr|rrr|rrr}
\hline\hline
Line			 	& $z$-range		& \multicolumn{12}{c}{\textbf{Objects within full $z$-range ($1.5<z<6.4$)}}		\\															
				&				& \multicolumn{3}{c}{All fields}	& \multicolumn{3}{|c}{MUSE-Wide} 	& \multicolumn{3}{|c}{UDF mosaic} 	& \multicolumn{3}{|c}{UDF10} 	\\	
\hline
Any detection		& 1.5000 -- 6.4432	& 103 & 2052 &  5.02\% & 24 & 1100 &  2.18\% & 49 & 719 &  6.82\% & 30 & 233 & 12.88\% \\
\nv			     	&   2.8918 -- 6.4432	& 7 & 1997 &  0.35\% & 3 & 1094 &  0.27\% & 3 & 688 &  0.44\% & 1 & 215 &  0.47\% \\
\civ				&   2.1114 -- 4.9699	& 45 & 1710 &  2.63\% & 15 & 947 &  1.58\% & 17 & 590 &  2.88\% & 13 & 173 &  7.51\% \\
\heii               		&  1.9379 -- 4.6411 	& 16 & 1465 &  1.09\% & 7 & 817 &  0.86\% & 8 & 495 &  1.62\% & 1 & 153 &  0.65\% \\
\oiii				&   1.8969 -- 4.5632	& 18 & 1451 &  1.24\% & 2 & 803 &  0.25\% & 7 & 496 &  1.41\% & 9 & 152 &  5.92\% \\
\siiii				&   1.5514 -- 3.9067	& 13 & 1000 &  1.30\% & 0 & 530 &  0.00\% & 8 & 347 &  2.31\% & 5 & 123 &  4.07\% \\
\ciii				&  1.5241 -- 3.8548 	& 52 & 985 &  5.28\% & 5 & 520 &  0.96\% & 29 & 343 &  8.45\% & 18 & 122 & 14.75\% \\
\hline 
\hline
\multicolumn{14}{c}{}     \\
\hline
\hline
Line			 	& $z$-range		& \multicolumn{12}{|c}{\textbf{MUSE redshift desert ($1.5<z<2.9$)}}     		\\
				&				& \multicolumn{3}{|c}{All fields} 	& \multicolumn{3}{|c}{MUSE-Wide}	& \multicolumn{3}{|c}{UDF mosaic} 	& \multicolumn{3}{|c}{UDF10} \\
\hline
Any detection		& 1.5000 -- 6.4432	& 49 & 55 & 89.09\% & 5 & 6 & 83.33\% & 28 & 31 & 90.32\% & 16 & 18 & 88.89\% \\
\nv			     	&   2.8918 -- 6.4432	& 0  & 0 &  0.0\% & 0 & 0 &  0.0\% & 0 & 0 &  0.0\% & 0 & 0 &  0.0s\% \\
\civ				&   2.1114 -- 4.9699	& 11 & 28 & 39.29\% & 2 & 5 & 40.00\% & 5 & 14 & 35.71\% & 4 & 9 & 44.44\% \\
\heii               		&  1.9379 -- 4.6411 	 & 8 & 34 & 23.53\% & 2 & 5 & 40.00\% & 5 & 19 & 26.32\% & 1 & 10 & 10.00\% \\
\oiii				&   1.8969 -- 4.5632	& 12 & 38 & 31.58\% & 0 & 5 &  0.00\% & 7 & 22 & 31.82\% & 5 & 11 & 45.45\% \\
\siiii				&   1.5514 -- 3.9067	 & 11 & 53 & 20.75\% & 0 & 5 &  0.00\% & 7 & 31 & 22.58\% & 4 & 17 & 23.53\% \\
\ciii				&  1.5241 -- 3.8548 	& 37 & 54 & 68.52\% & 1 & 5 & 20.00\% & 22 & 31 & 70.97\% & 14 & 18 & 77.78\% \\
\hline     
\hline
\multicolumn{14}{c}{}     \\
\hline
\hline
Line			 	& $z$-range		& \multicolumn{12}{|c}{\textbf{Ly$\alpha$ emitters ($2.9<z<6.4$)}}     		\\
				&				& \multicolumn{3}{|c}{All fields} 	& \multicolumn{3}{|c}{MUSE-Wide}	& \multicolumn{3}{|c}{UDF mosaic} 	& \multicolumn{3}{|c}{UDF10} \\
\hline
Any detection		& 2.9 -- 6.4432		& 54 & 1997 &  2.70\% & 19 & 1094 &  1.74\% & 21 & 688 &  3.05\% & 14 & 215 &  6.51\% \\
\nv			     	&   2.9 -- 6.4432	& 7 & 1997 &  0.35\% & 3 & 1094 &  0.27\% & 3 & 688 &  0.44\% & 1 & 215 &  0.47\% \\
\civ				&   2.9 -- 4.9699	& 34 & 1682 &  2.02\% & 13 & 942 &  1.38\% & 12 & 576 &  2.08\% & 9 & 164 &  5.49\% \\
\heii               		&  2.9 -- 4.6411 	&  8 & 1431 &  0.56\% & 5 & 812 &  0.62\% & 3 & 476 &  0.63\% & 0 & 143 &  0.00\% \\
\oiii				&  2.9 -- 4.5632		&  6 & 1413 &  0.42\% & 2 & 798 &  0.25\% & 0 & 474 &  0.00\% & 4 & 141 &  2.84\% \\
\siiii				&  2.9 -- 3.9067		&  2 & 947 &  0.21\% & 0 & 525 &  0.00\% & 1 & 316 &  0.32\% & 1 & 106 &  0.94\% \\
\ciii				& 2.9 -- 3.8548 		& 15 & 931 &  1.61\% & 4 & 515 &  0.78\% & 7 & 312 &  2.24\% & 4 & 104 &  3.85\% \\
\hline     
\hline
\end{tabular}
}
\tablefoot{The number of reliable secondary UV emission lines detected in the objects from the MUSE-Wide, the UDF mosaic and UDF10.
The top part provides numbers when including all available sources studied in this work, 
the central part list the statistics for the object in the so-called MUSE redshift desert, and
the bottom part provides the statistics for LAEs.
Each subsample (all fields, MUSE-Wide, UDF mosaic and UDF10) include three numbers corresponding to the number of lines detected, the number of objects available in the effective redshift range, and the percentage of spectra with a detection this corresponds to.
The MUSE-Wide includes $\approx91$~arcmin$^2$ at 1 hour depth, the UDF mosaic $\approx9$~arcmin$^2$ at 10 hours depth, and the UDF10 $\approx1$~arcmin$^2$ at 31 hours depth.
Each 1D spectrum was searched for UV emission via template matching with the FELIS 
software (Appendix~\ref{sec:felis}) using the template sets listed in Table~\ref{tab:LinesAndTemplates}.
}
\end{table*}

%
%

\begin{figure}
\begin{center}
\includegraphics[width=0.49\textwidth]{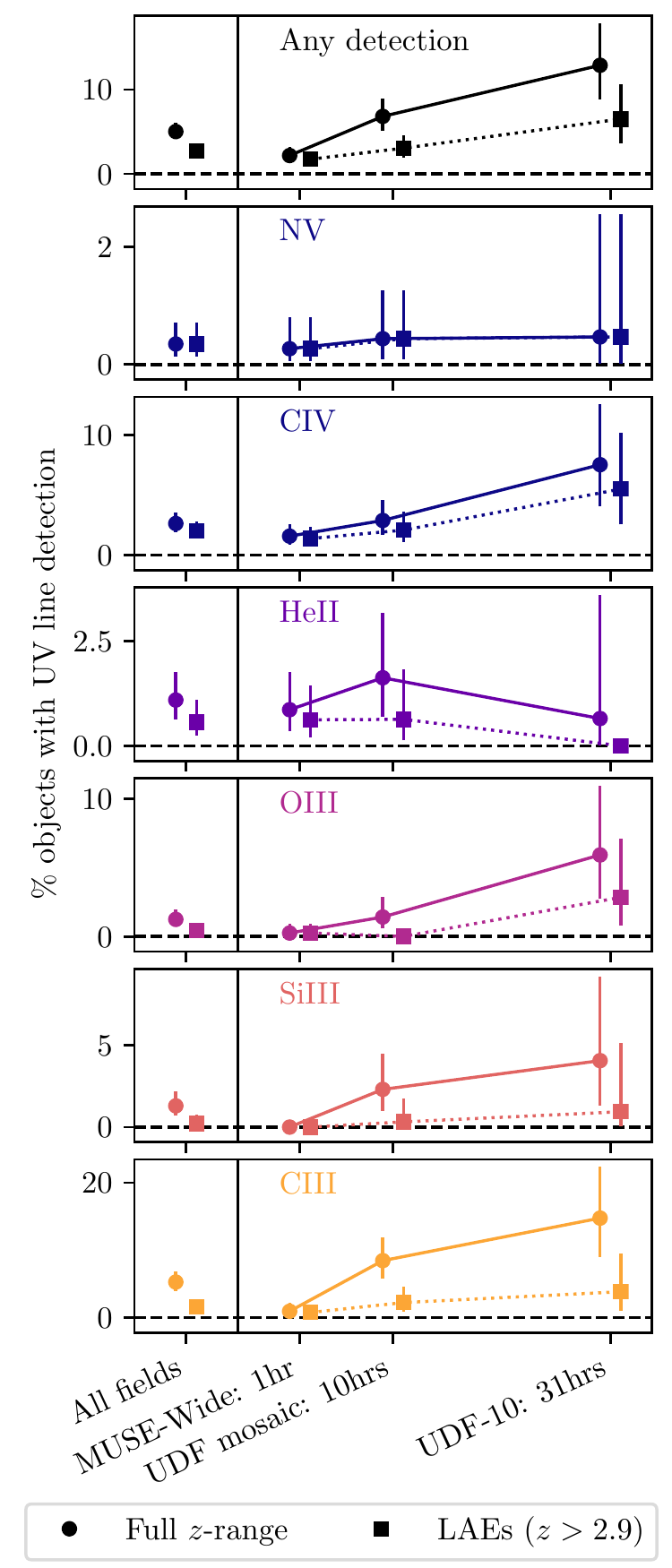}
\caption{Percentage of objects with detected secondary UV emission lines red-wards of \lya{} for the full sample (circles) and the subsample of LAEs (squares).
The percentages are provided for all fields studied, the MUSE-Wide data (1 hour depth), the UDF mosaic data (10 hours depth) and the UDF10 pointing (31 hours depth).
The error bars indicate the 95\% confidence Clopper-Pearson interval of the binomial distribution for $k$ UV line detections given the $n$ available objects searched (see Table~\ref{tab:UVESdetections}).
}
\label{fig:detfrac}
\end{center}
\end{figure}

Figure~\ref{fig:detfrac} shows the UV emission line detection fractions for each of the subsamples.
In general, there is an increase of the fraction of objects with detected UV emission lines red-wards of \lya{} as the depth of the data is increased (1, 10, and 31 hours depth for MUSE-Wide, UDF mosaic and UDF10, respectively).
In some cases the low number of detections might be responsible for the potential trends as indicated by the error bars showing the 95\% confidence intervals for the Clopper-Pearson interval \citep{CLOPPER:1934ee}.
These intervals are conservative as they are extracted based directly on the cumulative probability function of the binomial distribution but reflect the uncertainty of each fraction.
An increase in the fraction of faint emission lines as a function of survey depth will only occur if there is a relative change in the shape of the object's (emission line) luminosity function for the different samples. 
For example, if the truncated LAE luminosity function shape (slope) was the same as that of the UV emitter luminosity function at the various depths, the fraction of objects with detected UV emission lines above the 3$\sigma$ threshold should remain the same irrespective of survey depth.
Given the limited sample size and survey depth \cite{2017MNRAS.471.2575S} fixed the faint-end slope for their \ciii{} LF fit suggesting that it is close to a scaled-down version of the \lya{} LF.
This is not what we see in the MUSE data, implying that the shape of the luminosity functions of the subsamples of objects emitting the detected UV emission lines differ in shape compared to the luminosity function of the parent (LAE) sample. 
A steeper faint-end slope of the UV emission line luminosity function could replicate the observed increase in the fraction of UV emitters with increasing depth. 

\begin{figure*}
\begin{center}
\includegraphics[width=0.8\textwidth]{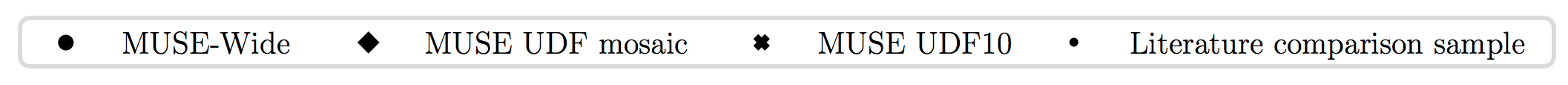}\\
\includegraphics[width=0.4\textwidth]{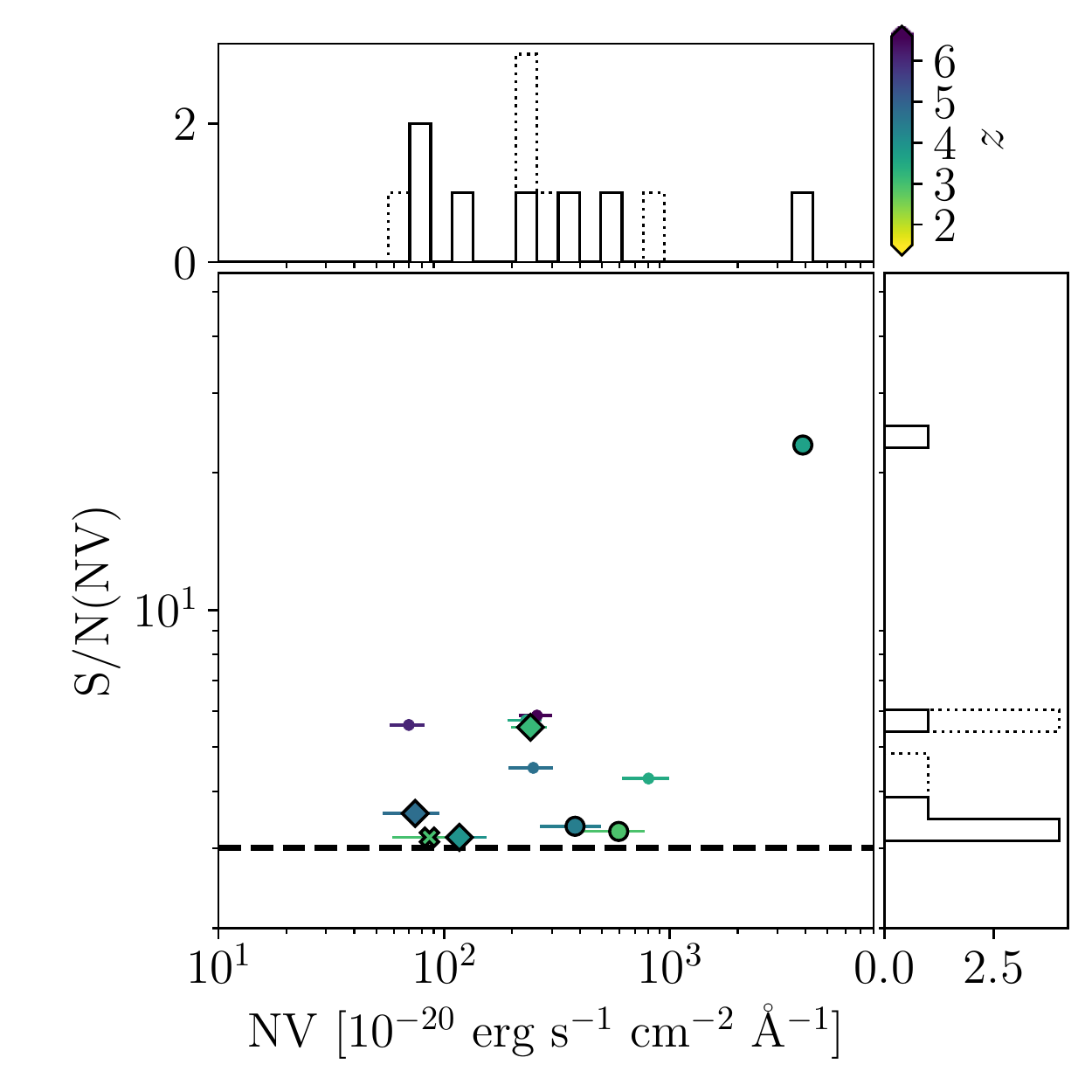}
\includegraphics[width=0.4\textwidth]{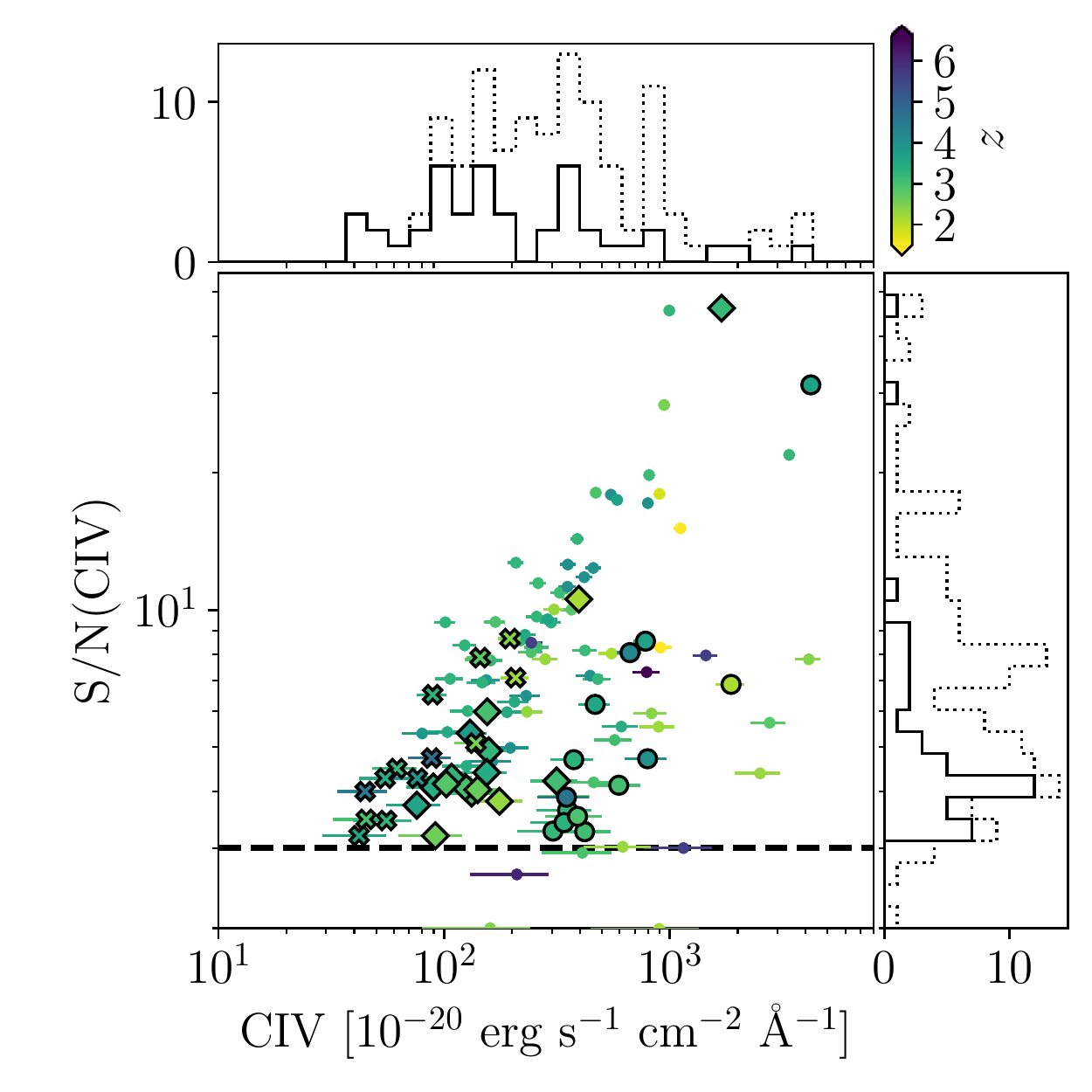}\\
\includegraphics[width=0.4\textwidth]{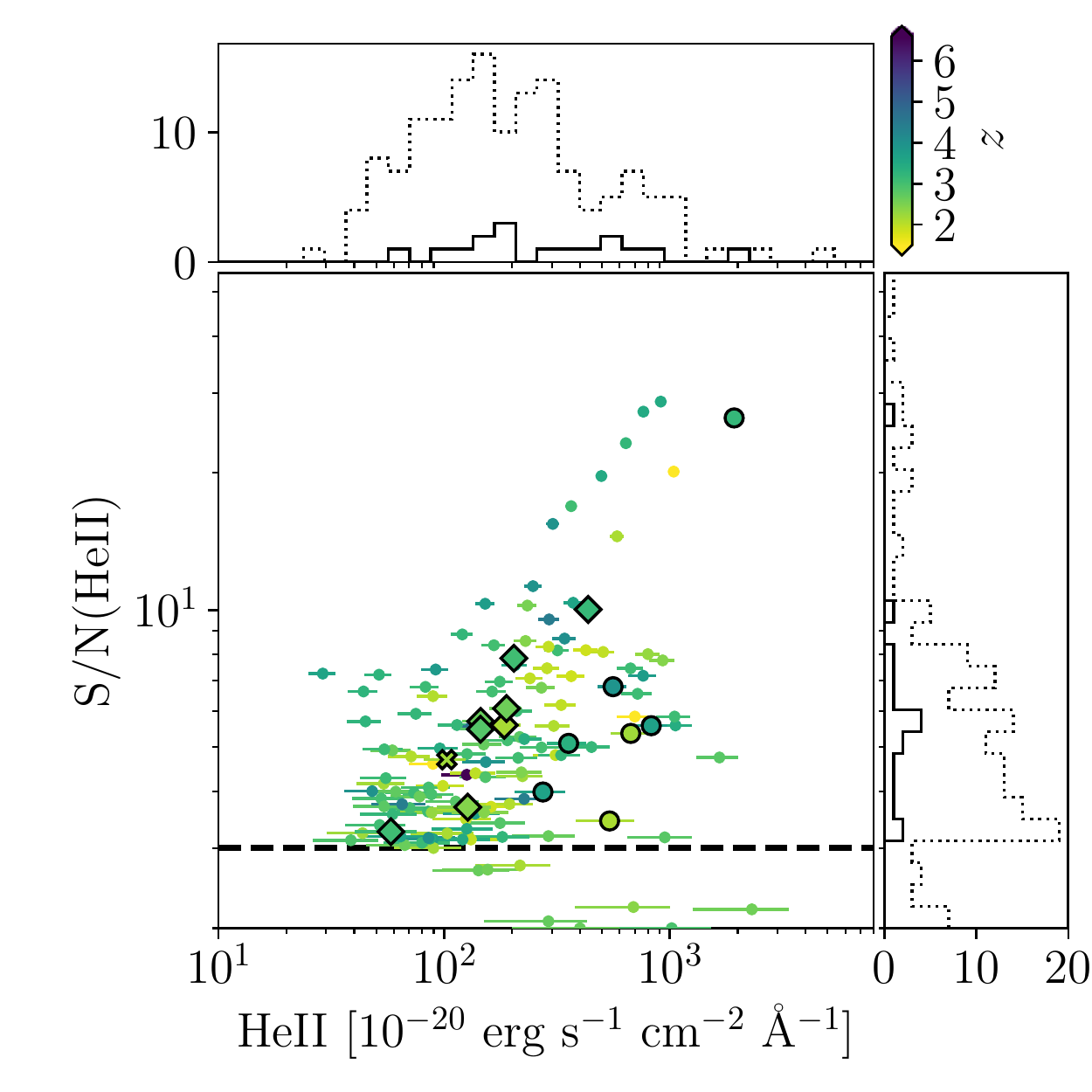}
\includegraphics[width=0.4\textwidth]{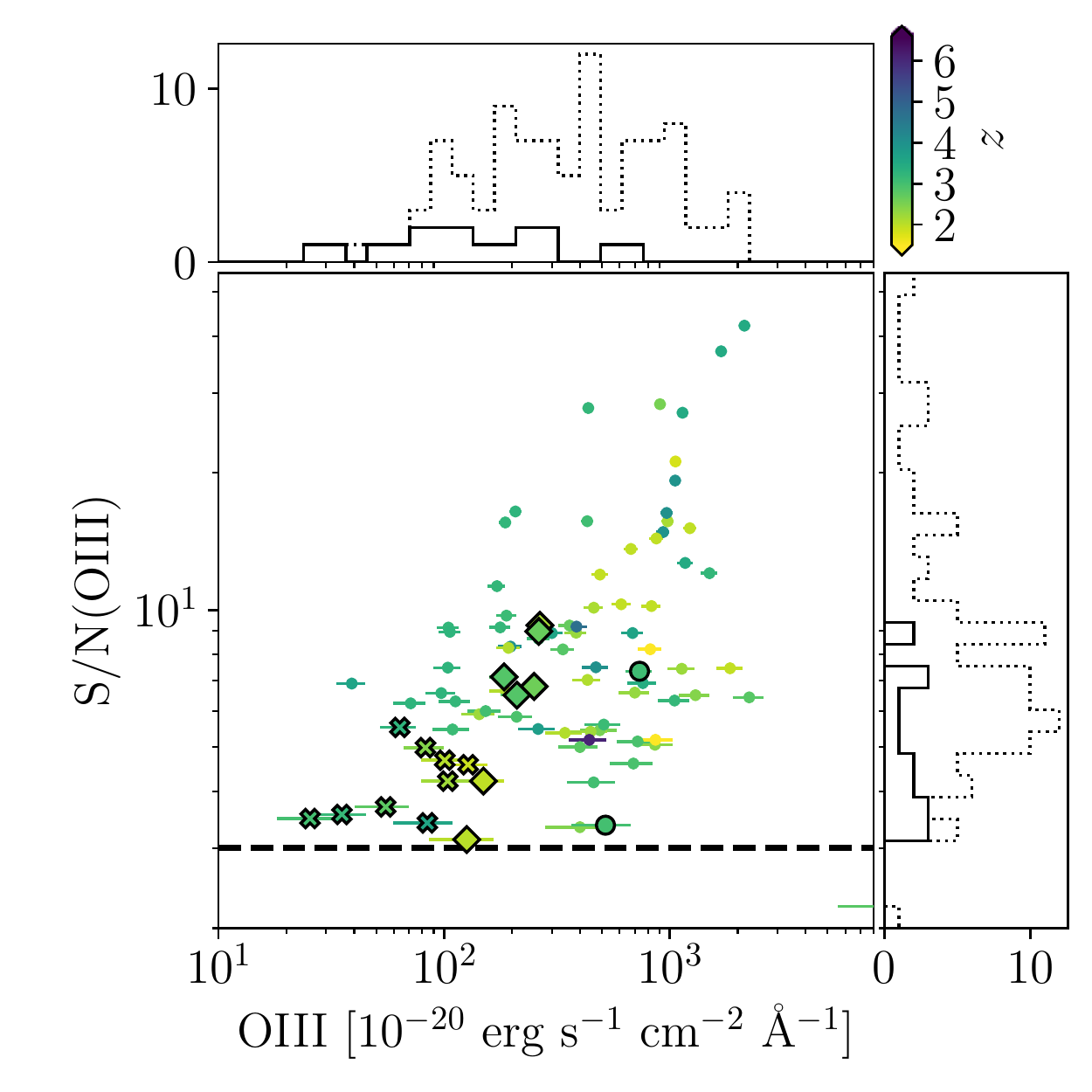}\\
\includegraphics[width=0.4\textwidth]{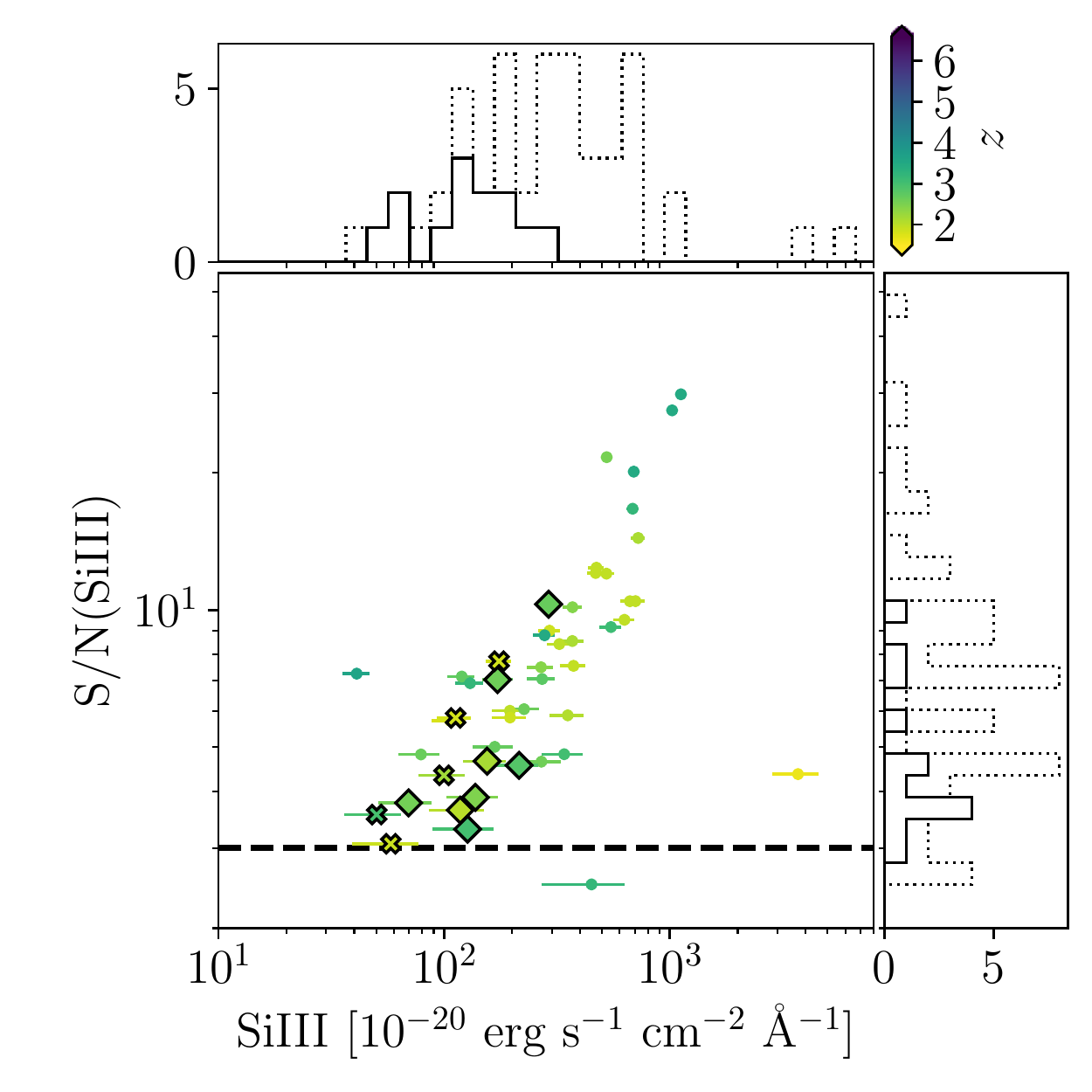}
\includegraphics[width=0.4\textwidth]{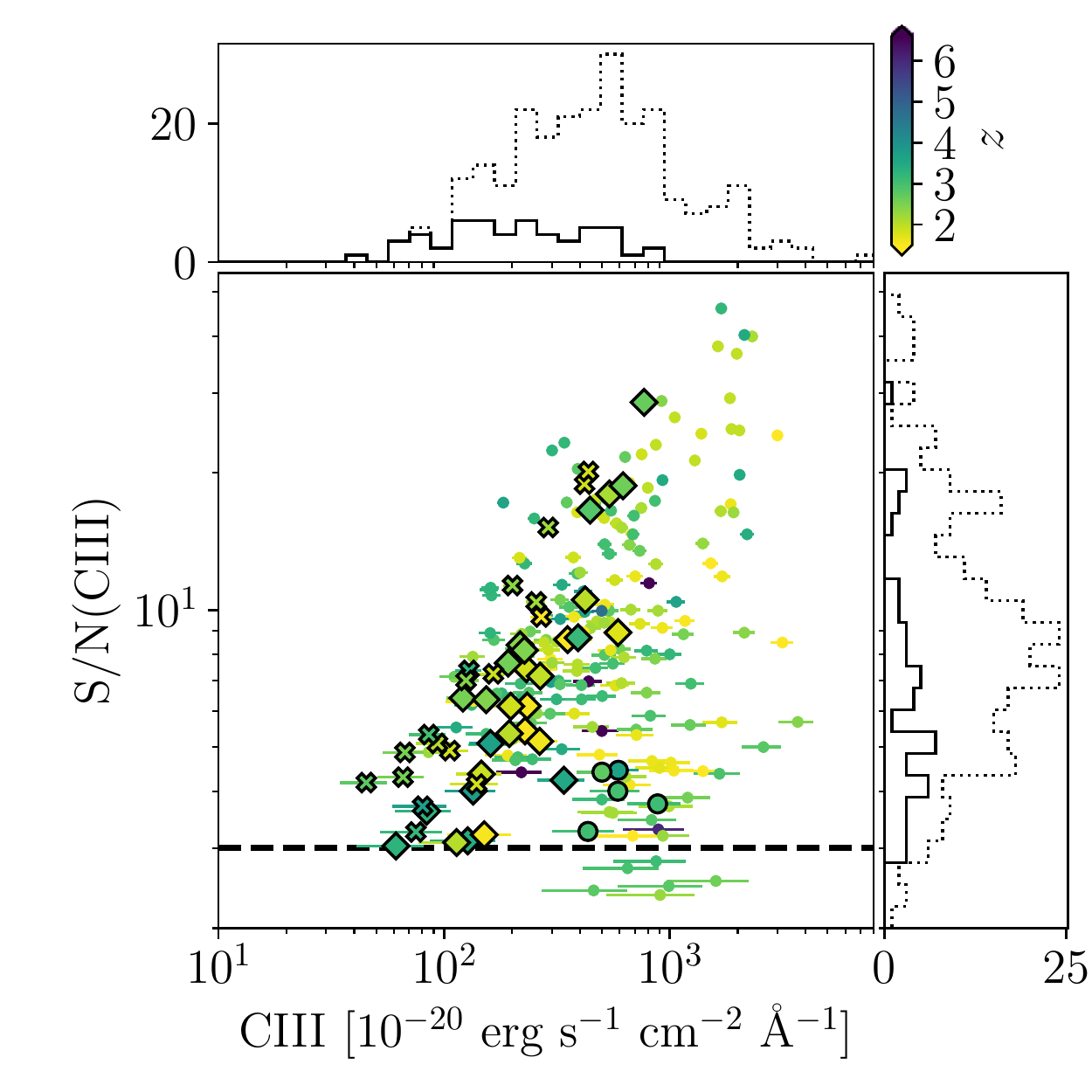}
\caption{Distribution of rest-frame UV emission line fluxes (x-axis) detected at a FELIS S/N (y-axis) above 3$\sigma$ (horizontal dashed line) in the MUSE-Wide (filled circles), the UDF mosaic (filled diamonds), and the UDF10 (filled x's) data sets studied in this paper.
These detections are compared to the collection of detections from the literature (small dots) described in Appendix~\ref{sec:litcol}. Here S/N represents the emission line S/N quoted in the original references. 
All points are color coded according to their redshift.
The solid histograms show the subset of the objects from this work.
The dotted histograms show the distribution of measurements when including the literature data.}
\label{fig:ELfluxes}
\end{center}
\end{figure*}

Figure~\ref{fig:ELfluxes} presents the line fluxes and their significance for the UV line detections together with a sample of literature values (small dots; see Appendix~\ref{sec:litcol} for details).
For the emission line doublets the total flux from combining both components is shown.
The uncertainties are determined by the variance of the template crossmatch defined in Equation~\ref{eq:CCvariance}.
The large number of upper limits provided by the FELIS template matching are not shown to prevent cluttering the figure, but are available in the public catalog published with this paper.
As a crosscheck, we compared the estimated \heii{} and \ciii{} emission line fluxes from FELIS to the PLATEFIT \citep{2004ApJ...613..898T,2004MNRAS.351.1151B} flux estimates for the sources in the UDF mosaic also studied by \cite{2019A&A...624A..89N} and \cite{2017A&A...608A...4M} and find good agreement between the fully independent detections and flux measures.

\begin{figure*}
\begin{center}
\includegraphics[width=0.80\textwidth]{mainlegend.png}\\
\includegraphics[width=0.41\textwidth]{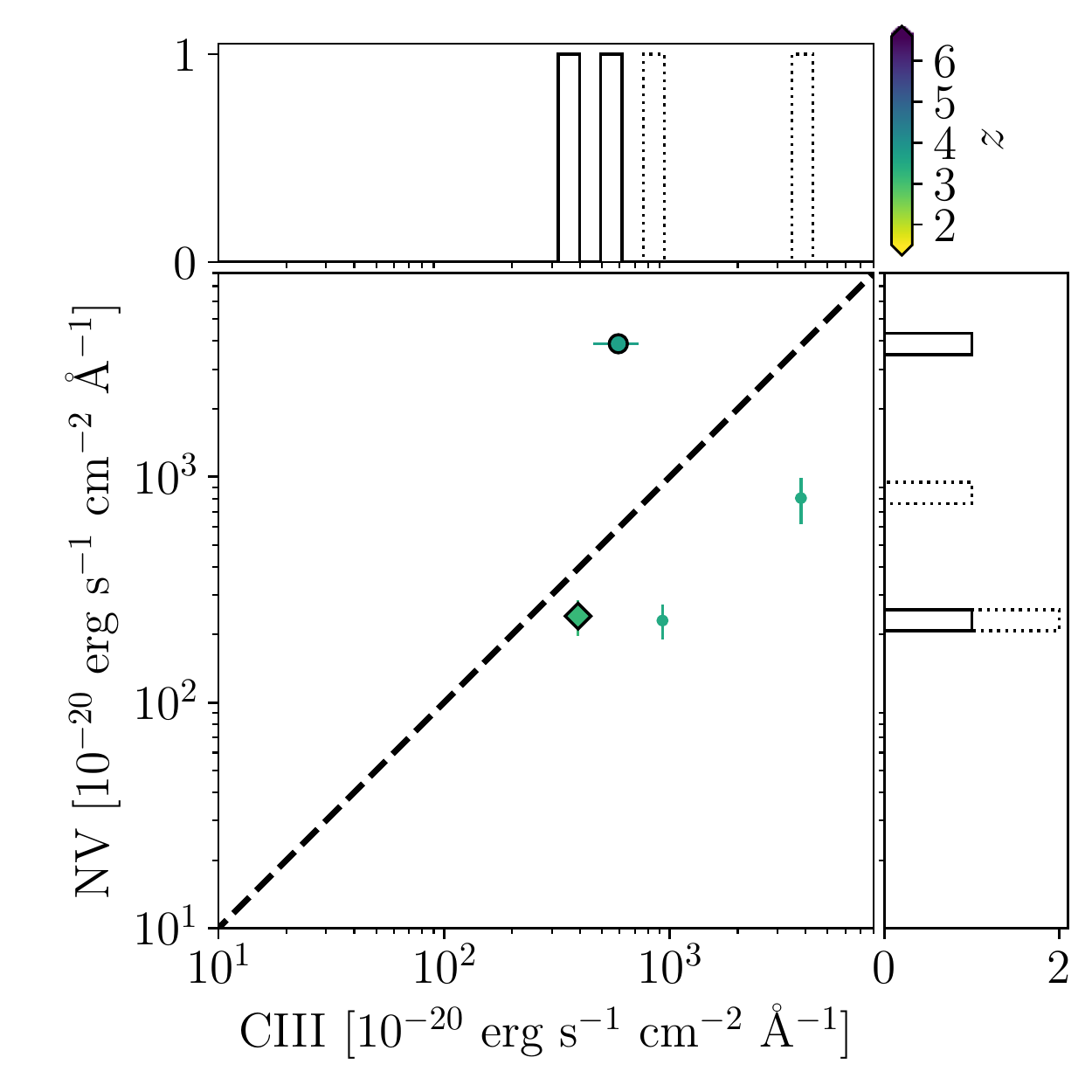}
\includegraphics[width=0.41\textwidth]{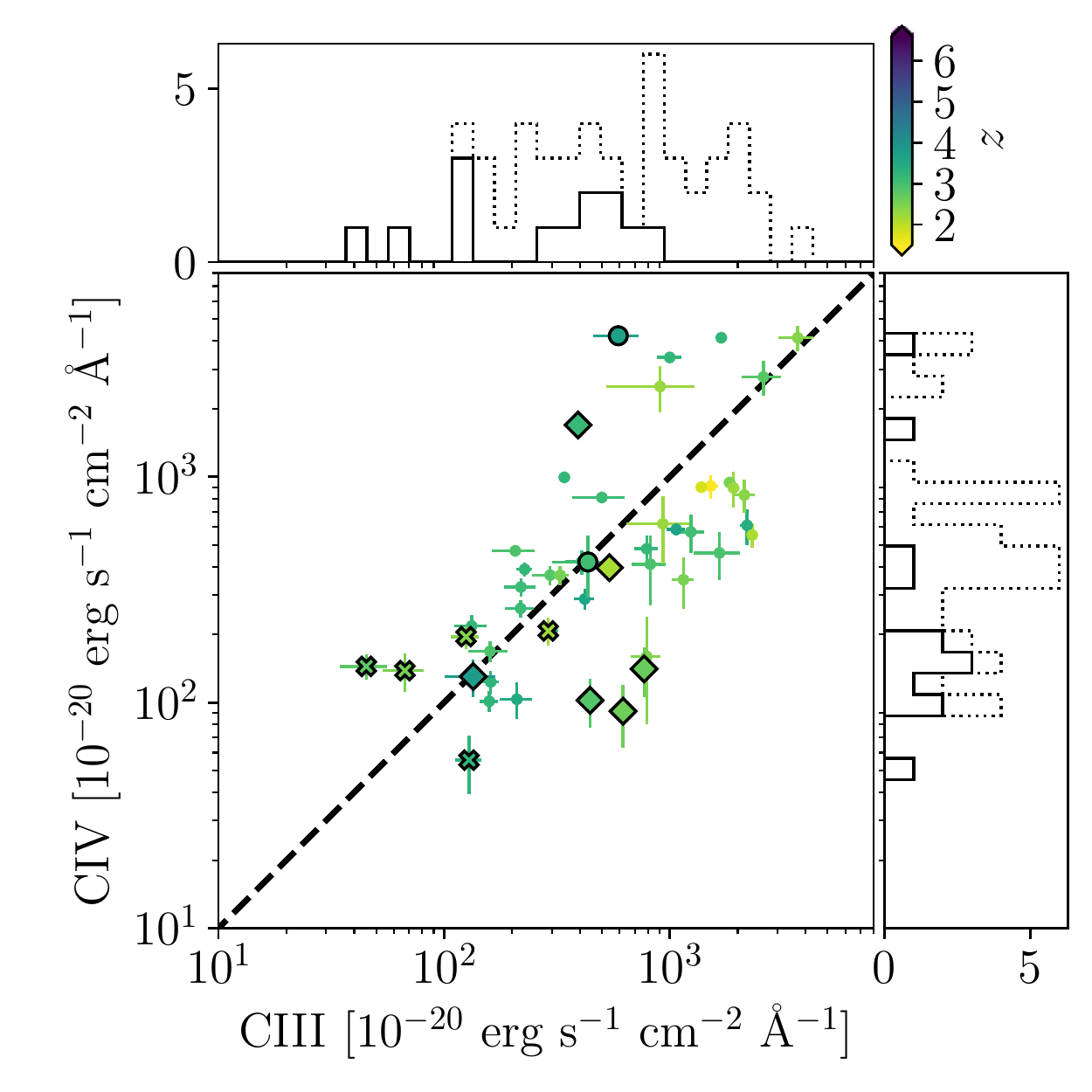}\\
\includegraphics[width=0.41\textwidth]{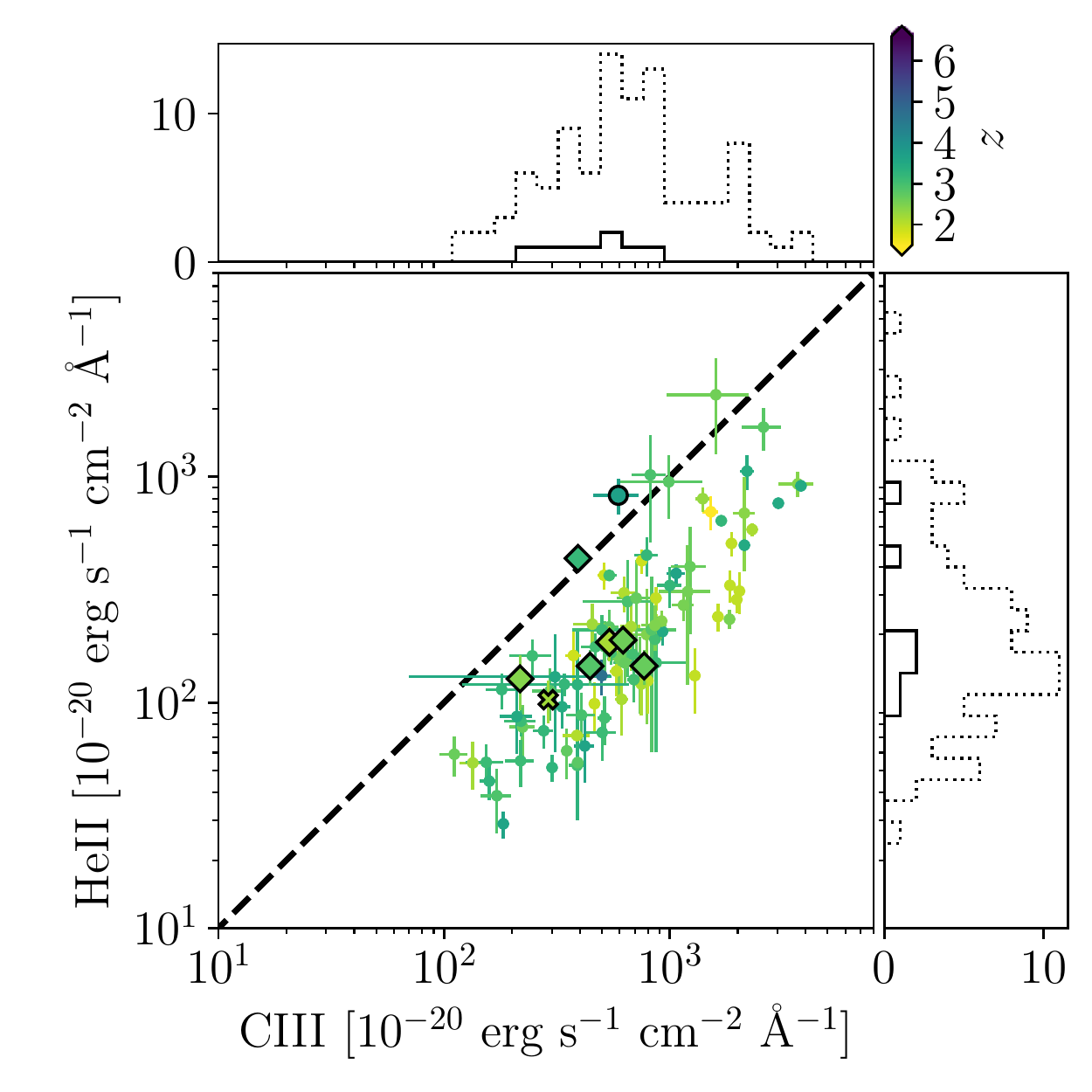}
\includegraphics[width=0.41\textwidth]{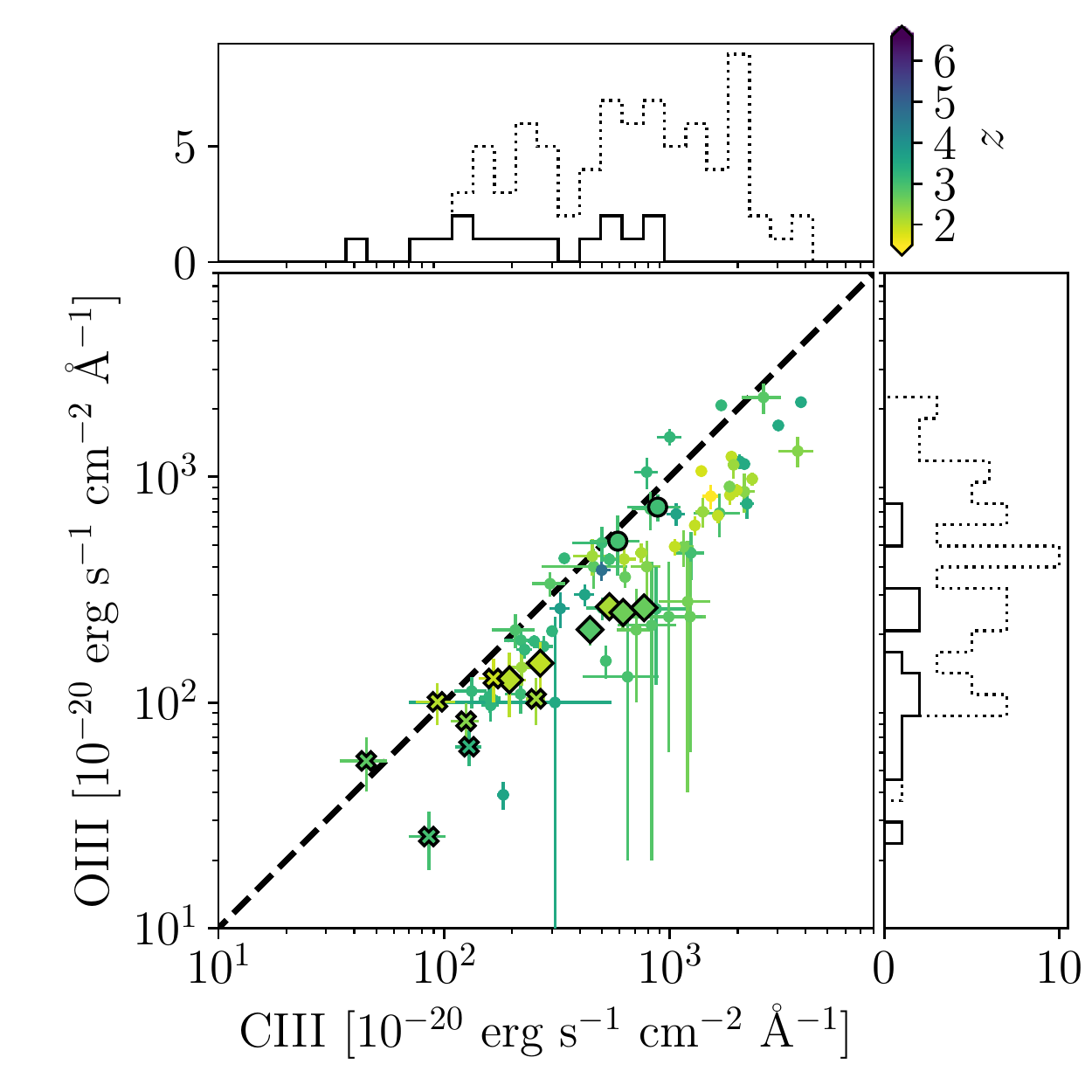}\\
\includegraphics[width=0.41\textwidth]{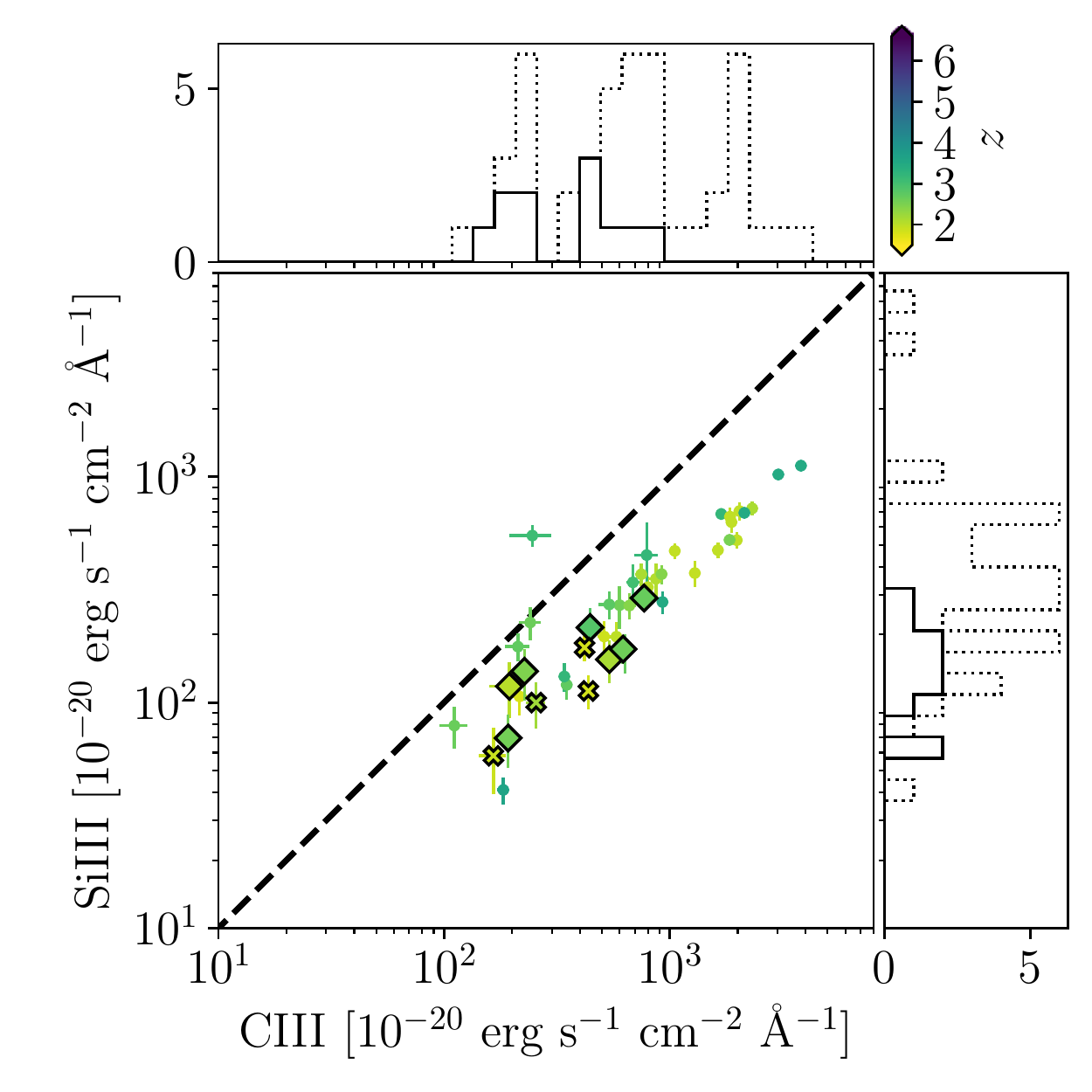}
\caption{Similar to Figure~\ref{fig:ELfluxes}, but showing correlations between the $>3\sigma$ \ciii{} detections and the other secondary UV emission line fluxes from MUSE-Wide, the UDF mosaic, and UDF10 shown as large symbols. The literature measurements described in Appendix~\ref{sec:litcol} are shown by the small dots.
The dashed lines show the one-to-one relations to guide the eye.}
\label{fig:fluxesVSciii}
\end{center}
\end{figure*}

Figure~\ref{fig:fluxesVSciii} presents comparisons between the UV emission line fluxes and the \ciii{} flux for objects with both \ciii{} and additional UV lines detected.
It is clear that several of the emission line fluxes correlate in an approximately linear fashion.
To quantify the strength and parametrize these correlations we estimate the 
Pearson correlation coefficient ($r_\textrm{P}$) and the Spearman's rank correlation coefficient ($r_\textrm{S}$).
The $r_\textrm{P}$ tests the strength of a linear relation, whereas $r_\textrm{S}$ only requires a monotonic correlation to provide a larger value.
The coefficients have values between -1 and 1 and as a rule of thumb absolute values of the correlation coefficients $|r|<0.3$, $0.3<|r|<0.5$, $0.5<|r|<0.7$, and $|r|>0.7$ can be considered very weak (nonexistent), weak, moderate, and strong correlations, respectively.
Standard linear regression usually does not account for uncertainties in both data sets. 
To obtain a parametric representation of the correlations, we therefore estimate the best-fit linear relations between the flux estimates via orthogonal distance regression (ODR) using Scipy's build-in version of this.
The correlation coefficients and resulting best fits for the emission line fluxes are presented as correlations 1-4 in Table~\ref{tab:paramcorr}.
\begin{table*}
\caption{\label{tab:paramcorr}Empirical linear correlations ($y = a x+b$) between measured quantities for the UV line emitters}
\centering
\begin{tabular}{crrrrrrc}
\hline\hline  
No.		&	$y$								&	$a$				&	$x$								&	$b$					&	$r_\textrm{P}$		&	$r_\textrm{S}$		&	Figure			\\ 	
\hline
1		&	$\log_{10}[F(\textrm{\civ})]$  			& $0.8387\pm0.0242$ 	& $\log_{10}[F(\textrm{\ciii})]$ 				& $0.5056\pm0.1167$		&	0.9680			&	0.8970			& \ref{fig:fluxesVSciii} 	\\ 
2		&	$\log_{10}[F(\textrm{\heii})]$			& $0.8698\pm0.0117$ 	& $\log_{10}[F(\textrm{\ciii})]$ 				& $-0.0988\pm0.0548$		&	0.9788			&	0.8901			& \ref{fig:fluxesVSciii} 	\\ 
3		&	$\log_{10}[F(\textrm{\oiii})]$  			& $0.9112\pm0.0092$ 	& $\log_{10}[F(\textrm{\ciii})]$ 				& $0.1238\pm0.0437$ 		&	0.9903			&	0.9529			& \ref{fig:fluxesVSciii} 	\\ 
4		&	$\log_{10}[F(\textrm{\siiii})]$ 			& $1.0111\pm0.0071$ 	& $\log_{10}[F(\textrm{\ciii})]$ 				& $-0.5164\pm0.0316$ 		&	0.9937			&	0.9548			& \ref{fig:fluxesVSciii} 	\\ 
5		&	$\log_{10}[F(\textrm{\nvtwo})] $			& $0.8437\pm0.0353$	& $\log_{10}[F(\textrm{\nvone})]$			& $0.5115\pm0.1080$		&	0.9556			&	0.9273			& \ref{fig:doubletratios}	\\  
6		&	$\log_{10}[F(\textrm{\civtwo})] $			& $1.0124\pm0.0255$	& $\log_{10}[F(\textrm{\civone})]$			& $0.0010\pm0.1050$		&	0.9754			&	0.8978			& \ref{fig:doubletratios}	\\  
7		&	$\log_{10}[F(\textrm{\oiiitwo})] $			& $1.0086\pm0.0118$	& $\log_{10}[F(\textrm{\oiiione})]$			& $0.2763\pm0.0513$		&	0.9893			&	0.9704			& \ref{fig:doubletratios}	\\  
8		&	$\log_{10}[F(\textrm{\siiiitwo})]$			& $1.0463\pm0.0085$	& $\log_{10}[F(\textrm{\siiiione})]$			& $-0.2825\pm0.0339$		&	0.9897			&	0.9418			& \ref{fig:doubletratios}	\\ 
9		&	$\log_{10}[F(\textrm{\ciiitwo})]$			& $1.0049\pm0.0063$	& $\log_{10}[F(\textrm{\ciiione})]$			& $-0.1647\pm0.0226$		&	0.9855			&	0.9269			& \ref{fig:doubletratios}	\\ 
10		&	$\log_{10}[\textrm{EW}_0(\textrm{\nv})]$	& $1.0390\pm0.3928$	& $\log_{10}[\textrm{EW}_0(\textrm{\ciii})]$	& $0.1618\pm0.3046$		&	0.0320			&	-0.0952			& \ref{fig:EWsciii}	\\  
11		&	$\log_{10}[\textrm{EW}_0(\textrm{\civ})]$	& $1.4381\pm0.2597$	& $\log_{10}[\textrm{EW}_0(\textrm{\ciii})]$	& $-0.5146\pm0.2271$		&	0.4621			&	0.4472			& \ref{fig:EWsciii}	\\ 
12		&	$\log_{10}[\textrm{EW}_0(\textrm{\heii})]$	& $1.1769\pm0.1866$	& $\log_{10}[\textrm{EW}_0(\textrm{\ciii})]$	& $-0.7772\pm0.1439$		&	0.6701			&	0.6534			& \ref{fig:EWsciii}	\\ 
13		&	$\log_{10}[\textrm{EW}_0(\textrm{\oiii})]$	& $0.8480\pm0.0774$	& $\log_{10}[\textrm{EW}_0(\textrm{\ciii})]$	& $-0.3058\pm0.0686$		&	0.8046			&	0.7781			& \ref{fig:EWsciii}	\\ 
14		&	$\log_{10}[\textrm{EW}_0(\textrm{\siiii})]$	& $0.8244\pm0.0578$	& $\log_{10}[\textrm{EW}_0(\textrm{\ciii})]$	& $-0.3243\pm0.0445$		&	0.9306			&	0.9019			& \ref{fig:EWsciii}	\\ 
15		&	$\log_{10}[\textrm{EW}_0(\nv)]$		& $-0.1085\pm0.2627$	& $\log_{10}[\textrm{EW}_0(\textrm{Ly}\alpha)]$& $1.3362\pm0.3791$		&	0.5471			&	0.3636			& \ref{fig:EWs} \\ 
16		&	$\log_{10}[\textrm{EW}_0(\civ)]$		& $1.0117\pm0.2477$	& $\log_{10}[\textrm{EW}_0(\textrm{Ly}\alpha)]$& $-0.8186\pm0.2693$		&	0.4479			&	0.5199			& \ref{fig:EWs} \\ 
17		&	$\log_{10}[\textrm{EW}_0(\heii)]$		& $0.8488\pm0.0697$	& $\log_{10}[\textrm{EW}_0(\textrm{Ly}\alpha)]$& $-1.0449\pm0.1029$		&	0.7014			&	0.7244			& \ref{fig:EWs} \\ 
18		&	$\log_{10}[\textrm{EW}_0(\oiii)]$		& $0.7140\pm0.0671$	& $\log_{10}[\textrm{EW}_0(\textrm{Ly}\alpha)]$& $-0.5714\pm0.0983$		&	0.4356			&	0.5915			& \ref{fig:EWs} \\ 
19		&	$\log_{10}[\textrm{EW}_0(\siiii)]$		& $0.7870\pm0.1190$	& $\log_{10}[\textrm{EW}_0(\textrm{Ly}\alpha)]$& $0.8464\pm0.1768$		&	0.7175			&	0.8846			& \ref{fig:EWs} \\ 
20		&	$\log_{10}[\textrm{EW}_0(\ciii)]$		& $0.5971\pm0.0459$	& $\log_{10}[\textrm{EW}_0(\textrm{Ly}\alpha)]$& $-0.1540\pm0.0697$		&	0.6248			&	0.7273			& \ref{fig:EWs} \\ 
\hline\hline
\end{tabular}
\tablefoot{
The best-fit parameters $a$ and $b$ were obtained with orthogonal distance regression (ODR) taking errors on both $x$ and $y$ into account, $r_\textrm{P}$ provides the 
Pearson correlation coefficient, and $r_\textrm{S}$ provides the Spearman's rank correlation coefficient.
The last columns provide references to figures of the different (non)correlations.
}
\end{table*}

The correlations between \ciii{} and \heii, \oiii, and \siiii{} are tighter (higher values of $r_\textrm{P}$) than the correlation with the resonant emission from \civ.
In all cases, the upper 3$\sigma$ limits resulting from nondetections of UV emission lines by the FELIS template matches follow the presented correlations. 
Hence, if the redshift evolution is insignificant (no indication of a redshift dependence is seen for the data in the redshift range probed by the MUSE samples) these correlations can be used to predict likely emission line fluxes for the fainter UV emission features based on \ciii{}.
This will be useful for targeting these fainter lines in the EoR where only the brightest UV lines (\lya{} and in some cases \ciii{} and \civ) have been recovered with current facilities. 

If we instead consider the individual components of the UV emission line doublets, we find that the flux ratios
$F$(\nvone)/$F$(\nvtwo) and $F$(\oiiione)/$F$(\oiiitwo) are both generally $<1$, whereas
$F$(\siiiione)/$F$(\siiiitwo) and $F$(\ciiione)/$F$(\ciiitwo) are $>1$.
The mean, median and standard deviation of these four flux ratios are 
(0.69,0.65,0.24), 
(0.49,0.44,0.25), 
(1.65,1.60,0.73) and 
(1.35,1.37,0.45).   
The data are shown in Figure~\ref{fig:doubletratios}.
The scatter for the $F$(\civone)/$F$(\civtwo) correlation is larger (lower $r_\textrm{P}$) than for the nebular emission lines and the values of this flux ratio have mean, median and standard deviation (1.34,1.30,0.83).
This can be attributed to the line profiles of the \civ{} emission which are poorly captured by the Gaussian FELIS line templates.
\civ{} is a resonant line that is scattered by high-ionization gas and is produced in the winds of young massive O and B stars as well as in the ISM \citep[e.g.;][]{2011AJ....141...37L,2016ApJ...829...64D,2019ApJ...878L...3B,2020A&A...641A.118F}.
The wind features can produce prominent P-Cygni profiles which, superimposed on the ISM emission, create line-profiles which are poorly approximated by a Gaussian.
Hence, when such a combination of emission processes is present the flux measures (and estimates of velocity offset with respect to systemic; see Section~\ref{sec:voffset}) are more uncertain for the \civ{} template matches and are therefore expected to scatter more.
Full modeling of the \civ{} line profiles is beyond the scope of this work, but provides valuable information about the underlying emission mechanisms, the metallicity and the initial mass function \cite{2011AJ....141...37L}.  

Even though, an attempt to fully model the stellar+nebular emission lines like, \heii, \civ, and \nv{} is beyond scope of this work, broader lines measured from the FELIS template matches, could be indicative of a predominantly stellar contribution to the emission (for non-AGN), as stellar emission is generally broader than the nebular emission \citep[e.g.,][]{,2012MNRAS.421.1043S,2006MNRAS.368..895C,2007ARA&A..45..177C,2019A&A...624A..89N}. 
The FWHM distributions from the FELIS matches for these lines span the range between roughly 50 km/s and 550 km/s with the majority of detections having FWHM~$\lesssim300$~km/s for all three lines. 
Thus, the FWHM of \heii, \civ, and \nv{} are generally modest, but could indicate contribution from stellar emission in some systems.
For details on the nature of the HeII emission and the emitter properties we refer to \cite{2019A&A...624A..89N}, who as mentioned studied the majority of the HeII emitters presented here in detail.   

By comparing and fitting literature compilations of the oscillator strength \cite{1976JPCRD...5..537M,1983A&A...122..335F} and \cite{1991ApJS...77..119M} predict the theoretical \civ{} doublet ratio to be $F$(\civone)/$F$(\civtwo) = 2 as listed in Table~\ref{tab:UVlines}. 
However, doublet ratios closer to one have also been found \citep[e.g.,][]{1983A&A...122..335F,2012MNRAS.427.1953C,2014MNRAS.445.3200S}. 
The $F$(\nvone)/$F$(\nvtwo) emissivity ratio is also expected to be two \citep{1976JPCRD...5..537M,1984RMxAA...9..107T,1991ApJS...77..119M}.
This is in disagreement with the \nv{} flux ratios measured by FELIS for the seven MUSE sources with potential \nv{} detections, where we generally see that $F$(\nvtwo)$>$$F$(\nvone). 
As \nv{} is also a resonant line arising from stellar winds and in the ISM \citep{2011AJ....141...37L}, profiles deviating from Gaussian could be part of the explanation for this discrepancy.
The best-fit linear correlations and correlation coefficients for these two emission doublets are presented as correlations number five and six in Table~\ref{tab:paramcorr}.

As listed in Table~\ref{tab:UVlines} \cite{1991ApJS...77..119M} estimates the \oiii{} doublet ratio to be $F$(\oiiione)/$F$(\oiiitwo)~$\approx0.7$ which is in agreement with recent findings where \oiiitwo{} tends to be strongest \citep[e.g.,][]{2017ApJ...836L..14M, 2016ApJ...821L..27V}.
This is also in agreement with what we find here.
Performing a linear fit to the measured \oiii{} doublet component fluxes from our sample and the literature and estimating the correlation coefficient we find the relation presented as correlation seven in Table~\ref{tab:paramcorr}.
Assuming an electron temperature of $10^4$~K the theoretically expected \siiii{} doublet ratio is $F$(\siiiione)/$F$(\siiiitwo)~$ \lesssim 1.7$, similar to the expected \ciii{} doublet ratio $F$(\ciiione)/$F$(\ciiitwo)~$ \lesssim 1.6$ \citep{2006agna.book.....O}.
As we show in Section~\ref{sec:neTe}, these two ratios depend somewhat on the assumed electron temperature and probe the electron density of the emitting gas.
As mentioned, we find that both of these flux ratios are $>1$ for the vast majority of sources.
The best-fit ODR linear empirical relations and correlation coefficients for these ratios are presented as correlations number eight and nine in Table~\ref{tab:paramcorr} and
are in agreement with the theoretical expectations.

\section{Rest-frame EW estimates}\label{sec:EWcoor}

\begin{figure}
\begin{center}
\includegraphics[width=0.49\textwidth]{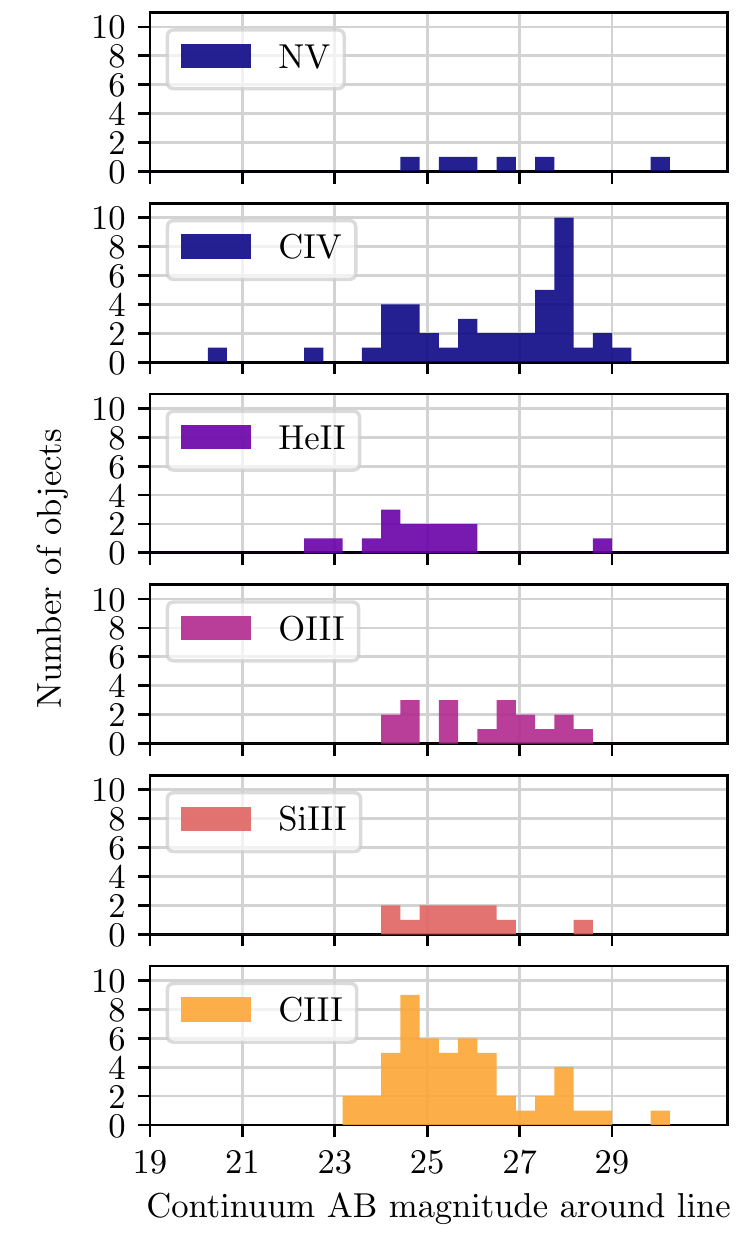}
\caption{AB magnitude distributions of the measured HST magnitudes used to infer the underlying continuum flux density when estimating the EW$_0$ of the UV emission lines targeted in this study.}
\label{fig:EWcontmag}
\end{center}
\end{figure}

In addition to the line flux measurements of the UV emission line detections from FELIS, we compute the rest-frame equivalent width (EW$_0$) using the expression
\begin{equation}\label{eq:EW0}
\textrm{EW}_0 = \frac{F_\textrm{UV emission, rest-frame}}{f_\textrm{continuum, observed} \times (1+z)}
\end{equation}
The UV emission line flux ($F_\textrm{UV emission, rest-frame}$) is provided directly from the FELIS template matching of the emission line templates ($\alpha$ in Equation~\ref{eq:felisflux}).
The observed continuum flux density ($f_\textrm{continuum, observed}$) is estimated from available photometric catalogs in the following way:
For each source we identify the HST broad-band nearest to the location of the detected UV emission line free of any emission line contamination. 
We then assume a power law continuum of the form $f(\lambda) \propto \lambda^\beta$ with a fixed spectral slope of $\beta=-1.97$, which is the median spectral slope for the LAEs studied here and by \cite{Kerutt:2021tr}.
If available, we use the photometry by \citet[][only estimated for LAEs]{Kerutt:2021tr} described in Section~\ref{sec:UVandLya} to predict the continuum flux density at the location of the relevant UV emission.
Otherwise we use the \cite{2015AJ....150...31R} or the \cite{2014ApJS..214...24S} HST broad-band fluxes to estimate the continuum level. 
Assuming $\beta=-1.97$ for the non-LAE ($z<2.9$) is also a good approximation of the average spectral slope based on the available HST photometry of these objects in the MUSE wavelength range.
Figure~\ref{fig:EWcontmag} shows the HST broad-band magnitudes used to infer the continuum flux density for each of the detected UV emission lines for sources with estimated EW$_0$. 

\begin{figure*}
\begin{center}
\includegraphics[width=0.55\textwidth]{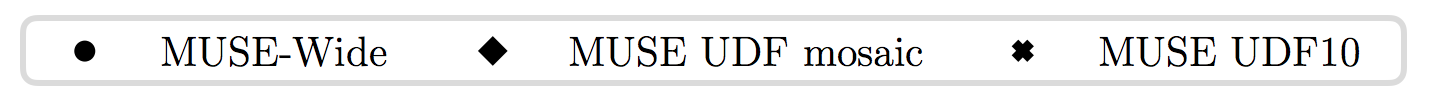}\\
\includegraphics[width=0.42\textwidth]{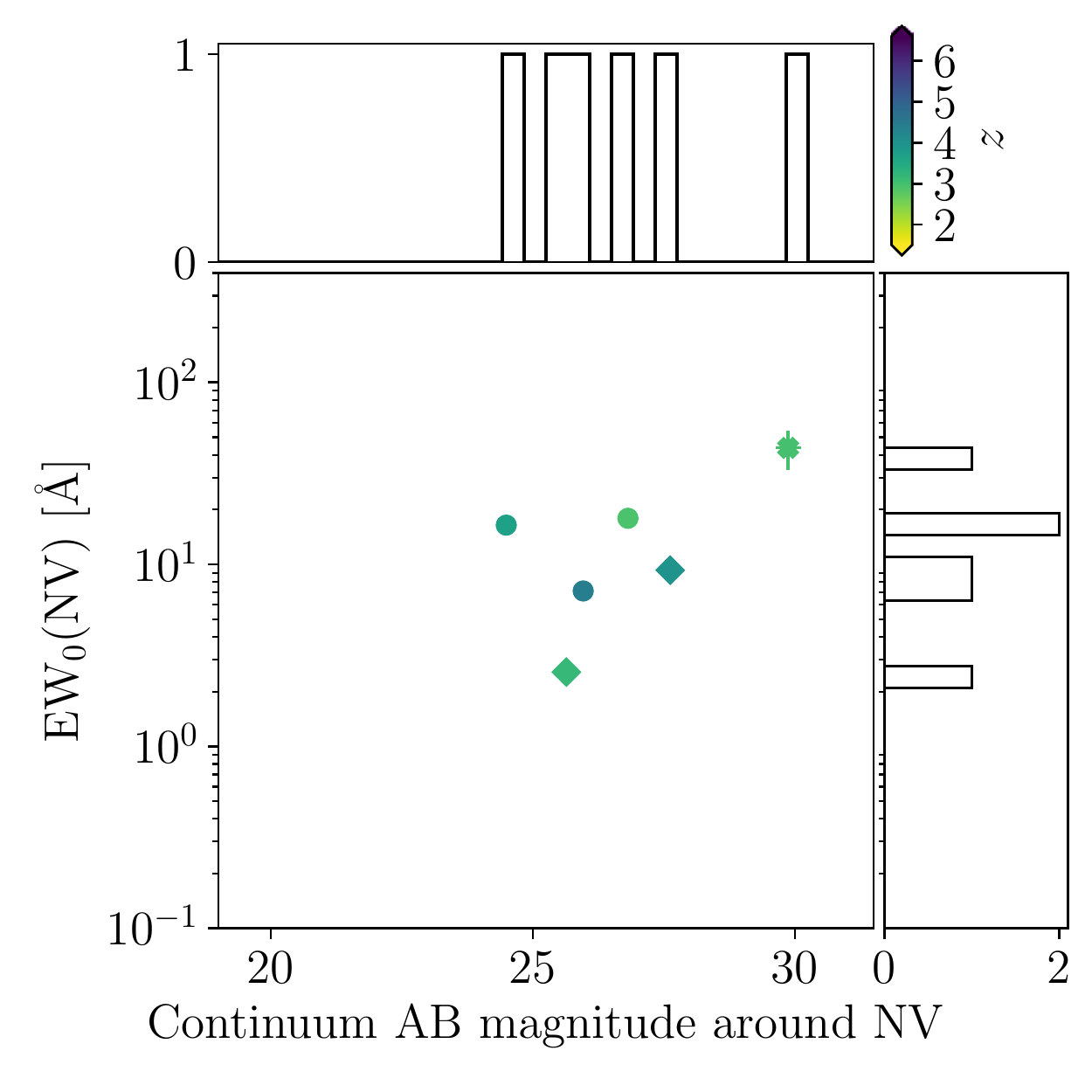}
\includegraphics[width=0.42\textwidth]{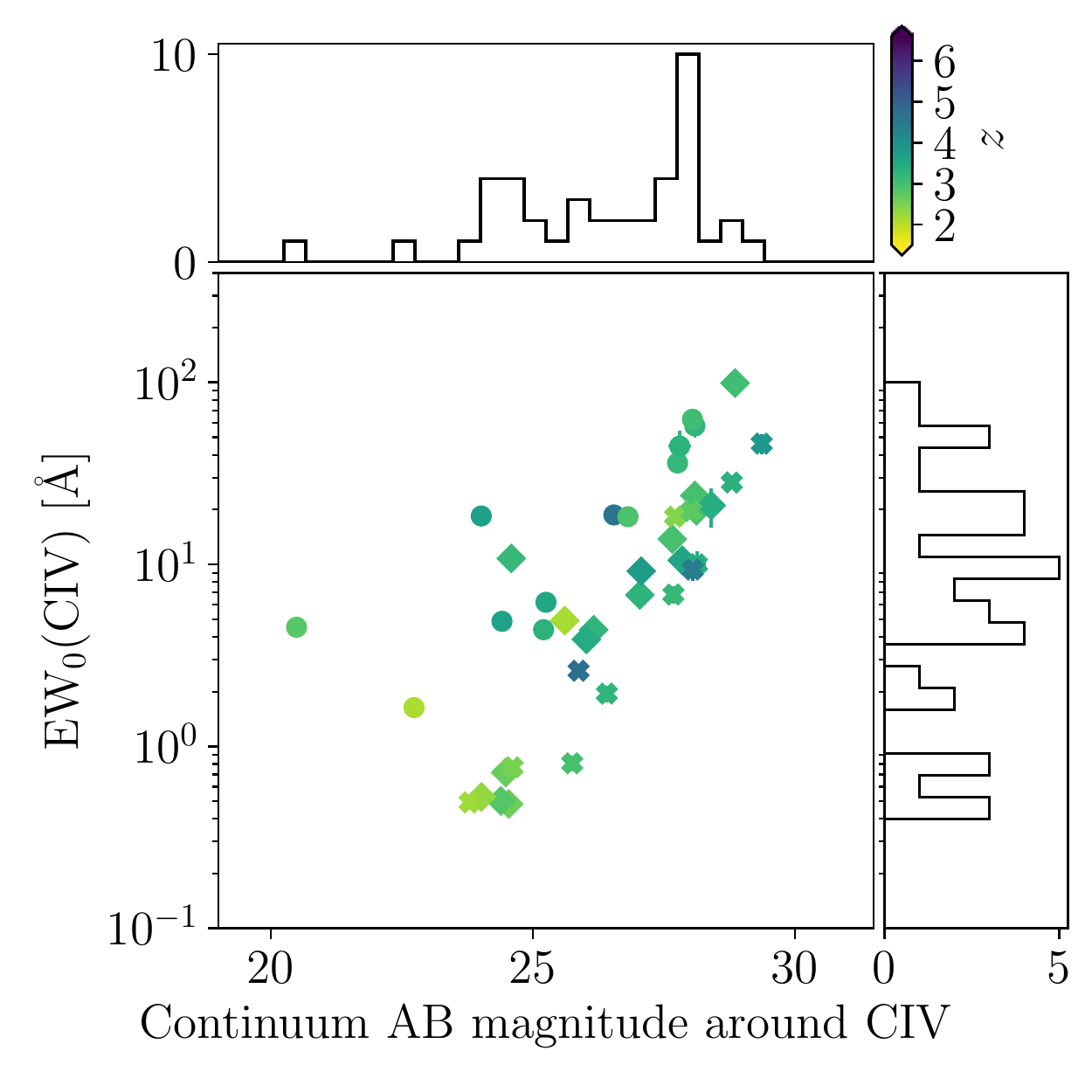}\\
\includegraphics[width=0.42\textwidth]{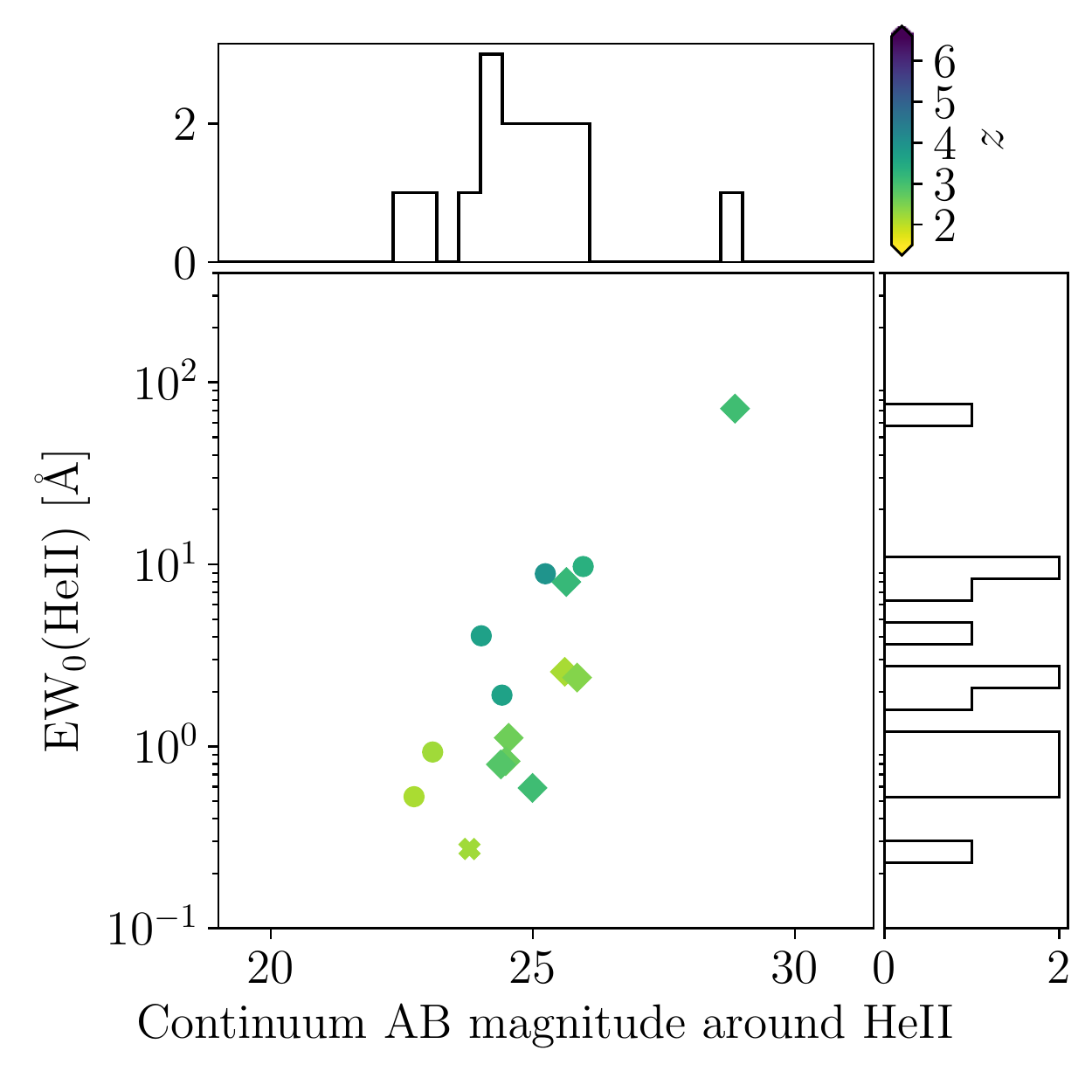}
\includegraphics[width=0.42\textwidth]{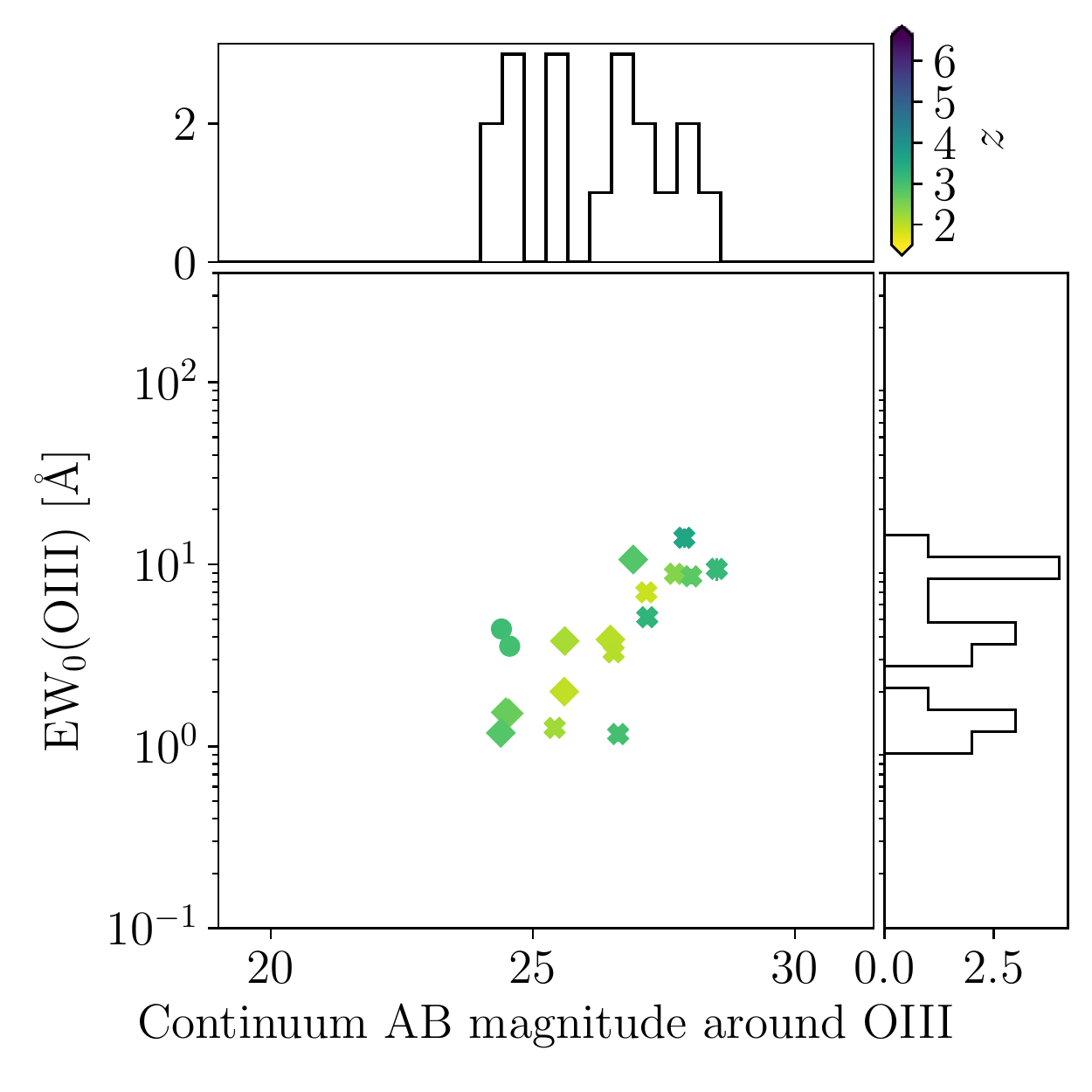}\\
\includegraphics[width=0.42\textwidth]{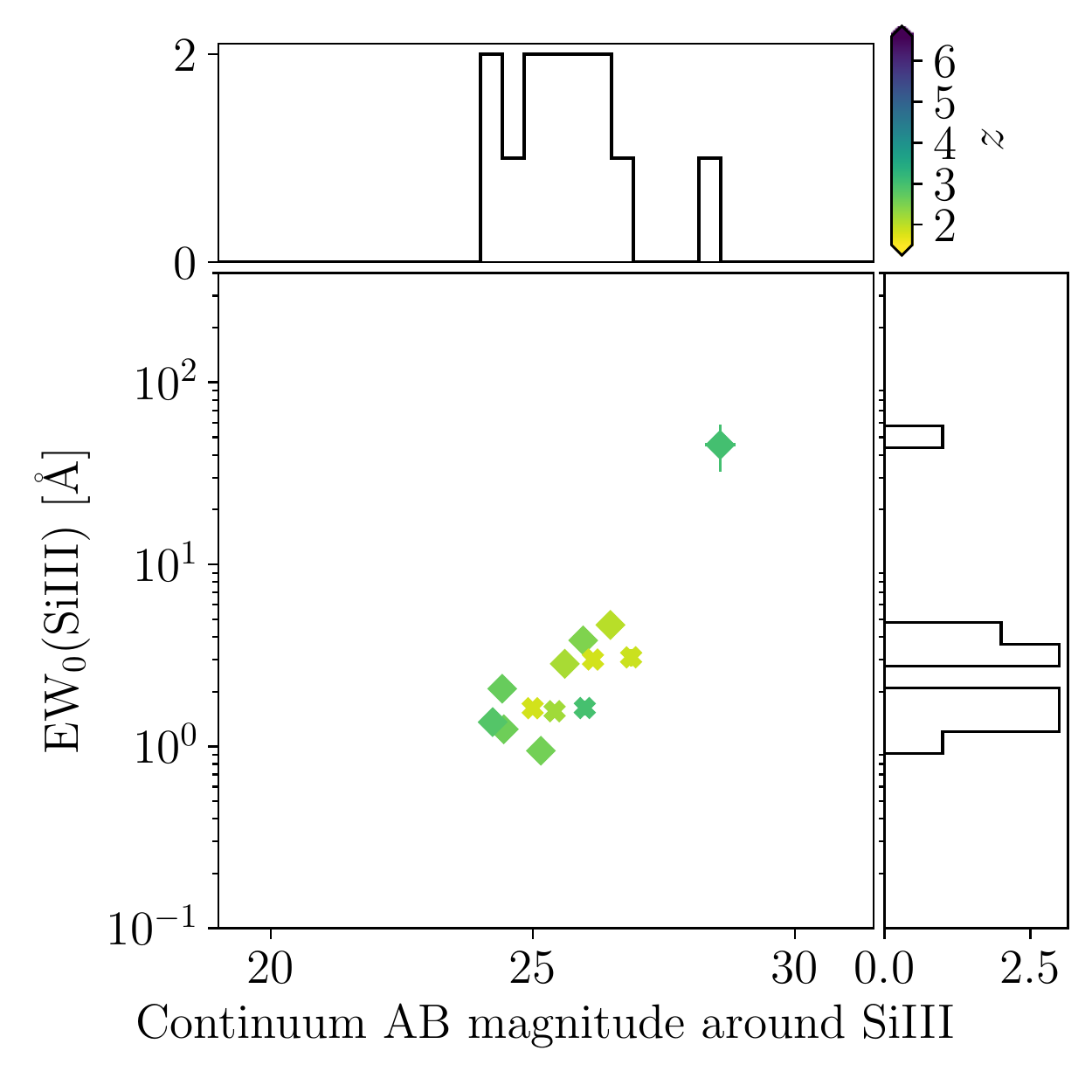}
\includegraphics[width=0.42\textwidth]{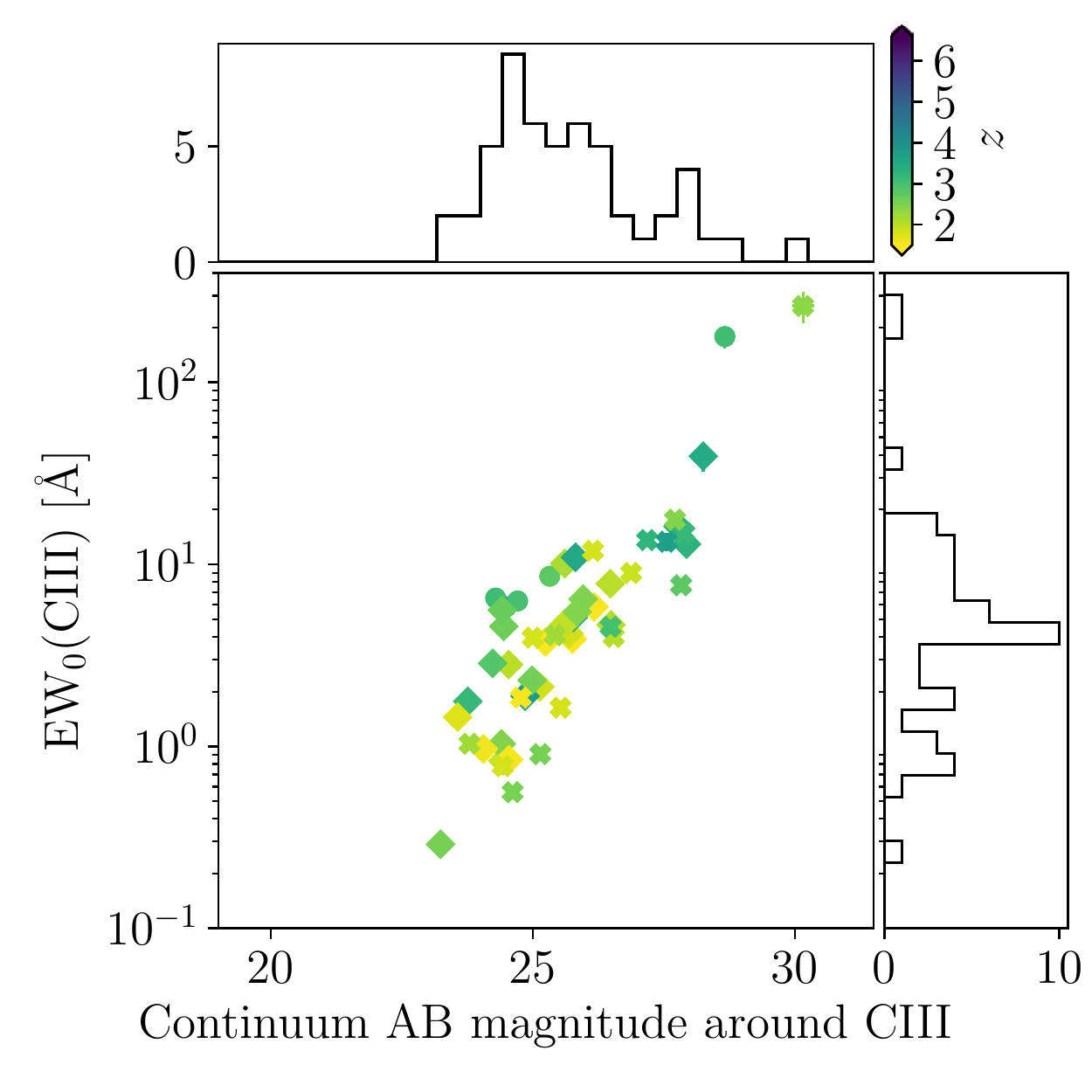}
\caption{Rest-frame equivalent width (EW$_0$) of the UV emission lines detected by FELIS as a function of the continuum magnitude used to estimate the flux density at the emission line location.
As in Figure~\ref{fig:ELfluxes}, filled circles, diamonds and x's correspond to objects from the MUSE-Wide, the UDF mosaic and the UDF10 samples, respectively.
Each point is color coded according to each object's redshift.}
\label{fig:EWvsContmag}
\end{center}
\end{figure*}

In Figure~\ref{fig:EWvsContmag} we show the EW$_0$ values resulting from a combination of these continuum flux density estimates and the emission line fluxes from the FELIS template matches.
We note that the apparent ``correlations'' seen in these panels are driven by the fact that the continuum magnitude shown on the x-axes is included in the definition of the EW$_0$ estimates plotted on the y-axes.
Taking out the dependence on the continuum magnitude of the EW$_0$, we see that the line fluxes as a function of continuum AB magnitude are fairly flat (with a large scatter), with a tendency for fainter lines to correspond to fainter objects.
The full sample of EW$_0$ estimates (and upper or lower limits; not shown in Figure~\ref{fig:EWvsContmag}) are available in the value added catalog described in Appendix~\ref{sec:mastercat}. 

As can be seen in Figure~\ref{fig:EWvsContmag}, a few sources show high EW estimates that are rarely seen in the literature, For example, we see objects predicted to have EW(\ciii) = 30-300\AA{} and EW(\civ) above 40\AA, and a single source with EW(\heii)$ \approx 100$\AA. 
If any of the high-EW measurements discussed below are confirmed, these systems will provide interesting extremes for further studies.
The high-EW(\ciii) systems are 603092083, 301006546, and 721870849. 
The potential detection from object 603092083 coincides with a sky feature and its residuals, which have likely affected the measured line fluxes.
Object 301006546 has a potential detection of both \ciii{} and \civ{}. Both of these detections are however low-significance detections and the large EW estimate questions whether these detections are real. 
Finally, the prominent \ciii{} emission from object 721870849 is present in a spectrum with no apparent continuum presumably arising from a low-luminosity system resulting in the extreme EW at $z=2.3796$.
The photoionization models discussed in Section~\ref{sec:pimodelinference} provide non-AGN solutions that are capable of reproducing the observed flux ratios of this system, though the parameter space is fairly limited.
The object with a potential \heii{} EW around 100\AA{} (object 601071350) also has EW(\civ) above 40\AA (which is however affected by sky residuals). As pointed out by \cite{2019A&A...624A..89N}, who also studied this object (ID=UDF 3621), a nearby AGN and an LAE can affect the measured HeII emission in this system. Our flux measurements agree with those from \cite{2019A&A...624A..89N}, whereas our EW estimate deviates, as they find EW(\heii)~$=6.8$~\AA{} for this object.
The three remaining objects with EW(\civ)~$>40$\AA{} not discussed above are 102014087, 210012237, and 721480767.
The estimated EW(\civ) of $58\pm8$\AA, $45\pm10$\AA, and $46\pm6$\AA{} from the spectra of these objects are generally of lower quality with S/N(FELIS) of 3.6, 3.4, and 4.3, which could indicate that the detected lines could be spurious or their strengths less certain. 

Considering correlations between the EWs of the various UV emission lines, similar to what is presented for the emission line fluxes in Figure~\ref{fig:fluxesVSciii}, reveals that EW$_0$(\siiii), EW$_0$(\heii), and EW$_0$(\oiii) generally follow the strength of EW$_0$(\ciii), though offset from the one-to-one relation (as shown in Figure~\ref{fig:EWsciii}).
The \civ{} resonant line does not correlate with EW$_0$(\ciii) to the same degree as the nonresonant UV lines. 
As there are only seven objects with both \ciii{} and \nv{} detected we cannot draw any firm conclusions but no clear correlation appears to be forming.
The linear ODR fits to the logarithmic distributions of the EW$_0$ estimates from the current study and the literature are presented together with the correlations coefficients in Table~\ref{tab:paramcorr} as correlations number 10-14.
Again, the limits for the nondetections are in agreement with the presented correlations.
Hence, similar to the correlations for the emission line fluxes found above, EW$_0$(\siiii), EW$_0$(\heii), EW$_0$(\oiii), and EW$_0$(\ciii) correlate with each other and can be used as predictive tools for estimating EW$_0$ of emission lines.
Based on photo-ionization models, a correlation between the EW$_0$ of \ciii{} and \oiii{} is expected as shown in Figure~9 of \cite{2016ApJ...833..136J}.

In line with these findings, multiple studies including \cite{2017A&A...608A...4M}, \cite{2020MNRAS.494..719M}, and \cite{2021MNRAS.501.3238T} have reported that high-EW \ciii{} emitters generally have high EWs of [\oiii]$\lambda$5007+H$\beta$ as also anticipated by the \cite{2016ApJ...833..136J} theoretical models.
\cite{2016ApJ...833..136J} stress that both EW$_0$([\oiii]$\lambda$5007+H$\beta$) and EW$_0$(\ciii) reach their largest values for young, high ionization parameter models, but confirm that \ciii{} emission is more sensitive to metallicity due to its temperature dependence.
\cite{2021MNRAS.501.3238T} show that EW$_0$(\ciii) depends strongly on the metallicity of the emitting system and that you generally need low metallicity to obtain large EW$_0$(\ciii).
They find that EW$_0$(\ciii) increases by a factor three when the metallicity changes from 0.3 to 0.1 of solar metallicity, whereas [\oiii]$\lambda$5007+H$\beta$ varies little with metallicity in their models. 
Hence, \cite{2021MNRAS.501.3238T} argue that to obtain $\textrm{EW}_0(\ciii)>10$~\AA{} requires an ionization parameter log$_{10}$(U) above $-2.5$ and a metallicity of 0.2 solar or lower. 
The fact that we see several such systems at redshift three (lower right panel in Figure~\ref{fig:EWvsContmag}) indicates that even at redshift 3, the large sample of MUSE sources includes low-metallicity objects 
that are typical for objects at redshifts approaching the EoR. 
Furthermore, it was pointed out by \cite{2018MNRAS.479.3264C} that systems with large EW$_0$([\oiii]~$\lambda$4959+[\oiii]~$\lambda$5007) and hence large EW$_0$(\ciii) tend to have a high production efficiency of \ion{H}-ionizing photons, $\xi_\textrm{ion}$.

\section{The UV emission of the \lya-emitter sample}\label{sec:UVandLya}

In this study we are focusing on rest-frame UV emission red-wards of the strongest UV emission line, \lya, partially motivated by the challenges of observing this line at high redshift ($z\gtrsim6$) where the significantly neutral CGM and IGM absorbs the \lya{} photons escaping the galaxy along the line of sight \citep[e.g.,][]{2011MNRAS.414.2139D,2011ApJ...728...52L,2019A&A...627A..84L,2017arXiv170403416D}.
Nevertheless, the \lya{} line itself has improved our understanding of star-forming galaxies in the (early) Universe. 
In particular, the asymmetric Ly$\alpha$ line profile has enabled redshift confirmations of large samples of sources at both $2<z<6$ 
\citep[e.g.,][]{
2014ApJ...795..165S,
2015A&A...576A..79L,
2017A&A...606A..12H,
2017A&A...608A...2I,
2019A&A...624A.141U}
and high redshift at $z>6$
\citep[e.g.,][]{
2013Natur.502..524F,
2015ApJ...804L..30O,
2016ApJ...818...38S,
2016ApJ...827L..14T,
2016ApJ...823L..14H,
2018A&A...619A.147P,
2020ApJ...896..156F}.
The resonant scattering of the photons and the resulting (occasional) multipeaked emission has been shown to relate closely to the column density and dynamics of the neutral hydrogen 
in the ISM and the CGM \citep{2015A&A...578A...7V,2018A&A...616A..29G,2020A&A...639A..85G}.
The fraction of galaxies with confirmed \lya{} emission has been used to probe the evolution (or lack thereof) of the fraction of LAEs among Lyman-break galaxies from low redshift to the EoR
\citep[e.g.,][]{
2013ApJ...775L..29T,
2014ApJ...793..113P,
2014ApJ...794....5T,
2017A&A...608A.123D,
2018MNRAS.473...30C,
2020A&A...638A..12K}.
Together with the observed velocity offset of the \lya{} line resulting from resonant scattering 
\citep{2013ApJ...777...67S,
2014ApJ...795...33E,
2015ApJ...812..157H,
2017MNRAS.464..469S,
2018MNRAS.478L..60V}, 
this has probed the amount of neutral gas in the IGM and has constrained the neutral fraction of the Universe during the EoR 
\citep{2010ApJ...723..869O,
2017MNRAS.466.4239G,
2018ApJ...857L..11M,
2018ApJ...856....2M,
2018Natur.553..473B,
2019ApJ...878...12H}.
Furthermore, comparisons between \lya{} and H$\alpha$ or UV emission line strengths have been used to study the production efficiency and escape of ionizing photons from LAEs \citep{2016ApJ...831L...9N,2017MNRAS.472..772M,2018ApJ...859...84H,2019A&A...627A.164L,2020MNRAS.tmp..584M}.
It is therefore of interest to relate and compare the measured rest-frame UV emission lines red-wards of \lya{} studied here, with the characteristics of the \lya{} line itself and the properties of the LAEs in our sample. 

For this comparison we rely on HST broad-band magnitudes, \lya{} line fluxes, \lya{} EW$_0$ estimates, spectral UV slopes $\beta$, continuum magnitudes, and \lya{} FWHM from the catalog that will be presented by \cite{Kerutt:2021tr}.
This LAE study is based on the same data and source identification (see Section~\ref{sec:objsel}) as the current study, but focuses on properties of the emanating \lya{} emission.
The \lya{} emission fluxes correspond to the measured flux within 3D apertures of three \cite{1980ApJS...43..305K} radii as measured by LSDCat when detecting sources in the MUSE data cubes.
We use these \lya{} fluxes as opposed to obtaining them directly from the TDOSE spectra, as the TDOSE extractions are based on the assumption that the morphological extent of the line emission follows the continuum morphology of the modeled HST images \citep[Section~\ref{sec:1Dspec} and][]{2019A&A...628A..91S}.
However, \lya{} emission is known to be extended beyond the continuum \citep{2011ApJ...736..160S,2014MNRAS.442..110M,2016A&A...587A..98W,2018Natur.562..229W,2017A&A...608A...8L,2020A&A...635A..82L} and fluxes based on the TDOSE spectra would therefore be biased. The LSDCat Kron radii fluxes therefore better represents the actual \lya{} flux emitted by the LAEs.
As for the secondary UV emission lines, the EW$_0$(\lya) values were calculated by comparing the fluxes to the continuum flux densities estimated from a continuum represented by a power law $f_\lambda\propto\lambda^\beta$. 
To obtain the spectral slope, \cite{Kerutt:2021tr} first determined the magnitudes from available ancillary broad-band HST photometry, by fitting the rest-frame UV morphology for each of the LAEs using GALFIT \citep{2010AJ....139.2097P,2002AJ....124..266P}.   
This provided morphological parameters for all LAEs including a measure of their effective radius. 
The estimated absolute UV magnitude at 1500~\AA{} is also based on these GALFIT models. 
The spectral slope $\beta$ was then obtained from fitting the continuum power law to the GALFIT-based HST magnitudes.
To avoid large scatter in the EW$_0$ measurements presented and analyzed by \cite{Kerutt:2021tr}, the EWs are based on the median $\beta$ for the full LAE sample of $\beta=-1.97$ similar to what was done for the secondary UV emission lines presented here.
The FWHM of the \lya{} emission was measured for each source by fitting a skewed Gaussian profile \citep[Equation~2 by][]{2014ApJ...788...74S} to the \lya{} line profiles in 1D spectral extractions weighted by the MUSE PSF to maximize S/N.
These fits also provide a \lya{} redshift which is more precise than the lead line redshits provided by the LSDCat source identification. We therefore use these redshifts for the analysis of the \lya{} velocity offsets described in Section~\ref{sec:voffset}.
Finally, \cite{Kerutt:2021tr}  provide estimates of the systemic redshifts based on the FWHM and peak separation between any double-peaked LAEs (identified through visual inspection of the 1D spectra) in the sample based on the empirical relations presented by \cite{2018MNRAS.478L..60V}.
We note that a handful of the $z>2.9$ objects studied here are not included in the \cite{Kerutt:2021tr} catalog, as their selection was based only on non-AGN objects with leading Ly$\alpha$ emission based on the LSDCat selections (Section~\ref{sec:objsel}). 
Hence, for $z>2.9$, the objects with IDs 121033078, 601381485, 720470421, 722551008, 722731033, and 723311101 are not included in the LAE parameter comparisons in the following.
For further details and for the full value-added catalog of \lya-related quantities we refer to \cite{Kerutt:2021tr}.

\begin{figure*}
\begin{center}
\includegraphics[width=0.80\textwidth]{mainlegend.png}\\
\includegraphics[width=0.4\textwidth]{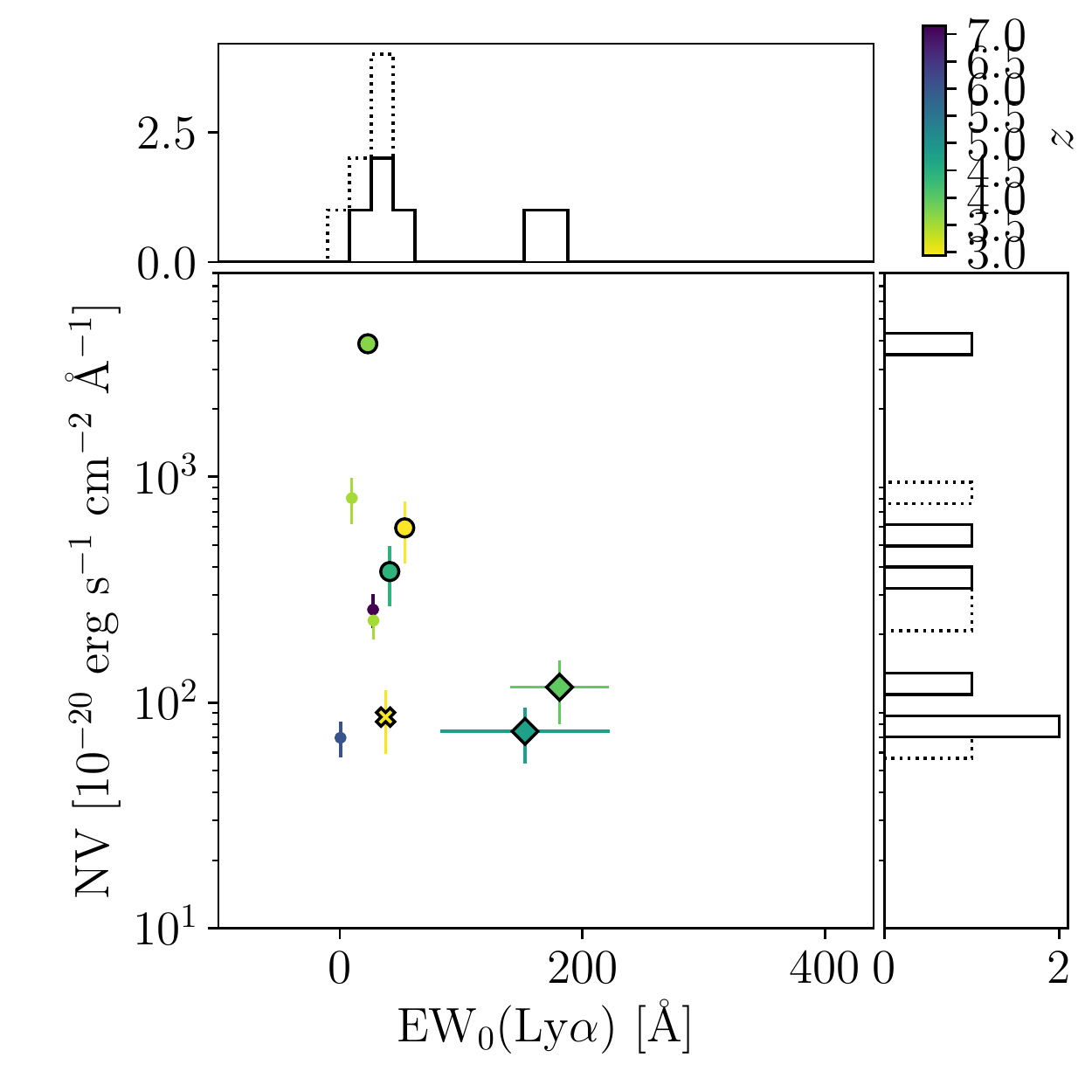}
\includegraphics[width=0.4\textwidth]{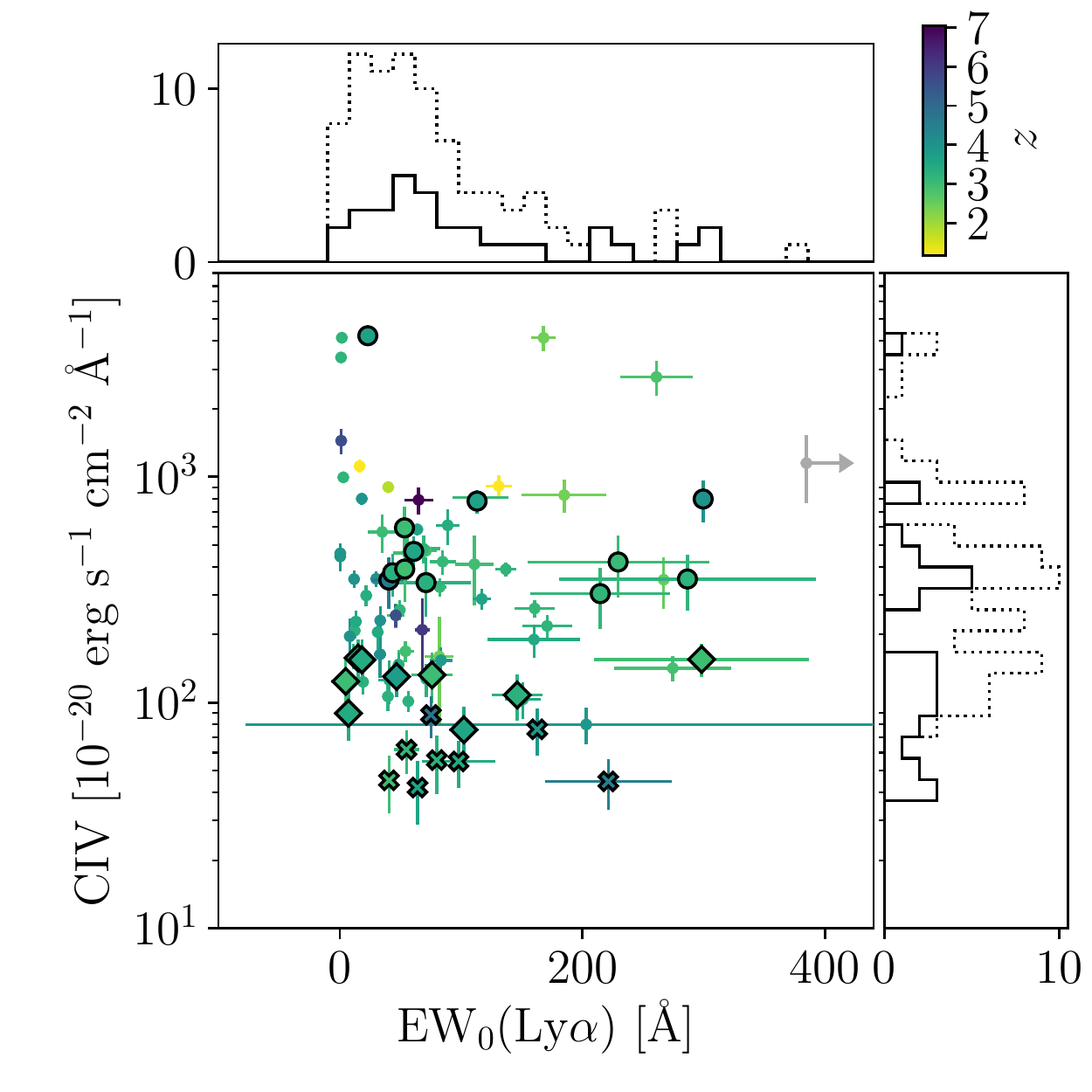}\\
\includegraphics[width=0.4\textwidth]{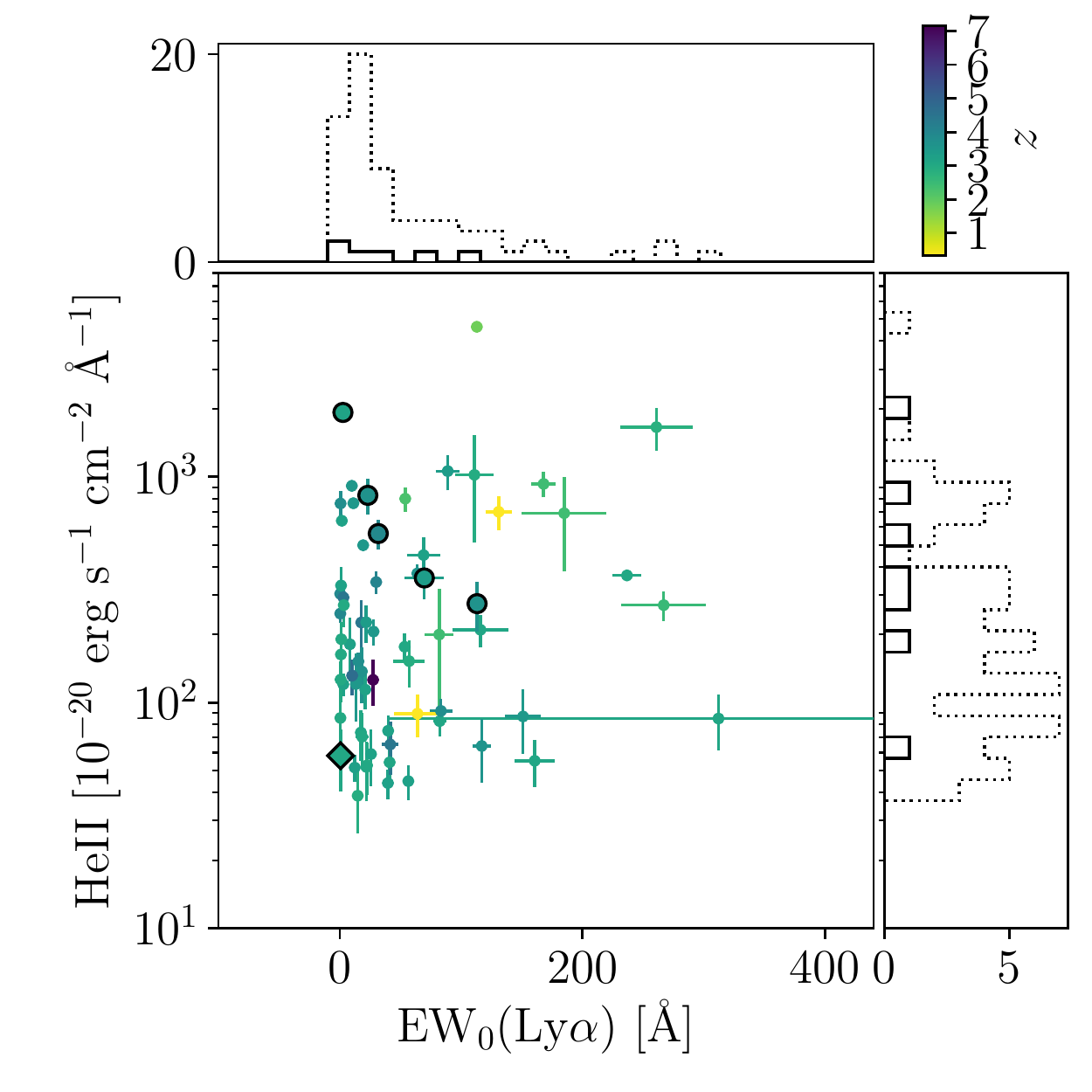}
\includegraphics[width=0.4\textwidth]{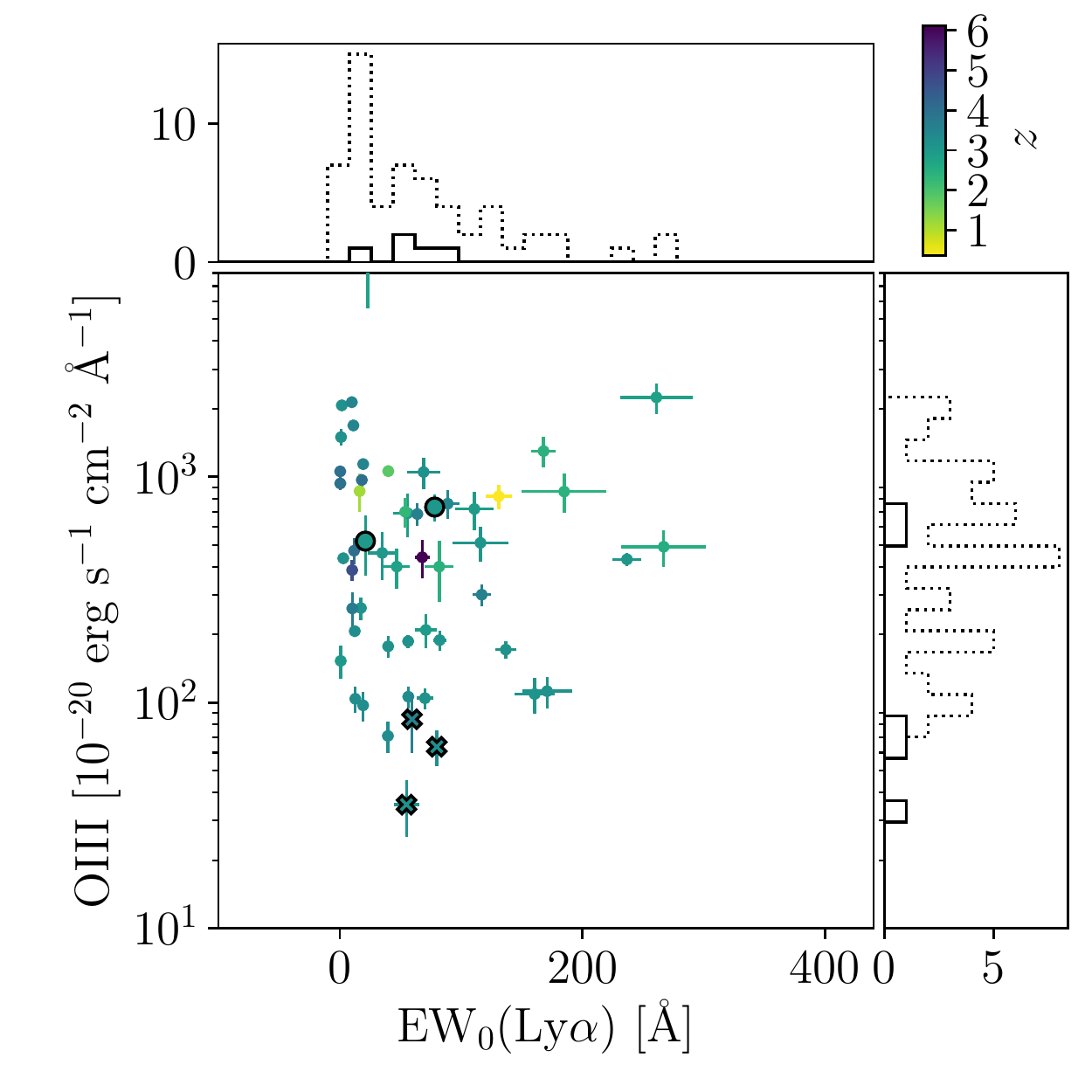}\\
\includegraphics[width=0.4\textwidth]{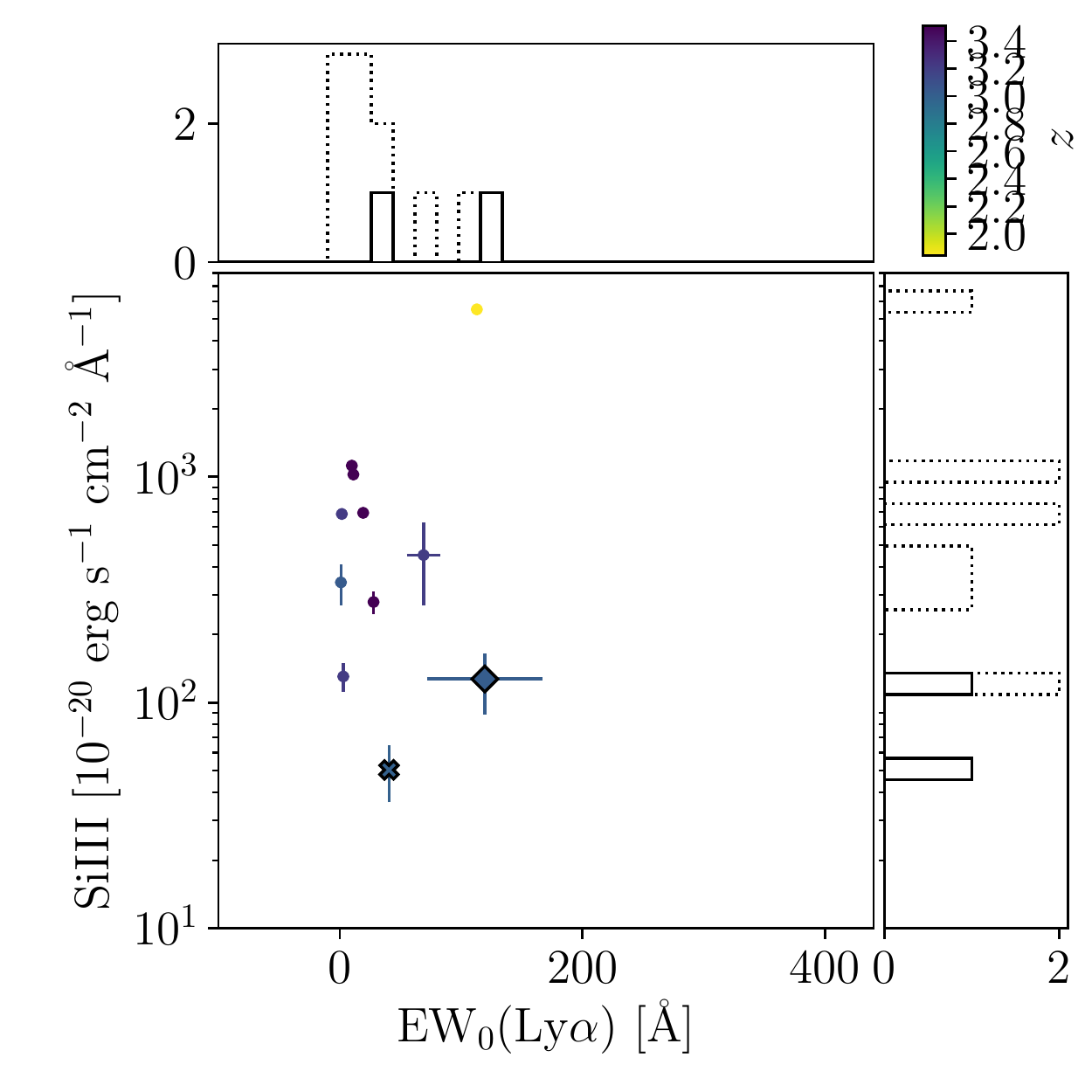}
\includegraphics[width=0.4\textwidth]{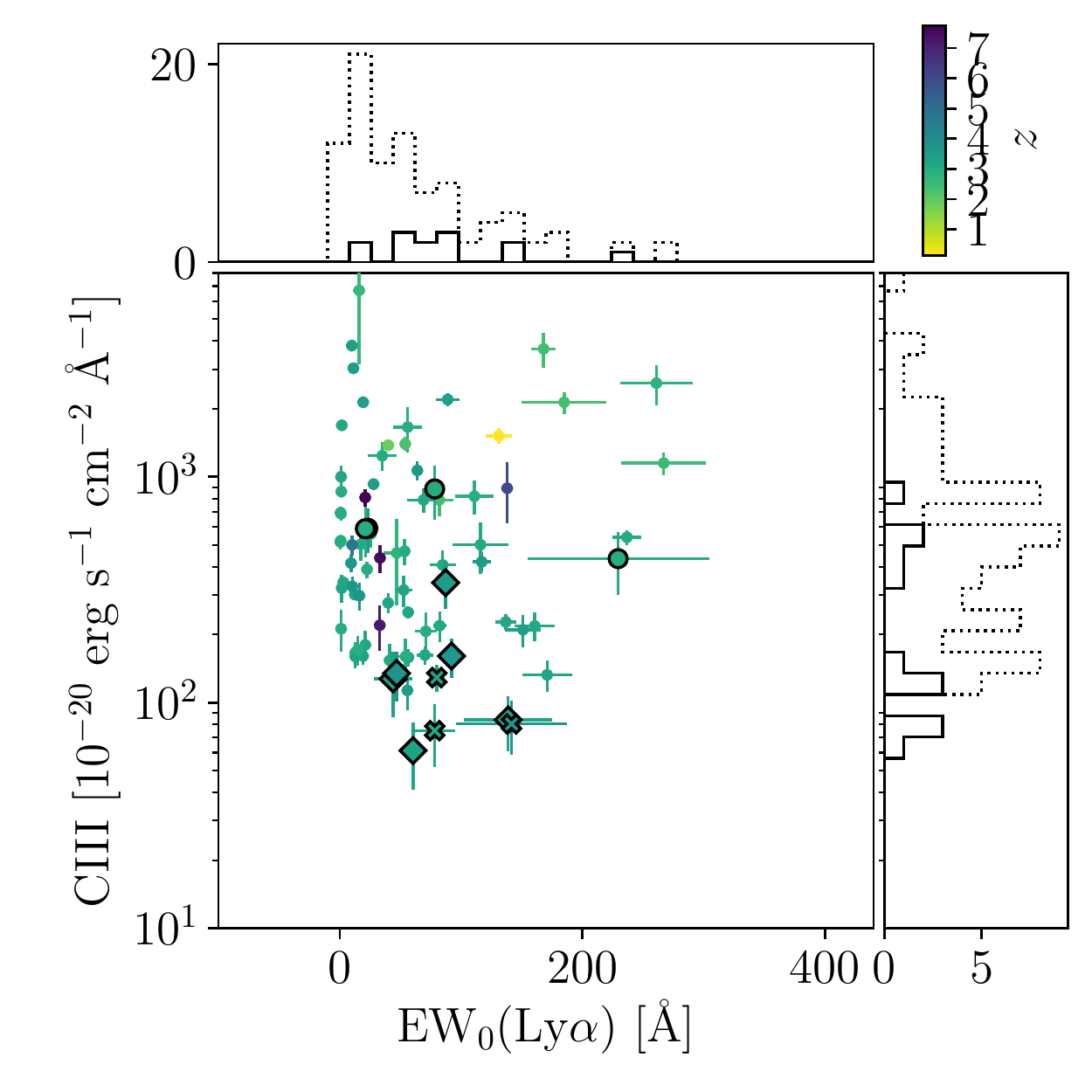}
\caption{Detected emission line fluxes as a function of rest-frame equivalent width of \lya{}.
The measurements presented here from MUSE-Wide, the UDF mosaic and the UDF10 are shown as large symbols.
Measurements from the literature are shown by small dots (see Appendix~\ref{sec:litcol}).
All symbols are color coded according to source redshifts and are shown with 1$\sigma$ error bars. 
The solid histograms show the subset of the objects from this work, whereas the dashed histograms include the estimates from the literature collection.}
\label{fig:UVvsEWlya}
\end{center}
\end{figure*}

In Figure~\ref{fig:UVvsEWlya} we compare EW$_0$(\lya{}) from the \cite{Kerutt:2021tr} catalog to the estimated UV emission line fluxes detected with FELIS in the 1997 LAEs ($z>2.9$).
We see no strong correlations in general. 
It does however appear that the objects with larger EW$_0$(\lya) mostly have \ciii{} or \oiii{} UV emission lines detected at larger fluxes, that is there are no EW$_0(\textrm{Ly}\alpha)>200$~\AA{} objects with \ciii{} or \oiii{} line detections below $\approx2\times10^{-18}$~erg~s$^{-1}$~cm$^{-2}$.
The few \siiii{} detections from the current study and available from the literature only arise in sources with EW$_0(\textrm{Ly}\alpha)\lesssim100$~\AA.

\begin{figure*}
\begin{center}
\includegraphics[width=0.8\textwidth]{mainlegend.png}\\
\includegraphics[width=0.4\textwidth]{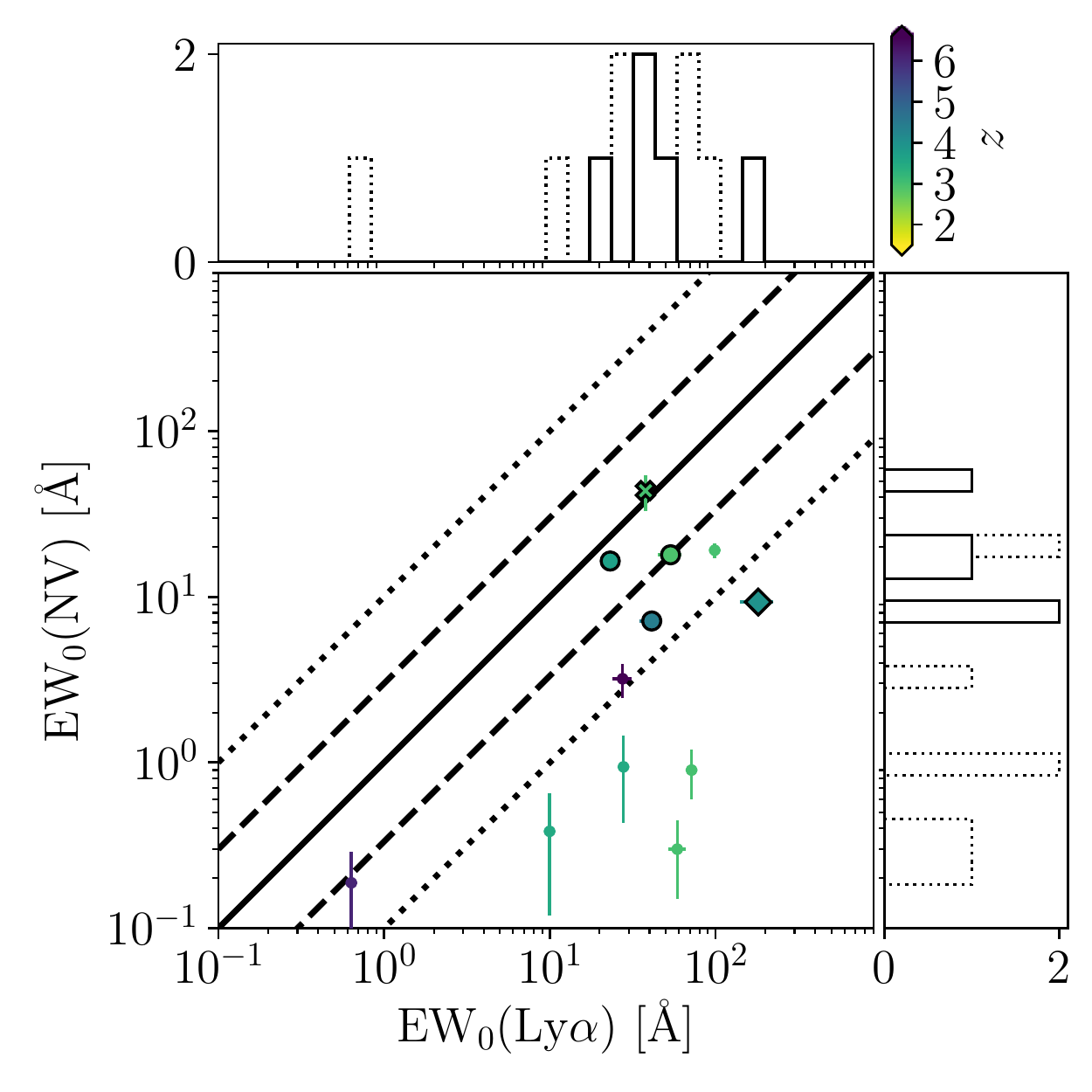}
\includegraphics[width=0.4\textwidth]{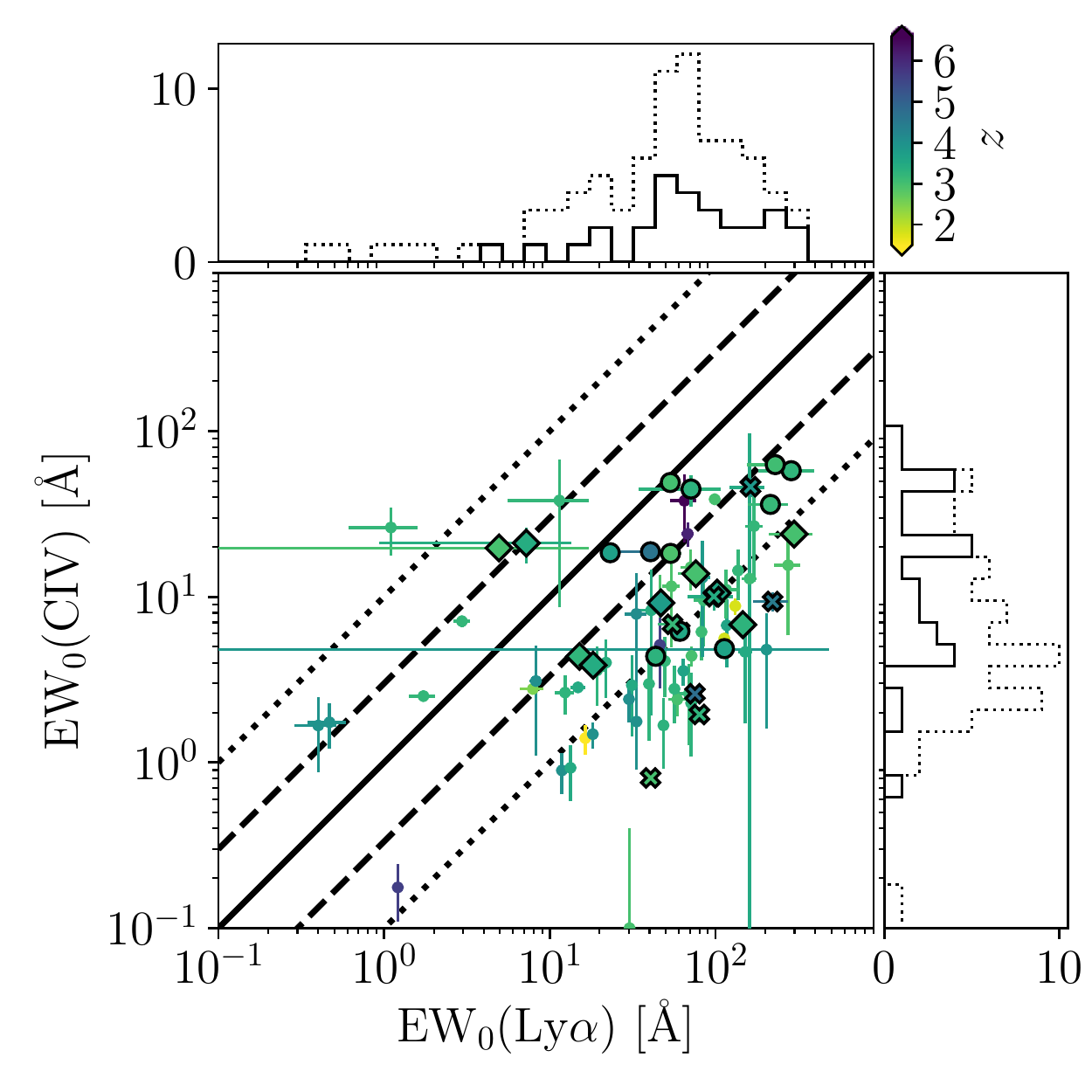}\\
\includegraphics[width=0.4\textwidth]{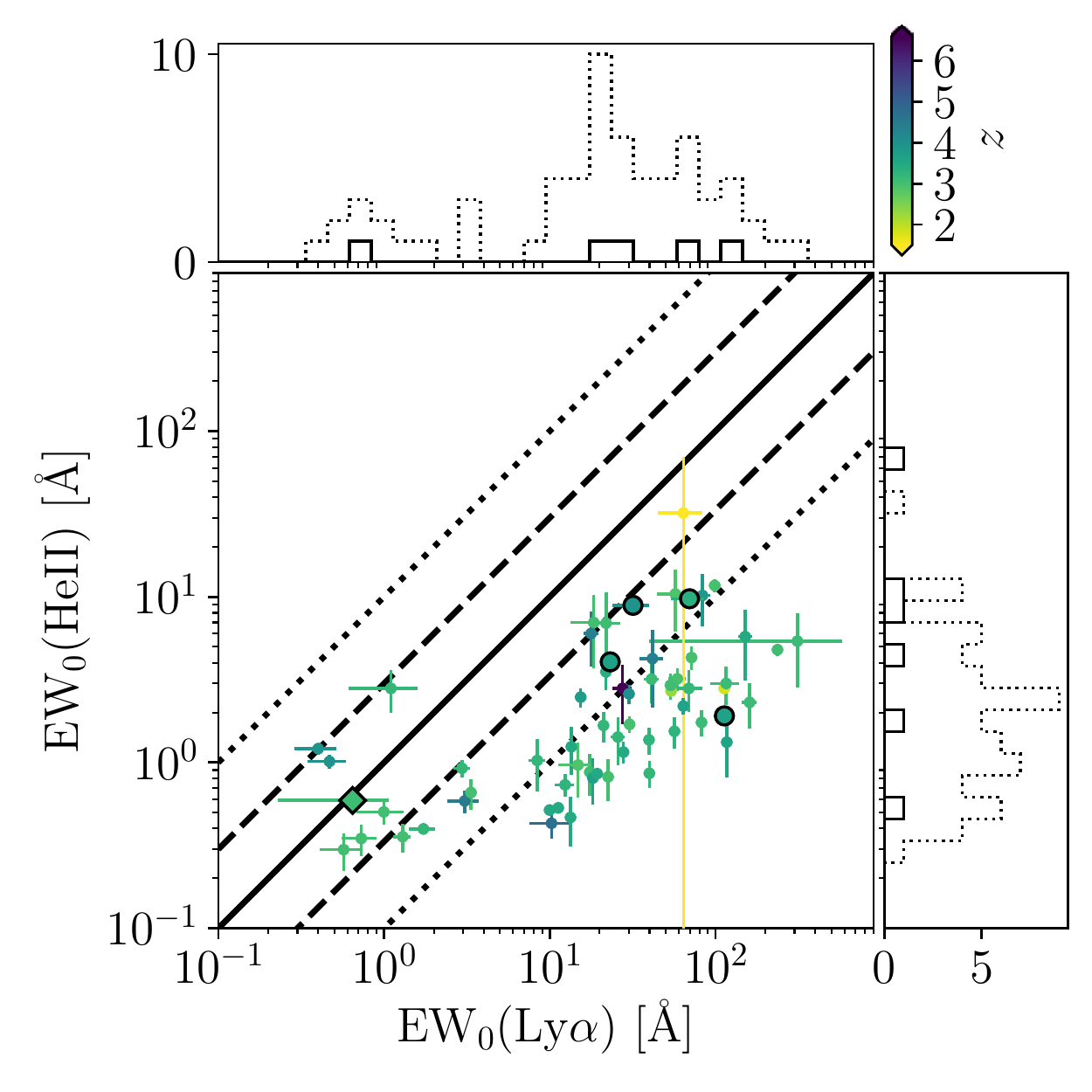}
\includegraphics[width=0.4\textwidth]{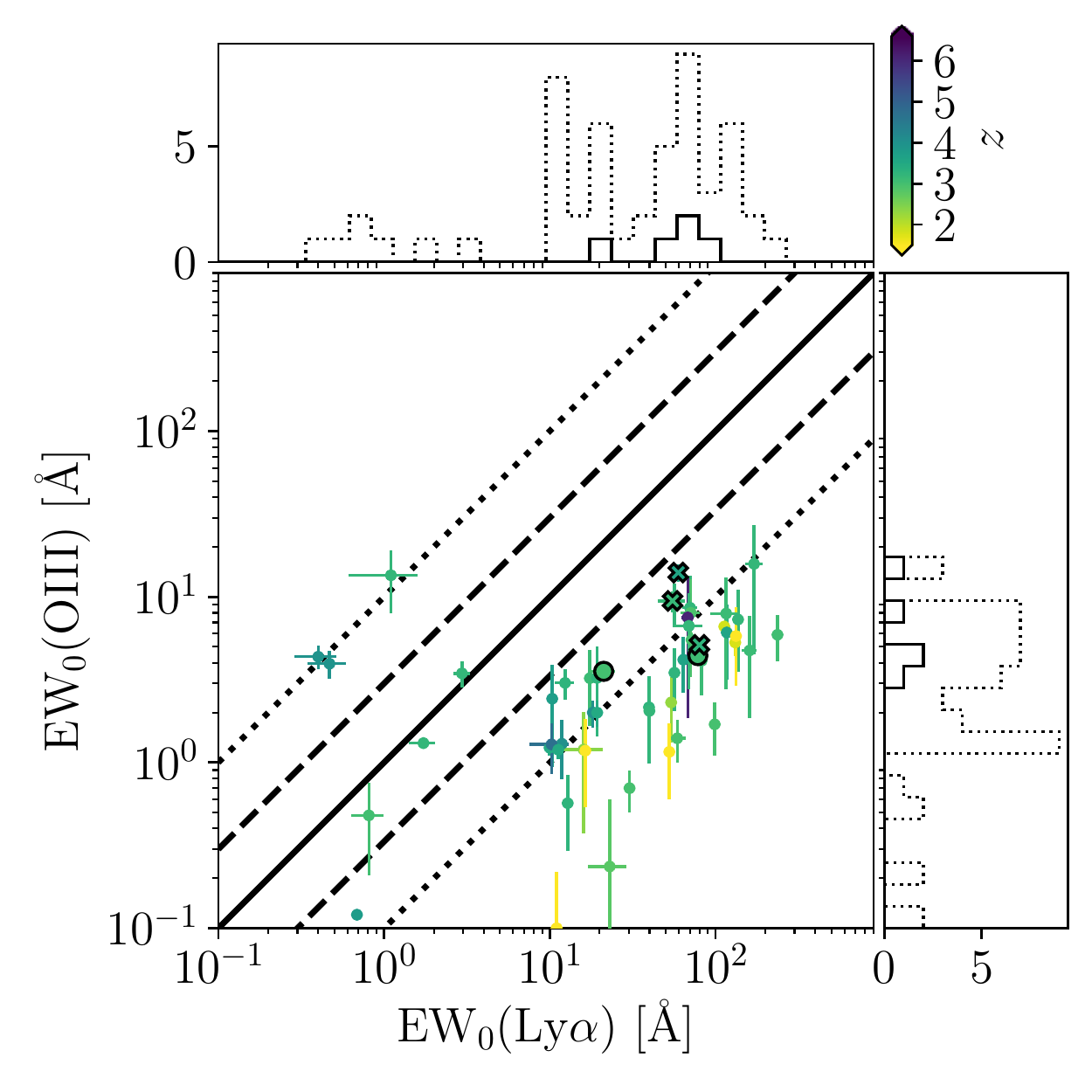}\\
\includegraphics[width=0.4\textwidth]{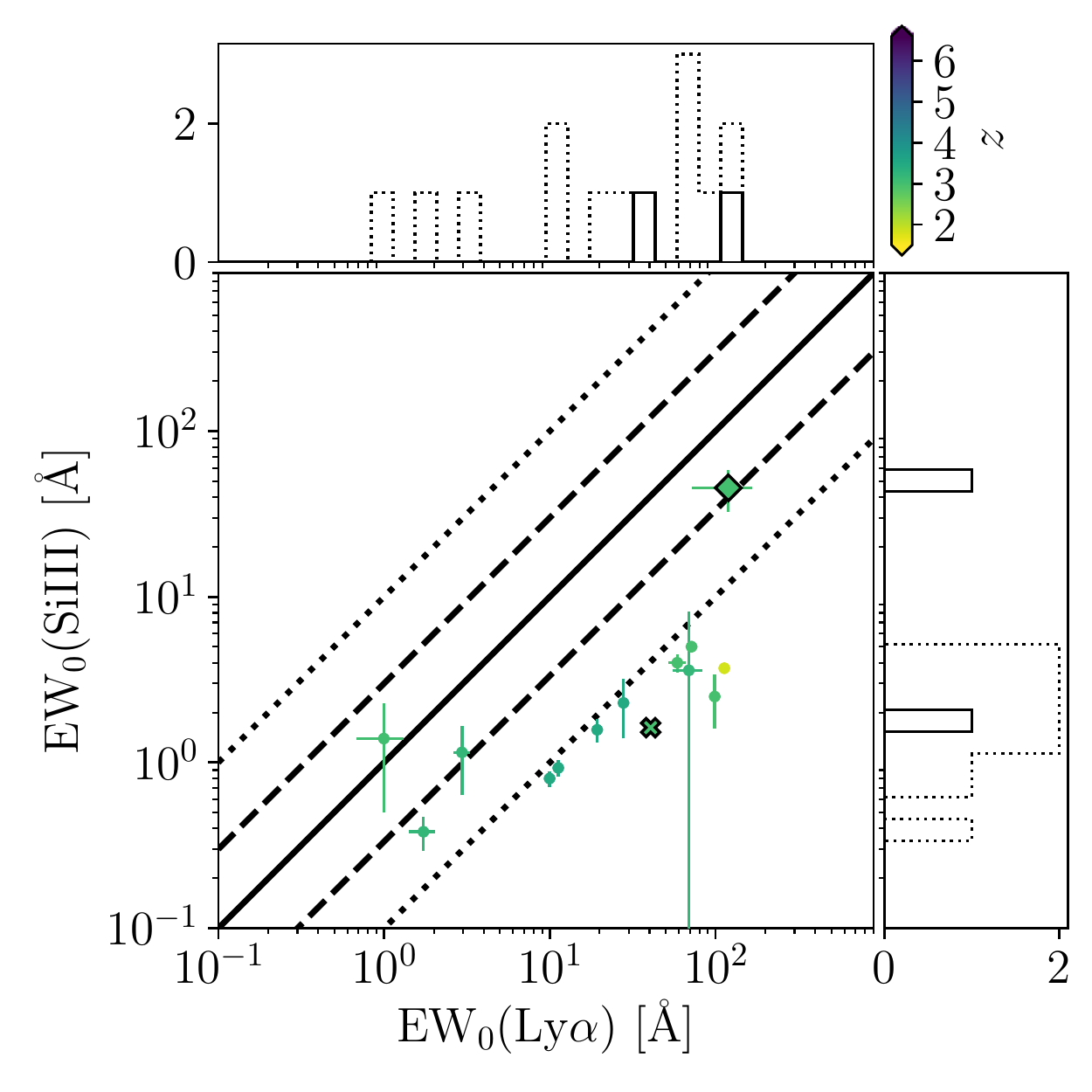}
\includegraphics[width=0.4\textwidth]{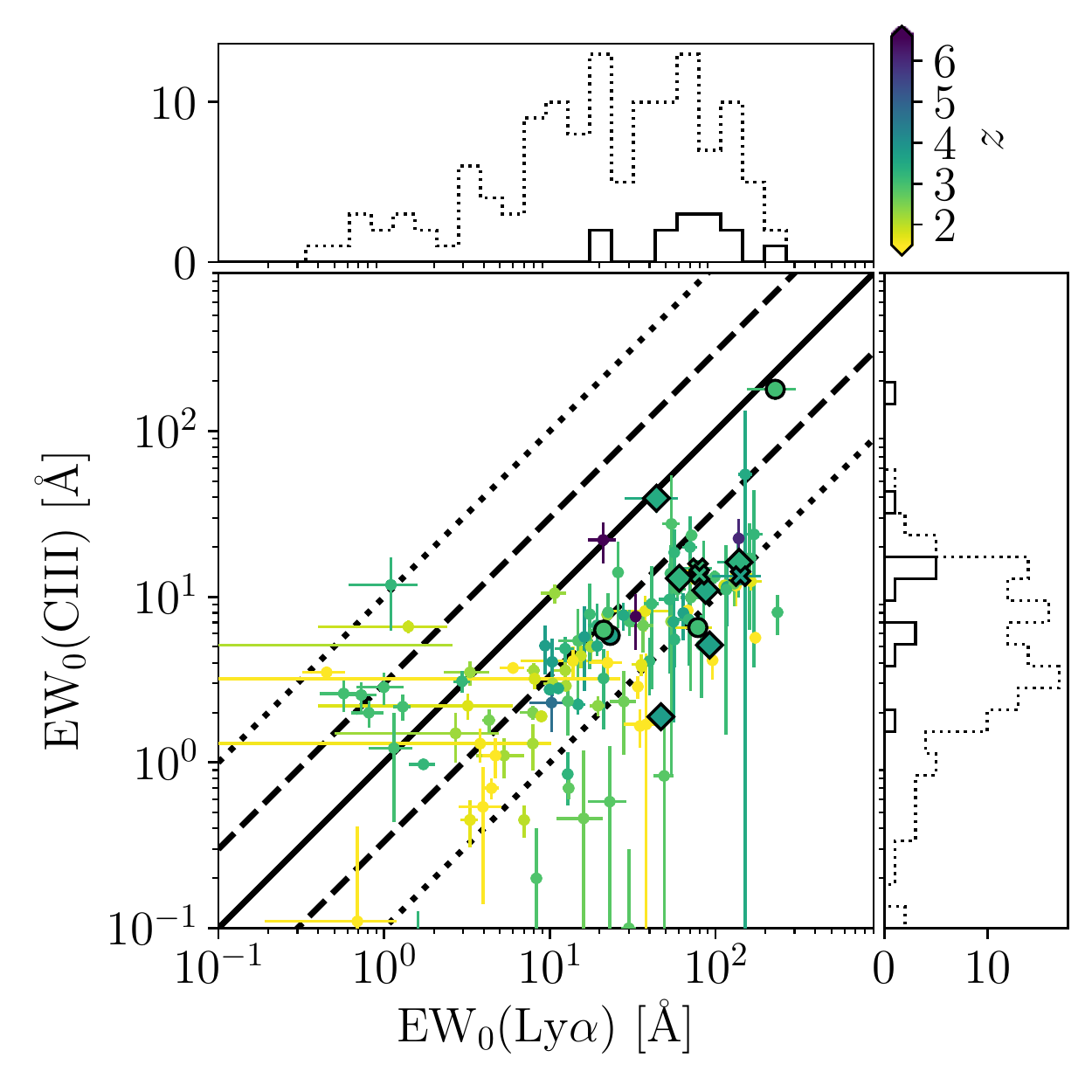}
\caption{UV emission line EW$_0$ estimates of LAEs as a function of EW$_0$(\lya).
The measurements from the current study are shown as large symbols (circles from MUSE-Wide, diamonds from the UDF mosaic, and x's from UDF10).
Measurements from the literature are shown by small dots (see Appendix~\ref{sec:litcol}).
All symbols are color coded according to source redshifts and are shown with 1$\sigma$ error bars. 
The diagonal curves correspond to the relations 10:1, 3:1, 1:1, 1:3, and 1:10 for reference.
Solid histograms show objects presented in this study, whereas dotted histograms include the literature samples.}
\label{fig:EWs}
\end{center}
\end{figure*}

Figure~\ref{fig:EWs} presents the current census of estimated UV emission line EW$_0$ from our study (large symbols) and the literature (small dots, see Appendix~\ref{sec:litcol}) for LAEs as a function of EW$_0$(\lya).
For reference, we show lines of fixed ratios between the UV emission line EW$_0$ and EW$_0$(\lya) estimates. 

Similar to the correlations of the EW$_0$ estimates for the UV emission lines probing systemic redshift reported in Section~\ref{sec:EWcoor}, we see that several of the UV lines have strengths that can be related to the \lya{} EW. 
Except for the resonant \civ{} emission, the \nv{} and \oiii{} emission, which all have correlation coefficients of roughly 0.5 or below, the UV lines show fair correlations with EW$_0$(\lya).
For completeness, in Table~\ref{tab:paramcorr} we list all six ODR fits to the data shown in Figure~\ref{fig:EWs}.
The majority of the EW$_0$(\ciii) measurements appear to align just below the dashed line marking EW$_0$(\ciii) of one-third the strength of the \lya{} lines for EW$_0$(\lya)~$>10$~\AA. 
If we focus on the subset of objects with EW$_0$(\ciii)~$<$~EW$_0$(\lya) the distribution of the ratio between the two EWs has a mean value of $0.22\pm0.18$.
For the full sample the mean ratio is $0.36\pm1.46$.
Similar trends, in some cases with equally large amounts of scatter, have been presented in various other studies
\citep[e.g.,][]{2003ApJ...588...65S,2014MNRAS.445.3200S,2015MNRAS.450.1846S,2015ApJ...814L...6R,2018ApJ...860...75D,2019A&A...625A..51L,2019A&A...631A..19M}.
As noted in Section~\ref{sec:EWcoor}, the strength of \ciii{} depends on the metallicity of the galaxy but is also affected by the hardness of the ionizing radiation \citep{2010ApJ...719.1168E,2017MNRAS.472.2608S}.
This has been shown to lead to expected correlations between EW$_0$(\ciii{}) and  EW$_0(\textrm{Ly}\alpha)$ in photoionization models \citep{2016ApJ...833..136J,2018ApJ...859...84H}.
Hence, despite that the strength and significance of the observed correlations between EW$_0$(\ciii) and EW$_0(\textrm{Ly}\alpha)$ are still being debated in the literature, the correlation presented here (correlation 20 in Table~\ref{tab:paramcorr}) appears to be in agreement with both previous observational results and theoretical expectations.

In addition to assessing potential correlations between the EW$_0$ and emission line fluxes derived in this work,
we also considered correlations with the LAE spectral slope $\beta$, \lya{} line peak separation, the apparent and absolute magnitudes at 1500~\AA{} (m$_\textrm{UV, 1500}$ and M$_\textrm{UV, 1500}$), LAE effective radius, (systemic) redshift, \lya{} flux, \lya{} luminosity and FWHM(\lya), all from the \cite{Kerutt:2021tr} LAE property catalog.
Overall, we did not find prominent trends between these parameters and line fluxes and EWs of the secondary UV emission line with the exception of tentative correlations with the \lya{} luminosity and the spectral slope $\beta$.

We find correlations between the \lya{} luminosity and the flux of the detected UV emission lines \ciii, \heii, \nv, and \oiii{} with $0.62 < r_\textrm{P} < 0.83$.
When accounting for the continuum level at the location of the emission features via the EW$_0$ estimates, 
the tentative correlations persist ($-0.72 < r_\textrm{P} < -0.64$) except for the \oiii{} measurements.
For the objects with detected \ciii{} a larger EW$_0$(\ciii) seems to imply a lower \lya{} luminosity when $ 41 < \log(L_\textrm{\lya}) < 43$.
The opposite appears to be true for objects with \heii{} and \nv{}. However, this is based on only a handful of objects in each case, and is therefore prone to scatter in the measurement and their uncertainties.
For all lines the UV emission line flux increases with the \lya{} luminosity.
Figures~\ref{fig:FvsLLya} and \ref{fig:EWvsLLya} present the \lya{} luminosity, the UV emission line fluxes and EW$_0$ estimates.

The $F$(\oiii) and EW$_0$(\oiii) of the LAEs where \oiii{} is detected correlates with the spectral slope $\beta$ with $r_\textrm{P} = 0.97$ and -0.82, respectively.
This indicates that \oiii-emitters with bluer ($\beta\approx-2.1$) spectral slopes have larger EW$_0$(\oiii) but smaller $F$(\oiii) than objects with redder slopes ($\beta\approx-1.6$).
A possible explanation for this trend is that bluer galaxies are likely younger and therefore have harder ionizing spectra leading to stronger UV lines.
This agrees with the trends seen with UV emission lines in the composite MUSE UDF mosaic spectra presented by \cite{2020A&A...641A.118F}.
For the other UV lines the correlation coefficients $r_\textrm{P}$ and $r_\textrm{P}$ are all below 0.46 and therefore provide no clear indication of potential trends.
For this comparison we used the individually measured $\beta$ values from \cite{Kerutt:2021tr}.
The potential correlations between \oiii{} and the spectral slope $\beta$ are shown in Figure~\ref{fig:FandEWvsbeta}.
%

\section{Emission line velocity offsets}\label{sec:voffset}

As noted, one of the key diagnostics obtainable from the \lya{} line is the velocity offset of the resonant \lya{} emission, $\Delta v_\textrm{\lya}$, from the systemic redshift.
In more general terms the offset of the \lya{} emission, or any other line, can be measured with respect to any reference (emission line) wavelength whether it is at systemic or not.
For instance, the relative velocity offset between the resonant \civ{} emission and the \lya{} line could potentially reveal similarities or differences between the neutral hydrogen that scatters \lya{} photons and the ionized gas that scatters the energetic \civ{} photons. 
Or the offset of \civ{} with respect to systemic can be used as a tracer of the ionized gas' velocity structure similar to what has been done for \lya{} and neutral hydrogen.
Following \cite{2014ApJ...795...33E} and \cite{2014ApJ...788...74S}, we therefore define the general emission line velocity offsets as 
\begin{equation}
\Delta v_\textrm{line}  = c \left(\frac{z_\textrm{line} - z_\textrm{reference}}{1+z_\textrm{reference}}\right)
\end{equation}
where $c$ is the speed of light in km~s$^{-1}$, ``line'' refers to the emission line for which the velocity offset is measured, and ``reference'' refers to the reference feature used to measure the offset.
Hence, the reference redshift, $z_\textrm{reference}$, is either the systemic redshift, or the redshift of the observed location of another line. 

\begin{figure*}
\begin{center}
\includegraphics[width=0.60\textwidth]{mainlegend_nolit.png}\\
\includegraphics[width=0.45\textwidth]{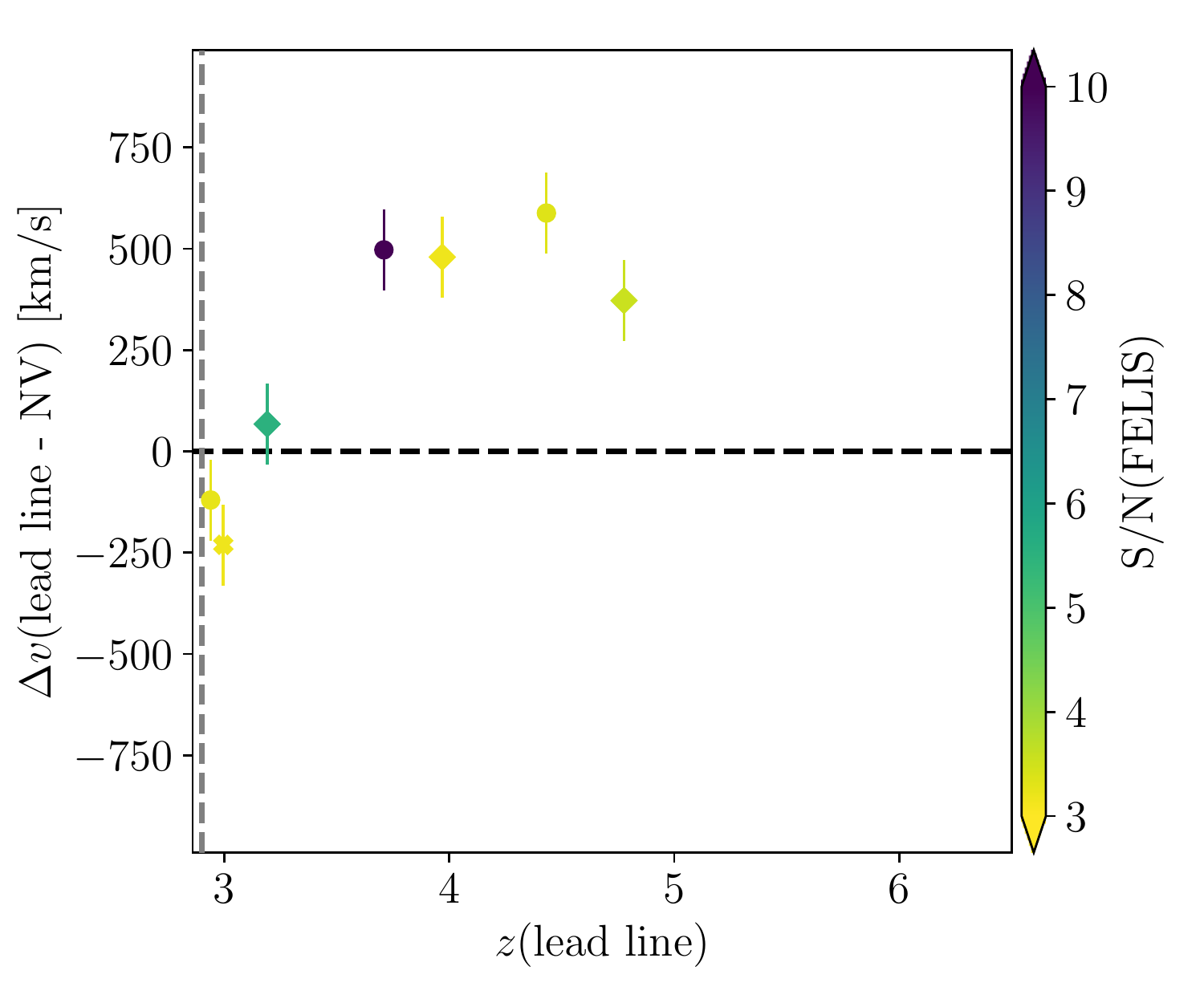}
\includegraphics[width=0.45\textwidth]{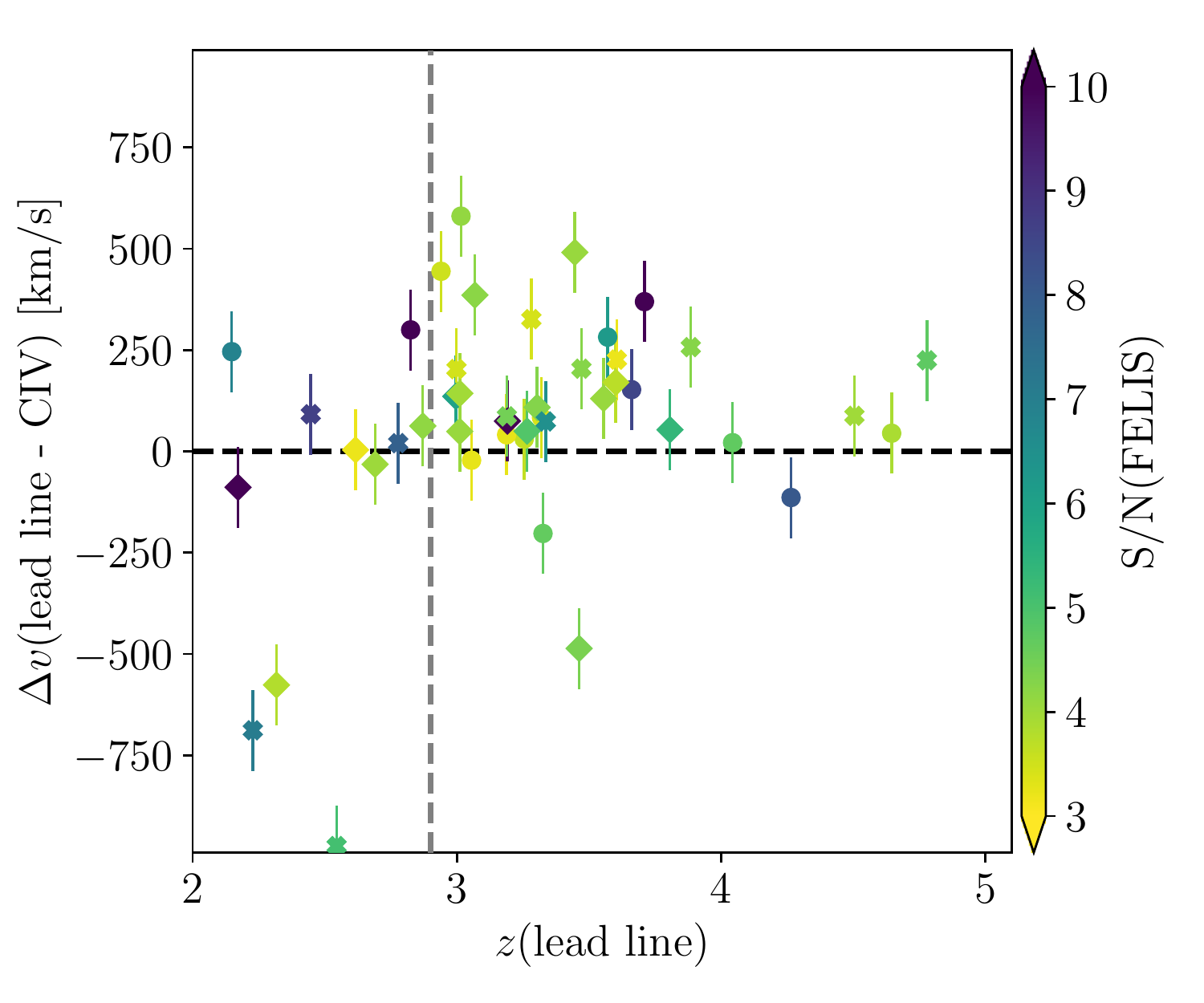}\\
\includegraphics[width=0.45\textwidth]{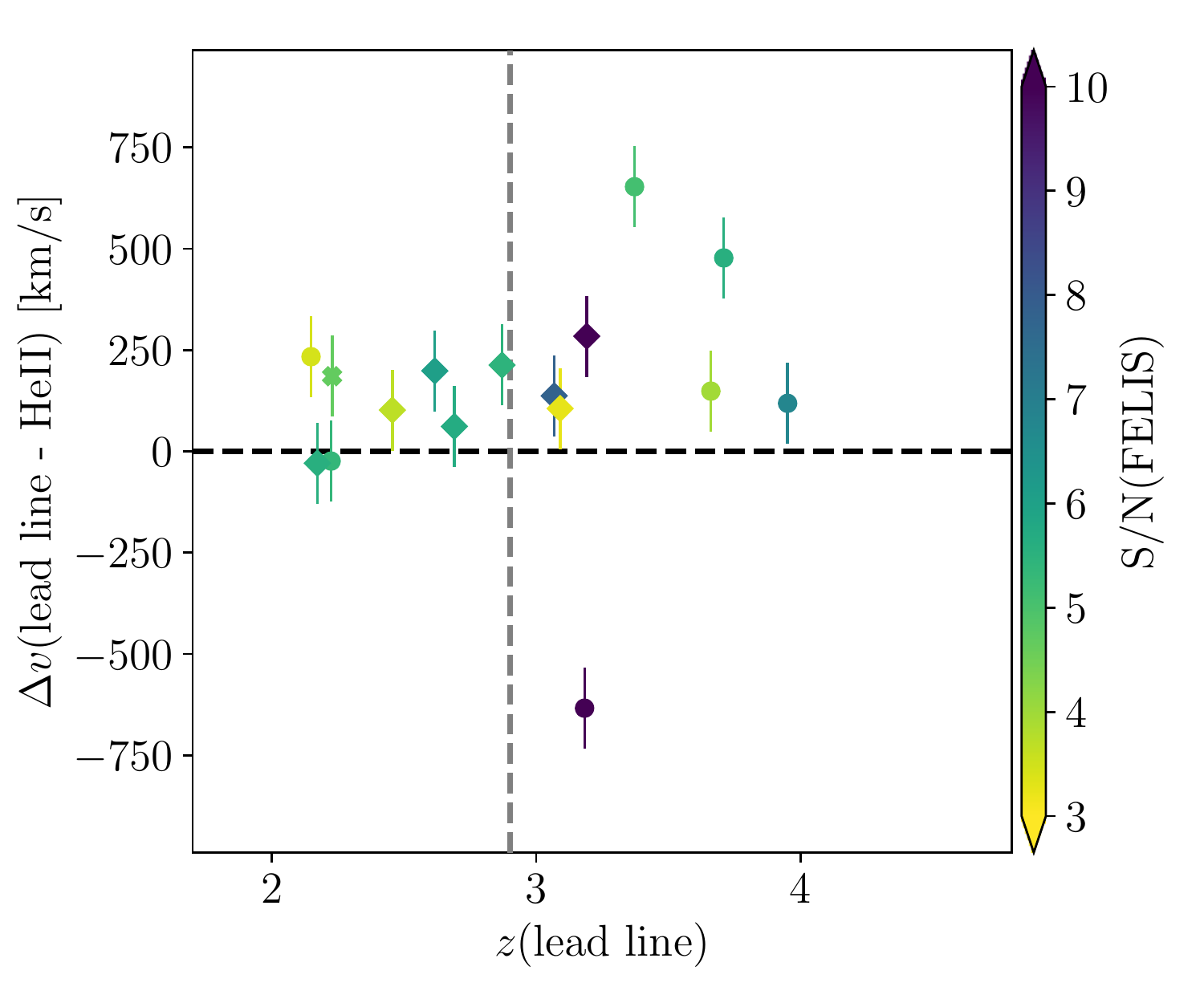}
\includegraphics[width=0.45\textwidth]{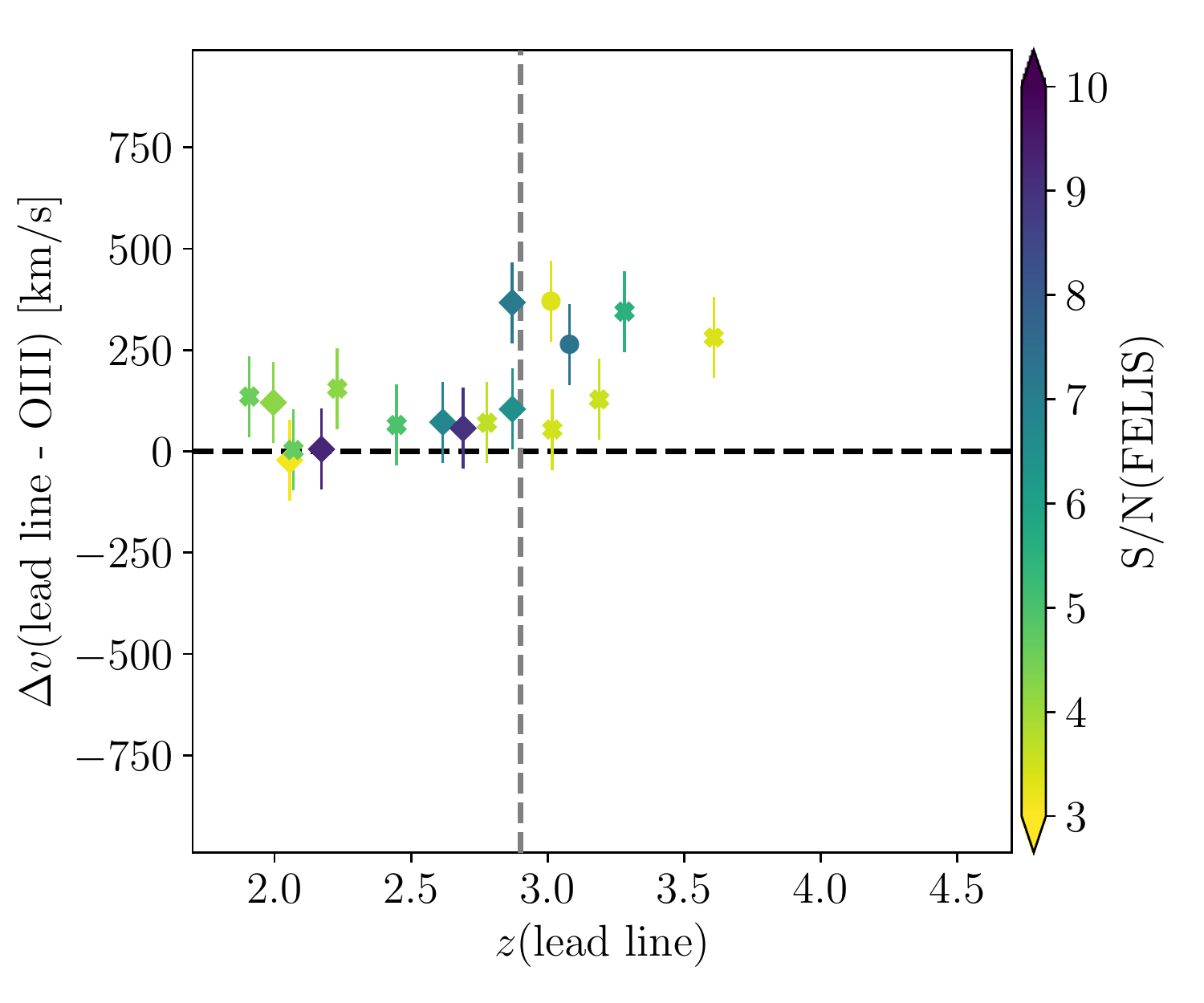}\\
\includegraphics[width=0.45\textwidth]{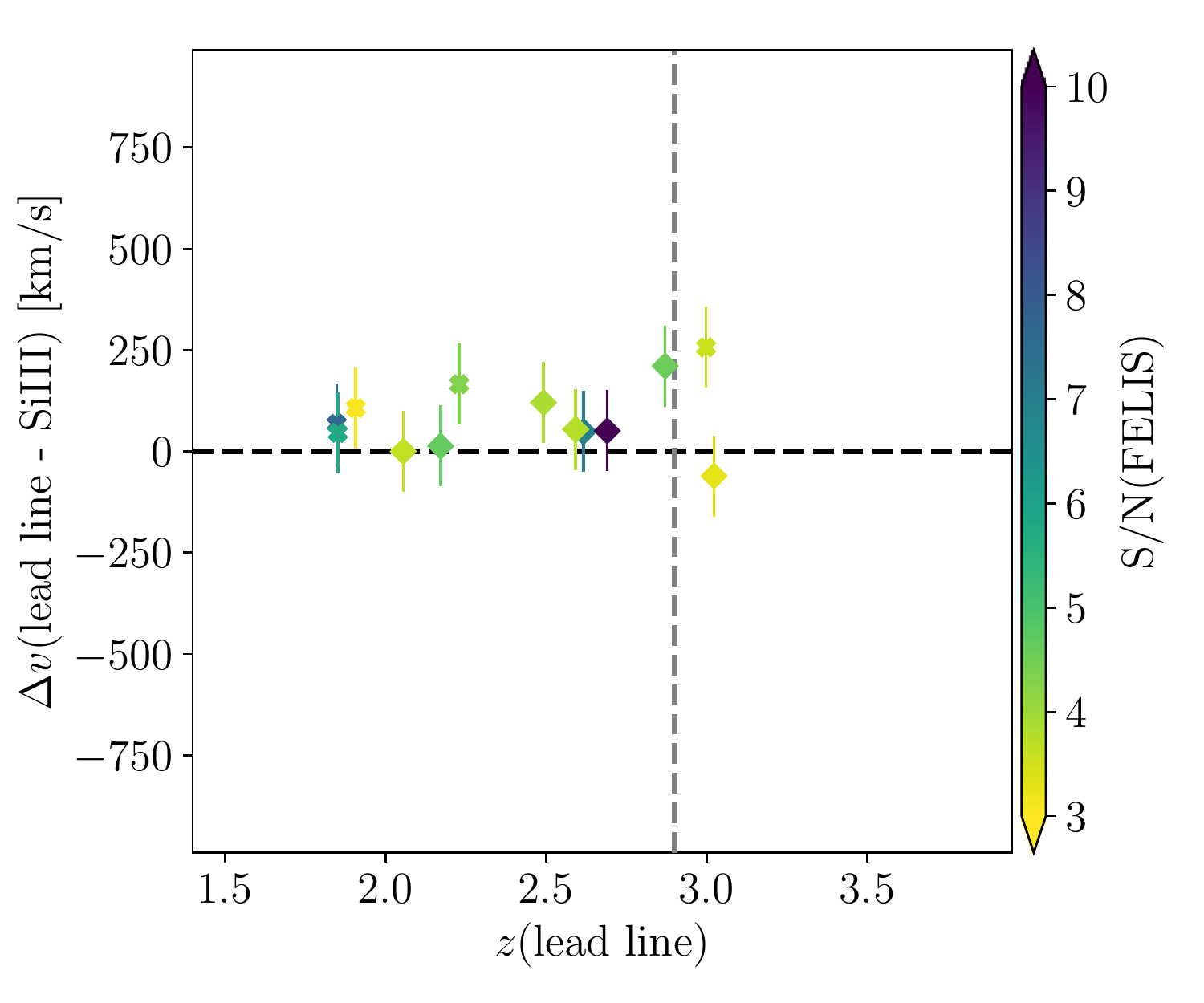}
\includegraphics[width=0.45\textwidth]{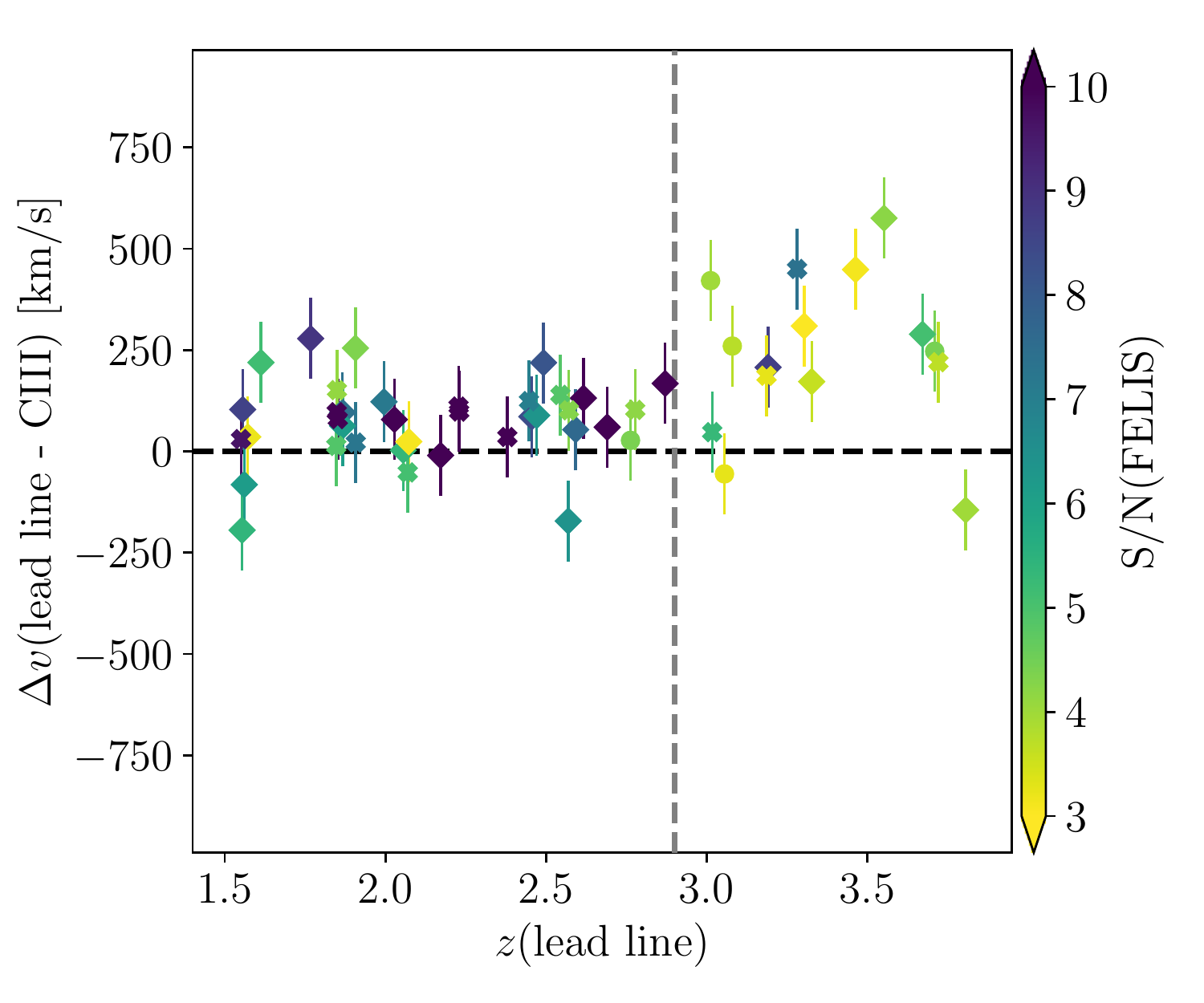}
\caption{Velocity offsets for the individual FELIS template matches of the detected UV emission lines with respect to the object redshift based on the ``lead line'' of the MUSE LSDCat redshift determinations. 
For LAEs the \lya{} redshift based on the line profile fits by \cite{Kerutt:2021tr} 
From top left to bottom right the velocity offset for \civ, \heii, \oiii, \siiii, and \ciii{}
with respect to the lead line are shown. 
For objects with $z\gtrsim2.9$ (vertical dashed line) the lead line of essentially all objects is Ly$\alpha$.
The color coding shows the significance of the FELIS detection in terms of the signal to noise ratio, S/N(FELIS).
All $\Delta v$ estimates were assigned an uncertainty of $\pm$100~km~s$^{-1}$ according to the precision uncertainty coming from the MUSE spectral resolution according to the discussion in Section~\ref{sec:FELIStest_mock}.
}
\label{fig:voffsetsall}
\end{center}
\end{figure*}

In Figure~\ref{fig:voffsetsall} we present the UV emission line velocity offsets from the FELIS detections between the LSDCat lead emission line and the respective UV emission lines. 
For $z\gtrsim2.9$, which is marked by the vertical dashed lines, this effectively means that the lead line is Ly$\alpha$ and hence this part of each panel shows $\Delta v_\textrm{\lya}$.
For the LAEs we replace the LSDCat redshift with the \lya{} redshift from the skewed Gaussian \lya{} profile fits when determining the velocity offsets.  
When estimating the velocity offsets, we did do not impose any cut on S/N(FELIS). As illustrated by the color coding in Figure~\ref{fig:voffsetsall} the S/N(FELIS) values for the detections used to estimate velocity offsets mostly fall around S/N~$\approx5$ (green). 
When visually inspecting the potential line detections from FELIS, we made an explicit cut to only consider emission features with velocity offsets below 1000~km~s$^{-1}$ with respect to the primary line of the object.
As AGN and quasars are known to emit rest-frame UV lines at velocity offsets up to several thousands km~s$^{-1}$ \citep[e.g.,][]{2020ApJ...898..105O}, our limitation to only consider candidate detections with $\Delta\textrm{v}<1000$~km~s$^{-1}$ potentially limits our ability to detect and recover emission from faint AGN in our sample.
However, this limitation does not prevent us from recovering several known AGN in the targeted fields with velocity offsets below 1000~km~s$^{-1}$  (see, for example, Figure~\ref{fig:ObjSpec04}).  
In cases where one component of a detected emission line doublet is not detected the solution with the highest S/N(FELIS) from the simultaneous doublet template match (see Section~\ref{sec:UVEmissionLineSearch}) is used to obtain the systemic redshift.
Of the ancillary spectroscopic redshift collected in support of the photometric redshifts presented by \cite{2014ApJS..214...24S}, none are of the sources presented with systemic redshifts estimates from MUSE presented here.   
%

Figure~\ref{fig:voffsetsall} shows that for the tracers of the systemic redshift with good statistics (\heii, \oiii, and \ciii) the \lya{} velocity offsets are 250-500~km~s$^{-1}$ for the majority of the objects. 
For non-LAEs on the other hand, most objects are consistent with no velocity offsets within 3$\sigma$.
These $\Delta v_\textrm{\lya}$ estimates agree well with what has been found in the literature at both low and high redshift \citep[e.g.,][]{2011MNRAS.414.3265R,2014ApJ...795...33E,2017MNRAS.464..469S,2017ApJ...836L..14M,2020A&A...643A...6C}.

There are two exceptions to this trend among these LAEs.
For object 604992563 the velocity offset between \lya{} and \ciii{} is estimated to be $-145$~km~s$^{-1}$. 
This either questions the potential detection of the \ciii{} doublet (the FELIS S/N for this detection is estimated to be 4.0) or could indicate infalling gas offsetting the \lya{} blue-wards of systemic \citep[e.g.,][]{2006A&A...460..397V,2006ApJ...649...14D,2021MNRAS.501.5757M}. 
However, within 2$\sigma$ the estimated offset is still consistent with zero or a small positive offset of the \lya{} line.
We show the spectrum of object 604992563 in Figure~\ref{fig:ObjSpec98}. 
The second LAE showing significant apparent blueshift of the \lya{} emission is object 219009247 which has a potential \heii{} emission detected $-634$~km~s$^{-1}$ offset from the \lya{} redshift. 
Upon further inspection, it turns out that the emission detected as \heii{} for object 219009247 is actually [\oiii]~$\lambda$5007 flux from a chance superposition (projected distance $<1\farcs0$) of a foreground line emitter at $z=0.3731$ contaminating the spectrum of object 219009247.
Hence, this detection presents a false positive detection of \heii{} emission from a LAE in our sample.  

\begin{figure}
\begin{center}
\includegraphics[width=0.49\textwidth]{mainlegend_nolit.png}\\
\includegraphics[width=0.49\textwidth]{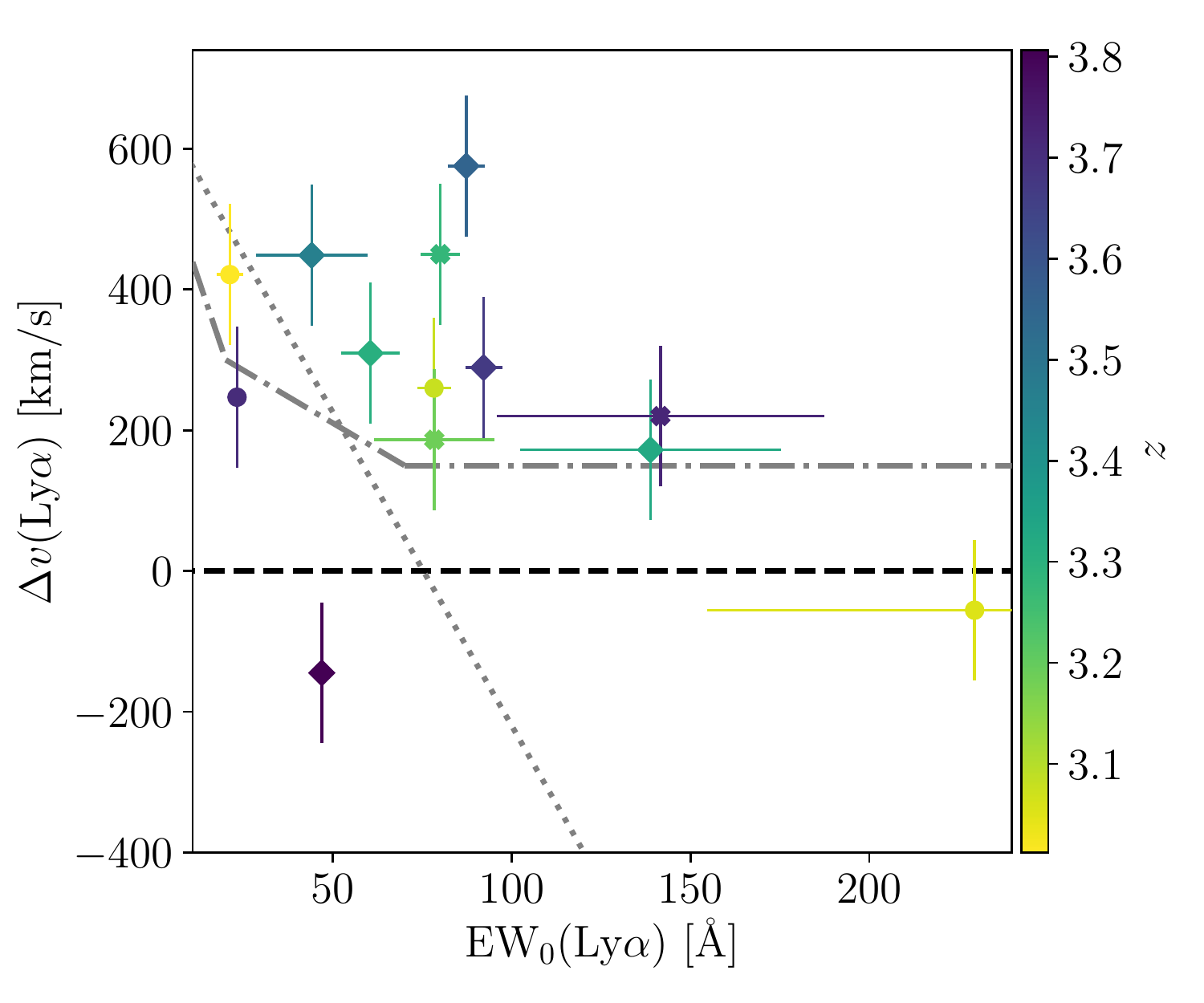}
\caption{Velocity offsets of Ly$\alpha$ determined with respect to the \ciii{} detections as a function of the Ly$\alpha$ rest-frame equivalent width estimated for a fixed UV slope of $\beta=-1.97$.
The gray dot-dashed and dotted lines show the empirical relationship between $\Delta v_\textrm{\lya}$ and equivalent width for $z\approx3$ objects presented by \cite{2018MNRAS.477.2098N} and \cite{2003ApJ...584...45A}, respectively.
The individual points are color coded according to redshift.
}
\label{fig:dvCIIIvsEWlya}
\end{center}
\end{figure}

Empirical correlations between $\Delta v_\textrm{\lya}$ and EW$_{0}$(\lya) \citep[see][for a recent collection]{2021MNRAS.503.4105T} have been parametrized through empirical relations by, for example, \cite{2018MNRAS.477.2098N} and \cite{2003ApJ...584...45A} for LAEs at $z\approx3$. 
In Figure~\ref{fig:dvCIIIvsEWlya} we show the $\Delta v_\textrm{\lya}$ estimates for our sample based on the \ciii{} detections together with these empirical relations.
The MUSE data appear to roughly follow the \citet[][dot-dashed]{2018MNRAS.477.2098N} relation though with large scatter. 
This amount of scatter is however comparable to the scatter of the collection of data that \cite{2018MNRAS.477.2098N} based their relation on.
%

\begin{figure}
\begin{center}
\includegraphics[width=0.40\textwidth]{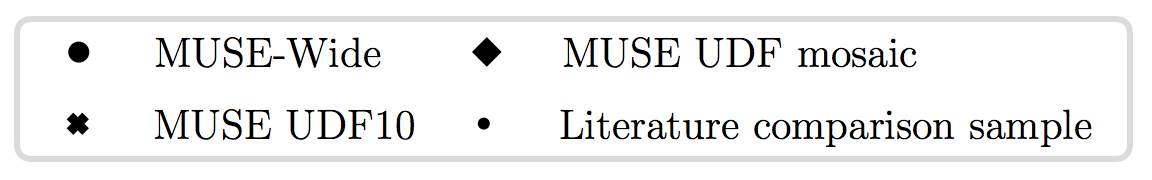}\\
\includegraphics[width=0.49\textwidth]{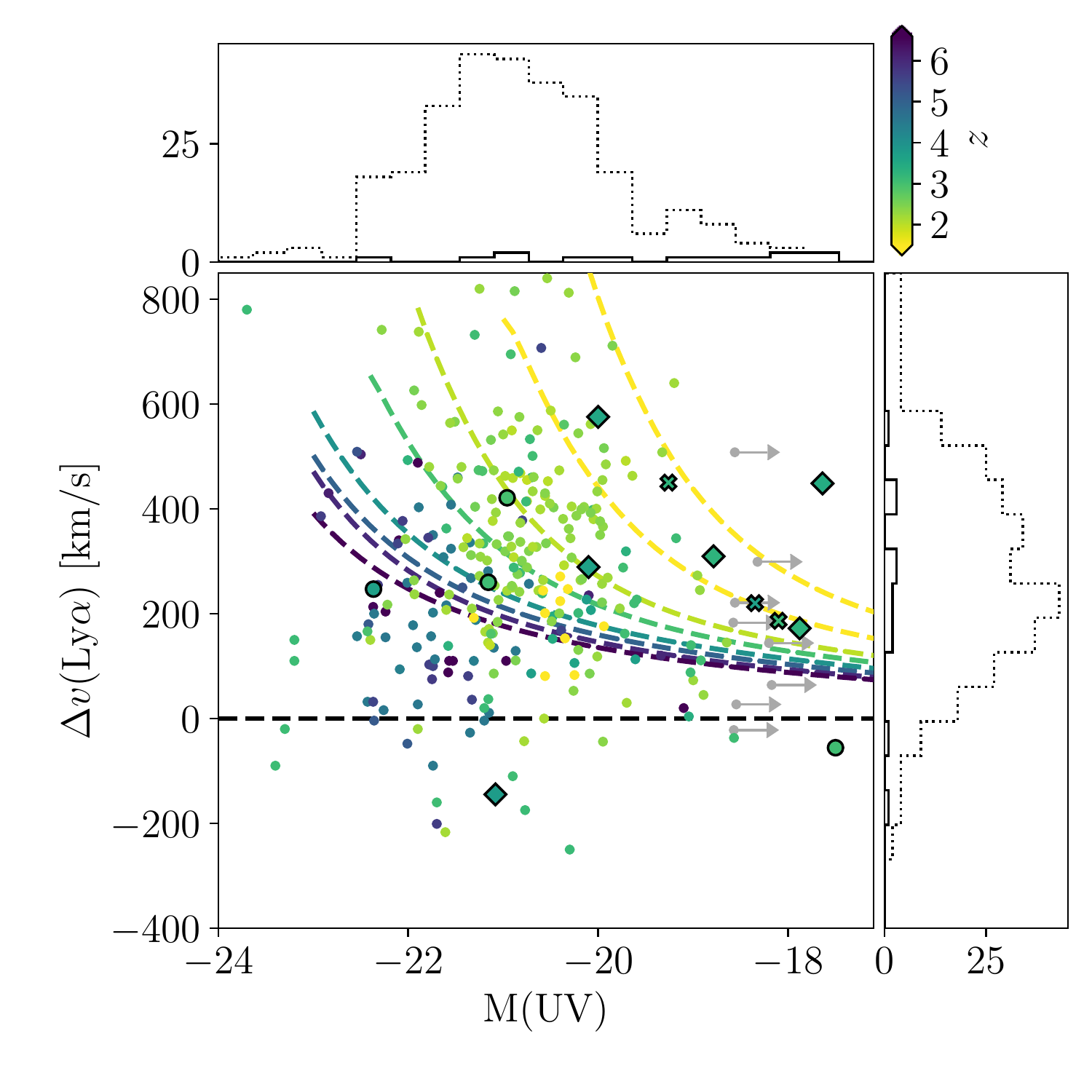}
\caption{Velocity offsets of Ly$\alpha$ as a function of the absolute UV magnitude for the MUSE LAEs with \ciii{} detections studied here shown by the large symbols and a collection of measurements at redshifts from $z\approx2$ to $z=7.73$ collected from
\cite{
  2012ApJ...745...33K, 
  2013ApJ...777...67S, 
  2014ApJ...795...33E,  
  2015MNRAS.450.1846S, 
  2017MNRAS.464..469S, 
  2015ApJ...809...19H, 
  2015ApJ...807..180W, 
  2016ApJ...829L..11P, 
  2016Sci...352.1559I, 
  2017ApJ...836L...2B, 
  2017ApJ...836L..14M, 
  2017A&A...597A..13V, 
  2018MNRAS.477.2817S, 
  2019ApJ...879...70H, 
  2020A&A...643A...6C,  
  2020MNRAS.498.3043M}, 
and \cite{2020MNRAS.492.1778M}
shown by the small dots.
To prevent cluttering the figure the estimated uncertainties are not shown. 
The median errors on the magnitudes and $\Delta v_\textrm{\lya}$ are 0.2~mag and 50~km~s$^{-1}$.
The dashed curves present the predicted median $\Delta v_\textrm{\lya}$ as a function of UV magnitude for a range of redshifts (according to the color coding) from \citet[][corresponding to the black curve in their Figure~2 from their Equation~2]{2018ApJ...856....2M}.
This is based on an abundance matching model relating the absolute UV magnitude to the galaxy halo mass \citep{2015ApJ...813...21M}.
Points and curves are color-coded according to object redshift and the color gradient highlights the importance of halo mass, that is available gas reservoir, in producing and determining $\Delta v_\textrm{\lya}$.  
The solid histograms show the distribution of objects from our study, whereas the dotted histograms include the sample of measurements from the literature.
}
\label{fig:dvCIIIvsMUV}
\end{center}
\end{figure}

Previous studies, among those \cite{2014ApJ...795...33E}, have noted a correlation between absolute UV magnitude and $\Delta v_\textrm{\lya}$.
In Figure~\ref{fig:dvCIIIvsMUV} we show the estimated UV magnitudes for the \lya{} velocity offsets of the \ciii{} emitters in our sample (large symbols) together with a collection of 
M$_\textrm{UV}$ and $\Delta v_\textrm{\lya}$ measurements from the literature shown as small dots (see figure caption for details).
All points have been color-coded according to the object redshift and they indicate that the highest redshift objects (blue and purple points) on average have smaller \lya{} velocity offsets and are brighter than objects at redshifts 2--3 (green points).
The latter could of course be affected by the \cite{1920MeLuS..22....3M,1922MeLuF.100....1M} bias, as it is generally harder to observe intrinsically faint objects at higher redshifts.
For comparison, we show a set of colored dashed curves predicting the correlation between the median $\Delta v_\textrm{\lya}$ and M$_\textrm{UV}$ from \cite{2018ApJ...856....2M}.
These curves correspond to the black curve in their Figure~2 at different redshifts (according to the color coding) which relates the velocity offsets to the galaxy halo mass (their Equation~2).
The galaxy halo masses are translated into absolute UV magnitudes through a simple abundance matching model.
Based on the relation between M$_\textrm{UV}$ and halo mass of the emitting systems by \cite{2015ApJ...813...21M}, \cite{2018ApJ...856....2M} suggest that $\Delta v_\textrm{\lya}$ probes the halo mass of the host galaxy (though with significant scatter), as the amount of neutral hydrogen scattering \lya{} is closely related to the galaxy mass.
Despite the significant scatter in the points plotted in Figure~\ref{fig:dvCIIIvsMUV} they appear to roughly follow the curves from \cite{2018ApJ...856....2M}. 
This could indicate that the halo mass, that is the reservoir of available gas to scatter the \lya{} photons, is indeed a more important quantity for determining the \lya{} velocity offsets than, for instance, star-formation rate (SFR), that is outflows from star formation.
If the opposite was true, at fixed M$_\textrm{UV}$ the high-redshift galaxies should show larger $\Delta v_\textrm{\lya}$ than the lower redshift galaxies as galaxies generally have lower mass for a fixed SFR at higher redshifts.
The lowest redshift targets from the literature ($z\lesssim1.5$), that is the yellow points, seem to be an exception to this apparent trend, potentially indicating that for these systems the hypothesis that the halo mass drives the size of the velocity offsets might not be true. Here star formation processes might be of higher importance.  

\cite{2019A&A...631A..19M} highlight the importance of ISM outflows in producing velocity offsets of \lya.
However, in line with the above, they argue that the key factor controlling the strength of the outflows, and hence the \lya{} velocity offsets, is the \ion{H}{i} column density.
Only in systems with low column density ($\approx10^{19}$cm$^{-2}$) can the ISM outflows produce velocity offsets, which are not expected to be larger than $\approx300$~km~s$^{-1}$. 
This appears to be in agreement with the literature $z\lesssim1.5$ objects which all have $\Delta v_\textrm{\lya}\lesssim300$~km~s$^{-1}$.
According to \cite{2019A&A...631A..19M}, to produce larger $\Delta v_\textrm{\lya}$ it appears that systems with larger gas reservoirs (halo masses) with higher \ion{H}{i} column densities are needed.
Based on samples of star forming galaxies at $z\approx0.2$ \cite{2014ApJ...788...74S} and \cite{2015ApJ...809...19H} arrive at similar conclusions, which are also supported by radiative transfer models \citep[e.g.,][]{2013ApJ...775...99C,2015A&A...578A...7V}. 

\cite{2020MNRAS.496.1013M} estimated the average velocity offset of \lya{} for a sample of LAEs from the MUSE Quasar-field Blind Emitters Survey \citep[MUSEQuBES; PI. Schaye;][]{2019MNRAS.484..431C} with respect to circumgalactic absorption lines, and found an average of $\Delta v_\textrm{\lya} = 171\pm 8$~km~s$^{-1}$, which appears to be somewhat lower than what we find here. 
Their sample generally contains objects with smaller L(\lya) than the UV line emitters providing $\Delta v_\textrm{\lya}$ here, which could explain part of this difference as brighter \lya{} implies larger halo mass and brighter M$_\textrm{UV}$ \citep[e.g.,][]{Kerutt:2021tr,2019MNRAS.489..555K} which again implies larger \lya{} velocity offsets at fixed redshift \citep{2018ApJ...856....2M}.

\begin{figure}
\begin{center}
\includegraphics[width=0.49\textwidth]{mainlegend_nolit.png}\\
\includegraphics[width=0.49\textwidth]{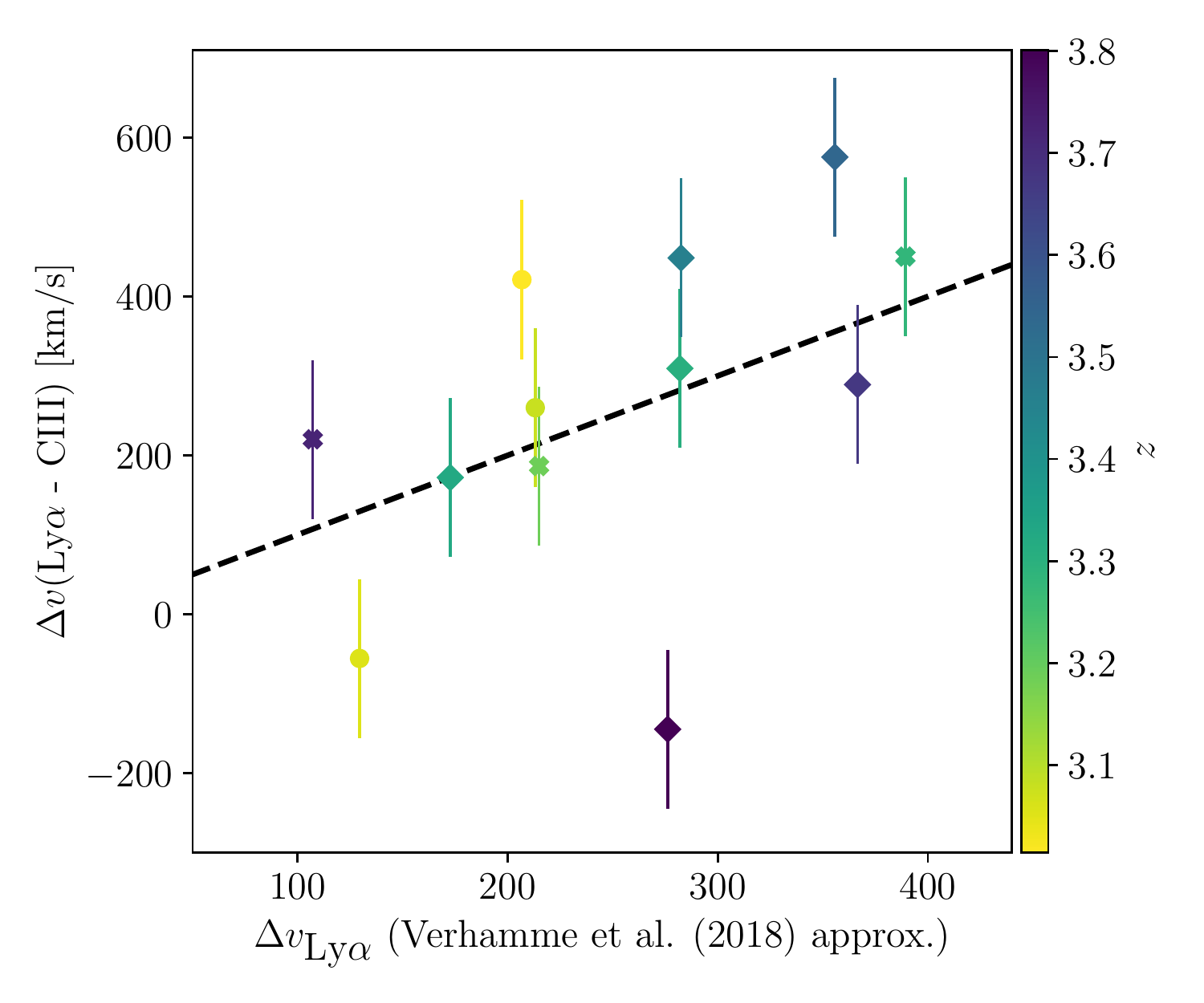}
\caption{Comparison of the velocity offset of Ly$\alpha$ with respect to the \ciii{} doublet detections and the Ly$\alpha$ velocity offset estimated based on the peak separation for double peaked LAEs and the \lya{} FWHM for single peaked LAEs following the empirical relations presented by \cite{2018MNRAS.478L..60V}.
The dashed line indicates the one-to-one relation.
}
\label{fig:dvCIIIverhamme}
\end{center}
\end{figure}

Based on another sample of \lya{} velocity offsets, \cite{2018MNRAS.478L..60V} present means of predicting the systemic redshifts based on the FWHM of the \lya{} line or the separation of the red and blue components of double-peaked \lya{} emission profiles.
Figure~\ref{fig:dvCIIIverhamme} presents a comparison of the \lya{} velocity offsets for the \ciii{} emitters among the LAEs in our sample and the predicted \lya{} velocity offset based on the \cite{2018MNRAS.478L..60V} relations.
Earlier measurements on five sources from our main sample were included in the work by \cite{2018MNRAS.478L..60V}.
If a \lya{} profile was classified as double peaked, the correlation between velocity offset and peak separation was used. Otherwise the correlation with \lya{} FWHM was used to predict the velocity offset.
Figure~\ref{fig:dvCIIIverhamme} shows that there is good agreement between the measured and predicted \lya{} velocity offsets.
Similarly, the \lya{} velocity offsets based on the \heii{} and \oiii{} detections also match the empirical predictions presented by \cite{2018MNRAS.478L..60V}.

\begin{figure}
\begin{center}
\includegraphics[width=0.40\textwidth]{mainlegend_tworow.png}\\
\includegraphics[width=0.49\textwidth]{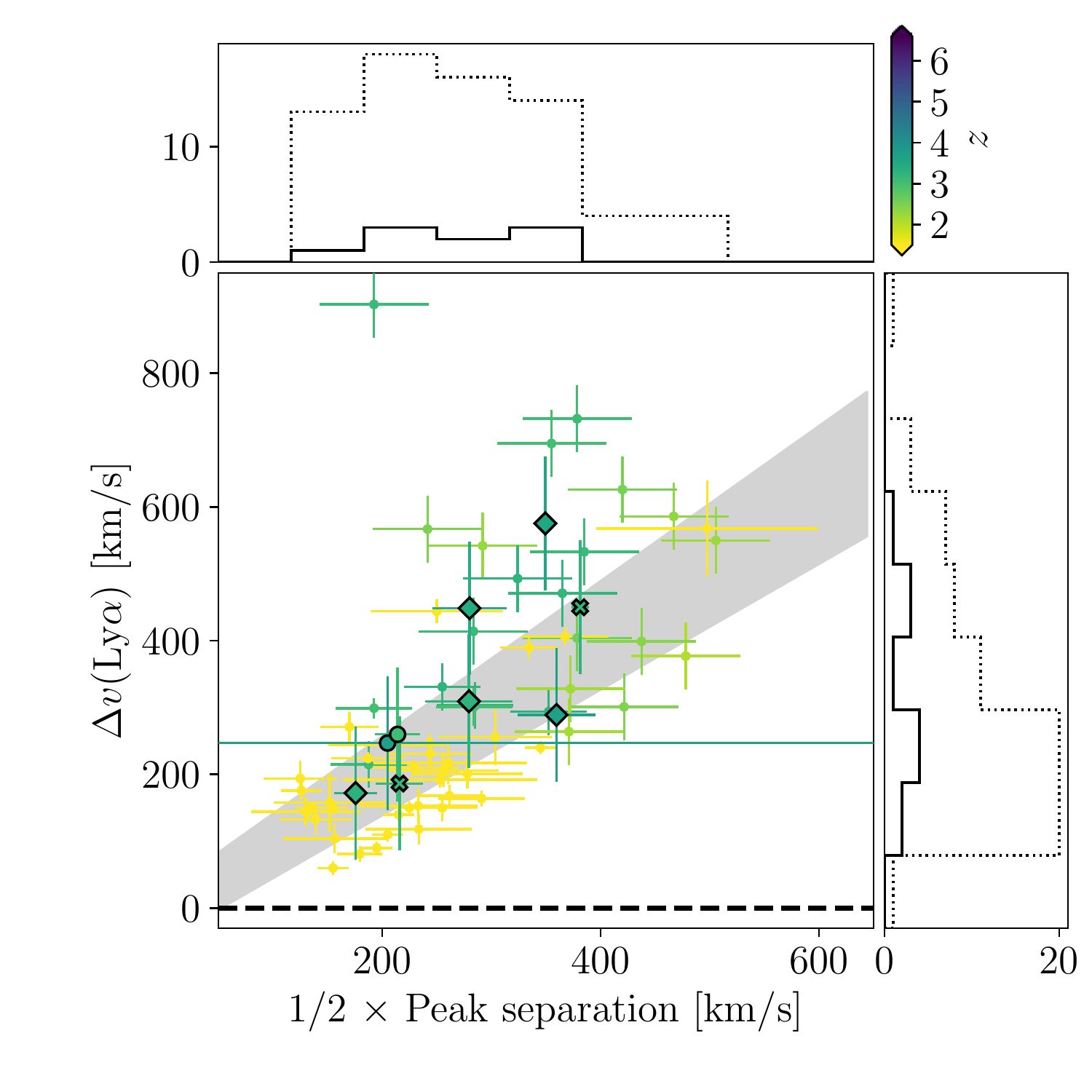}\\
\includegraphics[width=0.49\textwidth]{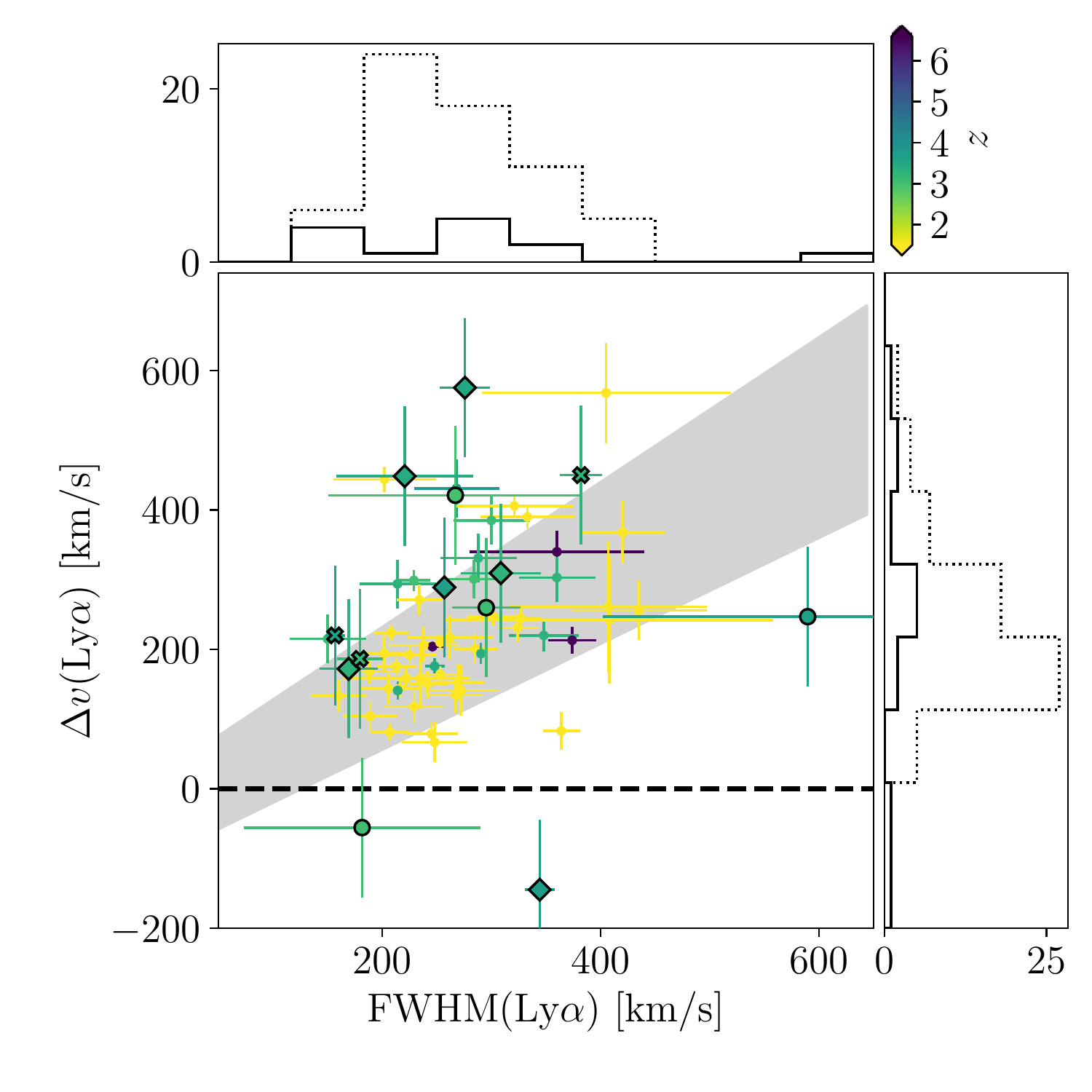}
\caption{Comparison of the velocity offset of Ly$\alpha$ and half the \lya{} peak separation for objects with \lya{} line profiles containing both a red peak and a blue bump (top) and the \lya{} FWHM (bottom).
These are the quantities used to predict $\Delta v_\textrm{\lya}$(\cite{2018MNRAS.478L..60V} approx.) on the x-axis in Figure~\ref{fig:dvCIIIverhamme}.
The gray bands show the empirical relations presented by \cite{2018MNRAS.478L..60V}.
The small dots present a collection of literature measurements from
\cite{2012ApJ...745...33K, 
2013A&A...553A.106L, 
2015ApJ...804L..30O, 
2015ApJ...812..157H, 
2015ApJ...809...19H, 
2016A&A...587A.133G, 
2016Natur.529..178I, 
2016MNRAS.461.3683I, 
2017MNRAS.464..469S, 
2017A&A...597A..13V, 
2018MNRAS.478L..60V, 
2017ApJ...844..171Y, 
2020MNRAS.492.1778M, 
2020MNRAS.496.1013M}, 
and Matthee et al. 2021 (submitted).
All points are color coded according to object redshift.
The solid histograms show the measurements presented in this study, whereas the dotted histograms include the measurements from the literature.  
}
\label{fig:dvCIIIpeaksep}
\end{center}
\end{figure}

If we consider the two individual predictions from the original study shown in Figure~\ref{fig:dvCIIIpeaksep} relating $\Delta v_\textrm{\lya}$ to the \lya{} peak separation (top) and FWHM (bottom), we see that our measurements (large symbols) are in good agreement with the empirical relations from \cite{2018MNRAS.478L..60V}:
\begin{eqnarray}
\Delta v_\textrm{\lya} &=& (1.05\pm0.11) \times \frac{1}{2}\textrm{PeakSep}(\textrm{\lya}) - (12\pm37)~\textrm{km~s}^{-1} 
\nonumber\\
r_\textrm{P} &=& 0.6122 \nonumber\\
r_\textrm{S} &=& 0.6906 \nonumber\\
\Delta v_\textrm{\lya} &=& (0.90\pm0.14) \times \textrm{FWHM}(\textrm{\lya}) - (34\pm60)~\textrm{km~s}^{-1} \nonumber \\
r_\textrm{P} &=& 0.2970  \nonumber \\
r_\textrm{S} &=& 0.4299\;. \nonumber 
\end{eqnarray}
These relations are shown as the gray bands in Figure~\ref{fig:dvCIIIpeaksep} and the correlation coefficients for each of them are estimated after including the collection of measurements from the literature shown as the small symbols in the figure. 
Our measurements appear to be offset slightly high with respect to the $\Delta v_\textrm{\lya}$--FWHM(\lya) relation, even though we also see that they present a scatter similar to what is found in the literature.
As an estimate of the scatter in the relations, we determine the perpendicular Euclidian distance between all points and the proposed linear relations.
The average distance from the peak separation (FWHM) relation is 70~km~s$^{-1}$ (68~km~s$^{-1}$) with a standard deviation of the distribution of distances of 77~km~s$^{-1}$ (67~km~s$^{-1}$).
From the data used to define the empirical relations \cite{2018MNRAS.478L..60V} quote a similar but tighter scatter of $53\pm9$~km~s$^{-1}$ and $72\pm12$~km~s$^{-1}$ for the two relations and determine an uncertainty on the systemic redshift correction of  $\lesssim\pm100$~km~s$^{-1}$ which is in agreement with the above estimates.

In addition to the correlations described above, we also checked for trends between $\Delta v_\textrm{\lya}$ and the remaining LAE properties described in Section~\ref{sec:UVandLya}, including $\beta$, LAE effective radius, \lya{} flux, and L(\lya).
We do not find any trends between these parameters and the \lya{} velocity offset with respect to the UV emission probes of systemic redshift ($|r_\textrm{P}| < 0.42$ and $|r_\textrm{S}| < 0.28$ for all  parameters).
The only exception to this is the potential correlation (shown in Figure~\ref{fig:dvVSRe}) between $\Delta v_\textrm{\lya}$ and the effective radius indicating that larger galaxies (R$_{e}\approx3.5$~kpc) have offsets of roughy 500~km~s$^{-1}$ as opposed to galaxies with R$_{e}\approx1$~kpc that have $\Delta v_\textrm{\lya}\approx300$~km~s$^{-1}$.
However, $r_\textrm{P} = 0.48$ and $r_\textrm{S} = 0.58$ for these data so at best the correlation is only tentative. 

Finally, to judge if any systemic offsets of the resonant \civ{} emission was detectable with respect to the estimated systemic redshift, we looked at the velocity offsets of \civ{} for all \ciii{} emitters.
Eleven of the thirteen $z>1.5$ \ciii{} emitters also showing \civ{} show estimated $\Delta v_\textrm{\civ}\lesssim250$~km~s$^{-1}$ within the error bars with a median of 92~km~s$^{-1}$ (including the two outliers the mean is 82~km~s$^{-1}$).
We have avoided an attempt to disentangle the stellar and nebular contribution to the \civ{} emission (see Section~\ref{sec:UVEmissionLineSearch}), but these velocity offsets are in agreement with the stellar absorption based $\Delta v_\textrm{\civ}$ estimates at lower redshift from \cite{2016ApJ...829...64D}.
Hence, the estimated $\Delta v_\textrm{\civ}$ with respect to systemic appears to be rather modest in the \ciii-\civ{} emitter sample.
It is therefore also likely that the large scatter of the \civ{} measurements seen in the upper right panel of Figure~\ref{fig:voffsetsall} can be contributed to mostly the \lya{} velocity offsets when larger than $\approx$250~km~s$^{-1}$.
The two objects with $\Delta v_\textrm{\civ}>250$~km~s$^{-1}$ (720320277 and 720830605) show larger offsets due to prominent \civ{} P-Cygni profiles as described in Section~\ref{sec:UVEmissionLineSearch}. 
Therefore, the estimated velocity offsets for these objects reflect the ``residual'' emission in the red part of the P-Cygni profile.  
As mentioned, full modeling of the \civ{} line profiles is beyond the scope of this work.

\section{Estimating electron density}\label{sec:neTe}

\begin{figure}
\begin{center}
\includegraphics[width=0.49\textwidth]{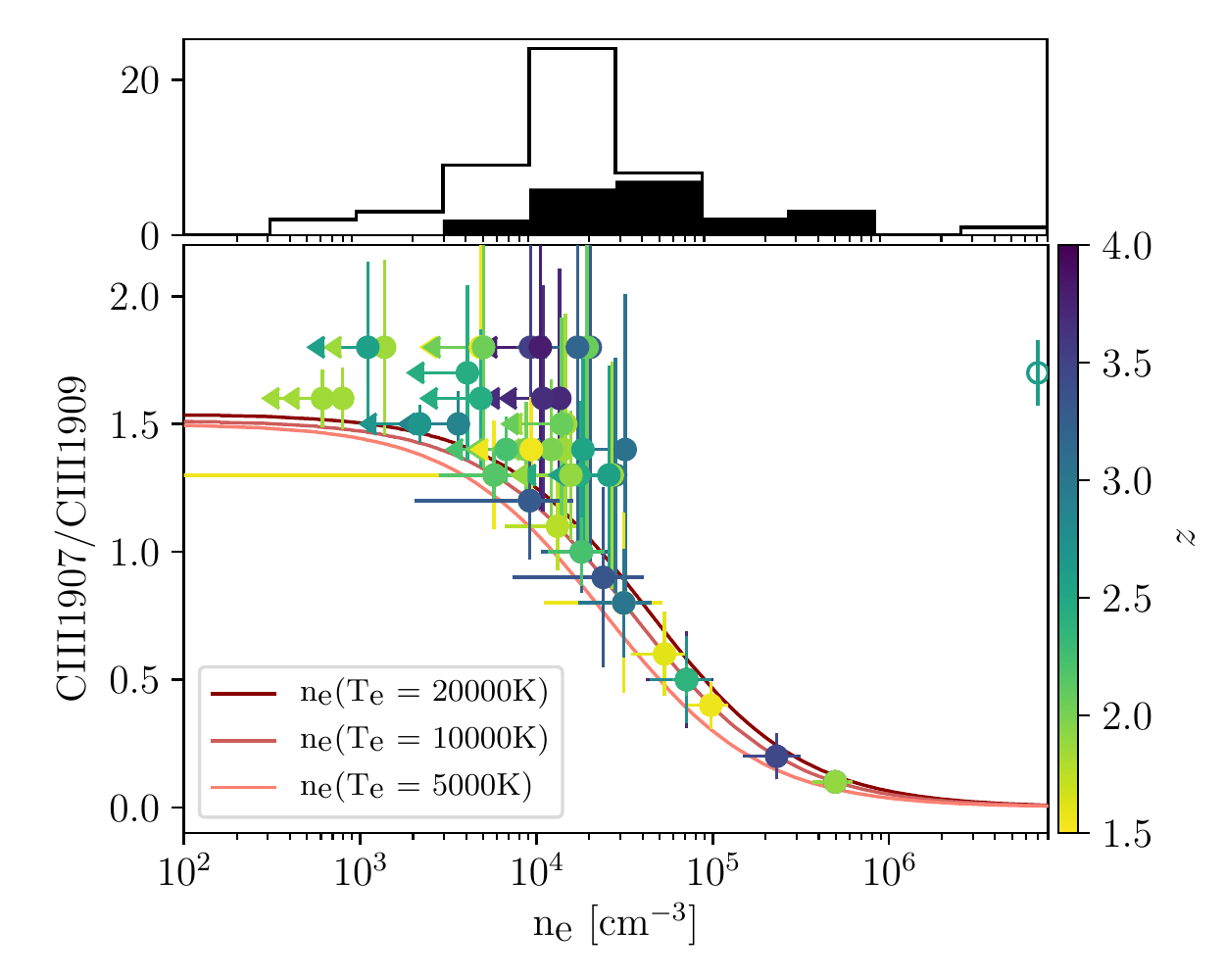}\\
\includegraphics[width=0.49\textwidth]{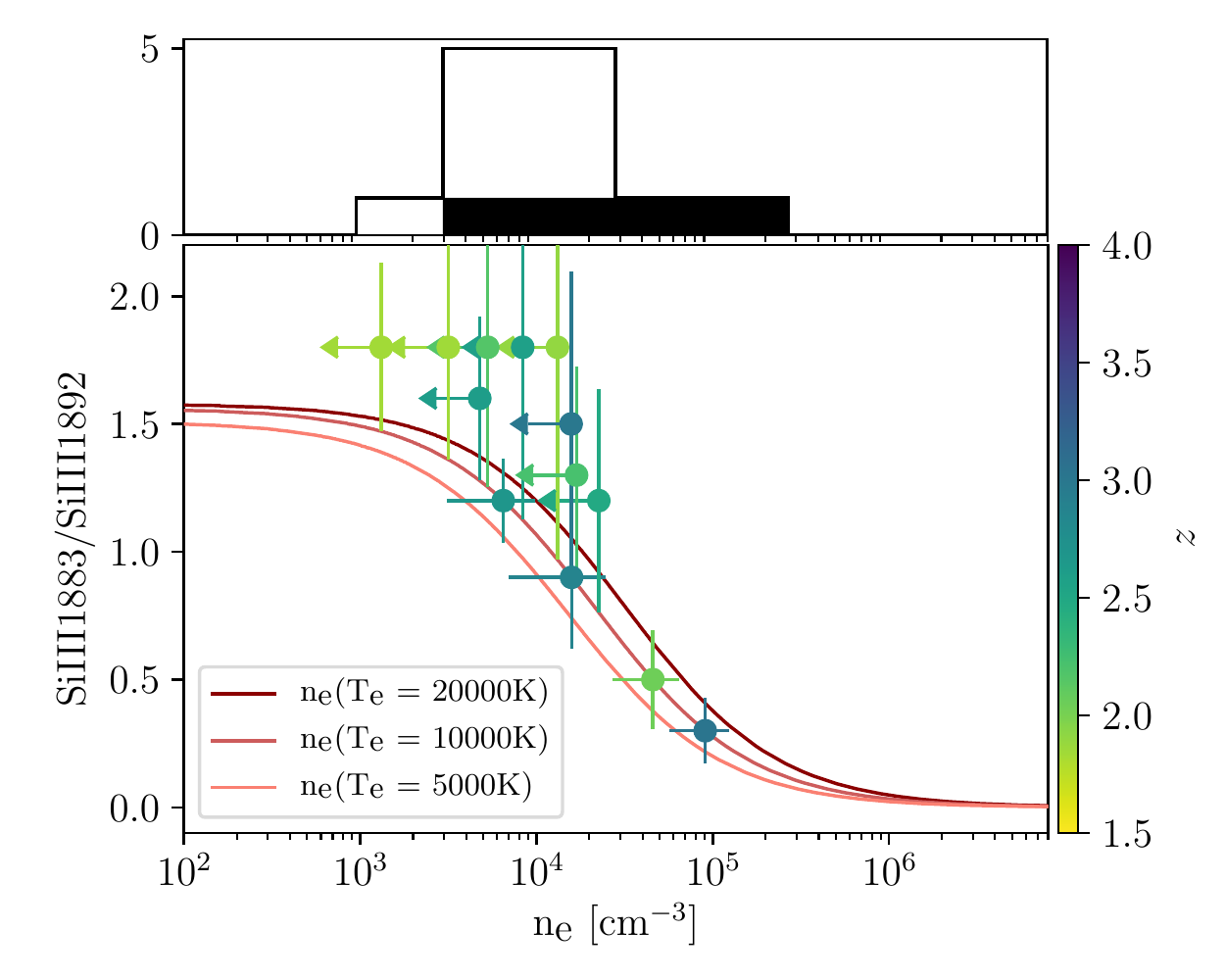}
\caption{Electron densities ($n_\textrm{e}$) estimated based on the \ciii{} (top) and \siiii{} (bottom) doublet line flux ratios for a fixed electron temperature of T$_\textrm{e}=10^4$~K.
The error bars (and limits) correspond to the 1$\sigma$ uncertainties on the doublet flux ratios.
For comparison, the theoretical curves \citep[e.g.,][]{2006agna.book.....O,2019ApJ...880...16K,2019ARA&A..57..511K} for electron temperatures of $5000$~K, $10000$~K and $20000$~K are shown as the solid red curves.
The filled histograms count the number of objects where $n_\textrm{e}$ could be determined from the observed flux ratio.
The open histograms add the subset of objects where the $n_\textrm{e}$ estimates are upper limits.
The single open symbol to the right in the top panel indicates the \ciii{} flux ratio of object 605172634 (Figure~\ref{fig:ObjSpec02}) which results in an unconstrained electron density estimate for T$_\textrm{e}=10^4$~K.
}
\label{fig:neestimates}
\end{center}
\end{figure}

The relative strengths of the \ciii{} and \siiii{} emission line doublet components are highly sensitive to the electron number density of the emitting gas and the ISM pressure \citep[e.g.,][]{1992ApJ...389..443K,2006agna.book.....O,2019ApJ...880...16K,2019ARA&A..57..511K}.
Using the PyNeb software \citep{2013ascl.soft04021L,2015A&A...573A..42L} we estimated the electron density, $n_\textrm{e}$, for the 52 and 13 objects with detected \ciii{} and \siiii{} emission, respectively (see Figure~\ref{fig:doubletratios}).
The doublet component flux ratios are determined directly from the FELIS template matches.
In 0/13 and 7/52 of the \siiii{} and \ciii{} doublet detections the fainter component is only marginally detected. However, for these few sources the flux ratio, and hence an estimate of the electron density, can still be inferred from the template matches, as the flux of each doublet component as mentioned varies independently.
In these cases the flux ratio will naturally be pushed to the edges of the allowed ranges for the templates (see Table~\ref{tab:LinesAndTemplates}).
The mean S/N(FELIS) of the doublet detections used to infer the electron density is 5.05 and 7.67 for the \siiii{} and \ciii{} detections, respectively. The S/N values span the ranges 3.07--10.30 and 3.03--28.58 for the two samples.
Example spectra and the corresponding FELIS template matches for a subsample of these sources are shown in Figures~\ref{fig:ObjSpec}, \ref{fig:ObjSpec98}, \ref{fig:ObjSpec03}, and \ref{fig:ObjSpec99}. 
Figure~\ref{fig:neestimates} shows the distribution of the $n_\textrm{e}$ estimates from our MUSE samples assuming a fixed electron temperature of 10$^4$~K.
Fixing the electron temperature to 5000K or $2\times10^4$~K instead would shift the data points onto the lighter or darker shaded red curves shown in Figure~\ref{fig:neestimates}, respectively.
The estimates sample the full range of electron densities where the \ciii{} and \siiii{} ratios are sensitive, but the majority of the estimates (and upper limits) have $n_\textrm{e}<10^5$cm$^{-3}$.
We checked for correlations between the inferred ISM electron densities and the estimated EW$_0$(\ciii) (and EW$_0$(\siiii)) values, as larger EW$_0$(\ciii) potentially indicates younger systems, but found no clear dependencies.

\begin{figure}
\begin{center}
\includegraphics[width=0.48\textwidth]{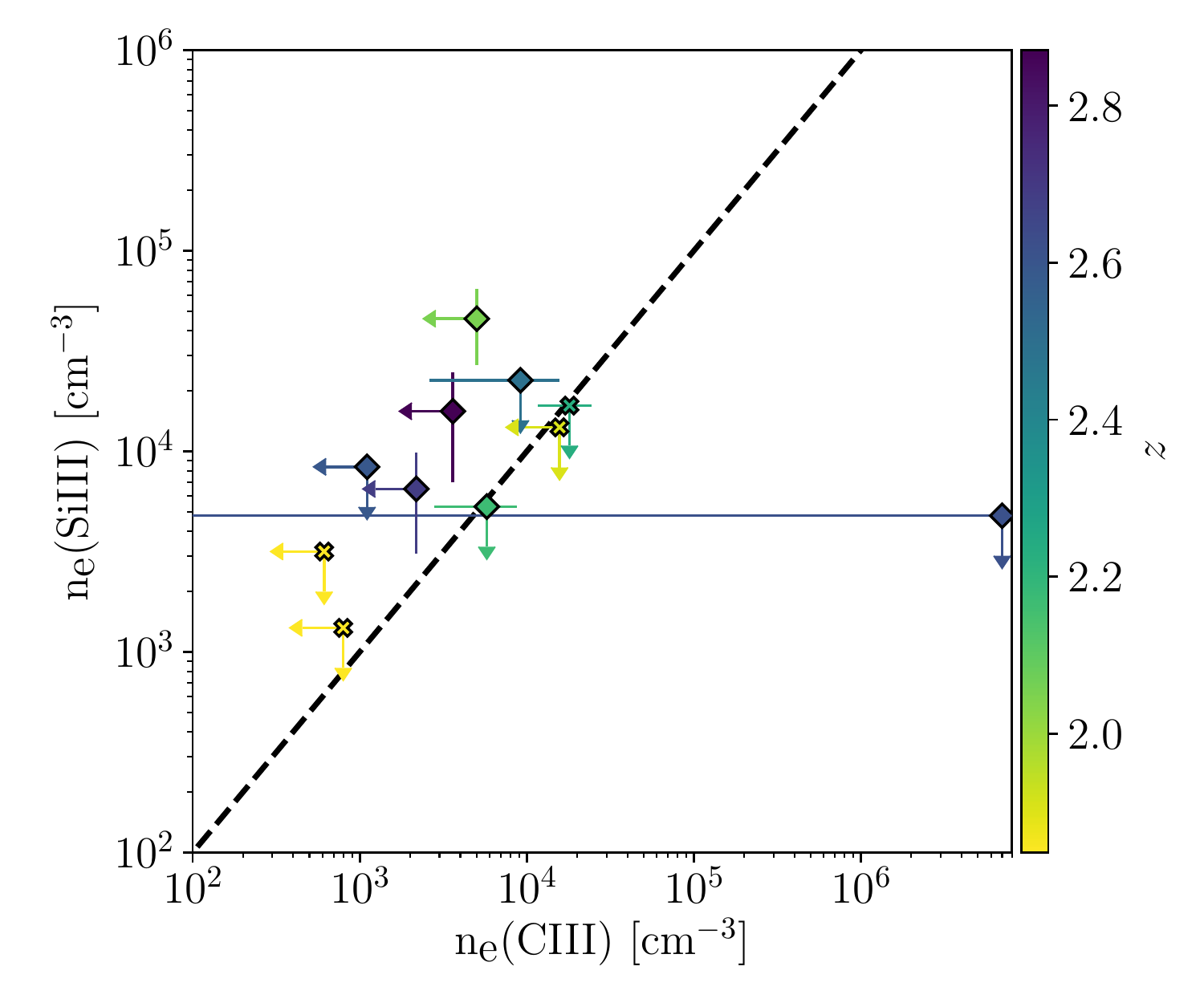}
\caption{Comparison of electron density estimates based on the \ciii{} (x-axis) and \siiii{} (y-axis) from Figure~\ref{fig:neestimates} for the 11 objects (all from the UDF mosaic and UDF10) with measurements from both doublets.
The shown electron densities were estimated for a fixed electron temperature of T$_\textrm{e}=10^4$~K.
The errorbars (and limits) corresponds to the 1$\sigma$ uncertainties on the doublet flux ratios.}
\label{fig:neCIIIVSneSiIII}
\end{center}
\end{figure}

%
Figure~\ref{fig:neCIIIVSneSiIII} shows a comparison of the electron density estimates for the objects with measurements from both \ciii{} and \siiii{} (11 sources all from the UDF mosaic and UDF10).
Within 3$\sigma$ the estimated electron densities generally agree, even though several objects only have upper limits on $n_\textrm{e}$(\ciii) and/or $n_\textrm{e}$(\siiii).
These 11 sources have mean S/N(FELIS) of 5.34 and 14.52 spanning the ranges 3.07--10.30 and 5.35--28.58 for the \siiii{} and \ciii{} detections, respectively. 

The upper limits on $n_\textrm{e}$ result from doublet flux ratios above the theoretically allowed values (solid red curves in Figure~\ref{fig:neestimates}) within the 1$\sigma$ error bars on the doublet component flux ratios.
The upper limits are then quoted as the electron density corresponding to the lowest value in the 1$\sigma$ uncertainty range on the flux ratio as shown by the vertical error bars on the upper limits in Figure~\ref{fig:neestimates}.
In some cases, if not just a result of large uncertainty on the measured flux ratios, these upper limits could indicate that the assumptions for the PyNeb calculations are incorrect. 
For instance, the outlier in the upper right corner of the top panel in Figure~\ref{fig:neestimates} (right-hand side of Figure~\ref{fig:neCIIIVSneSiIII}) is object 605172634 (open symbol). 
This object (manually put at $n_\textrm{e}=7\times10^6$cm$^{-3}$) has a well-constrained \ciii{} doublet component flux ratio of $1.7\pm0.1$ from the emission line template fit to the \ciii{} emission (shown in Figure~\ref{fig:ObjSpec02}).
This leads to an un-constrained estimate of $n_\textrm{e}$ as the 1$\sigma$ error range is outside the allowed theoretical range for $n_\textrm{e}$ given the assumption on the electron temperature.
A higher electron temperature (darker red curve) allows for larger flux ratios.
However, for a \ciii{} flux ratio of $1.7\pm0.1$ a very high electron temperature of roughly $10^5$~K is needed to get a constraint on the electron density ($n_\textrm{e}<5\times10^5$cm$^{-3}$).
In a similar fashion, some of the lower inferred UV emission line doublet flux ratios with large uncertainties quoted as upper limits in Figure~\ref{fig:neestimates} can also be turned into estimates of an allowed electron density range by considering higher electron temperatures (and vice versa).
A caveat of the classical density diagnostics that could also play an important role in explaining similar measurements is the existence of density (and temperature) inhomogeneities that challenges the simple interpretation of constant temperature and density throughout the emitting gas \citep{2017PASP..129h2001P}.

The electron densities estimated for the sample studied here are generally larger than the electron densities (obtained from \siifull) of the ``green peas'' and ``Lyman break analogs'' studied by \cite{2019ApJ...872..146J}, where they find $n_\textrm{e}\lesssim10^3$cm$^{-3}$. However, as shown by the curves in Figure~\ref{fig:neestimates} both \ciii{} and \siiii{} saturate and are not sensitive to these low densities. 
At the same time the [\ion{S}{ii}] flux ratio saturates at $n_\textrm{e}\approx10^4$cm$^{-3}$  \citep{2019ApJ...880...16K} providing little overlap between the UV and optical tracers.
The electron densities estimated via \siifull{} and \oiifull{} for the sample of more common $z\approx2.3$ galaxies described by \cite{2016ApJ...816...23S} show $n_\textrm{e}\lesssim3\times10^3$cm$^{-3}$. 
Similar estimates were presented at $z = 0.1-1$ by \cite{2016ApJS..226....5L}.
Again, the optical probes used in these studies are not sensitive to $n_\textrm{e}\gtrsim10^4$cm$^{-3}$ \citep{2019ApJ...880...16K}. 
Hence, there appear to be a few objects with relatively high densities in our sample \citep[similar to what is seen in the objects presented by][]{2014MNRAS.440.1794J,2018MNRAS.476.1726J}, whereas the large number of upper limits for the \siiii{} and \ciii{} estimates are fully consistent with electron densities of special but also more generic galaxy samples at lower redshifts.
However, as discussed by \cite{2017A&A...608A...4M}, considering the differences between electron density estimates obtained from lines with different ionization energies, one should keep in mind that these lines originate in physically different parts of the nebulae so different densities are to be expected even for the same parent galaxy. 
For example, the \ion{S}{ii} lines are generated in the outskirts, that is lowest density parts assuming a density stratification, of the ionized regions, while \ciii{} and \siiii{} originate in the inner denser parts.

\section{Estimating gas-phase abundances}\label{sec:logOH}

\begin{figure}
\begin{center}
\includegraphics[width=0.50\textwidth]{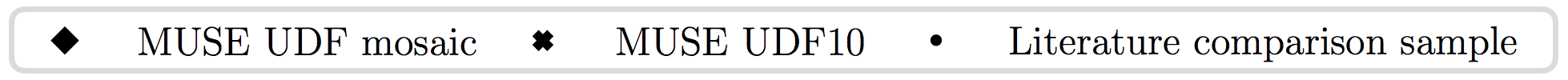}\\
\includegraphics[width=0.47\textwidth]{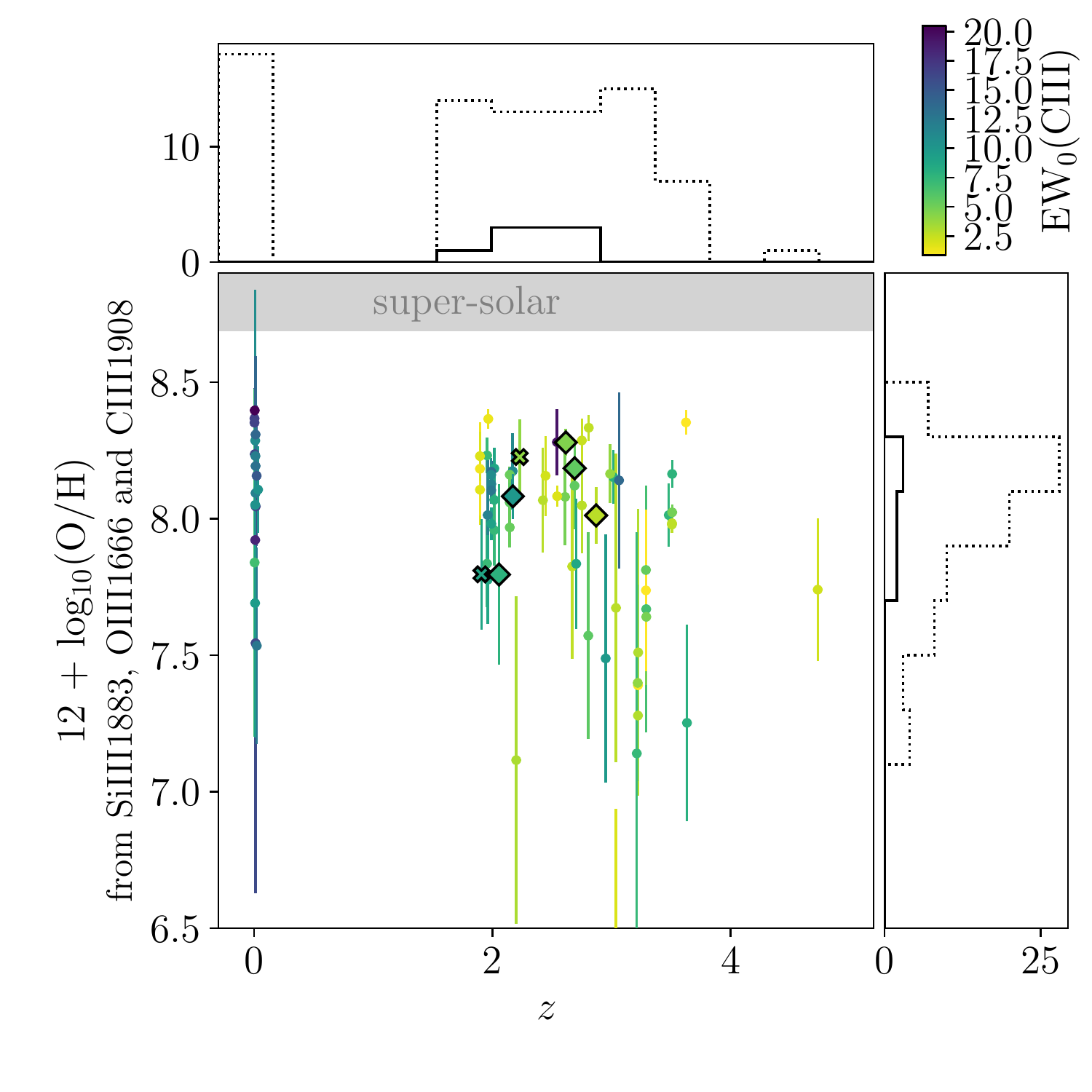}\\
\includegraphics[width=0.47\textwidth]{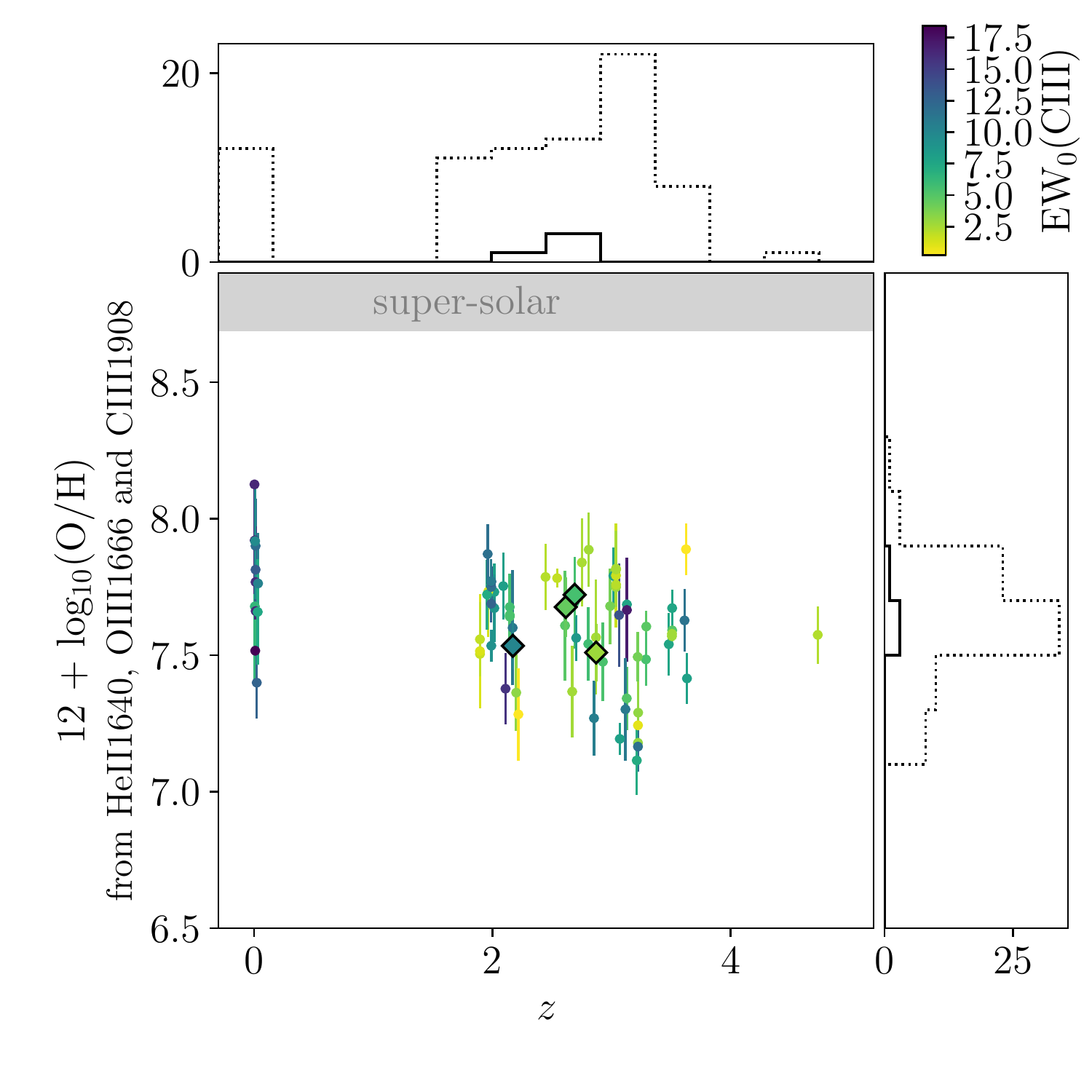}
\caption{Gas-phase abundances estimated using the third order polynomial fitting functions for the Si3-O3C3 (top panel) and He2-03C3 (bottom panel) diagnostics from \cite{2020ApJ...893....1B} as a function of source redshift. 
The large symbols are MUSE sources from the UDF and small dots are from the literature collection described in Appendix~\ref{sec:litcol}.
The solid histograms show the subset of the objects from this work, whereas the dotted histograms also include the sources from the literature.
All points are color coded according to EW$_0$(\ciii).
The gray bands mark regions of super-solar gas-phase abundances at 12+log$_{10}$(O/H)~$>8.69$.}
\label{fig:12logOH}
\end{center}
\end{figure}

The recent study by \cite{2020ApJ...893....1B} tested the diagnostics presented by \cite{2018ApJ...863...14B} using a sample of galaxies at redshift below 0.1 and at $z=2-3$.
They used predictions of UV emission fluxes from the Flexible Stellar Population Synthesis (FSPS) nebular emission model \citep[][]{2009ApJ...699..486C,2010ascl.soft10043C} to infer the connection between UV emission strengths and the gas-phase abundance parameterized as 12+log$_{10}$(O/H).
They compress the multidimensional space of predicted UV line fluxes from \siiii, \oiii, \heii, and \ciii{} into a set of fitting formulas that provide estimates of the gas-phase abundance.
The first is based on a combination of \siiii, \oiii, and \ciii{}, dubbed Si3-O3C3: 
\begin{eqnarray}
12+\log_{10}\textrm{(O/H)} = &3.09\;+0.09x -1.71x^2 -0.73x^{3} \nonumber\\
&-16.51y -19.84y^{2} -6.26y^{3} \nonumber\\
&4.79xy -0.28xy^{2} +1.67x^{2}y,
\end{eqnarray}
where $x$ corresponds to $\log_{10}$(\oiiitwo/\ciii) and $y$ is $\log_{10}$(\siiiione/\ciii). 
\cite{2020ApJ...893....1B} quote a typical statistical error for this relation of $\pm 0.14$\,dex.
The second fitting formula is based on a combination of \oiii, \heii, and \ciii{}, dubbed He2-03C3:
\begin{eqnarray}
12+\log_{10}\textrm{(O/H)} = &6.88 -1.13x -0.46x^2 -0.03x^{3} \nonumber\\
&-0.61y +0.02y^{2} -0.04y^{3}  \nonumber\\
&-0.32xy +0.03xy^{2} -0.21x^{2}y,
\end{eqnarray}
where $x$ is again $\log_{10}$(\oiiitwo/\ciii) and $y$ corresponds to $\log_{10}$(\heii/\ciii). The quoted typical statistical error for this fit is $\pm 0.08$\,dex.
We estimate the gas phase abundance following the Si3-O3C3 (He2-03C3) method for all seven (four) MUSE objects and 76 (82) objects from the collection of data from the literature described in Appendix~\ref{sec:litcol} which have all relevant lines detected above a S/N of 3.
Figure~\ref{fig:12logOH} shows the results applying the \cite{2020ApJ...893....1B} fitting formulas to these sources.
The MUSE objects from our study are shown  with large symbols; $z\approx2-3$ and literature sample as small dots. 

First, we see that all estimates of the gas-phase abundances from our study and the literature sample show subsolar abundances.
This indicates that sources with prominent UV emission lines are mostly systems with low gas-phase abundances, that is with low gas-phase metallicities.
This agrees with the notion that the emitting gas of star forming galaxies capable of producing strong rest-frame UV emission generally has low (subsolar) metallicity with higher temperatures and stronger ionization fields of gas surrounding young massive stars \cite[e.g.,][]{2014MNRAS.445.3200S,2017A&A...608A...4M,2019MNRAS.488.3492S,2020A&A...641A.118F}.
The fact that none of the UV emission line sources from the large collection of sources assembled from the literature show super-solar gas-phase abundances supports this.
There is no indication that the $z>1$ data would be particularly biased and only contain systems with low gas-phase abundances. In several cases the line emitters were detected in broader representative samples of galaxies. In the case of the MUSE sources studied here, the parent sample consists of line-emitters (mostly LAEs) from an unbiased untargeted emission line search in the full data cubes.

The $z\approx0$ observations from the literature do on average have slightly higher gas-phase abundances and EW$_0$(\ciii) than the measurements at higher redshift. 
However, as
the EW$_0$(\ciii) color coding in Figure~\ref{fig:12logOH} shows, these galaxies predominantly have strong \ciii{} emission which might cause a selection bias rather than illustrating an underlying intrinsic correlation between EW$_0$(\ciii) and 12+log$_{10}$(O/H).
This also results in an apparent slight decrease in gas-phase abundance with redshift, which is however fully consistent with a flat nonevolving gas-phase abundance as a function of redshift within the scatter and errors of the measurements.
From gas phase abundance measurements based on rest-frame optical emission lines, it has been suggested that there is a deficiency of EW$_0$(\ciii)~$\gtrsim10$~\AA{} objects at 12+log$_{10}$(O/H)~$\gtrsim8.5$ \citep{2017A&A...608A...4M,2017MNRAS.472.2608S}. 
\cite{2020arXiv200809780S} argues that this deficiency might even be for objects at 12+log$_{10}$(O/H)~$\gtrsim8.0$ (Z/Z$_\odot \gtrsim 0.2$).
In line with these findings the UV-based gas-phase abundances presented in Figure~\ref{fig:12logOH}, which are all below 8.5, do not present any clear evidence for a correlation between EW$_0$(\ciii) and 12+log$_{10}$(O/H) as just explained.
It is however worth noting that a comparison by \cite{2021ApJ...908..154R} between optical and the \cite{2020ApJ...893....1B} UV gas phase abundance diagnostics for a single gravitationally lensed source find that the UV estimates are 0.5--0.8 dex lower than the optical estimates. 
A correction of this size to the measurements presented in Figure~\ref{fig:12logOH} would shift the values to roughly 8.5, which is the scale where trends with EW$_0$(\ciii) starts to emerge when considering optical gas phase abundance diagnostics.

\begin{figure}
\begin{center}
\includegraphics[width=0.4\textwidth]{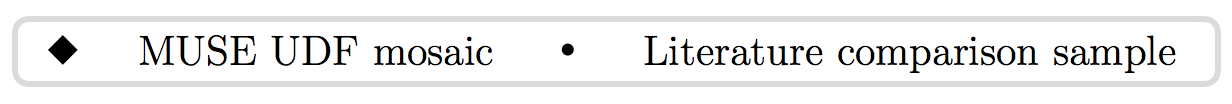}\\
\includegraphics[width=0.48\textwidth]{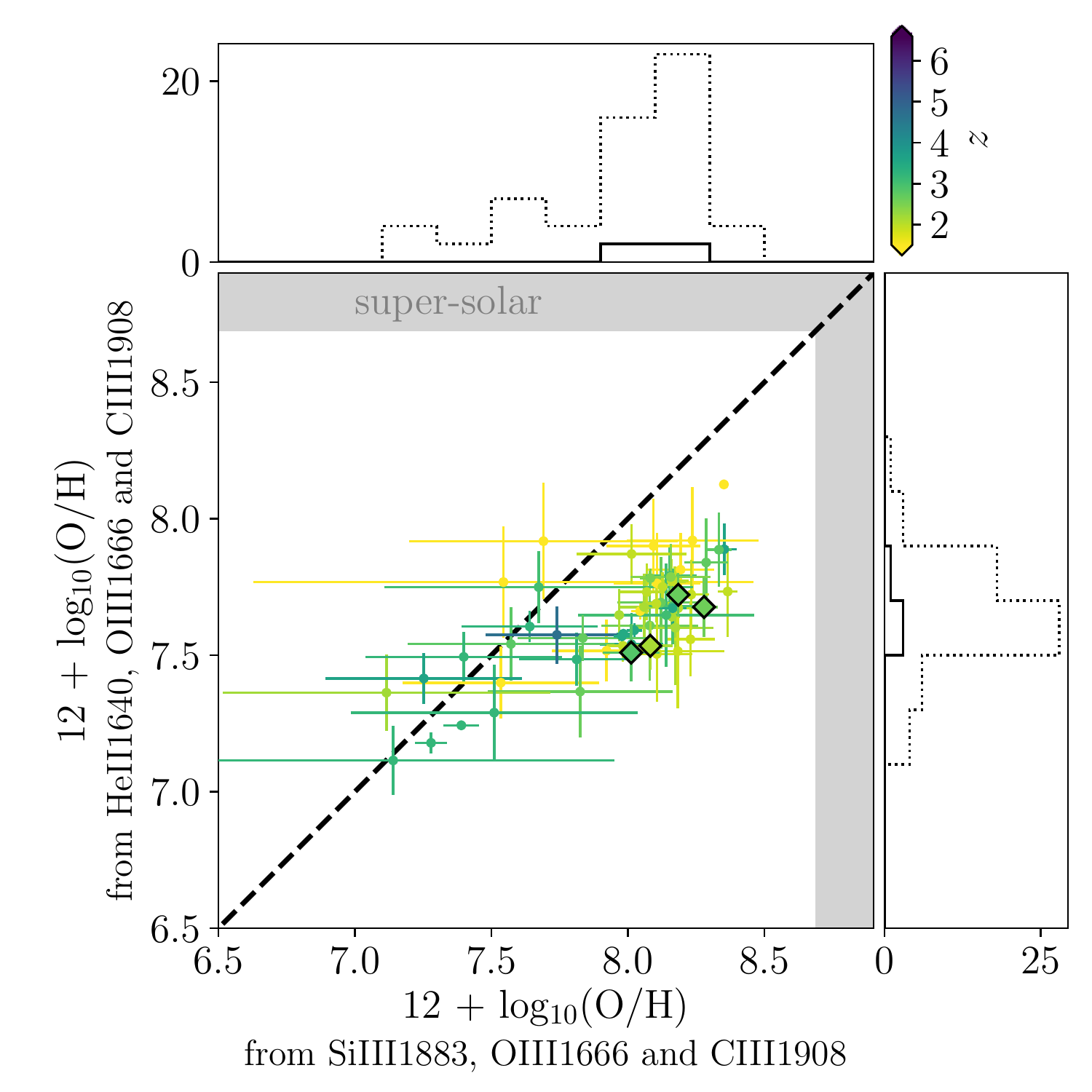}
\caption{Comparison of the gas-phase abundance estimates from Figure~\ref{fig:12logOH} for objects with all relevant emission features detected. 
The large symbols are MUSE sources from the UDF and small dots are objects from the literature collection described in Appendix~\ref{sec:litcol}.
The solid histograms show the subset of objects from this work and the dotted histograms include the sample from the literature.
The gray region marks values of super-solar abundances (12+log$_{10}$(O/H)~$>8.69$).
Points are color-coded according to object redshift.}
\label{fig:12logOHcomp}
\end{center}
\end{figure}

Figure~\ref{fig:12logOHcomp} presents a direct comparison of the four MUSE sources and 57 objects from the literature where both estimates could be performed.
All EW$_0$ and flux measurements for these subsamples of sources with close to the full suite of rest-frame UV emission lines detected are available for the MUSE objects in the catalog described in Table~\ref{tab:mastercatcol} and for the literature sample in the catalog described in Table~\ref{tab:litcatcol}.
In Figure~\ref{fig:12logOHcomp} we see that the He2-03C3 gas-phase abundance estimator predicts lower values than the Si3-O3C3 estimator for Si3-O3C3 abundances of 12+log$_{10}$(O/H) $\gtrsim8$.
This confirms the results from the similar comparison performed by \citet[][their figure 11]{2020ApJ...893....1B}.
Also the average gas-phase abundance predicted for the collected source samples for the two predictors (12+log$_{10}$(O/H) $\approx7.6$ and $\approx8.2$ for He2-03C3 and Si3-O3C3, respectively) agrees with the estimates by \cite{2020ApJ...893....1B}.
%
We caution that it has been pointed out that the \heii{} emission line is potentially a problematic tracer of the gas-phase abundances, as it likely includes both nebular emission and emission from stellar winds as described by \citet[][]{2018ApJ...863...14B,2020ApJ...893....1B} among others. 
They however stress that He2-03C3 is a reliable metallicity tracer, particularly at low gas-phase abundances (12+log$_{10}$(O/H) $\lesssim 8$), where stellar contributions are minimal.
\cite{2020ApJ...893....1B} furthermore find that the ability for He2-03C3 to reproduce the abundance estimates based on optical lines depends on the photoionization grid used for individual sources, which also indicates that indeed multiple radiation processes contribute to the \heii{} emission.

\section{Physical parameter inference from photoionization models}\label{sec:pimodelinference}

Photoionization models are not only useful for probing the gas-phase abundances as just described and exemplified.
They also present predictions for a range of physical galaxy properties given theoretically predicted emission line fluxes.
Thus, by constraining the emission line fluxes and ratios between individual line species observationally, model comparisons can provide constraints on the emitting galaxies and their environment's likely physical properties. 
To infer the characteristics of the observed galaxy sample from MUSE studied here, 
we consider three suites of photoionization models, taken from the literature, that reproduce the nebular emission
from different ionizing sources (massive single and binary stars and AGN) obtained combining the
ionizing radiation field of these sources with the CLOUDY photoionization code
\citep[c13.03;][]{2013RMxAA..49..137F}. 
Specifically, the models we consider here are spectral models of
the nebular emission from gas ionized by single young and massive stars by \cite{2016MNRAS.462.1757G}, models
that include the contribution from binary stars by \cite{2018MNRAS.477..904X}, and the \cite{2016MNRAS.456.3354F} models which describe the emission of the gas ionized by an AGN. 
In the following we summarize the main features of these models relevant for our work.
For further details we refer to the individual papers describing the models.

\begin{table*}
\caption{\label{tab:photoionnizationparam}Sampled parameters of the suites of photoionization models used to produce PIM-PDFs}
\begin{center}
\resizebox{\textwidth}{!}{ 
\begin{tabular}{lcccc}
\hline\hline
Model 					& 								&NEOGAL AGN narrow-line regions.	& NEOGAL star-forming galaxies.   	& BPASS star-forming galaxies. 	\\ 
parameter					&								&Power law accretion disk.		& Single stars.					& Single stars and binaries.		\\
						& 								& \cite{2016MNRAS.456.3354F} 	& \cite{2016MNRAS.462.1757G} 	& \cite{2018MNRAS.477..904X}	\\
\hline
Ionization parameter 		& logU 							& $-5.0, -4.5, -4.0, -3.5, -3.0, $ 		&  $-4.0, -3.5, -3.0, -2.5, -2.0,$		& $-3.5$ to $-1.5$ in steps of 0.1 dex	\\
\vspace{0.2cm}
 						& 	 							& $-2.5, -2.0, -1.5, -1.0$ 			&  $ -1.5, -1.0$					& 				 			\\ 
\vspace{0.2cm}
Hydrogen number density 	& $\log_{10}(n_{\rm H}/{\rm cm}^{-3})$ 	& 2.0, 3.0, 4.0 					& 2.0, 3.0, 4.0					& 0.0, 0.5, 1.0, 1.5, 2.0, 2.5, 3.0		\\  
Relative gas metallicity 		& $Z$ 							& 0.0001, 0.0002, 0.0005, 0.001,  	& 0.0001, 0.0002, 0.0005, 0.001, 	& 1e$-5$, 1e$-4$, 0.001, 0.002,    		\\
(Z$_\odot$ = 0.01524)		& 								& 0.002, 0.004, 0.006, 0.008, 		& 0.002, 0.004, 0.006, 0.008, 		& 0.003, 0.004, 0.006, 0.008, 	 	\\
						& 								& 0.014, 0.01774, 0.03, 0.04, 		& 0.010, 0.014, 0.017, 0.020, 		& 0.010, 0.014, 0.020, 0.030, 	 	\\
\vspace{0.2cm}
						& 								& 0.05, 0.06, 0.07				& 0.030, 0.040					& 0.040						\\
\vspace{0.2cm}
Dust-to-metal mass ratio		& $\xi_\textrm{d}$ 					& 0.1, 0.3, 0.5 					& 0.1, 0.3, 0.5					& N/A	 					\\
Carbon to Oxygen ratio		& (C/O) / (C/O)$_\odot$				& N/A						& 0.10, 0.14, 0.20, 0.27, 0.38, 		& N/A						\\
\vspace{0.2cm}
$(C/O)_\odot=0.44$			&								& 							& 0.52, 0.72, 1.00, 1.40 			& 							\\
\vspace{0.2cm}
Upper mass cutoff of IMF		& $m_\textrm{up}/\mathrm{M}_{\sun}$ 	& N/A						&100, 300						& N/A						\\
\vspace{0.2cm}
Spectrum power law index 	& $\alpha$ 						& $-2.0, -1.7, -1.4, -1.2$ 			& N/A 						& N/A			 			\\
Stellar population age	 	& log$_{10}$(Age) [yr] 				& N/A			 			& 8.0$^\star$ 						& 6.0 to 8.0 in steps of 0.1 dex 		\\
\hline\hline
\end{tabular}
}
\tablefoot{The columns list the physical parameters sampled by the NEOGAL AGN narrow-line region, NEOGAL star-forming, and BPASS-based star-forming photoionization models used to infer the `PIM-PDFs' (PhotoIonization Model Probability Density Functions) shown in Figure~\ref{fig:pimodelPDFsObj} and described in Section~\ref{sec:pimodelinference}. 
The sampled values for each model are listed in each of the columns. $^\star$This is the effective age of the most recent episode of star formation in the NEOGAL galaxies, where the actual stellar population ages are between 0 and 10 Myr \citep{2016MNRAS.462.1757G}.
}
\end{center}
\end{table*}


The \cite{2016MNRAS.462.1757G} models, 
developed as part of the NEw frOntiers in Galaxy spectrAl modeLing (NEOGAL\footnote{\href{http://www.iap.fr/neogal/}{http://www.iap.fr/neogal/}}) project, 
compute the nebular emission from the gas in \ion{H}{ii} regions using
the latest version of the stellar population synthesis models of \cite{2003MNRAS.344.1000B} to be presented by Charlot
\& Bruzual (in prep.). These models incorporate updated stellar evolutionary tracks
\citep{2012MNRAS.427..127B}, including new prescriptions for the evolution of the most massive stars
in the Wolf-Rayet phase. 
The \cite{2016MNRAS.462.1757G} models provide the nebular emission from a whole galaxy,
parametrized in terms of ``galaxy-wide'' parameters, by convolving the spectral evolution of single,
ionization bound \ion{H}{ii} regions with a constant star formation history.
Here we use the publicly available photoionization models, each computed at 90 stellar population ages between 0 and 10~Myr, assuming a constant star formation rate for 100~Myr, resulting in this being the effective age of the most recent episode of star formation in the models \citep{2016MNRAS.462.1757G}.

To explore the effect of binary interactions, we consider the \cite{2018MNRAS.477..904X} photoionization models,
which provide the emission of  \ion{H}{ii} regions ionized by single or binary stellar populations treated as a single instantaneous starburst with a given age. 
These models are based on the Binary Population and Spectral Synthesis (BPASS\footnote{\href{https://bpass.auckland.ac.nz}{https://bpass.auckland.ac.nz}}) v2.1 models
\citep[][the latter referring to v2.2]{2016MNRAS.456..485S, 2017PASA...34...58E,2018MNRAS.479...75S}.  
We consider the available models for binary stellar populations of ages from $10^6$ to $10^8$ years.
By including the BPASS-based models we attempt to accommodate the influence and likely important effect binary stellar populations have on the amount and strength of the ionizing photons being produced by star formation.
In particular, for high-ionization lines like \civ{} and \heii{} it has been argued that binary stellar populations are needed, though not always enough, to produce the observed emission from non-AGN galaxies \citep[e.g.,][]{2016ApJ...826..159S,2018ApJ...869..123S,2019ApJ...878L...3B,2019A&A...624A..89N,2020arXiv200809780S}.

To also account for the potential contribution of an AGN, we consider the emission from the narrow-line
gas emitting regions in AGN computed by \cite{2016MNRAS.456.3354F} (also part of NEOGAL), where the ionizing radiation field is represented by a broken power law, with the UV spectral index $\alpha$ for $F_{\nu}\propto
\nu^{\alpha}$ between five and 1000 eV. 

Each of the models are parametrized in terms of physical quantities like ionization
parameter (log$_{10}$U) at the edge of the Str\"omgren sphere (i.e., the dimensionless ratio between the number of
ionizing photons and the total number of hydrogen atoms computed at the Str\"omgren radius), the
hydrogen density of the gas cloud ($\log_{10} n_{\rm H}$), the carbon-to-oxygen abundance ratio \cite[C/O; only sampled in the models by][]{2016MNRAS.462.1757G}, and the metallicity ($Z$), that is the mass fraction of all elements heavier than helium.
Hence, this metallicity estimate, which therefore includes both gas-phase and interstellar (dust-phase) metals, differs from the  gas-phase abundance metallicity (12+log$_{10}$(O/H)) discussed earlier.

The NEOGAL\footnote{\href{http://www.iap.fr/neogal/}{http://www.iap.fr/neogal/}} \citep{2016MNRAS.462.1757G, 2016MNRAS.456.3354F} models also include dust physics \citep[e.g.,][for grain physics in CLOUDY]{2004MNRAS.350.1330V} 
and a self-consistent treatment of metal abundances and dust depletion. 
The depletion of refractory metals onto dust grains is parametrized by means of the
dust-to-metal mass ratio ($\xi_{\rm d}$). 
We note that the treatment of dust and, therefore, of all
the related effects, like photon absorption and scattering, radiation pressure, collisional cooling,
photo-electric heating of the gas, and metal depletion, are not included in the \cite{2018MNRAS.477..904X} BPASS
models \citep[see, e.g.,][for the impact of dust on the emergent nebular
emission]{1995ApJ...454..807S, 2002ApJ...572..753D}. 
A subgrid of models of the nebular emission from gas ionized by binary stars (using the BPASSv2.2) which includes dust physics and depletion is presented by \cite{2019MNRAS.490..978P}.
As an overview of the various parameters used for the three suites of models considered here and to ease comparison between them, we summarize the sampled parameter spaces and their ranges in Table~\ref{tab:photoionnizationparam}. 
This table summarizes the grid of models that predict the line intensities of galaxies and AGN based on stellar populations with both single and binary stars.
As we searched the MUSE objects for \nv, \civ, \heii, \oiii, \siiii, and \ciii{} and therefore have constraints on all of these lines (when they fall in the observed rest-frame wavelength range), in principle the predictions of the photoionization models span a fifteen-dimensional space of possible flux ratios consisting of the line ratios 
\nv/\civ, 
\nv/\heii, 
\nv/\oiii, 
\nv/\siiii, 
\nv/\ciii, 
\civ/\heii, 
\civ/\oiii, 
\civ/\siiii, 
\civ/\ciii, 
\heii/\oiii, 
\heii/\siiii, 
\heii/\ciii, 
\oiii/\siiii, 
\oiii/\ciii, and
\siiii/\ciii.
Here we have combined the fluxes of the individual components of the UV emission line doublets, for instance, F(\ciii)~=~F(\ciiione)~+~F(\ciiitwo), to limit the number of dimensions spanned. 

\begin{figure*}
\begin{center}
\includegraphics[width=0.97\textwidth]{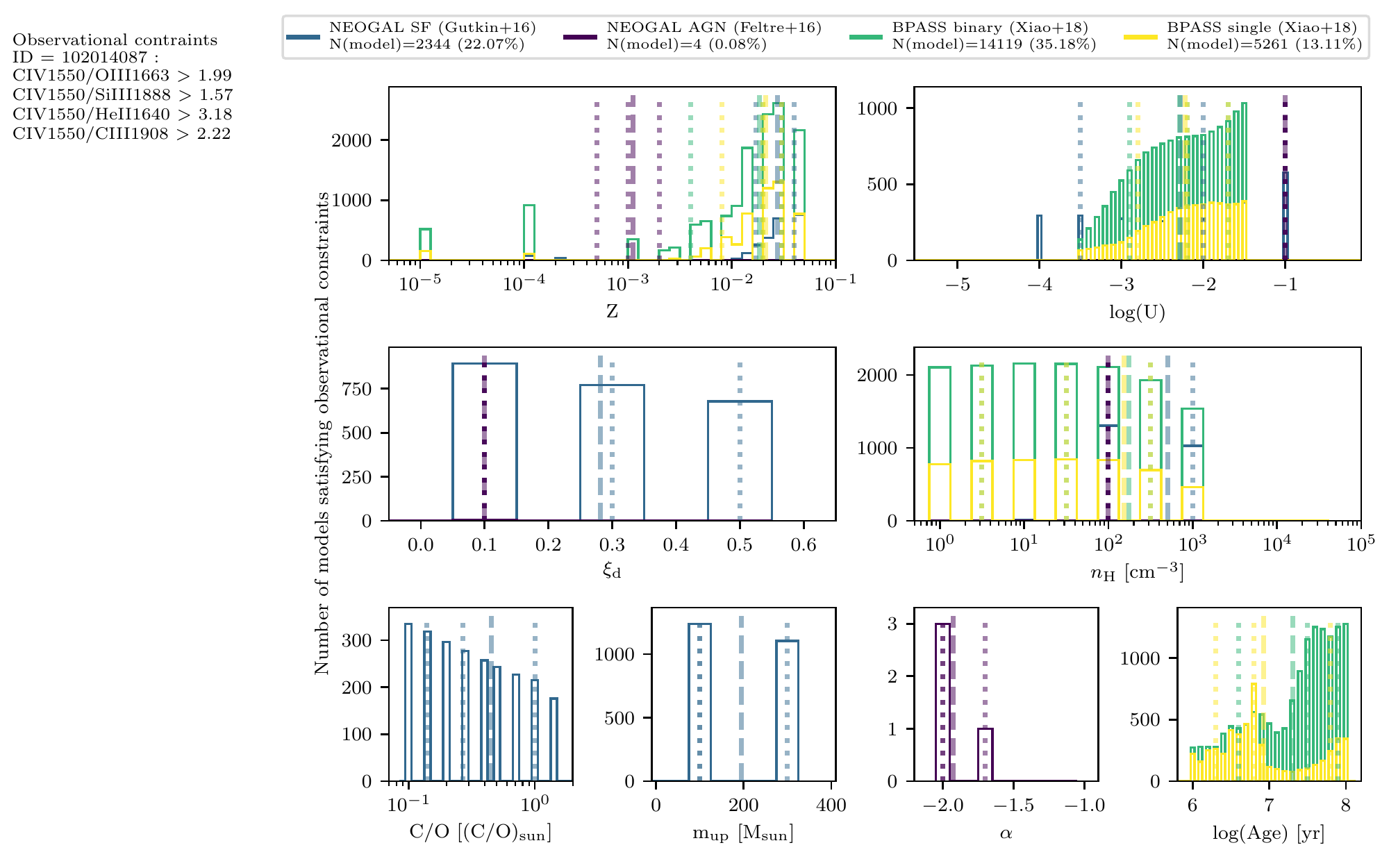}\\
\includegraphics[width=0.97\textwidth]{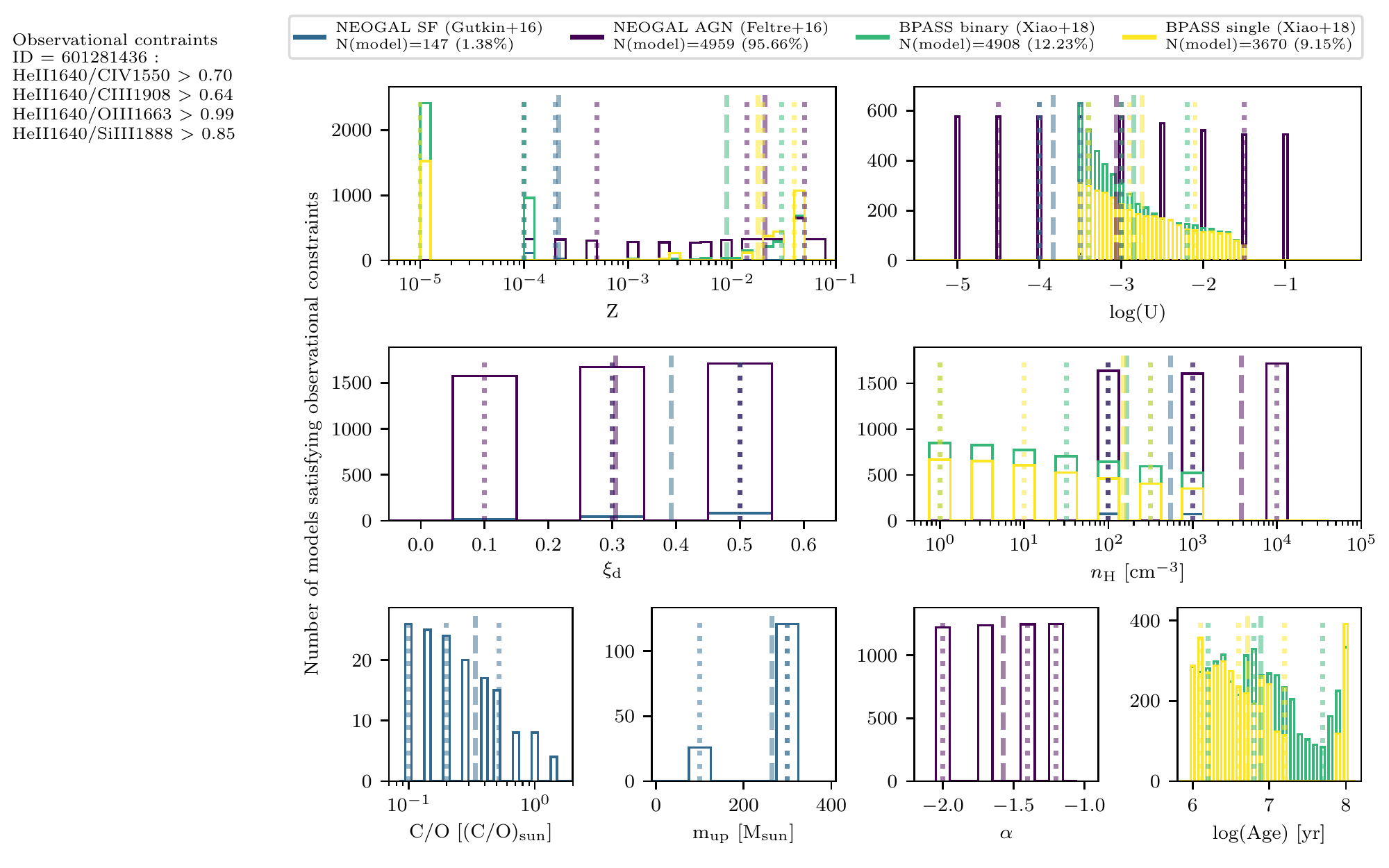}
\caption{Examples of PhotoIonization Model Probability Density Functions (PIM-PDFs) for the objects 102014087 (top) and 601281436 (bottom).
The histograms in the different panels show the distribution of AGN \citep[purple;][]{2016MNRAS.456.3354F}, 
star forming \citep[blue;][]{2016MNRAS.462.1757G},  
BPASS binary (green) and single star \citep[yellow;][]{2018MNRAS.477..904X} photoionization model parameters satisfying the observational emission line flux ratio constraints listed in the left margin. 
The spacing of the parameter histograms' bars indicate the discreteness of the sampling of each parameter space sampled by the models as listed in Table~\ref{tab:photoionnizationparam}.
The spectra of the two objects are shown in Figure~\ref{fig:ObjSpec02}.
The limits on the observed flux ratios are all 3$\sigma$.
The vertical dotted lines show the median values and the 68\% confidence intervals of the distributions. 
The vertical dashed lines mark the mean values.
For comparison the full unconstrained set of PIM-PDFs are shown in Figure~\ref{fig:pimodelPDFsNoConstraints}.}
\label{fig:pimodelPDFsObj}
\end{center}
\end{figure*}

Being projections of the multidimensional model grid, 2D explorations and projections of line ratios do not provide a robust way of inferring the physical parameters of the objects considered and fail to convey the full information and constraints available for them.
This can to some extent be remedied by ``marginalizing'' over individual parameters to lower the dimensionality of the model parameter space explored, for instance, by assuming a fixed carbon-to-oxygen ratio, hydrogen number density, or ionization parameter of the models to display.    
However, to condense the full information of the photoionization models we here present a new approach, where we explore the distributions of the model parameters of the full multidimensional photoionization model grid points (see Table~\ref{tab:photoionnizationparam}) and their flux ratio predictions.
We assume flat priors on the individual parameters for all galaxies, and let the data tell us what the most likely best-fit models are.
The sample of best-fit models agreeing within, for instance, 3$\sigma$ of the measured emission line fluxes (limits) provide a distribution of the model parameters sampled by this subset of models.
We refer to these parameter distributions as PIM-PDFs (PhotoIonization Model Probability Density Functions) indicating that they resemble the ``probability density functions'' of the actual physical parameters of the individual objects given the measured constraints on the observed UV emission line flux ratios. 
Two examples of PIM-PDFs are shown in Figure~\ref{fig:pimodelPDFsObj} (for reference and comparison Appendix~\ref{sec:pimodels} shows the ``full'' PIM-PDFs for the NEOGAL and BPASS-based photoionization models in Figure~\ref{fig:pimodelPDFsNoConstraints}). 
The PIM-PDF approach is different from the inferences performed by codes like the BayEsian Analysis of GaLaxy sEds \citep[BEAGLE][]{2016MNRAS.462.1415C} and Prospector \citep{2017ApJ...837..170L,2021ApJS..254...22J}. These codes sample the true PDF through Bayesian inference and Markov chain Monte Carlo for each object analyzed. Hence, these codes are optimizing the parameter grid as opposed to the PIM-PDFs which by construction just select among the available precomputed models and base the resulting distribution on these estimates. This approach is what makes PIM-PDFs more efficient when evaluating large samples of objects, as each object is not independently forced through a time-consuming optimization process.

The first object (102014087; top panel) is a case where the observational constraints (listed in the left margin) are more easily reproduced by models of ionizing photons from star formation as only 0.08\% of the available AGN models can reproduce the observed emission line flux ratios.
The set of models including contribution from binary stars (green PIM-PDFs) have a higher fraction of solutions that are capable of reproducing the observations.
The resulting PIM-PDFs predict that the object 102014087 likely has a (relative) metallicity of roughly a few times 10$^{-2}$ \citep[$Z_\odot=0.01524$;][]{2016MNRAS.462.1757G,2016MNRAS.456.3354F} and an ionization parameter that is $\gtrsim10^{-3}$. 
The stellar populations of the BPASS models prefer ages above $\approx10^7$ years.
The hydrogen number density and the additional parameters sampled by the NEOGAL model grids are essentially  unconstrained for the models that are able to reproduce the observations of 102014087, even though the preferred C/O ratio appears to be small rather than large. 

The observational constraints for the second object shown in Figure~\ref{fig:pimodelPDFsObj} (601281436; bottom panel) have a larger set of models with an AGN as ionizing source capable of reproducing the observational constraints, as essentially all the NEOGAL AGN models (purple) can reproduce the observations (and therefore poorly constrain the physical parameters), whereas the NEOGAL star formation model grids (blue) are struggling as only 1.38\% of them can reproduce the measured flux ratios. The few models that can have low C/O and a high mass-function cutoff.
A subsample of roughly 10\% of the BPASS-based models with preferentially small $n_\textrm{H}$ and age, a relative  metallicity of $10^{-5}$ and log$_{10}$(U)~$\lesssim-2.5$ are also capable of reproducing the observational constraints. 
The spectra of both 102014087 and 601281436 are shown in Figure~\ref{fig:ObjSpec02}.

\begin{figure*}
\begin{center}
\includegraphics[width=0.60\textwidth]{mainlegend_nolit.png}\\
\includegraphics[width=0.99\textwidth]{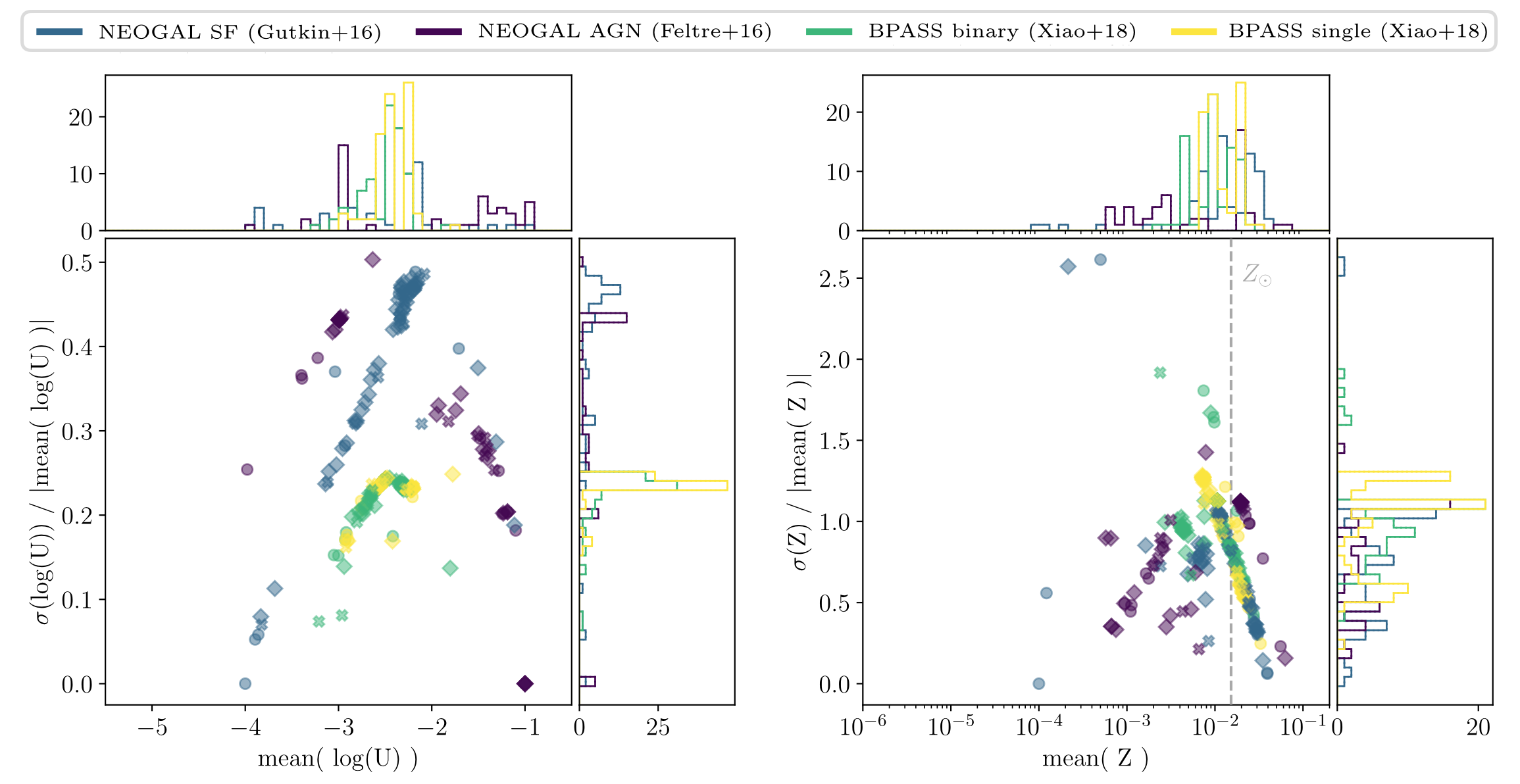}
\caption{Mean value (x-axis) and normalized standard deviation (y-axis) of the ionization parameter (log$_{10}$(U)) and the metallicity (Z) PIM-PDFs for the MUSE-Wide, the UDF mosaic, and the UDF10 objects studied in this work.
The blue points represent the distribution of star forming models \citep{2016MNRAS.462.1757G} matching the observations, the purple points indicate AGN models \citep{2016MNRAS.456.3354F}, the green points show models based on BPASS binaries \citep{2018MNRAS.477..904X}, and the yellow points represent models of BPASS single stars \citep{2018MNRAS.477..904X}.
The solid histograms show the distribution of the individual parameters.
The vertical dashed line in the right panel marks the solar metallicity $Z_\odot=0.01524$.}
\label{fig:pimodelPDFsDist}
\end{center}
\end{figure*}

We produced the PIM-PDFs for all studied objects with flux ratio constraints, that is the 103 objects with at least one UV emission line detected. Among these objects 62 (for 23 of these sources the line has S/N(FELIS)~$ >5$) only have a single line detected, resulting in an inference based on flux ratio limits only.  
By recording the mean values and the standard deviation of the PIM-PDFs for all objects in our sample, we can estimate the distribution of the most likely physical (model) parameters for the objects detected in our study.
Figure~\ref{fig:pimodelPDFsDist} presents these statistics for the ionization parameter (log$_{10}$(U), left panel) and  metallicity ($Z$, right panel) for the full sample.
Points at the top are generally unconstrained given the large relative width of the PIM-PDFs ($\sigma/\left|\textrm{mean}\right|$), whereas points towards the lower parts of each panel provide reliable estimates of the sampled parameter.
For the distributions of the ionization parameter in the left panel the preferred values appear to scatter around $\log_{10}(\textrm{U})\approx -2.5$.
In the right panel we see that the full sample of PIM-PDFs predicts systems (of star-forming models) that span a range in metallicity from roughly $5\times10^{-3}$ to $3\times10^{-2}$ (corresponding to $0.32 \lesssim Z/Z_\odot \lesssim 1.97$) with a mean value of approximately solar at $Z_\odot=0.01524$ indicated by the vertical gray dashed line.
As mentioned earlier, \cite{2021MNRAS.501.3238T} argue that $\textrm{EW}(\ciii)>10$~\AA{} mostly requires $\log_{10}(\textrm{U}) \lesssim -2.5$ and metallicity of $0.2~Z_{\odot}$ or lower.
If we add the results from the PIM-PDFs based on the emission line fluxes from the collection of UV line emitters from the literature, the distribution of the best-fit log$_{10}$(U) stays roughly the same, whereas the $Z$ distribution shifts to slightly lower values with a mean of subsolar metallicity at $Z\approx10^{-2}$ ($Z/Z_\odot\approx0.7$).
Hence, the PIM-PDFs predict the larger sample of MUSE galaxies studied here to have higher metallicities compared to metallicity estimates of samples at similar redshifts in the literature \citep[e.g.,][]{2008A&A...488..463M,2020MNRAS.491.1427S}.

We performed similar comparisons for the remaining parameters sampled by the photoionization models listed in Table~\ref{tab:photoionnizationparam}. 
The MUSE sample as well as the sample from the literature (Appendix~\ref{sec:litcol}) prefer values of $m_\text{up}=200~M_{\odot}$, $\xi_\textrm{d}\approx0.3$ and $\alpha\approx-1.6$ from the best-fit NEOGAL models.
The BPASS-based model's best-fit age distributions peak at the central value of $10^7$ years for both the MUSE and literature sample.
The C/O abundance ratio from the \cite{2016MNRAS.462.1757G} models spans the full range of sampled values for the MUSE objects with a peak around the solar value $(\textrm{C/O})_\odot = 0.44$. The sample of emitters from the literature more strongly favors C/O ratios around solar.
The distributions of the neutral hydrogen number density for the combined sample PIM-PDFs of the BPASS and NEOGAL star formation models peak at values between 10$^2$~cm$^{-3}$ and 10$^3$~cm$^{-3}$.
The solutions from the BPASS-based models generally predict densities lower than the NEOGAL models, which reflects the different sampling ranges of the two different suites of models as the means of the sampled ranges are 10$^{1.5}$~cm$^{-3}$ and 10$^3$~cm$^{-3}$ for the BPASS and NEOGAL models, respectively.

Even though the inferences on the physical parameters of the individual objects and the galaxy samples using the PIM-PDFs presented here estimates the characteristics of the UV line emitters, it should be stressed that these predictions should be considered only indicative of the true intrinsic values.
First, there are uncertainties and assumptions in producing the photoionization models resulting from the sampled parameter grids.
Second, the recent study by \cite{2020arXiv200611387F} presents spatially resolved estimates of physical parameters based on comparisons between rest-frame optical lines and photoionization grids of two lensed (and hence highly resolved) sources from HST grism spectroscopy.
They show and discuss the challenges in inferring metallicity and ionization parameters based on such grids, both in terms of uncertainties in the models, but also in terms of intra-object differences between, for instance, individual star-forming regions and integrated estimates of line ratios.
This of course also applies to the inferences on the photoionization model parameters obtained for the objects studied here, where all quantities are integrated over the whole galaxy given the nonlensed nature of the objects.

Another limitation of this approach is the nonuniform and in some cases coarse sampling of the physical parameters considered. 
This is done to limit the number of models to generate when calculating the emission line flux predictions when producing the photoionization model outputs.
However, this means that well-constrained line fluxes (with small uncertainties) translate into small regions of parameter space being allowed by observations. In those cases, it might very well be that no models are able to reproduce the observations to within 3$\sigma$ (as used here) of the measured line flux ratios. 
This does not mean that the models fail, but rather that they are not sampled finely enough to reflect the precision of the measurement. This issue of course scales with the number of well-constrained lines. A way to solve this is to either increase the considered multidimensional region of parameter space used to generate the PIM-PDFs, produce new more finely sampled models, or consider only the single best-fit model (in terms of $\chi^2$). 
Alternatively, dedicated machinery to infer physical parameters of individual objects, like for instance BEAGLE, can be used.
A detailed comparison between the PIM-PDF approach presented here and other methods, is beyond the scope of the current work, but will likely prove useful and insightful when exploring further the physical properties of individual galaxies from the MUSE and literature samples in the future.

Lastly, more fundamental challenges with the considered photoionization models also provide empty PIM-PDFs when attempting to match observational constraints. 
An example of one such challenge or limitation to the models in general, and therefore also to the PIM-PDF approach, is \heii{} emitted in star-forming galaxies.
As we noted in Section~\ref{sec:logOH}, \heii{} emission can contain contributions from both nebular emission and stellar winds \citep[e.g.,][]{2012MNRAS.421.1043S,2019A&A...624A..89N,2018ApJ...863...14B,2020ApJ...893....1B}, and producing enough high-energy photons to produce the \heii{} emission has been proven challenging.
Binaries (for example, as implemented in the BPASS models) have in some cases relieved part of the tension \citep{2016ApJ...826..159S}, but they are often not enough. 
It has therefore been proposed and debated that the inclusion of X-ray binaries \citep{2019A&A...622L..10S,2020MNRAS.494..941S} or energetic shocks \citep{2008ApJS..178...20A,2017MNRAS.472.2608S,2019MNRAS.490..978P} could provide means for reproducing the observed \heii{} fluxes and other lines produced by high-energy ionizing photons like \nv{} and \civ{} within photoionization model frameworks.

Despite these limitations and challenges with inferring physical properties from photoionization models based on a set of observation constraints, the PIM-PDFs presented here provide a simple and fast way to obtain photoionization model predictions for large samples of objects including all available information and avoiding marginalization over individual parameters or projections onto lower-dimensional parameter spaces.

\section{Conclusions}
\label{sec:conc}

We have presented a comprehensive search for rest-frame UV emission lines in a parent sample of 2052 emission line galaxies detected with the MUSE integral field spectrograph in the COSMOS, GOODS-South, and HUDF as part of the MUSE-Wide and MUSE-Deep GTO surveys (Section~\ref{sec:data}). 
The studied objects were selected to have redshifts above 1.5 based on the emission feature(s) identified during the untargeted emission line search in the MUSE data cubes.
This resulted in 1997 LAEs at $z>2.9$ and 55 objects in the so-called MUSE redshift desert at $1.5<z<2.9$ where neither \oiifull{} nor \lya{} are available for source redshift identification (Section~\ref{sec:objsel}). 
For each of the objects in the parent sample, we extracted 1D spectra from the 3D MUSE data cubes optimized in both flux and S/N using the software TDOSE (Section~\ref{sec:1Dspec}). 
These spectra were searched for rest-frame UV emission lines red-wards of \lya{} using Gaussian emission line template matching using the tool FELIS (presented as part of this paper in Section~\ref{sec:UVEmissionLineSearch} and Appendix~\ref{sec:felis}).
We visually vetted all potential detections of the UV emission lines \nv, \civ, \heii, \oiii, \siiii, and \ciii{} and found 54 line emitters with 3$\sigma$ detections among the 1997 LAEs.
Including the sample in the redshift desert this number increases to 103 line emitters.
Table~\ref{tab:UVESdetections} summarizes the complete sample of rest-frame UV detections from our search which are made publicly available with this paper (Appendix~\ref{sec:mastercat}).
To further improve the statistics of the assessments and for comparison purposes, we complemented the main sample with an extensive collection of UV emission line flux measurements from the literature which are also made available with this work (Appendix~\ref{sec:litcol}).

Based on the UV emission lines and complementary measurements of LAE characteristics from \cite{Kerutt:2021tr}, we explored the range of physical parameters of the galaxies in our main sample.
Our main conclusions can be summarized as follows:
\begin{itemize}
\item The fraction of objects with detected rest-frame UV emission lines grows with increasing depth of the data. 
This implies a relative change in the shapes (slopes) of the luminosity function of the parent sample and the luminosity function of the UV line emitter subsamples (Section~\ref{sec:UVEmissionLineSearch}).

\item The strengths of \heii, \oiii, and \siiii{} correlate with the flux of the \ciii{} emission as parametrized by the correlations presented in Table~\ref{tab:paramcorr}.

\item Based on the FELIS Gaussian template matches we determine the relative strength of the UV emission line doublet component flux ratios. Here we find that $F$(\oiiione)/$F$(\oiiitwo), $F$(\siiiione)/$F$(\siiiitwo), and $F$(\ciiione)/$F$(\ciiitwo) are all in agreement with expected theoretical values for the vast majority of objects (Section~\ref{sec:UVEmissionLineSearch} and Table~\ref{tab:paramcorr}).

\item We determine EW$_0$ for all identified UV emission lines, and similar to the flux measurements we find significant correlations between them for the \heii, \oiii, \siiii, and \ciii{} emission lines (Section~\ref{sec:EWcoor}).

\item We find correlations between EW$_0$(\heii), EW$_0$(\siiii), and EW$_0$(\ciii) and EW$_0$(\lya).
The majority of the LAEs with \ciii{} detected have $\textrm{EW}_0(\textrm{\ciii})\approx 0.22\pm0.18\times\textrm{EW}_0(\textrm{\lya})$ for $10\lesssim\textrm{EW}_0(\textrm{\lya})/\textrm{\AA}\lesssim100$ which is in agreement with previous studies (Section~\ref{sec:UVandLya}).

\item Considering the subsample of LAEs only, we explored correlations between multiple LAE characteristics including EW$_0$(\lya), M$_\textrm{UV}$, spectral slope, and LAE effective radius and found no prominent relationships. 
We do however see potential correlations between \lya{} luminosity and the UV emission lines indicating that objects with lower \lya{} luminosity have larger UV emission line EW$_0$ estimates (Section~\ref{sec:UVandLya}).

\item The detection of multiple rest-frame emission lines enabled us to assess the velocity offset $\Delta v$ of resonant lines like \lya{} and \civ{} with respect to the systemic redshift probed by, for instance, \ciii{} (Section~\ref{sec:voffset}).
In agreement with previous measurements we find $\Delta v_\textrm{\lya}$ of 250-500~km~s$^{-1}$.
Again checking for correlations with LAE characteristics we find broad agreement between theoretical and empirical relations between $\Delta v_\textrm{\lya}$, M$_\textrm{UV}$, and $\textrm{EW}_0(\textrm{\lya})$, despite a large scatter in the parameters for the objects studied here.
In addition we confirm the empirical relation between the \lya{} line width, \lya{} peak separation, and $\Delta v_\textrm{\lya}$ even though a large scatter in the individual correlations is present. 
For the resonant emission of \civ{} we find that $\Delta v_\textrm{\civ} \lesssim 250$~km~s$^{-1}$ with a few outliers caused by the limitations of trying to model \civ{} P-Cygni profiles from combined absorption and emission by a pair of Gaussians.

\item For objects with detected \ciii{} and \siiii{} the estimated electron density from the doublet flux ratios is generally $n_e<10^5$~cm$^{-3}$ (Section~\ref{sec:neTe}).

\item Using the fitting formulas by \cite{2020ApJ...893....1B}, we show that the sample of objects with simultaneous detections of \heii, \oiii, \siiii, and \ciii{} have subsolar gas-phase abundances of 12+log$_{10}$(O/H)~$\approx 8$. 
We do not find any trends with redshift or EW$_0$(\ciii) for the estimated gas-phase abundances. 
In agreement with \cite{2020ApJ...893....1B} we find that the tracer including \heii, \oiii, and \ciii{} in most cases results in lower metallicity estimates than the tracer based on \oiii, \siiii, and \ciii{} (Section~\ref{sec:logOH}).

\item Finally, we present a new approach to condense information from physical parameter grids of photoionization models taking the full amount of information into account without marginalizing over individual parameters. 
We refer to the resulting distribution of model parameters that are able to reproduce the observational constraints from the UV emission lines as PIM-PDFs (PhotoIonization Model Probability Density Functions; Section~\ref{sec:pimodelinference}).
The PIM-PDFs provide the distribution of the best-fit (integrated) model parameters of individual sources given the observational constraints on the UV emission line fluxes and flux ratios obtained from the FELIS template matches.
We show that the general sample of emitters recovered from the MUSE data have an average ionization parameter $\log_{10}(\textrm{U})\approx-2.5$ and a mean metallicity of order solar, though individual objects span the range  $5\times10^{-3} \lesssim Z \lesssim 3\times10^{-2}$ corresponding to roughly $0.32 \lesssim Z/Z_\odot \lesssim 1.97$ which is at the high end when comparing to previous estimates from the literature.

\end{itemize}
In summary, with the large sample of emission line sources recovered from the MUSE GTO surveys, we have demonstrated the wealth of information and physical properties that rest-frame UV emission features red-wards of \lya{} probe.
Apart from gaining insight into the individual galaxies and samples themselves, the rest-frame UV emission lines also provide promising probes of high-redshift galaxies at epochs where the \lya{} emission is strongly affected and absorbed by the increasing neutral CGM and IGM.
Especially in light of upcoming near-infrared missions like the James Webb Space Telescope (JWST) and the Nancy Grace Roman Space Telescope (formerly known as WFIRST), further exploration of the characteristics of rest-frame UV emission line sources can serve as benchmarks, links to lower redshift, and means of comparison and redshift evolution assessment.

\begin{acknowledgements}
We would like to thank Charlotte Mason for useful discussions and for providing the data for the curves shown in Figure~\ref{fig:dvCIIIvsMUV} and Dawn Erb for providing the observational data for the comparison sample studied by \cite{2014ApJ...795..165S}, also shown in Figure~\ref{fig:dvCIIIvsMUV}.
%
%
This work has been supported by the BMBF grant 05A14BAC 
and we acknowledge support by the Competitive Fund of the Leibniz Association through grant SAW-2015-AIP-2.
AF acknowledges the support from grant PRIN MIUR2017-20173ML3WW\_001.
JS acknowledges the support from Vici grant 639.043.409 from the Dutch Research Council (NWO).
GM received funding from the European Union's Horizon 2020 research and innovation programme under the Marie Sklodowska-Curie grant agreement No MARACAS - DLV-896778.
%
%
This paper is
based on observations collected at the European Organisation for Astronomical Research in the Southern Hemisphere under ESO programmes 
094.A-0289(B),   
095.A-0010(A),   
096.A-0045(A),   
096.A-0045(B),   
094.A-0205,  
095.A-0240,  
096.A-0090,   
097.A-0160,   
and 
098.A-0017.  
This paper also makes use of observations made with the NASA/ESA Hubble Space Telescope obtained at STScI.

This research made use of the following programs and open-source packages for Python and we are thankful to their developers:
DS9 \citep{2003ASPC..295..489J},
Astropy \citep{2013A&A...558A..33A,2018AJ....156..123A},
APLpy \citep{2012ascl.soft08017R},
iPython \citep{Perez:2007hy},
numpy \citep{vanderWalt:2011dp},
matplotlib \citep{Hunter:2007ux}, 
and
SciPy \citep{SciPyOpensources:tUcReTVZ}
%
\end{acknowledgements}

%
%
\bibliographystyle{aa}
\bibliography{paperslibrary}

\begin{thebibliography}{236}
\expandafter\ifx\csname natexlab\endcsname\relax\def\natexlab#1{#1}\fi

\bibitem[{Adelberger {et~al.}(2003)Adelberger, Steidel, Shapley, \&
  Pettini}]{2003ApJ...584...45A}
Adelberger, K.~L., Steidel, C.~C., Shapley, A.~E., \& Pettini, M. 2003, The
  Astrophysical Journal, 584, 45

\bibitem[{Allen {et~al.}(2008)Allen, Groves, Dopita, Sutherland, \&
  Kewley}]{2008ApJS..178...20A}
Allen, M.~G., Groves, B.~A., Dopita, M.~A., Sutherland, R.~S., \& Kewley, L.~J.
  2008, The Astrophysical Journal Supplement Series, 178, 20

\bibitem[{Amor{\'\i}n {et~al.}(2017)Amor{\'\i}n, Fontana, P{\'e}rez-Montero,
  Castellano, Guaita, Grazian, F{\`e}vre, Ribeiro, Schaerer, Tasca, Thomas,
  Bardelli, Cassar{\`a}, Cassata, Cimatti, Contini, Barros, Garilli,
  Giavalisco, Hathi, Koekemoer, Le~Brun, Lemaux, Maccagni, Pentericci, Pforr,
  Talia, Tresse, Vanzella, Vergani, Zamorani, Zucca, \&
  Merlin}]{2017NatAs...1E..52A}
Amor{\'\i}n, R., Fontana, A., P{\'e}rez-Montero, E., {et~al.} 2017, Nature
  Astronomy, 1, 0052

\bibitem[{Astropy~Collaboration {et~al.}(2018)Astropy~Collaboration,
  Price-Whelan, Sip{\H o}cz, G{\"u}nther, Lim, Crawford, Conseil, Shupe, Craig,
  Dencheva, Ginsburg, VanderPlas, Bradley, P{\'e}rez-Su{\'a}rez, de~Val-Borro,
  Aldcroft, Cruz, Robitaille, Tollerud, Ardelean, Babej, Bach, Bachetti,
  Bakanov, Bamford, Barentsen, Barmby, Baumbach, Berry, Biscani, Boquien,
  Bostroem, Bouma, Brammer, Bray, Breytenbach, Buddelmeijer, Burke, Calderone,
  Cano~Rodr{\'\i}guez, Cara, Cardoso, Cheedella, Copin, Corrales, Crichton,
  D'Avella, Deil, Depagne, Dietrich, Donath, Droettboom, Earl, Erben, Fabbro,
  Ferreira, Finethy, Fox, Garrison, Gibbons, Goldstein, Gommers, Greco,
  Greenfield, Groener, Grollier, Hagen, Hirst, Homeier, Horton, Hosseinzadeh,
  Hu, Hunkeler, Ivezi{\'c}, Jain, Jenness, Kanarek, Kendrew, Kern, Kerzendorf,
  Khvalko, King, Kirkby, Kulkarni, Kumar, Lee, Lenz, Littlefair, Ma, Macleod,
  Mastropietro, McCully, Montagnac, Morris, Mueller, Mumford, Muna, Murphy,
  Nelson, Nguyen, Ninan, N{\"o}the, Ogaz, Oh, Parejko, Parley, Pascual, Patil,
  Patil, Plunkett, Prochaska, Rastogi, Reddy~Janga, Sabater, Sakurikar,
  Seifert, Sherbert, Sherwood-Taylor, Shih, Sick, Silbiger, Singanamalla,
  {Singer, L. P.}, Sladen, Sooley, Sornarajah, Streicher, Teuben, Thomas,
  Tremblay, Turner, Terr{\'o}n, van Kerkwijk, de~la Vega, Watkins, Weaver,
  Whitmore, Woillez, Zabalza, \& Contributors}]{2018AJ....156..123A}
Astropy~Collaboration, T., Price-Whelan, A.~M., Sip{\H o}cz, B.~M., {et~al.}
  2018, The Astronomical Journal, 156, 123

\bibitem[{Astropy~Collaboration {et~al.}(2013)Astropy~Collaboration,
  Robitaille, Tollerud, Greenfield, Droettboom, Bray, Aldcroft, Davis,
  Ginsburg, Price-Whelan, Kerzendorf, Conley, Crighton, Barbary, Muna,
  Ferguson, Grollier, Parikh, Nair, Unther, Deil, Woillez, Conseil, Kramer,
  Turner, Singer, Fox, Weaver, Zabalza, Edwards, Azalee~Bostroem, Burke, Casey,
  Crawford, Dencheva, Ely, Jenness, Labrie, Lim, Pierfederici, Pontzen, Ptak,
  Refsdal, Servillat, \& Streicher}]{2013A&A...558A..33A}
Astropy~Collaboration, T., Robitaille, T.~P., Tollerud, E.~J., {et~al.} 2013,
  Astronomy and Astrophysics, 558, A33

\bibitem[{Bacon {et~al.}(2010)Bacon, Accardo, Adjali, Anwand, Bauer, Biswas,
  Blaizot, Boudon, Brau-Nogue, Brinchmann, Caillier, Capoani, Carollo, Contini,
  Couderc, Daguis{\'e}, Deiries, Delabre, Dreizler, Dubois, Dupieux, Dupuy,
  Emsellem, Fechner, Fleischmann, Fran{\c c}ois, Gallou, Gharsa, Glindemann,
  Gojak, Guiderdoni, Hansali, Hahn, Jarno, Kelz, Koehler, Kosmalski, Laurent,
  Le~Floch, Lilly, Lizon, Loupias, Manescau, Monstein, Nicklas, Olaya, Pares,
  Pasquini, Pecontal-Rousset, Pello, Petit, Popow, Reiss, Remillieux, Renault,
  Roth, Rupprecht, Serre, Schaye, Soucail, Steinmetz, Streicher, Stuik,
  Valentin, Vernet, Weilbacher, Wisotzki, \& Yerle}]{2010SPIE.7735E..08B}
Bacon, R., Accardo, M., Adjali, L., {et~al.} 2010, in Proceedings of the SPIE,
  ed. I.~S. McLean, S.~K. Ramsay, \& H.~Takami, Ctr. de Recherche Astrophysique
  de Lyon, CNRS, Univ. Claude-Bernard Lyon I, France (SPIE), 773508

\bibitem[{Bacon {et~al.}(2017)Bacon, Conseil, Mary, Brinchmann, Shepherd,
  Akhlaghi, Weilbacher, Piqueras, Wisotzki, Lagattuta, Epinat, Guerou, Inami,
  Cantalupo, Courbot, Contini, Richard, Maseda, Bouwens, Bouch{\'e},
  Kollatschny, Schaye, Marino, Pell{\'o}, Herenz, Guiderdoni, \&
  Carollo}]{2017A&A...608A...1B}
Bacon, R., Conseil, S., Mary, D., {et~al.} 2017, Astronomy and Astrophysics,
  608, A1

\bibitem[{Bacon {et~al.}(2021)Bacon, Mary, Garel, Blaizot, Maseda, Schaye,
  Wisotzki, Conseil, Brinchmann, Leclercq, Abril-Melgarejo, Boogaard,
  Bouch{\'e}, Contini, Feltre, Guiderdoni, Herenz, Kollatschny, Kusakabe,
  Matthee, Michel-Dansac, Nanayakkara, Richard, Roth, Schmidt, Steinmetz,
  Tresse, Urrutia, Verhamme, Weilbacher, Zabl, \&
  Zoutendijk}]{2021A&A...647A.107B}
Bacon, R., Mary, D., Garel, T., {et~al.} 2021, Astronomy and Astrophysics, 647,
  A107

\bibitem[{Banados {et~al.}(2018)Banados, Venemans, Mazzucchelli, Farina,
  Walter, Wang, Decarli, Stern, Fan, Davies, Hennawi, Simcoe, Turner, Rix,
  Yang, Kelson, Rudie, \& Winters}]{2018Natur.553..473B}
Banados, E., Venemans, B.~P., Mazzucchelli, C., {et~al.} 2018, Nature, 553, 473

\bibitem[{Bayliss {et~al.}(2014)Bayliss, Rigby, Sharon, Wuyts, Florian,
  Gladders, Johnson, \& Oguri}]{2014ApJ...790..144B}
Bayliss, M.~B., Rigby, J.~R., Sharon, K., {et~al.} 2014, The Astrophysical
  Journal, 790, 144

\bibitem[{Beckwith {et~al.}(2006)Beckwith, Stiavelli, Koekemoer, Caldwell,
  Ferguson, Hook, Lucas, Bergeron, Corbin, Jogee, Panagia, Robberto, Royle,
  Somerville, \& Sosey}]{2006AJ....132.1729B}
Beckwith, S. V.~W., Stiavelli, M., Koekemoer, A.~M., {et~al.} 2006, The
  Astronomical Journal, 132, 1729

\bibitem[{Berg {et~al.}(2019{\natexlab{a}})Berg, Chisholm, Erb, Pogge, Henry,
  \& Olivier}]{2019ApJ...878L...3B}
Berg, D.~A., Chisholm, J., Erb, D.~K., {et~al.} 2019{\natexlab{a}}, The
  Astrophysical Journal Letters, 878, L3

\bibitem[{Berg {et~al.}(2018)Berg, Erb, Auger, Pettini, \&
  Brammer}]{2018ApJ...859..164B}
Berg, D.~A., Erb, D.~K., Auger, M.~W., Pettini, M., \& Brammer, G.~B. 2018, The
  Astrophysical Journal, 859, 164

\bibitem[{Berg {et~al.}(2019{\natexlab{b}})Berg, Erb, Henry, Skillman, \&
  McQuinn}]{2019ApJ...874...93B}
Berg, D.~A., Erb, D.~K., Henry, R. B.~C., Skillman, E.~D., \& McQuinn, K. B.~W.
  2019{\natexlab{b}}, The Astrophysical Journal, 874, 93

\bibitem[{Berg {et~al.}(2016)Berg, Skillman, Henry, Erb, \&
  Carigi}]{2016ApJ...827..126B}
Berg, D.~A., Skillman, E.~D., Henry, R. B.~C., Erb, D.~K., \& Carigi, L. 2016,
  The Astrophysical Journal, 827, 126

\bibitem[{Bian {et~al.}(2018)Bian, Kewley, \& Dopita}]{2018ApJ...859..175B}
Bian, F., Kewley, L.~J., \& Dopita, M.~A. 2018, The Astrophysical Journal, 859,
  175

\bibitem[{Brada{\v c} {et~al.}(2017)Brada{\v c}, Garcia-Appadoo, Huang,
  Vallini, Quinn~Finney, Hoag, Lemaux, Borello~Schmidt, Treu, Carilli,
  Dijkstra, Ferrara, Fontana, Jones, Ryan, Wagg, \&
  Gonzalez}]{2017ApJ...836L...2B}
Brada{\v c}, M., Garcia-Appadoo, D., Huang, K.-H., {et~al.} 2017, The
  Astrophysical Journal Letters, 836, L2

\bibitem[{Bressan {et~al.}(2012)Bressan, Marigo, Girardi, Salasnich, Dal~Cero,
  Rubele, \& Nanni}]{2012MNRAS.427..127B}
Bressan, A., Marigo, P., Girardi, L., {et~al.} 2012, Monthly Notices of the
  Royal Astronomical Society, 427, 127

\bibitem[{Brinchmann {et~al.}(2004)Brinchmann, Charlot, White, Tremonti,
  Kauffmann, Heckman, \& Brinkmann}]{2004MNRAS.351.1151B}
Brinchmann, J., Charlot, S., White, S. D.~M., {et~al.} 2004, Monthly Notices of
  the Royal Astronomical Society, 351, 1151

\bibitem[{Bruzual \& Charlot(2003)}]{2003MNRAS.344.1000B}
Bruzual, G. \& Charlot, S. 2003, Monthly Notices of the Royal Astronomical
  Society, 344, 1000

\bibitem[{Byler {et~al.}(2017)Byler, Dalcanton, Conroy, \&
  Johnson}]{2017ApJ...840...44B}
Byler, N., Dalcanton, J.~J., Conroy, C., \& Johnson, B.~D. 2017, The
  Astrophysical Journal, 840, 44

\bibitem[{Byler {et~al.}(2018)Byler, Dalcanton, Conroy, Johnson, Levesque, \&
  Berg}]{2018ApJ...863...14B}
Byler, N., Dalcanton, J.~J., Conroy, C., {et~al.} 2018, The Astrophysical
  Journal, 863, 14

\bibitem[{Byler {et~al.}(2020)Byler, Kewley, Rigby, Acharyya, Berg, Bayliss, \&
  Sharon}]{2020ApJ...893....1B}
Byler, N., Kewley, L.~J., Rigby, J.~R., {et~al.} 2020, The Astrophysical
  Journal, 893, 1

\bibitem[{Caruana {et~al.}(2018)Caruana, Wisotzki, Herenz, Kerutt, Urrutia,
  Schmidt, Bouwens, Brinchmann, Cantalupo, Carollo, Diener, Drake, Garel,
  Marino, Richard, Saust, Schaye, \& Verhamme}]{2018MNRAS.473...30C}
Caruana, J., Wisotzki, L., Herenz, E.~C., {et~al.} 2018, Monthly Notices of the
  Royal Astronomical Society, 473, 30

\bibitem[{Cassata {et~al.}(2020)Cassata, Morselli, Faisst, Ginolfi,
  B{\'e}thermin, Capak, Le~Fevre, Schaerer, Silverman, Yan, Lemaux, Romano,
  Talia, Bardelli, Boquien, Cimatti, Dessauges-Zavadsky, Fudamoto, Fujimoto,
  Giavalisco, Hathi, Ibar, Jones, Koekemoer, Mendez-Hernandez, Mancini, Oesch,
  Pozzi, Riechers, Rodighiero, Vergani, Zamorani, \&
  Zucca}]{2020A&A...643A...6C}
Cassata, P., Morselli, L., Faisst, A., {et~al.} 2020, Astronomy and
  Astrophysics, 643, A6

\bibitem[{Chen {et~al.}(2019)Chen, Johnson, Straka, Zahedy, Schaye, Muzahid,
  Bouch{\'e}, Cantalupo, Marino, \& Wendt}]{2019MNRAS.484..431C}
Chen, H.-W., Johnson, S.~D., Straka, L.~A., {et~al.} 2019, Monthly Notices of
  the Royal Astronomical Society, 484, 431

\bibitem[{Chevallard \& Charlot(2016)}]{2016MNRAS.462.1415C}
Chevallard, J. \& Charlot, S. 2016, Monthly Notices of the Royal Astronomical
  Society, 462, 1415

\bibitem[{Chevallard {et~al.}(2018)Chevallard, Charlot, Senchyna, Stark,
  Vidal-Garcia, Feltre, Gutkin, Jones, Mainali, \&
  Wofford}]{2018MNRAS.479.3264C}
Chevallard, J., Charlot, S., Senchyna, P., {et~al.} 2018, Monthly Notices of
  the Royal Astronomical Society, 479, 3264

\bibitem[{Chonis {et~al.}(2013)Chonis, Blanc, Hill, Adams, Finkelstein,
  Gebhardt, Kollmeier, Ciardullo, Drory, Gronwall, Hagen, Overzier, Song, \&
  Zeimann}]{2013ApJ...775...99C}
Chonis, T.~S., Blanc, G.~A., Hill, G.~J., {et~al.} 2013, The Astrophysical
  Journal, 775, 99

\bibitem[{Christensen {et~al.}(2012)Christensen, Richard, Hjorth,
  Milvang-Jensen, Laursen, Limousin, Dessauges-Zavadsky, Grillo, \&
  Ebeling}]{2012MNRAS.427.1953C}
Christensen, L., Richard, J., Hjorth, J., {et~al.} 2012, Monthly Notices of the
  Royal Astronomical Society, 427, 1953

\bibitem[{Clopper \& Pearson(1934)}]{CLOPPER:1934ee}
Clopper, C.~J. \& Pearson, E.~S. 1934, Biometrika, 26, 404

\bibitem[{Conroy \& Gunn(2010)}]{2010ascl.soft10043C}
Conroy, C. \& Gunn, J.~E. 2010, Astrophysics Source Code Library, ascl:1010.043

\bibitem[{Conroy {et~al.}(2009)Conroy, Gunn, \& White}]{2009ApJ...699..486C}
Conroy, C., Gunn, J.~E., \& White, M. 2009, The Astrophysical Journal, 699, 486

\bibitem[{Conseil {et~al.}(2016)Conseil, Bacon, Piqueras, \&
  Shepherd}]{2016arXiv161205308C}
Conseil, S., Bacon, R., Piqueras, L., \& Shepherd, M. 2016, arXiv.org,
  arXiv:1612.05308

\bibitem[{Crowther(2007)}]{2007ARA&A..45..177C}
Crowther, P.~A. 2007, Annual Review of Astronomy and Astrophysics, 45, 177

\bibitem[{Crowther {et~al.}(2006)Crowther, Prinja, Pettini, \&
  Steidel}]{2006MNRAS.368..895C}
Crowther, P.~A., Prinja, R.~K., Pettini, M., \& Steidel, C.~C. 2006, Monthly
  Notices of the Royal Astronomical Society, 368, 895

\bibitem[{de~Barros {et~al.}(2017)de~Barros, Pentericci, Vanzella, Castellano,
  Fontana, Grazian, Conselice, Yan, Koekemoer, Cristiani, Dickinson,
  Finkelstein, \& Maiolino}]{2017A&A...608A.123D}
de~Barros, S., Pentericci, L., Vanzella, E., {et~al.} 2017, Astronomy and
  Astrophysics, 608, A123

\bibitem[{Dickinson {et~al.}(2003)Dickinson, Giavalisco, \&
  Team}]{2003mglh.conf..324D}
Dickinson, M., Giavalisco, M., \& Team, G. 2003, in The Mass of Galaxies at Low
  and High Redshift: Proceedings of the European Southern Observatory and
  Universit{\"a}ts-Sternwarte M{\"u}nchen Workshop Held in Venice, Space
  Telescope Science Institute, Baltimore MD 21218, USA (Berlin/Heidelberg:
  Springer-Verlag), 324--

\bibitem[{Dijkstra(2017)}]{2017arXiv170403416D}
Dijkstra, M. 2017, arXiv.org, arXiv:1704.03416

\bibitem[{Dijkstra {et~al.}(2006)Dijkstra, Haiman, \&
  Spaans}]{2006ApJ...649...14D}
Dijkstra, M., Haiman, Z., \& Spaans, M. 2006, The Astrophysical Journal, 649,
  14

\bibitem[{Dijkstra {et~al.}(2011)Dijkstra, Mesinger, \&
  Wyithe}]{2011MNRAS.414.2139D}
Dijkstra, M., Mesinger, A., \& Wyithe, J. S.~B. 2011, Monthly Notices of the
  Royal Astronomical Society, 414, 2139

\bibitem[{Ding {et~al.}(2017)Ding, Cai, Fan, Stark, Bian, Jiang, McGreer,
  Robertson, \& Siana}]{2017ApJ...838L..22D}
Ding, J., Cai, Z., Fan, X., {et~al.} 2017, The Astrophysical Journal Letters,
  838, L22

\bibitem[{Dopita {et~al.}(2002)Dopita, Groves, Sutherland, Binette, \&
  Cecil}]{2002ApJ...572..753D}
Dopita, M.~A., Groves, B.~A., Sutherland, R.~S., Binette, L., \& Cecil, G.
  2002, The Astrophysical Journal, 572, 753

\bibitem[{Du {et~al.}(2016)Du, Shapley, Martin, \& Coil}]{2016ApJ...829...64D}
Du, X., Shapley, A.~E., Martin, C.~L., \& Coil, A.~L. 2016, The Astrophysical
  Journal, 829, 64

\bibitem[{Du {et~al.}(2018)Du, Shapley, Reddy, Jones, Stark, Steidel, Strom,
  Rudie, Erb, Ellis, \& Pettini}]{2018ApJ...860...75D}
Du, X., Shapley, A.~E., Reddy, N.~A., {et~al.} 2018, The Astrophysical Journal,
  860, 75

\bibitem[{Du {et~al.}(2020)Du, Shapley, Tang, Stark, Martin, Mobasher, Topping,
  \& Chevallard}]{2020ApJ...890...65D}
Du, X., Shapley, A.~E., Tang, M., {et~al.} 2020, The Astrophysical Journal,
  890, 65

\bibitem[{Eldridge {et~al.}(2017)Eldridge, Stanway, Xiao, McClelland, Taylor,
  Ng, Greis, \& Bray}]{2017PASA...34...58E}
Eldridge, J.~J., Stanway, E.~R., Xiao, L., {et~al.} 2017, Publications of the
  Astronomical Society of Australia, 34, e058

\bibitem[{Erb {et~al.}(2010)Erb, Pettini, Shapley, Steidel, Law, \&
  Reddy}]{2010ApJ...719.1168E}
Erb, D.~K., Pettini, M., Shapley, A.~E., {et~al.} 2010, The Astrophysical
  Journal, 719, 1168

\bibitem[{Erb {et~al.}(2014)Erb, Steidel, Trainor, Bogosavljevi{\'c}, Shapley,
  Nestor, Kulas, Law, Strom, Rudie, Reddy, Pettini, Konidaris, Mace, Matthews,
  \& McLean}]{2014ApJ...795...33E}
Erb, D.~K., Steidel, C.~C., Trainor, R.~F., {et~al.} 2014, The Astrophysical
  Journal, 795, 33

\bibitem[{Feibelman(1983)}]{1983A&A...122..335F}
Feibelman, W.~A. 1983, Astronomy and Astrophysics (ISSN 0004-6361), 122, 335

\bibitem[{Feltre {et~al.}(2016)Feltre, Charlot, \&
  Gutkin}]{2016MNRAS.456.3354F}
Feltre, A., Charlot, S., \& Gutkin, J. 2016, Monthly Notices of the Royal
  Astronomical Society, 456, 3354

\bibitem[{Feltre {et~al.}(2020)Feltre, Maseda, Bacon, Pradeep, Leclercq,
  Kusakabe, Wisotzki, Hashimoto, Schmidt, Blaizot, Brinchmann, Boogaard,
  Cantalupo, Carton, Inami, Kollatschny, Marino, Matthee, Nanayakkara, Richard,
  Schaye, Tresse, Urrutia, Verhamme, \& Weilbacher}]{2020A&A...641A.118F}
Feltre, A., Maseda, M.~V., Bacon, R., {et~al.} 2020, Astronomy and
  Astrophysics, 641, A118

\bibitem[{Ferland {et~al.}(2013)Ferland, Porter, van Hoof, Williams, Abel,
  Lykins, Shaw, Henney, \& {Stancil, P. C.}}]{2013RMxAA..49..137F}
Ferland, G.~J., Porter, R.~L., van Hoof, P. A.~M., {et~al.} 2013, Revista
  Mexicana de Astronom{\'\i}a y Astrof{\'\i}sica Vol. 49, 49, 137

\bibitem[{Finkelstein {et~al.}(2013)Finkelstein, Papovich, Dickinson, Song,
  Tilvi, Koekemoer, Finkelstein, Mobasher, Ferguson, Giavalisco, Reddy, Ashby,
  Dekel, Fazio, Fontana, Grogin, Huang, Kocevski, Rafelski, Weiner, \&
  Willner}]{2013Natur.502..524F}
Finkelstein, S.~L., Papovich, C., Dickinson, M., {et~al.} 2013, Nature, 502,
  524

\bibitem[{Florian {et~al.}(2020)Florian, Rigby, Acharyya, Sharon, Gladders,
  Kewley, Khullar, Gozman, Brammer, Momcheva, Nicholls, LaMassa, Dahle,
  Bayliss, Wuyts, Johnson, \& Whitaker}]{2020arXiv200611387F}
Florian, M.~K., Rigby, J.~R., Acharyya, A., {et~al.} 2020, arXiv.org,
  arXiv:2006.11387

\bibitem[{Fuller {et~al.}(2020)Fuller, Lemaux, Bradac, Hoag, Schmidt, Huang,
  Strait, Mason, Treu, Pentericci, Trenti, Henry, \&
  Malkan}]{2020ApJ...896..156F}
Fuller, S., Lemaux, B.~C., Bradac, M., {et~al.} 2020, The Astrophysical
  Journal, 896, 156

\bibitem[{Gazagnes {et~al.}(2020)Gazagnes, Chisholm, Schaerer, Verhamme, \&
  Izotov}]{2020A&A...639A..85G}
Gazagnes, S., Chisholm, J., Schaerer, D., Verhamme, A., \& Izotov, Y. 2020,
  Astronomy and Astrophysics, 639, A85

\bibitem[{Gazagnes {et~al.}(2018)Gazagnes, Chisholm, Schaerer, Verhamme, Rigby,
  \& Bayliss}]{2018A&A...616A..29G}
Gazagnes, S., Chisholm, J., Schaerer, D., {et~al.} 2018, Astronomy and
  Astrophysics, 616, A29

\bibitem[{Giavalisco {et~al.}(2004)Giavalisco, Ferguson, Koekemoer, Dickinson,
  Alexander, Bauer, Bergeron, Biagetti, Brandt, Casertano, Cesarsky,
  Chatzichristou, Conselice, Cristiani, Da~Costa, Dahlen, de~Mello, Eisenhardt,
  Erben, Fall, Fassnacht, Fosbury, Fruchter, Gardner, Grogin, Hook,
  Hornschemeier, Idzi, Jogee, Kretchmer, Laidler, Lee, Livio, Lucas, Madau,
  Mobasher, Moustakas, Nonino, Padovani, Papovich, Park, Ravindranath, Renzini,
  Richardson, Riess, Rosati, Schirmer, Schreier, Somerville, Spinrad, Stern,
  Stiavelli, Strolger, Urry, Vandame, Williams, \& Wolf}]{2004ApJ...600L..93G}
Giavalisco, M., Ferguson, H.~C., Koekemoer, A.~M., {et~al.} 2004, The
  Astrophysical Journal, 600, L93

\bibitem[{Greig {et~al.}(2017)Greig, Mesinger, Haiman, \&
  Simcoe}]{2017MNRAS.466.4239G}
Greig, B., Mesinger, A., Haiman, Z., \& Simcoe, R.~A. 2017, Monthly Notices of
  the Royal Astronomical Society, 466, 4239

\bibitem[{Grogin {et~al.}(2011)Grogin, Kocevski, Faber, Ferguson, Koekemoer,
  Riess, Acquaviva, Alexander, Almaini, Ashby, Barden, Bell, Bournaud, Brown,
  Caputi, Casertano, Cassata, Castellano, Challis, Chary, Cheung, Cirasuolo,
  Conselice, Roshan~Cooray, Croton, Daddi, Dahlen, Dav{\'e}, de~Mello, Dekel,
  Dickinson, Dolch, Donley, Dunlop, Dutton, Elbaz, Fazio, Filippenko,
  Finkelstein, Fontana, Gardner, Garnavich, Gawiser, Giavalisco, Grazian, Guo,
  Hathi, H{\"a}ussler, Hopkins, Huang, Huang, Jha, Kartaltepe, Kirshner, Koo,
  Lai, Lee, Li, Lotz, Lucas, Madau, McCarthy, McGrath, McIntosh, Mclure,
  Mobasher, Moustakas, Mozena, Nandra, Newman, Niemi, Noeske, Papovich,
  Pentericci, Pope, Primack, Rajan, Ravindranath, Reddy, Renzini, Rix, Robaina,
  Rodney, Rosario, Rosati, Salimbeni, Scarlata, Siana, Simard, Smidt,
  Somerville, Spinrad, Straughn, Strolger, Telford, Teplitz, Trump, Van
  Der~Wel, Villforth, Wechsler, Weiner, Wiklind, Wild, Wilson, Wuyts, Yan, \&
  Yun}]{2011ApJS..197...35G}
Grogin, N.~A., Kocevski, D.~D., Faber, S.~M., {et~al.} 2011, The Astrophysical
  Journal Supplement, 197, 35

\bibitem[{Guaita {et~al.}(2016)Guaita, Pentericci, Grazian, Vanzella, Nonino,
  Giavalisco, Zamorani, Bongiorno, Cassata, Castellano, Garilli, Gawiser,
  Le~Brun, Le~Fevre, Lemaux, Maccagni, Merlin, Santini, Tasca, Thomas, Zucca,
  de~Barros, Hathi, Amor{\'\i}n, Bardelli, \& Fontana}]{2016A&A...587A.133G}
Guaita, L., Pentericci, L., Grazian, A., {et~al.} 2016, Astronomy and
  Astrophysics, 587, A133

\bibitem[{Guo {et~al.}(2013)Guo, Ferguson, Giavalisco, Barro, Willner, Ashby,
  Dahlen, Donley, Faber, Fontana, Galametz, Grazian, Huang, Kocevski,
  Koekemoer, Koo, McGrath, Peth, Salvato, Wuyts, Castellano, Cooray, Dickinson,
  Dunlop, Fazio, Gardner, Gawiser, Grogin, Hathi, Hsu, Lee, Lucas, Mobasher,
  Nandra, Newman, \& Van Der~Wel}]{2013ApJS..207...24G}
Guo, Y., Ferguson, H.~C., Giavalisco, M., {et~al.} 2013, The Astrophysical
  Journal Supplement, 207, 24

\bibitem[{Gutkin {et~al.}(2016)Gutkin, Charlot, \&
  Bruzual}]{2016MNRAS.462.1757G}
Gutkin, J., Charlot, S., \& Bruzual, G. 2016, Monthly Notices of the Royal
  Astronomical Society, 462, 1757

\bibitem[{Harikane {et~al.}(2018)Harikane, Ouchi, Shibuya, Kojima, Zhang, Itoh,
  Ono, Higuchi, Inoue, Chevallard, Capak, Nagao, Onodera, Faisst, Martin,
  Rauch, Bruzual, Charlot, Davidzon, Fujimoto, Hilmi, Ilbert, Lee, Matsuoka,
  Silverman, \& Toft}]{2018ApJ...859...84H}
Harikane, Y., Ouchi, M., Shibuya, T., {et~al.} 2018, The Astrophysical Journal,
  859, 84

\bibitem[{Hashimoto {et~al.}(2015)Hashimoto, Verhamme, Ouchi, Shimasaku,
  Schaerer, Nakajima, Shibuya, Rauch, Ono, \& Goto}]{2015ApJ...812..157H}
Hashimoto, T., Verhamme, A., Ouchi, M., {et~al.} 2015, The Astrophysical
  Journal, 812, 157

\bibitem[{Henry {et~al.}(2015)Henry, Scarlata, Martin, \&
  Erb}]{2015ApJ...809...19H}
Henry, A., Scarlata, C., Martin, C.~L., \& Erb, D. 2015, The Astrophysical
  Journal, 809, 19

\bibitem[{Herenz {et~al.}(2020)Herenz, Hayes, \&
  Scarlata}]{2020A&A...642A..55H}
Herenz, E.~C., Hayes, M., \& Scarlata, C. 2020, Astronomy and Astrophysics,
  642, A55

\bibitem[{Herenz {et~al.}(2017)Herenz, Urrutia, Wisotzki, Kerutt, Saust,
  Werhahn, Schmidt, Caruana, Diener, Bacon, Brinchmann, Schaye, Maseda, \&
  Weilbacher}]{2017A&A...606A..12H}
Herenz, E.~C., Urrutia, T., Wisotzki, L., {et~al.} 2017, Astronomy and
  Astrophysics, 606, A12

\bibitem[{Herenz \& Wisotzki(2017)}]{2017A&A...602A.111H}
Herenz, E.~C. \& Wisotzki, L. 2017, Astronomy and Astrophysics, 602, A111

\bibitem[{Herenz \& Wisotzki(2021)}]{2021A&A...649.C5H}
Herenz, E.~C. \& Wisotzki, L. 2021, Astronomy and Astrophysics, 649, C5

\bibitem[{Herenz {et~al.}(2019)Herenz, Wisotzki, Saust, Kerutt, Urrutia,
  Diener, Schmidt, Marino, de~la Vieuville, Boogaard, Schaye, Guiderdoni,
  Richard, \& Bacon}]{2019A&A...621A.107H}
Herenz, E.~C., Wisotzki, L., Saust, R., {et~al.} 2019, Astronomy and
  Astrophysics, 621, A107

\bibitem[{Hirschmann {et~al.}(2017)Hirschmann, Charlot, Feltre, Naab, Choi,
  Ostriker, \& Somerville}]{2017MNRAS.472.2468H}
Hirschmann, M., Charlot, S., Feltre, A., {et~al.} 2017, Monthly Notices of the
  Royal Astronomical Society, 472, 2468

\bibitem[{Hirschmann {et~al.}(2019)Hirschmann, Charlot, Feltre, Naab,
  Somerville, \& Choi}]{2019MNRAS.487..333H}
Hirschmann, M., Charlot, S., Feltre, A., {et~al.} 2019, Monthly Notices of the
  Royal Astronomical Society, 487, 333

\bibitem[{Hoag {et~al.}(2019)Hoag, Bradac, Huang, Mason, Treu, Schmidt, Trenti,
  Strait, Lemaux, Finney, \& Paddock}]{2019ApJ...878...12H}
Hoag, A., Bradac, M., Huang, K., {et~al.} 2019, The Astrophysical Journal, 878,
  12

\bibitem[{Huang {et~al.}(2016)Huang, Lemaux, Schmidt, Hoag, Brada{\v c}, Treu,
  Dijkstra, Fontana, Henry, Malkan, Mason, Morishita, Pentericci, Ryan, Trenti,
  \& Wang}]{2016ApJ...823L..14H}
Huang, K.-H., Lemaux, B.~C., Schmidt, K.~B., {et~al.} 2016, The Astrophysical
  Journal Letters, 823, L14

\bibitem[{Hunter(2007)}]{Hunter:2007ux}
Hunter, J.~D. 2007, Computing in Science and Engineering

\bibitem[{Hutchison {et~al.}(2019)Hutchison, Papovich, Finkelstein, Dickinson,
  Jung, Zitrin, Ellis, Malhotra, Rhoads, Roberts-Borsani, Song, \&
  Tilvi}]{2019ApJ...879...70H}
Hutchison, T.~A., Papovich, C., Finkelstein, S.~L., {et~al.} 2019, The
  Astrophysical Journal, 879, 70

\bibitem[{Illingworth {et~al.}(2016)Illingworth, Magee, Bouwens, Oesch,
  Labb{\'e}, Van~Dokkum, Whitaker, Holden, Franx, \&
  Gonzalez}]{2016arXiv160600841I}
Illingworth, G., Magee, D., Bouwens, R., {et~al.} 2016, arXiv.org,
  arXiv:1606.00841

\bibitem[{Inami {et~al.}(2017)Inami, Bacon, Brinchmann, Richard, Contini,
  Conseil, Hamer, Akhlaghi, Bouch{\'e}, Cl{\'e}ment, Desprez, Drake, Hashimoto,
  Leclercq, Maseda, Michel-Dansac, Paalvast, Tresse, Ventou, Kollatschny,
  Boogaard, Finley, Marino, Schaye, \& Wisotzki}]{2017A&A...608A...2I}
Inami, H., Bacon, R., Brinchmann, J., {et~al.} 2017, Astronomy and
  Astrophysics, 608, A2

\bibitem[{Inoue {et~al.}(2016)Inoue, Tamura, Matsuo, Mawatari, Shimizu,
  Shibuya, Ota, Yoshida, Zackrisson, Kashikawa, Kohno, Umehata, Hatsukade, Iye,
  Matsuda, Okamoto, \& Yamaguchi}]{2016Sci...352.1559I}
Inoue, A.~K., Tamura, Y., Matsuo, H., {et~al.} 2016, Science, 352, 1559

\bibitem[{Izotov {et~al.}(2016{\natexlab{a}})Izotov, Orlitov{\'a}, Schaerer,
  Thuan, Verhamme, Guseva, \& Worseck}]{2016Natur.529..178I}
Izotov, Y.~I., Orlitov{\'a}, I., Schaerer, D., {et~al.} 2016{\natexlab{a}},
  Nature, 529, 178

\bibitem[{Izotov {et~al.}(2016{\natexlab{b}})Izotov, Schaerer, Thuan, Worseck,
  Guseva, Orlitov{\'a}, \& Verhamme}]{2016MNRAS.461.3683I}
Izotov, Y.~I., Schaerer, D., Thuan, T.~X., {et~al.} 2016{\natexlab{b}}, Monthly
  Notices of the Royal Astronomical Society, 461, 3683

\bibitem[{Izotov {et~al.}(2018)Izotov, Schaerer, Worseck, Guseva, Thuan,
  Verhamme, Orlitov{\'a}, \& Fricke}]{2018MNRAS.474.4514I}
Izotov, Y.~I., Schaerer, D., Worseck, G., {et~al.} 2018, Monthly Notices of the
  Royal Astronomical Society, 474, 4514

\bibitem[{James {et~al.}(2018)James, Auger, Pettini, Stark, Belokurov, \&
  Carniani}]{2018MNRAS.476.1726J}
James, B.~L., Auger, M., Pettini, M., {et~al.} 2018, Monthly Notices of the
  Royal Astronomical Society, 476, 1726

\bibitem[{James {et~al.}(2014)James, Pettini, Christensen, Auger, Becker, King,
  Quider, Shapley, \& Steidel}]{2014MNRAS.440.1794J}
James, B.~L., Pettini, M., Christensen, L., {et~al.} 2014, Monthly Notices of
  the Royal Astronomical Society, 440, 1794

\bibitem[{Jaskot \& Ravindranath(2016)}]{2016ApJ...833..136J}
Jaskot, A.~E. \& Ravindranath, S. 2016, The Astrophysical Journal, 833, 136

\bibitem[{Jiang {et~al.}(2020)Jiang, Kashikawa, Wang, Walth, Ho, Cai, Egami,
  Fan, Ito, Liang, Schaerer, \& Stark}]{2020NatAs.tmp..246J}
Jiang, L., Kashikawa, N., Wang, S., {et~al.} 2020, Nature Astronomy

\bibitem[{Jiang {et~al.}(2019)Jiang, Malhotra, Yang, \&
  Rhoads}]{2019ApJ...872..146J}
Jiang, T., Malhotra, S., Yang, H., \& Rhoads, J.~E. 2019, The Astrophysical
  Journal, 872, 146

\bibitem[{Johnson {et~al.}(2021)Johnson, Leja, Conroy, \&
  Speagle}]{2021ApJS..254...22J}
Johnson, B.~D., Leja, J., Conroy, C., \& Speagle, J.~S. 2021, Astrophysical
  Journal Supplement, 254, 22

\bibitem[{Jones {et~al.}(2001)Jones, Oliphant, \&
  Peterson}]{SciPyOpensources:tUcReTVZ}
Jones, E., Oliphant, T., \& Peterson, P. 2001

\bibitem[{Joye \& Mandel(2003)}]{2003ASPC..295..489J}
Joye, W.~A. \& Mandel, E. 2003, Astronomical Data Analysis Software and Systems
  XII ASP Conference Series, 295, 489

\bibitem[{Jung {et~al.}(2019)Jung, Finkelstein, Dickinson, Hutchison, Larson,
  Papovich, Pentericci, Song, Ferguson, Guo, Malhotra, Mobasher, Rhoads, Tilvi,
  \& Wold}]{2019ApJ...877..146J}
Jung, I., Finkelstein, S.~L., Dickinson, M., {et~al.} 2019, The Astrophysical
  Journal, 877, 146

\bibitem[{Kamann(2018)}]{2018ascl.soft05021K}
Kamann, S. 2018, Astrophysics Source Code Library, ascl:1805.021

\bibitem[{Kamann {et~al.}(2013)Kamann, Wisotzki, \& Roth}]{2013A&A...549A..71K}
Kamann, S., Wisotzki, L., \& Roth, M.~M. 2013, Astronomy and Astrophysics, 549,
  A71

\bibitem[{Keenan {et~al.}(1992)Keenan, Feibelman, \&
  Berrington}]{1992ApJ...389..443K}
Keenan, F.~P., Feibelman, W.~A., \& Berrington, K.~A. 1992, Astrophysical
  Journal, 389, 443

\bibitem[{Kerutt(2017)}]{2017ascl.soft03011K}
Kerutt, J. 2017, Astrophysics Source Code Library, ascl:1703.011

\bibitem[{Kerutt {et~al.}(2021)Kerutt, Wisotzki, Verhamme, Schmidt, Leclercq,
  \& Herenz}]{Kerutt:2021tr}
Kerutt, J., Wisotzki, L., Verhamme, A., {et~al.} 2021, Astronomy and
  Astrophysics, Submitted

\bibitem[{Kewley {et~al.}(2019{\natexlab{a}})Kewley, Nicholls, Sutherland,
  Rigby, Acharya, Dopita, \& Bayliss}]{2019ApJ...880...16K}
Kewley, L.~J., Nicholls, D.~C., Sutherland, R., {et~al.} 2019{\natexlab{a}},
  The Astrophysical Journal, 880, 16

\bibitem[{Kewley {et~al.}(2019{\natexlab{b}})Kewley, Nicholls, \&
  Sutherland}]{2019ARA&A..57..511K}
Kewley, L.~J., Nicholls, D.~C., \& Sutherland, R.~S. 2019{\natexlab{b}}, Annual
  Review of Astronomy and Astrophysics, 57, 511

\bibitem[{Khostovan {et~al.}(2019)Khostovan, Sobral, Mobasher, Matthee,
  Cochrane, Chartab, Jafariyazani, Paulino-Afonso, Santos, \&
  Calhau}]{2019MNRAS.489..555K}
Khostovan, A.~A., Sobral, D., Mobasher, B., {et~al.} 2019, Monthly Notices of
  the Royal Astronomical Society, 489, 555

\bibitem[{Koekemoer {et~al.}(2011)Koekemoer, Faber, Ferguson, Grogin, Kocevski,
  Koo, Lai, Lotz, Lucas, McGrath, Ogaz, Rajan, Riess, Rodney, Strolger,
  Casertano, Castellano, Dahlen, Dickinson, Dolch, Fontana, Giavalisco,
  Grazian, Guo, Hathi, Huang, Van Der~Wel, Yan, Acquaviva, Alexander, Almaini,
  Ashby, Barden, Bell, Bournaud, Brown, Caputi, Cassata, Challis, Chary,
  Cheung, Cirasuolo, Conselice, Roshan~Cooray, Croton, Daddi, Dav{\'e},
  de~Mello, De~Ravel, Dekel, Donley, Dunlop, Dutton, Elbaz, Fazio, Filippenko,
  Finkelstein, Frazer, Gardner, Garnavich, Gawiser, Gruetzbauch, Hartley,
  H{\"a}ussler, Herrington, Hopkins, Huang, Jha, Johnson, Kartaltepe,
  Khostovan, Kirshner, Lani, Lee, Li, Madau, McCarthy, McIntosh, Mclure,
  Mcpartland, Mobasher, Moreira, Mortlock, Moustakas, Mozena, Nandra, Newman,
  Nielsen, Niemi, Noeske, Papovich, Pentericci, Pope, Primack, Ravindranath,
  Reddy, Renzini, Rix, Robaina, Rosario, Rosati, Salimbeni, Scarlata, Siana,
  Simard, Smidt, Snyder, Somerville, Spinrad, Straughn, Telford, Teplitz,
  Trump, Vargas, Villforth, Wagner, Wandro, Wechsler, Weiner, Wiklind, Wild,
  Wilson, Wuyts, \& Yun}]{2011ApJS..197...36K}
Koekemoer, A.~M., Faber, S.~M., Ferguson, H.~C., {et~al.} 2011, The
  Astrophysical Journal Supplement, 197, 36

\bibitem[{Kramida {et~al.}(2019)Kramida, Ralchenko, \&
  Reader}]{NISTAtomicSpectra:uk}
Kramida, A., Ralchenko, Y., \& Reader, J. 2019, {NIST Atomic Spectra Database}

\bibitem[{Kron(1980)}]{1980ApJS...43..305K}
Kron, R.~G. 1980, Astrophysical Journal Supplement Series, 43, 305

\bibitem[{Kulas {et~al.}(2012)Kulas, Shapley, Kollmeier, Zheng, Steidel, \&
  Hainline}]{2012ApJ...745...33K}
Kulas, K.~R., Shapley, A.~E., Kollmeier, J.~A., {et~al.} 2012, The
  Astrophysical Journal, 745, 33

\bibitem[{Kusakabe {et~al.}(2020)Kusakabe, Blaizot, Garel, Verhamme, Bacon,
  Richard, Hashimoto, Inami, Conseil, Guiderdoni, Drake, Christian~Herenz,
  Schaye, Oesch, Matthee, Anna~Marino, Borello~Schmidt, Pell{\'o}, Maseda,
  Leclercq, Kerutt, \& Mahler}]{2020A&A...638A..12K}
Kusakabe, H., Blaizot, J., Garel, T., {et~al.} 2020, Astronomy and
  Astrophysics, 638, A12

\bibitem[{Laigle {et~al.}(2016)Laigle, McCracken, Ilbert, Hsieh, Davidzon,
  Capak, Hasinger, Silverman, Pichon, Coupon, Aussel, Le~Borgne, Caputi,
  Cassata, Chang, Civano, Dunlop, Fynbo, Kartaltepe, Koekemoer, Le~Fevre,
  Le~Floc'h, Leauthaud, Lilly, Lin, Marchesi, Milvang-Jensen, Salvato, Sanders,
  Scoville, Smol{\v c}i{\'c}, Stockmann, Taniguchi, Tasca, Toft, Vaccari, \&
  Zabl}]{2016ApJS..224...24L}
Laigle, C., McCracken, H.~J., Ilbert, O., {et~al.} 2016, The Astrophysical
  Journal Supplement Series, 224, 24

\bibitem[{Lam {et~al.}(2019)Lam, Bouwens, Labb{\'e}, Schaye, Schmidt, Maseda,
  Bacon, Boogaard, Nanayakkara, Richard, Mahler, \&
  Urrutia}]{2019A&A...627A.164L}
Lam, D., Bouwens, R.~J., Labb{\'e}, I., {et~al.} 2019, Astronomy and
  Astrophysics, 627, A164

\bibitem[{Laporte {et~al.}(2017)Laporte, Nakajima, Ellis, Zitrin, Stark,
  Mainali, \& Roberts-Borsani}]{2017ApJ...851...40L}
Laporte, N., Nakajima, K., Ellis, R.~S., {et~al.} 2017, The Astrophysical
  Journal, 851, 40

\bibitem[{Laursen {et~al.}(2019)Laursen, Sommer-Larsen, Milvang-Jensen, Fynbo,
  \& Razoumov}]{2019A&A...627A..84L}
Laursen, P., Sommer-Larsen, J., Milvang-Jensen, B., Fynbo, J. P.~U., \&
  Razoumov, A.~O. 2019, Astronomy and Astrophysics, 627, A84

\bibitem[{Laursen {et~al.}(2011)Laursen, Sommer-Larsen, \&
  Razoumov}]{2011ApJ...728...52L}
Laursen, P., Sommer-Larsen, J., \& Razoumov, A.~O. 2011, The Astrophysical
  Journal, 728, 52

\bibitem[{Le~Fevre {et~al.}(2019)Le~Fevre, Lemaux, Nakajima, Schaerer, Talia,
  Zamorani, Cassata, Garilli, Maccagni, Pentericci, Tasca, Zucca, Amor{\'\i}n,
  Bardelli, Cimatti, Giavalisco, Guaita, Hathi, Marchi, Vanzella, Vergani, \&
  Dunlop}]{2019A&A...625A..51L}
Le~Fevre, O., Lemaux, B.~C., Nakajima, K., {et~al.} 2019, Astronomy and
  Astrophysics, 625, A51

\bibitem[{Le~Fevre {et~al.}(2015)Le~Fevre, Tasca, Cassata, Garilli, Le~Brun,
  Maccagni, Pentericci, Thomas, Vanzella, Zamorani, Zucca, Amor{\'\i}n,
  Bardelli, Capak, Cassara, Castellano, Cimatti, Cuby, Cucciati, de~la Torre,
  Durkalec, Fontana, Giavalisco, Grazian, Hathi, Ilbert, Lemaux, Moreau,
  Paltani, Ribeiro, Salvato, Schaerer, Scodeggio, Sommariva, Talia, Taniguchi,
  Tresse, Vergani, Wang, Charlot, Contini, Fotopoulou, L{\'o}pez-Sanjuan,
  Mellier, \& Scoville}]{2015A&A...576A..79L}
Le~Fevre, O., Tasca, L. A.~M., Cassata, P., {et~al.} 2015, Astronomy and
  Astrophysics, 576, A79

\bibitem[{Leclercq {et~al.}(2020)Leclercq, Bacon, Verhamme, Garel, Blaizot,
  Brinchmann, Cantalupo, Claeyssens, Conseil, Contini, Hashimoto, Herenz,
  Kusakabe, Marino, Maseda, Matthee, Mitchell, Pezzuli, Richard, Schmidt, \&
  Wisotzki}]{2020A&A...635A..82L}
Leclercq, F., Bacon, R., Verhamme, A., {et~al.} 2020, Astronomy and
  Astrophysics, 635, A82

\bibitem[{Leclercq {et~al.}(2017)Leclercq, Bacon, Wisotzki, Mitchell, Garel,
  Verhamme, Blaizot, Hashimoto, Herenz, Conseil, Cantalupo, Inami, Contini,
  Richard, Maseda, Schaye, Marino, Akhlaghi, Brinchmann, \&
  Carollo}]{2017A&A...608A...8L}
Leclercq, F., Bacon, R., Wisotzki, L., {et~al.} 2017, Astronomy and
  Astrophysics, 608, A8

\bibitem[{Leitet {et~al.}(2013)Leitet, Bergvall, Hayes, Linn{\'e}, \&
  Zackrisson}]{2013A&A...553A.106L}
Leitet, E., Bergvall, N., Hayes, M., Linn{\'e}, S., \& Zackrisson, E. 2013,
  Astronomy and Astrophysics, 553, A106

\bibitem[{Leitherer {et~al.}(2011)Leitherer, Tremonti, Heckman, \&
  Calzetti}]{2011AJ....141...37L}
Leitherer, C., Tremonti, C.~A., Heckman, T.~M., \& Calzetti, D. 2011, The
  Astronomical Journal, 141, 37

\bibitem[{Leja {et~al.}(2017)Leja, Johnson, Conroy, van Dokkum, \&
  Byler}]{2017ApJ...837..170L}
Leja, J., Johnson, B.~D., Conroy, C., van Dokkum, P.~G., \& Byler, N. 2017, The
  Astrophysical Journal, 837, 170

\bibitem[{Luridiana {et~al.}(2013)Luridiana, Morisset, \&
  Shaw}]{2013ascl.soft04021L}
Luridiana, V., Morisset, C., \& Shaw, R.~A. 2013, Astrophysics Source Code
  Library, ascl:1304.021

\bibitem[{Luridiana {et~al.}(2015)Luridiana, Morisset, \&
  Shaw}]{2015A&A...573A..42L}
Luridiana, V., Morisset, C., \& Shaw, R.~A. 2015, Astronomy and Astrophysics,
  573, A42

\bibitem[{Ly {et~al.}(2016)Ly, Malhotra, Malkan, Rigby, Kashikawa, de~los
  Reyes, \& Rhoads}]{2016ApJS..226....5L}
Ly, C., Malhotra, S., Malkan, M.~A., {et~al.} 2016, The Astrophysical Journal
  Supplement Series, 226, 5

\bibitem[{Mainali {et~al.}(2017)Mainali, Kollmeier, Stark, Simcoe, Walth,
  Newman, \& Miller}]{2017ApJ...836L..14M}
Mainali, R., Kollmeier, J.~A., Stark, D.~P., {et~al.} 2017, The Astrophysical
  Journal Letters, 836, L14

\bibitem[{Mainali {et~al.}(2020)Mainali, Stark, Tang, Chevallard, Charlot,
  Sharon, Coe, Salmon, Bradley, Johnson, Frye, Avila, Ogaz, Zitrin, Brada{\v
  c}, Lemaux, Mahler, Paterno-Mahler, Strait, \&
  Andrade-Santos}]{2020MNRAS.494..719M}
Mainali, R., Stark, D.~P., Tang, M., {et~al.} 2020, Monthly Notices of the
  Royal Astronomical Society, 494, 719

\bibitem[{Mainali {et~al.}(2018)Mainali, Zitrin, Stark, Ellis, Richard, Tang,
  Laporte, Oesch, \& Mcgreer}]{2018MNRAS.479.1180M}
Mainali, R., Zitrin, A., Stark, D.~P., {et~al.} 2018, Monthly Notices of the
  Royal Astronomical Society, 479, 1180

\bibitem[{Maiolino {et~al.}(2008)Maiolino, Nagao, Grazian, Cocchia, Marconi,
  Mannucci, Cimatti, Pipino, Ballero, Calura, Chiappini, Fontana, Granato,
  Matteucci, Pastorini, Pentericci, Risaliti, Salvati, \&
  Silva}]{2008A&A...488..463M}
Maiolino, R., Nagao, T., Grazian, A., {et~al.} 2008, Astronomy and
  Astrophysics, 488, 463

\bibitem[{Malkan {et~al.}(1996)Malkan, Teplitz, \&
  McLean}]{1996ApJ...468L...9M}
Malkan, M.~A., Teplitz, H., \& McLean, I.~S. 1996, Astrophysical Journal
  Letters v.468, 468, L9

\bibitem[{Malmquist(1920)}]{1920MeLuS..22....3M}
Malmquist, G.~K. 1920, Meddelanden fran Lunds Astronomiska Observatorium Series
  II, 22, 3

\bibitem[{Malmquist(1922)}]{1922MeLuF.100....1M}
Malmquist, K.~G. 1922, Meddelanden fran Lunds Astronomiska Observatorium Series
  I, 100, 1

\bibitem[{Marchi {et~al.}(2019)Marchi, Pentericci, Guaita, Talia, Castellano,
  Hathi, Schaerer, Amor{\'\i}n, Bolzonella, Carnall, Charlot, Chevallard,
  Cullen, Finkelstein, Fontana, Fontanot, Garilli, Hibon, Koekemoer, Maccagni,
  McLure, Papovich, Pozzetti, \& Saxena}]{2019A&A...631A..19M}
Marchi, F., Pentericci, L., Guaita, L., {et~al.} 2019, Astronomy and
  Astrophysics, 631, A19

\bibitem[{Marques-Chaves {et~al.}(2020)Marques-Chaves, Perez-Fournon, Shu,
  Colina, Bolton, {\'A}lvarez-M{\'a}rquez, Brownstein, Cornachione, Geier,
  Jim{\'e}nez-{\'A}ngel, Kojima, Mao, Montero-Dorta, Oguri, Ouchi, Poidevin,
  Shirley, \& Zheng}]{2020MNRAS.492.1257M}
Marques-Chaves, R., Perez-Fournon, I., Shu, Y., {et~al.} 2020, Monthly Notices
  of the Royal Astronomical Society, 492, 1257

\bibitem[{Martin \& Wiese(1976)}]{1976JPCRD...5..537M}
Martin, G.~A. \& Wiese, W.~L. 1976, Journal of Physical and Chemical Reference
  Data, 5, 537

\bibitem[{Mary {et~al.}(2020)Mary, Bacon, Conseil, Piqueras, \&
  Schutz}]{2020A&A...635A.194M}
Mary, D., Bacon, R., Conseil, S., Piqueras, L., \& Schutz, A. 2020, Astronomy
  and Astrophysics, 635, A194

\bibitem[{Maseda {et~al.}(2020)Maseda, Bacon, Lam, Matthee, Brinchmann, Schaye,
  Labb{\'e}, Schmidt, Boogaard, Bouwens, Cantalupo, Franx, Hashimoto, Inami,
  Kusakabe, Mahler, Nanayakkara, Richard, \& Wisotzki}]{2020MNRAS.tmp..584M}
Maseda, M.~V., Bacon, R., Lam, D., {et~al.} 2020, Monthly Notices of the Royal
  Astronomical Society, 493, 5120

\bibitem[{Maseda {et~al.}(2017)Maseda, Brinchmann, Franx, Bacon, Bouwens,
  Schmidt, Boogaard, Contini, Feltre, Inami, Kollatschny, Marino, Richard,
  Verhamme, \& Wisotzki}]{2017A&A...608A...4M}
Maseda, M.~V., Brinchmann, J., Franx, M., {et~al.} 2017, Astronomy and
  Astrophysics, 608, A4

\bibitem[{Mason {et~al.}(2015)Mason, Trenti, \& Treu}]{2015ApJ...813...21M}
Mason, C.~A., Trenti, M., \& Treu, T. 2015, The Astrophysical Journal, 813, 21

\bibitem[{Mason {et~al.}(2018{\natexlab{a}})Mason, Treu, de~Barros, Dijkstra,
  Fontana, Mesinger, Pentericci, Trenti, \& Vanzella}]{2018ApJ...857L..11M}
Mason, C.~A., Treu, T., de~Barros, S., {et~al.} 2018{\natexlab{a}}, The
  Astrophysical Journal Letters, 857, L11

\bibitem[{Mason {et~al.}(2018{\natexlab{b}})Mason, Treu, Dijkstra, Mesinger,
  Trenti, Pentericci, de~Barros, \& Vanzella}]{2018ApJ...856....2M}
Mason, C.~A., Treu, T., Dijkstra, M., {et~al.} 2018{\natexlab{b}}, The
  Astrophysical Journal, 856, 2

\bibitem[{Matthee {et~al.}(2020{\natexlab{a}})Matthee, Pezzulli, Mackenzie,
  Cantalupo, Kusakabe, Leclercq, Sobral, Richard, Wisotzki, Lilly, Boogaard,
  Marino, Maseda, \& Nanayakkara}]{2020MNRAS.498.3043M}
Matthee, J., Pezzulli, G., Mackenzie, R., {et~al.} 2020{\natexlab{a}}, Monthly
  Notices of the Royal Astronomical Society, 498, 3043

\bibitem[{Matthee {et~al.}(2017)Matthee, Sobral, Darvish, Santos, Mobasher,
  Paulino-Afonso, R{\"o}ttgering, \& Alegre}]{2017MNRAS.472..772M}
Matthee, J., Sobral, D., Darvish, B., {et~al.} 2017, Monthly Notices of the
  Royal Astronomical Society, 472, 772

\bibitem[{Matthee {et~al.}(2020{\natexlab{b}})Matthee, Sobral, Gronke,
  Pezzulli, Cantalupo, R{\"o}ttgering, Darvish, \&
  Santos}]{2020MNRAS.492.1778M}
Matthee, J., Sobral, D., Gronke, M., {et~al.} 2020{\natexlab{b}}, Monthly
  Notices of the Royal Astronomical Society, 492, 1778

\bibitem[{Mitchell {et~al.}(2021)Mitchell, Blaizot, Cadiou, Dubois, Garel, \&
  Rosdahl}]{2021MNRAS.501.5757M}
Mitchell, P.~D., Blaizot, J., Cadiou, C., {et~al.} 2021, Monthly Notices of the
  Royal Astronomical Society, 501, 5757

\bibitem[{Momose {et~al.}(2014)Momose, Ouchi, Nakajima, Ono, Shibuya,
  Shimasaku, Yuma, Mori, \& Umemura}]{2014MNRAS.442..110M}
Momose, R., Ouchi, M., Nakajima, K., {et~al.} 2014, Monthly Notices of the
  Royal Astronomical Society, 442, 110

\bibitem[{Morton(1991)}]{1991ApJS...77..119M}
Morton, D.~C. 1991, Astrophysical Journal Supplement Series (ISSN 0067-0049),
  77, 119

\bibitem[{Muzahid {et~al.}(2020)Muzahid, Schaye, Marino, Cantalupo, Brinchmann,
  Contini, Wendt, Wisotzki, Zabl, Bouch{\'e}, Akhlaghi, Chen, Claeyssens,
  Johnson, Leclercq, Maseda, Matthee, Richard, Urrutia, \&
  Verhamme}]{2020MNRAS.496.1013M}
Muzahid, S., Schaye, J., Marino, R.~A., {et~al.} 2020, Monthly Notices of the
  Royal Astronomical Society, 496, 1013

\bibitem[{Nakajima {et~al.}(2016)Nakajima, Ellis, Iwata, Inoue, Kusakabe,
  Ouchi, \& Robertson}]{2016ApJ...831L...9N}
Nakajima, K., Ellis, R.~S., Iwata, I., {et~al.} 2016, The Astrophysical Journal
  Letters, 831, L9

\bibitem[{Nakajima {et~al.}(2018)Nakajima, Fletcher, Ellis, Robertson, \&
  Iwata}]{2018MNRAS.477.2098N}
Nakajima, K., Fletcher, T., Ellis, R.~S., Robertson, B.~E., \& Iwata, I. 2018,
  Monthly Notices of the Royal Astronomical Society, 477, 2098

\bibitem[{Nanayakkara {et~al.}(2019)Nanayakkara, Brinchmann, Boogaard, Bouwens,
  Cantalupo, Feltre, Kollatschny, Marino, Maseda, Matthee, Paalvast, Richard,
  \& Verhamme}]{2019A&A...624A..89N}
Nanayakkara, T., Brinchmann, J., Boogaard, L., {et~al.} 2019, Astronomy and
  Astrophysics, 624, A89

\bibitem[{Norman {et~al.}(2002)Norman, Hasinger, Giacconi, Gilli, Kewley,
  Nonino, Rosati, Szokoly, Tozzi, Wang, Zheng, Zirm, Bergeron, Gilmozzi,
  Grogin, Koekemoer, \& Schreier}]{2002ApJ...571..218N}
Norman, C., Hasinger, G., Giacconi, R., {et~al.} 2002, The Astrophysical
  Journal, 571, 218

\bibitem[{Oesch {et~al.}(2015)Oesch, van Dokkum, Illingworth, Bouwens,
  Momcheva, Holden, Roberts-Borsani, Smit, Franx, Labbe, Gonzalez, \&
  Magee}]{2015ApJ...804L..30O}
Oesch, P.~A., van Dokkum, P.~G., Illingworth, G.~D., {et~al.} 2015, The
  Astrophysical Journal Letters, 804, L30

\bibitem[{Oke \& Gunn(1983)}]{1983ApJ...266..713O}
Oke, J.~B. \& Gunn, J.~E. 1983, Astrophysical Journal, 266, 713

\bibitem[{Onoue {et~al.}(2020)Onoue, Banados, Mazzucchelli, Venemans,
  Schindler, Walter, Hennawi, Andika, Davies, Decarli, Farina, Jahnke, Nagao,
  Tominaga, \& Wang}]{2020ApJ...898..105O}
Onoue, M., Banados, E., Mazzucchelli, C., {et~al.} 2020, The Astrophysical
  Journal, 898, 105

\bibitem[{Osterbrock \& Ferland(2006)}]{2006agna.book.....O}
Osterbrock, D.~E. \& Ferland, G.~J. 2006, Astrophysics of gaseous nebulae and
  active galactic nuclei

\bibitem[{Ouchi {et~al.}(2010)Ouchi, Shimasaku, Furusawa, Saito, Yoshida,
  Akiyama, Ono, Yamada, Ota, Kashikawa, Iye, Kodama, Okamura, Simpson, \&
  Yoshida}]{2010ApJ...723..869O}
Ouchi, M., Shimasaku, K., Furusawa, H., {et~al.} 2010, The Astrophysical
  Journal, 723, 869

\bibitem[{Patr{\'\i}cio {et~al.}(2016)Patr{\'\i}cio, Richard, Verhamme,
  Wisotzki, Brinchmann, Turner, Christensen, Weilbacher, Blaizot, Bacon,
  Contini, Lagattuta, Cantalupo, Clement, \& Soucail}]{2016MNRAS.456.4191P}
Patr{\'\i}cio, V., Richard, J., Verhamme, A., {et~al.} 2016, Monthly Notices of
  the Royal Astronomical Society, 456, 4191

\bibitem[{Peimbert {et~al.}(2017)Peimbert, Peimbert, \&
  Delgado-Inglada}]{2017PASP..129h2001P}
Peimbert, M., Peimbert, A., \& Delgado-Inglada, G. 2017, Publications of the
  Astronomical Society of the Pacific, 129, 082001

\bibitem[{Peng {et~al.}(2002)Peng, Ho, Impey, \& Rix}]{2002AJ....124..266P}
Peng, C.~Y., Ho, L.~C., Impey, C.~D., \& Rix, H.~W. 2002, The Astronomical
  Journal, 124, 266

\bibitem[{Peng {et~al.}(2010)Peng, Ho, Impey, \& Rix}]{2010AJ....139.2097P}
Peng, C.~Y., Ho, L.~C., Impey, C.~D., \& Rix, H.~W. 2010, The Astronomical
  Journal, 139, 2097

\bibitem[{Pentericci {et~al.}(2016)Pentericci, Carniani, Castellano, Fontana,
  Maiolino, Guaita, Vanzella, Grazian, Santini, Yan, Cristiani, Conselice,
  Giavalisco, Hathi, \& Koekemoer}]{2016ApJ...829L..11P}
Pentericci, L., Carniani, S., Castellano, M., {et~al.} 2016, The Astrophysical
  Journal Letters, 829, L11

\bibitem[{Pentericci {et~al.}(2018)Pentericci, Vanzella, Castellano, Fontana,
  de~Barros, Grazian, Marchi, Bradac, Conselice, Cristiani, Dickinson,
  Finkelstein, Giallongo, Guaita, Koekemoer, Maiolino, Santini, \&
  Tilvi}]{2018A&A...619A.147P}
Pentericci, L., Vanzella, E., Castellano, M., {et~al.} 2018, Astronomy and
  Astrophysics, 619, A147

\bibitem[{Pentericci {et~al.}(2014)Pentericci, Vanzella, Fontana, Castellano,
  Treu, Mesinger, Dijkstra, Grazian, Bradac, Conselice, Cristiani, Dunlop,
  Galametz, Giavalisco, Giallongo, Koekemoer, McLure, Maiolino, Paris, \&
  Santini}]{2014ApJ...793..113P}
Pentericci, L., Vanzella, E., Fontana, A., {et~al.} 2014, The Astrophysical
  Journal, 793, 113

\bibitem[{P{\'e}rez \& Granger(2007)}]{Perez:2007hy}
P{\'e}rez, F. \& Granger, B.~E. 2007, Computing in Science and Engineering, 9,
  21

\bibitem[{Piqueras {et~al.}(2017)Piqueras, Conseil, Shepherd, Bacon, Leclercq,
  \& Richard}]{2017arXiv171003554P}
Piqueras, L., Conseil, S., Shepherd, M., {et~al.} 2017, arXiv.org,
  arXiv:1710.03554

\bibitem[{Plat {et~al.}(2019)Plat, Charlot, Bruzual, Feltre, Vidal-Garc{\'\i}a,
  Morisset, Chevallard, \& Todt}]{2019MNRAS.490..978P}
Plat, A., Charlot, S., Bruzual, G., {et~al.} 2019, Monthly Notices of the Royal
  Astronomical Society, 490, 978

\bibitem[{Rafelski {et~al.}(2015)Rafelski, Teplitz, Gardner, Coe, Bond,
  Koekemoer, Grogin, Kurczynski, McGrath, Bourque, Atek, Brown, Colbert,
  Codoreanu, Ferguson, Finkelstein, Gawiser, Giavalisco, Gronwall, Hanish, Lee,
  Mehta, de~Mello, Ravindranath, Ryan, Scarlata, Siana, Soto, \&
  Voyer}]{2015AJ....150...31R}
Rafelski, M., Teplitz, H.~I., Gardner, J.~P., {et~al.} 2015, The Astronomical
  Journal, 150, 31

\bibitem[{Rakic {et~al.}(2011)Rakic, Schaye, Steidel, \&
  Rudie}]{2011MNRAS.414.3265R}
Rakic, O., Schaye, J., Steidel, C.~C., \& Rudie, G.~C. 2011, Monthly Notices of
  the Royal Astronomical Society, 414, 3265

\bibitem[{Ravindranath {et~al.}(2020)Ravindranath, Monroe, Jaskot, Ferguson, \&
  Tumlinson}]{2020ApJ...896..170R}
Ravindranath, S., Monroe, T., Jaskot, A., Ferguson, H.~C., \& Tumlinson, J.
  2020, The Astrophysical Journal, 896, 170

\bibitem[{Richard {et~al.}(2021)Richard, Claeyssens, Lagattuta, Guaita, Bauer,
  Pell{\'o}, Carton, Bacon, Soucail, Lyon, Kneib, Mahler, Clement, Mercier,
  Variu, Tamone, Ebeling, Schmidt, Nanayakkara, Maseda, Weilbacher, Bouch{\'e},
  Bouwens, Wisotzki, de~la Vieuville, Martinez, \&
  Patr{\'\i}cio}]{2021A&A...646A..83R}
Richard, J., Claeyssens, A., Lagattuta, D., {et~al.} 2021, Astronomy and
  Astrophysics, 646, A83

\bibitem[{Rigby {et~al.}(2018{\natexlab{a}})Rigby, Bayliss, Chisholm, Bordoloi,
  Sharon, Gladders, Johnson, Paterno-Mahler, Wuyts, Dahle, \&
  Acharyya}]{2018ApJ...853...87R}
Rigby, J.~R., Bayliss, M.~B., Chisholm, J., {et~al.} 2018{\natexlab{a}}, The
  Astrophysical Journal, 853, 87

\bibitem[{Rigby {et~al.}(2014)Rigby, Bayliss, Gladders, Sharon, Wuyts, \&
  Dahle}]{2014ApJ...790...44R}
Rigby, J.~R., Bayliss, M.~B., Gladders, M.~D., {et~al.} 2014, The Astrophysical
  Journal, 790, 44

\bibitem[{Rigby {et~al.}(2015)Rigby, Bayliss, Gladders, Sharon, Wuyts, Dahle,
  Johnson, \& Pe{\~n}a-Guerrero}]{2015ApJ...814L...6R}
Rigby, J.~R., Bayliss, M.~B., Gladders, M.~D., {et~al.} 2015, The Astrophysical
  Journal Letters, 814, L6

\bibitem[{Rigby {et~al.}(2018{\natexlab{b}})Rigby, Bayliss, Sharon, Gladders,
  Chisholm, Dahle, Johnson, Paterno-Mahler, Wuyts, \&
  Kelson}]{2018AJ....155..104R}
Rigby, J.~R., Bayliss, M.~B., Sharon, K., {et~al.} 2018{\natexlab{b}}, The
  Astronomical Journal, 155, 104

\bibitem[{Rigby {et~al.}(2021)Rigby, Florian, Acharyya, Bayliss, Gladders,
  Sharon, Brammer, Momcheva, LaMassa, Bian, Dahle, Johnson, Kewley, Murray,
  Whitaker, \& Wuyts}]{2021ApJ...908..154R}
Rigby, J.~R., Florian, M., Acharyya, A., {et~al.} 2021, The Astrophysical
  Journal, 908, 154

\bibitem[{Robitaille \& Bressert(2012)}]{2012ascl.soft08017R}
Robitaille, T. \& Bressert, E. 2012, Astrophysics Source Code Library,
  ascl:1208.017

\bibitem[{Sanders {et~al.}(2016{\natexlab{a}})Sanders, Shapley, Kriek, Reddy,
  Freeman, Coil, Siana, Mobasher, Shivaei, Price, \&
  de~Groot}]{2016ApJ...825L..23S}
Sanders, R.~L., Shapley, A.~E., Kriek, M., {et~al.} 2016{\natexlab{a}}, The
  Astrophysical Journal Letters, 825, L23

\bibitem[{Sanders {et~al.}(2016{\natexlab{b}})Sanders, Shapley, Kriek, Reddy,
  Freeman, Coil, Siana, Mobasher, Shivaei, Price, \&
  de~Groot}]{2016ApJ...816...23S}
Sanders, R.~L., Shapley, A.~E., Kriek, M., {et~al.} 2016{\natexlab{b}}, The
  Astrophysical Journal, 816, 23

\bibitem[{Sanders {et~al.}(2020)Sanders, Shapley, Reddy, Kriek, Siana, Coil,
  Mobasher, Shivaei, Freeman, Azadi, Price, Leung, Fetherolf, de~Groot, Zick,
  Fornasini, \& Barro}]{2020MNRAS.491.1427S}
Sanders, R.~L., Shapley, A.~E., Reddy, N.~A., {et~al.} 2020, Monthly Notices of
  the Royal Astronomical Society, 491, 1427

\bibitem[{Saxena {et~al.}(2020)Saxena, Pentericci, Mirabelli, Schaerer,
  Schneider, Cullen, Amor{\'\i}n, Bolzonella, Bongiorno, Carnall, Castellano,
  Cucciati, Fontana, Fynbo, Garilli, Gargiulo, Guaita, Hathi, Hutchison,
  Koekemoer, Marchi, McLeod, McLure, Papovich, Pozzetti, Talia, \&
  Zamorani}]{2020A&A...636A..47S}
Saxena, A., Pentericci, L., Mirabelli, M., {et~al.} 2020, Astronomy and
  Astrophysics, 636, A47

\bibitem[{Schaerer {et~al.}(2019)Schaerer, Fragos, \&
  Izotov}]{2019A&A...622L..10S}
Schaerer, D., Fragos, T., \& Izotov, Y.~I. 2019, Astronomy and Astrophysics,
  622, L10

\bibitem[{Schaerer {et~al.}(2018)Schaerer, Izotov, Nakajima, Worseck, Chisholm,
  Verhamme, Thuan, \& de~Barros}]{2018A&A...616L..14S}
Schaerer, D., Izotov, Y.~I., Nakajima, K., {et~al.} 2018, Astronomy and
  Astrophysics, 616, L14

\bibitem[{Schenker {et~al.}(2013)Schenker, Ellis, Konidaris, \&
  Stark}]{2013ApJ...777...67S}
Schenker, M.~A., Ellis, R.~S., Konidaris, N.~P., \& Stark, D.~P. 2013, The
  Astrophysical Journal, 777, 67

\bibitem[{Schmidt(2021)}]{kasper_schmidt_2021_5131705}
Schmidt, K.~B. 2021, https://doi.org/10.5281/zenodo.5131705

\bibitem[{Schmidt {et~al.}(2017)Schmidt, Huang, Treu, Hoag, Bradac, Henry,
  Jones, Mason, Malkan, Morishita, Pentericci, Trenti, Vulcani, \&
  Wang}]{2017ApJ...839...17S}
Schmidt, K.~B., Huang, K.~H., Treu, T., {et~al.} 2017, The Astrophysical
  Journal, 839, 17

\bibitem[{Schmidt {et~al.}(2016)Schmidt, Treu, Bradac, Vulcani, Huang, Hoag,
  Maseda, Guaita, Pentericci, Brammer, Dijkstra, Dressler, Fontana, Henry,
  Jones, Mason, Trenti, \& Wang}]{2016ApJ...818...38S}
Schmidt, K.~B., Treu, T., Bradac, M., {et~al.} 2016, The Astrophysical Journal,
  818, 38

\bibitem[{Schmidt {et~al.}(2019)Schmidt, Wisotzki, Urrutia, Kerutt, Krajnovic,
  Herenz, Saust, Contini, Epinat, Inami, \& Maseda}]{2019A&A...628A..91S}
Schmidt, K.~B., Wisotzki, L., Urrutia, T., {et~al.} 2019, Astronomy and
  Astrophysics, 628, A91

\bibitem[{Scoville {et~al.}(2007)Scoville, Aussel, Brusa, Capak, Carollo,
  Elvis, Giavalisco, Guzzo, Hasinger, Impey, Kneib, Lefevre, Lilly, Mobasher,
  Renzini, Rich, Sanders, Schinnerer, Schminovich, Shopbell, Taniguchi, \&
  Tyson}]{2007ApJS..172....1S}
Scoville, N., Aussel, H., Brusa, M., {et~al.} 2007, The Astrophysical Journal
  Supplement Series, 172, 1

\bibitem[{Senchyna {et~al.}(2019)Senchyna, Stark, Chevallard, Charlot, Jones,
  \& Vidal-Garcia}]{2019MNRAS.488.3492S}
Senchyna, P., Stark, D.~P., Chevallard, J., {et~al.} 2019, Monthly Notices of
  the Royal Astronomical Society, 488, 3492

\bibitem[{Senchyna {et~al.}(2020)Senchyna, Stark, Mirocha, Reines, Charlot,
  Jones, \& Mulchaey}]{2020MNRAS.494..941S}
Senchyna, P., Stark, D.~P., Mirocha, J., {et~al.} 2020, Monthly Notices of the
  Royal Astronomical Society, 494, 941

\bibitem[{Senchyna {et~al.}(2021)Senchyna, Stark, of, \&
  {2021}}]{2020arXiv200809780S}
Senchyna, P., Stark, D.~P., of, S. C. M.~N., \& {2021}. 2021, Monthly Notice of
  the Royal Astronomical Society, 503, 6112

\bibitem[{Senchyna {et~al.}(2017)Senchyna, Stark, Vidal-Garcia, Chevallard,
  Charlot, Mainali, Jones, Wofford, Feltre, \& Gutkin}]{2017MNRAS.472.2608S}
Senchyna, P., Stark, D.~P., Vidal-Garcia, A., {et~al.} 2017, Monthly Notices of
  the Royal Astronomical Society, 472, 2608

\bibitem[{Shapley {et~al.}(2003)Shapley, Steidel, Pettini, \&
  Adelberger}]{2003ApJ...588...65S}
Shapley, A.~E., Steidel, C.~C., Pettini, M., \& Adelberger, K.~L. 2003, The
  Astrophysical Journal, 588, 65

\bibitem[{Shibuya {et~al.}(2018)Shibuya, Ouchi, Harikane, Rauch, Ono, Mukae,
  Higuchi, Kojima, Yuma, Lee, Furusawa, Konno, Martin, Shimasaku, Taniguchi,
  Kobayashi, Kajisawa, Nagao, Goto, Kashikawa, Komiyama, Kusakabe, Momose,
  Nakajima, Tanaka, \& Wang}]{2018PASJ...70S..15S}
Shibuya, T., Ouchi, M., Harikane, Y., {et~al.} 2018, Publications of the
  Astronomical Society of Japan, 70, S15

\bibitem[{Shibuya {et~al.}(2014)Shibuya, Ouchi, Nakajima, Hashimoto, Ono,
  Rauch, Gauthier, Shimasaku, Goto, Mori, \& Umemura}]{2014ApJ...788...74S}
Shibuya, T., Ouchi, M., Nakajima, K., {et~al.} 2014, The Astrophysical Journal,
  788, 74

\bibitem[{Shields \& Kennicutt(1995)}]{1995ApJ...454..807S}
Shields, J.~C. \& Kennicutt, R. C.~J. 1995, Astrophysical Journal v.454, 454,
  807

\bibitem[{Shirazi \& Brinchmann(2012)}]{2012MNRAS.421.1043S}
Shirazi, M. \& Brinchmann, J. 2012, Monthly Notices of the Royal Astronomical
  Society, 421, 1043

\bibitem[{Skelton {et~al.}(2014)Skelton, Whitaker, Momcheva, Brammer, van
  Dokkum, Labb{\'e}, Franx, Van Der~Wel, Bezanson, da~Cunha, Fumagalli,
  F{\"o}rster~Schreiber, Kriek, Leja, Lundgren, Magee, Marchesini, Maseda,
  Nelson, Oesch, Pacifici, Patel, Price, Rix, Tal, Wake, \&
  Wuyts}]{2014ApJS..214...24S}
Skelton, R.~E., Whitaker, K.~E., Momcheva, I.~G., {et~al.} 2014, The
  Astrophysical Journal Supplement Series, 214, 24

\bibitem[{Smit {et~al.}(2017)Smit, Swinbank, Massey, Richard, Smail, \&
  Kneib}]{2017MNRAS.467.3306S}
Smit, R., Swinbank, A.~M., Massey, R., {et~al.} 2017, Monthly Notices of the
  Royal Astronomical Society, 467, 3306

\bibitem[{Sobral {et~al.}(2018)Sobral, Matthee, Darvish, Smail, Best, Alegre,
  R{\"o}ttgering, Mobasher, Paulino-Afonso, Stroe, \&
  Oteo}]{2018MNRAS.477.2817S}
Sobral, D., Matthee, J., Darvish, B., {et~al.} 2018, Monthly Notices of the
  Royal Astronomical Society, 477, 2817

\bibitem[{Soto {et~al.}(2016)Soto, Lilly, Bacon, Richard, \&
  Conseil}]{2016MNRAS.458.3210S}
Soto, K.~T., Lilly, S.~J., Bacon, R., Richard, J., \& Conseil, S. 2016, Monthly
  Notices of the Royal Astronomical Society, 458, 3210

\bibitem[{Stanway \& Eldridge(2018)}]{2018MNRAS.479...75S}
Stanway, E.~R. \& Eldridge, J.~J. 2018, Monthly Notices of the Royal
  Astronomical Society, 479, 75

\bibitem[{Stanway {et~al.}(2016)Stanway, Eldridge, \&
  Becker}]{2016MNRAS.456..485S}
Stanway, E.~R., Eldridge, J.~J., \& Becker, G.~D. 2016, Monthly Notices of the
  Royal Astronomical Society, 456, 485

\bibitem[{Stark {et~al.}(2017)Stark, Ellis, Charlot, Chevallard, Tang, Belli,
  Zitrin, Mainali, Gutkin, Vidal-Garcia, Bouwens, \&
  Oesch}]{2017MNRAS.464..469S}
Stark, D.~P., Ellis, R.~S., Charlot, S., {et~al.} 2017, Monthly Notices of the
  Royal Astronomical Society, 464, 469

\bibitem[{Stark {et~al.}(2015{\natexlab{a}})Stark, Richard, Charlot, Clement,
  Ellis, Siana, Robertson, Schenker, Gutkin, \& Wofford}]{2015MNRAS.450.1846S}
Stark, D.~P., Richard, J., Charlot, S., {et~al.} 2015{\natexlab{a}}, Monthly
  Notices of the Royal Astronomical Society, 450, 1846

\bibitem[{Stark {et~al.}(2014)Stark, Richard, Siana, Charlot, Freeman, Gutkin,
  Wofford, Robertson, Amanullah, Watson, \&
  Milvang-Jensen}]{2014MNRAS.445.3200S}
Stark, D.~P., Richard, J., Siana, B., {et~al.} 2014, Monthly Notices of the
  Royal Astronomical Society, 445, 3200

\bibitem[{Stark {et~al.}(2015{\natexlab{b}})Stark, Walth, Charlot, Clement,
  Feltre, Gutkin, Richard, Mainali, Robertson, Siana, Tang, \&
  Schenker}]{2015MNRAS.454.1393S}
Stark, D.~P., Walth, G., Charlot, S., {et~al.} 2015{\natexlab{b}}, Monthly
  Notices of the Royal Astronomical Society, 454, 1393

\bibitem[{Steidel {et~al.}(2011)Steidel, Bogosavljevi{\'c}, Shapley, Kollmeier,
  Reddy, Erb, \& Pettini}]{2011ApJ...736..160S}
Steidel, C.~C., Bogosavljevi{\'c}, M., Shapley, A.~E., {et~al.} 2011, The
  Astrophysical Journal, 736, 160

\bibitem[{Steidel {et~al.}(2018)Steidel, Bogosavljevi{\'c}, Shapley, Reddy,
  Rudie, Pettini, Trainor, \& Strom}]{2018ApJ...869..123S}
Steidel, C.~C., Bogosavljevi{\'c}, M., Shapley, A.~E., {et~al.} 2018, The
  Astrophysical Journal, 869, 123

\bibitem[{Steidel {et~al.}(2014)Steidel, Rudie, Strom, Pettini, Reddy, Shapley,
  Trainor, Erb, Turner, Konidaris, Kulas, Mace, Matthews, \&
  McLean}]{2014ApJ...795..165S}
Steidel, C.~C., Rudie, G.~C., Strom, A.~L., {et~al.} 2014, The Astrophysical
  Journal, 795, 165

\bibitem[{Steidel {et~al.}(2016)Steidel, Strom, Pettini, Rudie, Reddy, \&
  Trainor}]{2016ApJ...826..159S}
Steidel, C.~C., Strom, A.~L., Pettini, M., {et~al.} 2016, The Astrophysical
  Journal, 826, 159

\bibitem[{Stroe {et~al.}(2017)Stroe, Sobral, Matthee, Calhau, \&
  Oteo}]{2017MNRAS.471.2575S}
Stroe, A., Sobral, D., Matthee, J., Calhau, J., \& Oteo, I. 2017, Monthly
  Notices of the Royal Astronomical Society, 471, 2575

\bibitem[{Tang {et~al.}(2021{\natexlab{a}})Tang, Stark, Chevallard, Charlot,
  Endsley, \& Congiu}]{2021MNRAS.503.4105T}
Tang, M., Stark, D., Chevallard, J., {et~al.} 2021{\natexlab{a}}, Monthly
  Notice of the Royal Astronomical Society, 503, 4105

\bibitem[{Tang {et~al.}(2021{\natexlab{b}})Tang, Stark, Chevallard, Charlot,
  Endsley, \& Congiu}]{2021MNRAS.501.3238T}
Tang, M., Stark, D.~P., Chevallard, J., {et~al.} 2021{\natexlab{b}}, Monthly
  Notices of the Royal Astronomical Society, 501, 3238

\bibitem[{Tilvi {et~al.}(2014)Tilvi, Papovich, Finkelstein, Long, Song,
  Dickinson, Ferguson, Koekemoer, Giavalisco, \&
  Mobasher}]{2014ApJ...794....5T}
Tilvi, V., Papovich, C., Finkelstein, S.~L., {et~al.} 2014, The Astrophysical
  Journal, 794, 5

\bibitem[{Tilvi {et~al.}(2016)Tilvi, Pirzkal, Malhotra, Finkelstein, Rhoads,
  Windhorst, Grogin, Koekemoer, Zakamska, Ryan, Christensen, Hathi, Pharo,
  Joshi, Yang, Gronwall, Cimatti, Walsh, OConnell, Straughn, {\"O}stlin,
  Rothberg, Livermore, Hibon, \& Gardner}]{2016ApJ...827L..14T}
Tilvi, V., Pirzkal, N., Malhotra, S., {et~al.} 2016, The Astrophysical Journal
  Letters, 827, L14

\bibitem[{Torres-Peimbert \& Pena(1984)}]{1984RMxAA...9..107T}
Torres-Peimbert, S. \& Pena, M. 1984, Revista Mexicana de Astronomia y
  Astrofisica, 9, 107

\bibitem[{Tremonti {et~al.}(2004)Tremonti, Heckman, Kauffmann, Brinchmann,
  Charlot, White, Seibert, Peng, Schlegel, Uomoto, Fukugita, \&
  Brinkmann}]{2004ApJ...613..898T}
Tremonti, C.~A., Heckman, T.~M., Kauffmann, G., {et~al.} 2004, The
  Astrophysical Journal, 613, 898

\bibitem[{Treu {et~al.}(2013)Treu, Schmidt, Trenti, Bradley, \&
  Stiavelli}]{2013ApJ...775L..29T}
Treu, T., Schmidt, K.~B., Trenti, M., Bradley, L.~D., \& Stiavelli, M. 2013,
  The Astrophysical Journal Letters, 775, L29

\bibitem[{Urrutia {et~al.}(2019)Urrutia, Wisotzki, Kerutt, Schmidt, Herenz,
  Klar, Saust, Werhahn, Diener, Caruana, Krajnovic, Bacon, Boogaard,
  Brinchmann, Enke, Maseda, Nanayakkara, Richard, Steinmetz, \&
  Weilbacher}]{2019A&A...624A.141U}
Urrutia, T., Wisotzki, L., Kerutt, J., {et~al.} 2019, Astronomy and
  Astrophysics, 624, A141

\bibitem[{van~der Walt {et~al.}(2011)van~der Walt, Colbert, \&
  Varoquaux}]{vanderWalt:2011dp}
van~der Walt, S., Colbert, S.~C., \& Varoquaux, G. 2011, Computing in Science
  and Engineering, 13, 22

\bibitem[{van Hoof {et~al.}(2004)van Hoof, Weingartner, Martin, Volk, \&
  Ferland}]{2004MNRAS.350.1330V}
van Hoof, P. A.~M., Weingartner, J.~C., Martin, P.~G., Volk, K., \& Ferland,
  G.~J. 2004, Monthly Notices of the Royal Astronomical Society, 350, 1330

\bibitem[{Vanzella {et~al.}(2020)Vanzella, Caminha, Calura, Cupani, Meneghetti,
  Castellano, Rosati, Mercurio, Sani, Grillo, Gilli, Mignoli, Comastri, Nonino,
  Cristiani, Giavalisco, \& Caputi}]{2020MNRAS.491.1093V}
Vanzella, E., Caminha, G.~B., Calura, F., {et~al.} 2020, Monthly Notices of the
  Royal Astronomical Society, 491, 1093

\bibitem[{Vanzella {et~al.}(2017)Vanzella, Castellano, Meneghetti, Mercurio,
  Caminha, Cupani, Calura, Christensen, Merlin, Rosati, Gronke, Dijkstra,
  Mignoli, Gilli, de~Barros, Caputi, Grillo, Balestra, Cristiani, Nonino,
  Giallongo, Grazian, Pentericci, Fontana, Comastri, Vignali, Zamorani, Brusa,
  Bergamini, \& Tozzi}]{2017ApJ...842...47V}
Vanzella, E., Castellano, M., Meneghetti, M., {et~al.} 2017, The Astrophysical
  Journal, 842, 47

\bibitem[{Vanzella {et~al.}(2016)Vanzella, de~Barros, Cupani, Karman, Gronke,
  Balestra, Coe, Mignoli, Brusa, Calura, Caminha, Caputi, Castellano,
  Christensen, Comastri, Cristiani, Dijkstra, Fontana, Giallongo, Giavalisco,
  Gilli, Grazian, Grillo, Koekemoer, Meneghetti, Nonino, Pentericci, Rosati,
  Schaerer, Verhamme, Vignali, \& Zamorani}]{2016ApJ...821L..27V}
Vanzella, E., de~Barros, S., Cupani, G., {et~al.} 2016, The Astrophysical
  Journal Letters, 821, L27

\bibitem[{Verhamme {et~al.}(2018)Verhamme, Garel, Ventou, Contini, Bouch{\'e},
  Herenz, Richard, Bacon, Schmidt, Maseda, Marino, Brinchmann, Cantalupo,
  Caruana, Cl{\'e}ment, Diener, {Drake, AB}, Hashimoto, Inami, Kerutt,
  Kollatschny, Leclercq, Patricio, Schaye, Wisotzki, \&
  Zabl}]{2018MNRAS.478L..60V}
Verhamme, A., Garel, T., Ventou, E., {et~al.} 2018, Monthly Notices of the
  Royal Astronomical Society: Letters, 478, L60

\bibitem[{Verhamme {et~al.}(2015)Verhamme, Orlitov{\'a}, Schaerer, \&
  Hayes}]{2015A&A...578A...7V}
Verhamme, A., Orlitov{\'a}, I., Schaerer, D., \& Hayes, M. 2015, Astronomy and
  Astrophysics, 578, A7

\bibitem[{Verhamme {et~al.}(2017)Verhamme, Orlitov{\'a}, Schaerer, Izotov,
  Worseck, Thuan, \& Guseva}]{2017A&A...597A..13V}
Verhamme, A., Orlitov{\'a}, I., Schaerer, D., {et~al.} 2017, Astronomy and
  Astrophysics, 597, A13

\bibitem[{Verhamme {et~al.}(2006)Verhamme, Schaerer, \&
  Maselli}]{2006A&A...460..397V}
Verhamme, A., Schaerer, D., \& Maselli, A. 2006, Astronomy and Astrophysics,
  460, 397

\bibitem[{Vidal-Garc{\'\i}a {et~al.}(2017)Vidal-Garc{\'\i}a, Charlot, Bruzual,
  \& Hubeny}]{2017MNRAS.470.3532V}
Vidal-Garc{\'\i}a, A., Charlot, S., Bruzual, G., \& Hubeny, I. 2017, Monthly
  Notices of the Royal Astronomical Society, 470, 3532

\bibitem[{Weilbacher {et~al.}(2020)Weilbacher, Palsa, Streicher, Bacon,
  Urrutia, Wisotzki, Conseil, Husemann, Jarno, Kelz, P{\'e}contal-Rousset,
  Richard, Roth, Selman, \& Vernet}]{2020A&A...641A..28W}
Weilbacher, P.~M., Palsa, R., Streicher, O., {et~al.} 2020, Astronomy and
  Astrophysics, 641, A28

\bibitem[{Weilbacher {et~al.}(2014)Weilbacher, Streicher, Urrutia,
  Pecontal-Rousset, Jarno, \& Bacon}]{2014ASPC..485..451W}
Weilbacher, P.~M., Streicher, O., Urrutia, T., {et~al.} 2014, in Astronomical
  Data Analysis Software and Systems XXIII. Proceedings of a meeting held 29
  September - 3 October 2013 at Waikoloa Beach Marriott, 451--

\bibitem[{Whitaker {et~al.}(2019)Whitaker, Ashas, Illingworth, Magee, Leja,
  Oesch, Van~Dokkum, Mowla, Bouwens, Franx, Holden, Labb{\'e}, Rafelski,
  Teplitz, \& Gonzalez}]{2019ApJS..244...16W}
Whitaker, K.~E., Ashas, M., Illingworth, G., {et~al.} 2019, The Astrophysical
  Journal Supplement Series, 244, 16

\bibitem[{Willott {et~al.}(2015)Willott, Carilli, Wagg, \&
  Wang}]{2015ApJ...807..180W}
Willott, C.~J., Carilli, C.~L., Wagg, J., \& Wang, R. 2015, The Astrophysical
  Journal, 807, 180

\bibitem[{Wisotzki {et~al.}(2016)Wisotzki, Bacon, Blaizot, Brinchmann, Herenz,
  Schaye, Bouch{\'e}, Cantalupo, Contini, Carollo, Caruana, Courbot, Emsellem,
  Kamann, Kerutt, Leclercq, Lilly, Patricio, Sandin, Steinmetz, Straka,
  Urrutia, Verhamme, Weilbacher, \& Wendt}]{2016A&A...587A..98W}
Wisotzki, L., Bacon, R., Blaizot, J., {et~al.} 2016, Astronomy and
  Astrophysics, 587, A98

\bibitem[{Wisotzki {et~al.}(2018)Wisotzki, Bacon, Brinchmann, Cantalupo,
  Richter, Schaye, Schmidt, Urrutia, Weilbacher, Akhlaghi, Bouch{\'e}, Contini,
  Guiderdoni, Herenz, Inami, Kerutt, Leclercq, Marino, Maseda, Monreal-Ibero,
  Nanayakkara, Richard, Saust, Steinmetz, \& Wendt}]{2018Natur.562..229W}
Wisotzki, L., Bacon, R., Brinchmann, J., {et~al.} 2018, Nature, 562, 229

\bibitem[{Wofford {et~al.}(2021)Wofford, Vidal-Garcia, Feltre, Chevallard,
  Charlot, Stark, Herenz, \& Hayes}]{2021MNRAS.500.2908W}
Wofford, A., Vidal-Garcia, A., Feltre, A., {et~al.} 2021, Monthly Notices of
  the Royal Astronomical Society, 500, 2908

\bibitem[{Xiao {et~al.}(2018)Xiao, Stanway, \& Eldridge}]{2018MNRAS.477..904X}
Xiao, L., Stanway, E.~R., \& Eldridge, J.~J. 2018, Monthly Notices of the Royal
  Astronomical Society, 477, 904

\bibitem[{Yang {et~al.}(2017)Yang, Malhotra, Gronke, Rhoads, Leitherer,
  Wofford, Jiang, Dijkstra, Tilvi, \& Wang}]{2017ApJ...844..171Y}
Yang, H., Malhotra, S., Gronke, M., {et~al.} 2017, The Astrophysical Journal,
  844, 171

\end{thebibliography}
\begin{appendix}

\section{Finding Emission Lines In Spectra (FELIS)}\label{sec:felis}

FELIS (Finding Emission Lines In Spectra) is a publicly available\footnote{\href{https://github.com/kasperschmidt/FELIS}{https://github.com/kasperschmidt/FELIS}, \cite{kasper_schmidt_2021_5131705}} Python tool enabling the search for emission lines and other spectral features in 1D spectra. 
FELIS was developed specifically for our particular science project but offers a general tool suitable for general detection of (weak) features in 1D spectra.
As described below, FELIS provides tools to build spectral templates and mock spectra (Section~\ref{sec:FELIStemplates}) which can be matched to observed spectra providing S/N estimates of potential emission features (Section~\ref{sec:runFELIS}).
In Section~\ref{sec:FELIStest_mock} we describe the results from testing FELIS on a set of idealized MUSE-Wide mock spectra with UV emission lines similar to the spectra analyzed in the study described in this paper.

\begin{figure}
\begin{center}
\includegraphics[trim=0 0 0 0.5cm,clip,width=0.49\textwidth]{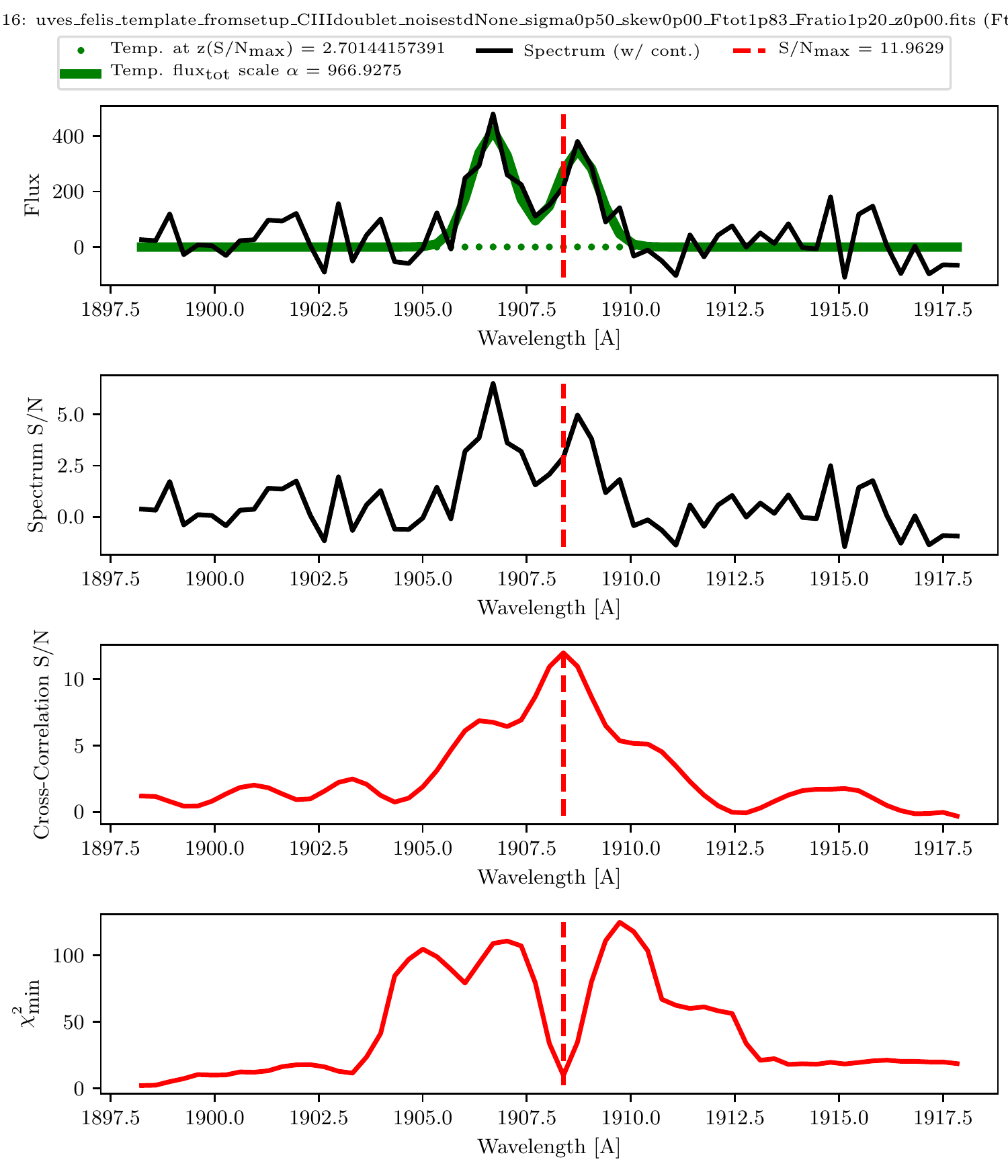}
\caption{Overview of the FELIS template fit to a simulated spectrum of a \ciii{} emitter with a doublet flux ratio of F(\ciiione)/F(\ciiitwo)~=~1.4 corresponding to an electron density $<10^4$cm$^{-3}$ \citep[][]{2006agna.book.....O}. 
The top panel shows the simulated spectrum (black) with the best-fit scaled template plotted on top (green).
The second panel shows the S/N spectrum, that is $\textrm{flux} / \sqrt{\textrm{variance}}$. 
The third panel from the top shows the template cross-correlation S/N calculated with Equation~\ref{eq:CCs2n}. 
The maximum value is indicated by the vertical dashed red line in all four panels.
The bottom panel shows the $\chi^2$ estimate of the template cross-correlation.}
\label{fig:FELIStempfit}
\end{center}
\end{figure}

The search for emission features with FELIS is performed via standard template matching by minimizing the $\chi^2$ between the input data (spectrum) and the model (idealized template of spectral feature). 
FELIS provides an estimate of the significance of the model template match by providing the S/N for the minimized $\chi^2$. 
Hence, FELIS matches the defined templates to the observed spectra by minimizing the $\chi^2$ expression
\begin{equation}
\chi^2         = \sum_i \frac{(\mathcal{D}_i - \alpha \mathcal{T}_i)^2}{\sigma_i^2} \; .
\end{equation}
Here the index $i$ runs over the individual pixels of the spectrum,
$\mathcal{D}$ represents the data, that is the flux measurements of the spectrum, 
$\sigma^2$ is the variance on the pixel fluxes, 
$\mathcal{T}$ indicates the template (normalized to an integrated flux of 1) to match to the spectrum,
and $\alpha$ is the flux scaling to apply to the template to obtain the best match.
Using that 
\begin{eqnarray}\label{eq:minicrit}
\frac{\partial \chi^2}{\partial \alpha} &=& -2\sum_i \left( \frac{\mathcal{D}_i-\alpha\mathcal{T}_i}{\sigma_i^2} \right) \frac{\partial}{\partial \alpha} \alpha\mathcal{T}_i  \\ 
\Downarrow  \quad \quad & & \nonumber \\
0 &=& \sum_i \frac{ \mathcal{D}_i - \alpha\mathcal{T}_i}{ \sigma_i^2}\mathcal{T}_i = \sum_i \frac{\mathcal{D}_i \mathcal{T}_i}{\sigma_i^2} -  \alpha\sum_i \frac{\mathcal{T}_i^2}{\sigma_i^2}
\end{eqnarray}
for the minimum $\chi^2$ value implies
\begin{equation}\label{eq:felisflux}
\alpha          = \frac{\sum_i \mathcal{D}_i\mathcal{T}_i/\sigma_i^2 }{ \sum_i \mathcal{T}_i^2 / \sigma_i^2 }
\end{equation}
with an estimated variance on $\alpha$ given by 
\begin{equation}\label{eq:CCvariance}
\sigma_\alpha^2 = \frac{1}{\sum_i \mathcal{T}_i^2 / \sigma_i^2 } \; .
\end{equation}
Hence, the best match of the normalized template $\mathcal{T}$ is obtained by scaling it by the flux $\alpha$.
The S/N of this template match can be estimated as 
\begin{equation}\label{eq:CCs2n}
\textrm{S/N} = \frac{\alpha}{\sigma_\alpha} \, .
\end{equation}
Cross-correlating each template with the spectrum, that is shifting the template over the spectrum pixel-by-pixel while minimizing $\chi^2$ provides a $\chi^2$ and S/N curve for each template and spectrum pair.
Examples of these curves are shown in the two lower panels of Figure~\ref{fig:FELIStempfit}).
The maximum of the S/N curve indicates the optimal alignment of the template with the spectrum.
Comparing the (S/N)$_{\mathcal{T}, \textrm{max}}$ with the maximum S/N values for a range of templates matched to the same spectrum provides the overall best match to the measured fluxes in the spectrum, (S/N)$_\textrm{max}$, given the set of templates explored.  

\subsection{Building spectral templates with FELIS}\label{sec:FELIStemplates}
The templates used for the FELIS cross-correlation can be generated with the function
\tiny
\begin{lstlisting}
felis_build_template.build_template()
\end{lstlisting}
\normalsize
The input to this function is a dictionary defining the template components to combine to the final template and the wavelength range and resolution of the template to generate.
Various template components can be added to the dictionary including:
\begin{itemize}
\item A skewed Gaussian emission line profile of the form 
\begin{eqnarray}
\textrm{PDF}(\lambda) &=& \frac{1}{ \sqrt{2\pi\sigma^2}}  \exp\left(-\frac{(\lambda-\mu)^2}{2\sigma^2} \right)  \\
2\times\textrm{CDF}(\lambda) &=& 1 + \textrm{erf}\left(\frac{\beta(\lambda-\mu)}{\sqrt{2}\sigma}\right)  \\
G(\lambda) &=& \textrm{PDF} \times 2\times\textrm{CDF}
\end{eqnarray}
where PDF and CDF refers to the probability and cumulative distribution functions, respectively. 
The template dictionary parameters to provide are
\tiny
\begin{lstlisting}
['GAUSS', mean, sigma, skew, scaling, 'info']
\end{lstlisting}
\normalsize
where he parameters \verb+mean+ and \verb+sigma+ are represented by $\mu$ and $\sigma$. The $\beta$ provides the \verb+skew+ of the Gaussian distribution. 
If $\beta=0$ then the error function (erf) is 0 and $G(\lambda)$ is a regular Gaussian PDF. 

\item A delta function (single pixel flux) added at the wavelength nearest the \verb+position+ in the generated template.
The template dictionary parameters to provide are
\tiny
\begin{lstlisting}
['DELTA', position, total flux, 'info']
\end{lstlisting}
\normalsize

\item A Gaussian line spread function (LSF) which the generated template will be convolved with before it is returned.
The template dictionary parameters to provide for this are
\tiny
\begin{lstlisting}
['LSF', sigma, 'info']
\end{lstlisting}
\normalsize
For the study presented in this paper, no LSF was added to the templates. 
The templates are ``raw'' Gaussians representing the observed signal focusing on recovering line fluxes.

\item A linear continuum with the template dictionary parameters
\tiny
\begin{lstlisting}
['CONT', level, slope, lam0, 'info']
\end{lstlisting}
\normalsize
describing the continuum as 
\begin{equation}
C = \verb+level+ + \verb+slope+ \times (\lambda-\verb+lam0+) 
\end{equation}
where \verb+lam0+ provides a reference wavelength $\lambda_0$ for the slope. 
Alternatively, a predefined continuum can be aded as a ``flux feature''.

\item A predefined spectral flux feature defined by a wavelength and flux vector as provided in the template dictionary parameter list:
\tiny
\begin{lstlisting}
['FEATURE',wavelength,flux,'info']
\end{lstlisting}
\normalsize
Examples of such features could be spectral breaks, nongaussian emission line profiles, or a nonlinear continuum.
\end{itemize}
The ``info'' provided to each of the template components contains a string with information that will be written to the header of the fits file containing the template returned by FELIS.
As an example generating a template containing a \ciii{} emission line doublet, a delta-function spike at 1900~\AA, and a blue continuum can be done with the commands:
\tiny
\begin{lstlisting}
import felis_build_template as fbt
# --- INPUT ---
# fill template component dictionary
tcdic = {}
tcdic['D1900']	= ['DELTA', 1900.0, 10.0,           'Delta function at 1900A']
tcdic['CIII1']	= ['GAUSS', 1907.0, 0.5, 0.0, 10.0, 'CIII]1907A']
tcdic['CIII2']	= ['GAUSS', 1909.0, 0.5, 0.5, 5.0,  'CIII]1909A']
tcdic['CONT']	= ['CONT', 1.0, -0.03, 1908.0,       'Continuum with flux 1.0 at 1908 + slope -0.03']
# --- COMMAND ---
template_range   = [1870,1980,0.1]
fbt.build_template(template_range,tcdic,tempfile='./outputname.fits')
\end{lstlisting}
\normalsize
This template will be noise-free.
To enable the generation of mock spectra a \verb+noise+ description can be provided to 
\tiny
\begin{lstlisting}
felis_build_template.build_template()
\end{lstlisting}
\normalsize
This adds noise to the template spectrum according to one of the following prescriptions:
\begin{itemize}
\item Noise drawn for each template pixel from a Poisson distribution around a mean value can be done with
\tiny
\begin{lstlisting}
noise=['POISSON',mean]
\end{lstlisting}
\normalsize

\item Gaussian random noise of each template pixel from a Gaussian error spectrum defined by the \verb+mean+ and standard deviation \verb+sigma+ in each pixel can be obtained with 
\tiny
\begin{lstlisting}
noise=['GAUSS',mean,sigma]
\end{lstlisting}
\normalsize

\item Assigning a Gaussian random noise around a constant noise level with the value \verb+value+ to each template pixel is done with
\tiny
\begin{lstlisting}
noise=['CONSTANT',value]
\end{lstlisting}
\normalsize

\item A Gaussian random noise for each template pixel drawn from a provided error spectrum defined by the wavelength and flux vectors \verb+wave+ and \verb+flux+ after interpolating it to the template wavelength range can be obtained with
\tiny
\begin{lstlisting}
noise=['SPECTRUM',wave,flux]
\end{lstlisting}
\normalsize
This noise prescription can be used to include information about sky-residuals, poor pixels, etc. in the noise prescription.
\end{itemize}
In the last three noise prescriptions an error spectrum is defined (or explicitly provided) from which Gaussian noise is drawn at each pixel. 
In practice, the pixel values in the error spectrum (which are stored as the template's flux error) are assumed to represent the standard deviation of the noise to be applied to each pixel. 
The noise-free template flux of each pixel is modified by adding a flux value drawn from a Gaussian distribution with mean 0 and standard deviation corresponding to the error spectrum value for that pixel.
Hence, the resulting S/N of a featureless noisy template is ensured to be a Gaussian centered on 0 with a standard deviation of 1. 

\subsection{Searching spectra with FELIS templates}\label{sec:runFELIS}

Having build a set of spectral templates that represents the parameter space of the spectral features expected to occur in the observed spectra, the spectra can be searched with FELIS by executing the following command:
\tiny
\begin{lstlisting}
import felis
# --- INPUT ---
spectra = [spectra,to,match]
redshifts = [1.5,3.5,5.5]
templates = [templates,to,match]
picklefile = './NameOfOutput.pkl'
plotdir = './path/to/directory/with/plots/'
wavewindow = [50]*len(spectra)            # observed frame
windowcenter = [5007]*len(spectra) # rest-frame
# --- COMMAND ---
ccdic      = felis.match_templates2specs(templates, spectra, redshifts, picklefile, plotdir=plotdir, wavewindow=wavewindow, wavecen_restframe=windowcenter)
\end{lstlisting}
\normalsize
This will produce a dictionary (\verb+ccdic+) containing the results from the cross-correlation $\chi^2$ minimizations performed by FELIS. 
The dictionary is saved to a binary Python pickle file if an output filename is provided.
This pickle file can be loaded with 
\tiny
\begin{lstlisting}
felis.load_picklefile() 
\end{lstlisting}
\normalsize
To select spectra from the pickle file output that have template matches of a certain quality and characteristics the function 
\tiny
\begin{lstlisting}
felis.selection_from_picklefile(picklefile)
\end{lstlisting}
\normalsize
is provided. 
For instance,
\tiny
\begin{lstlisting}
speclist   = felis.selection_from_picklefile(picklefile,S2Nmaxrange=[3,5])
\end{lstlisting}
\normalsize
will return all spectra that have a template match with a maximum S/N between three and five.
When provided a FELIS output dictionary and a dictionary key (the name of a spectrum) the function 
\tiny
\begin{lstlisting}
felis.getresult4maxS2N() 
\end{lstlisting}
\normalsize
will return information (including best-fit redshift, maximum S/N value, and template name) of the template matching the observed spectrum best.  
The content of the output can be plotted with 
\tiny
\begin{lstlisting}
felis.plot_picklefilecontent()
\end{lstlisting}
\normalsize

Above, the \verb+redshifts+ are provided as input to perform the template matching by FELIS in rest-frame. 
Hence, the templates are assumed to be in rest-frame, and the observed spectra will be moved to rest-frame by scaling wavelength and fluxes such that:
\begin{eqnarray}
f_\textrm{rest} &=& f_\textrm{obs} \times (1+z) \\
\sigma_\textrm{rest} &=& \sigma_\textrm{obs} \times (1+z) \\
\lambda_\textrm{rest} &=& \lambda_\textrm{obs} / (1+z)
\end{eqnarray}
This ensures that integrated fluxes and S/N are conserved between the observed-frame and rest-frame spectra.

To avoid cross-matching a template over the whole spectral range, if for instance the redshift of the object is roughly known and the search is for a well-defined emission line, the \verb+wavewindow+ and \verb+wavecen_restframe+ can be used to define the window over which the template match (cross-correlation) is performed.
The window is defined as 
\begin{equation}
\verb+wavecen_restframe+ \times (1+z) \pm \verb+wavewindow+
\end{equation}
The redshifts can be both photometric or spectroscopic. Supplying more uncertain photometric redshifts of course requires searching for features over a larger wavelength range. 
To look for potential velocity shifts of emission lines with respect to each other or systemic redshift, a spectroscopic redshift is clearly preferred.

\subsection{Testing FELIS on MUSE-Wide mock spectra}\label{sec:FELIStest_mock}

To test the reliability of FELIS to recover line fluxes, line widths and doublet flux ratios at different emission line S/N we generated a series of mock spectra for regions around the main rest-frame UV emission lines studied in this work (see Table~\ref{tab:LinesAndTemplates}). 
The mock spectra were generated using the template creation capabilities of FELIS described in Section~\ref{sec:FELIStemplates}.
Each line was modeled as a Gaussian with a width $\sigma_\textrm{Gauss}$ and a total emission line flux in units of 10$^{-20}$~erg~s$^{-1}$~cm$^{-2}$~\AA$^{-1}$. For doublets the emission line flux ratios were set as an extra constraint.
For the testing we also generate mock \lya{} spectra where we represented the \lya{} emission by a single skewed Gaussian ignoring any blue bumps.
All spectra were generated with a wavelength resolution of 1.25~\AA{}  and generated in a wavelength range of $\pm$50~\AA{} around the central wavelength of the emission line (for doublets we used the central wavelength of the doublet).
Table~\ref{tab:mockspecparam} summarizes the template parameters for the mock spectra.

\begin{figure}
\begin{center}
\includegraphics[width=0.49\textwidth]{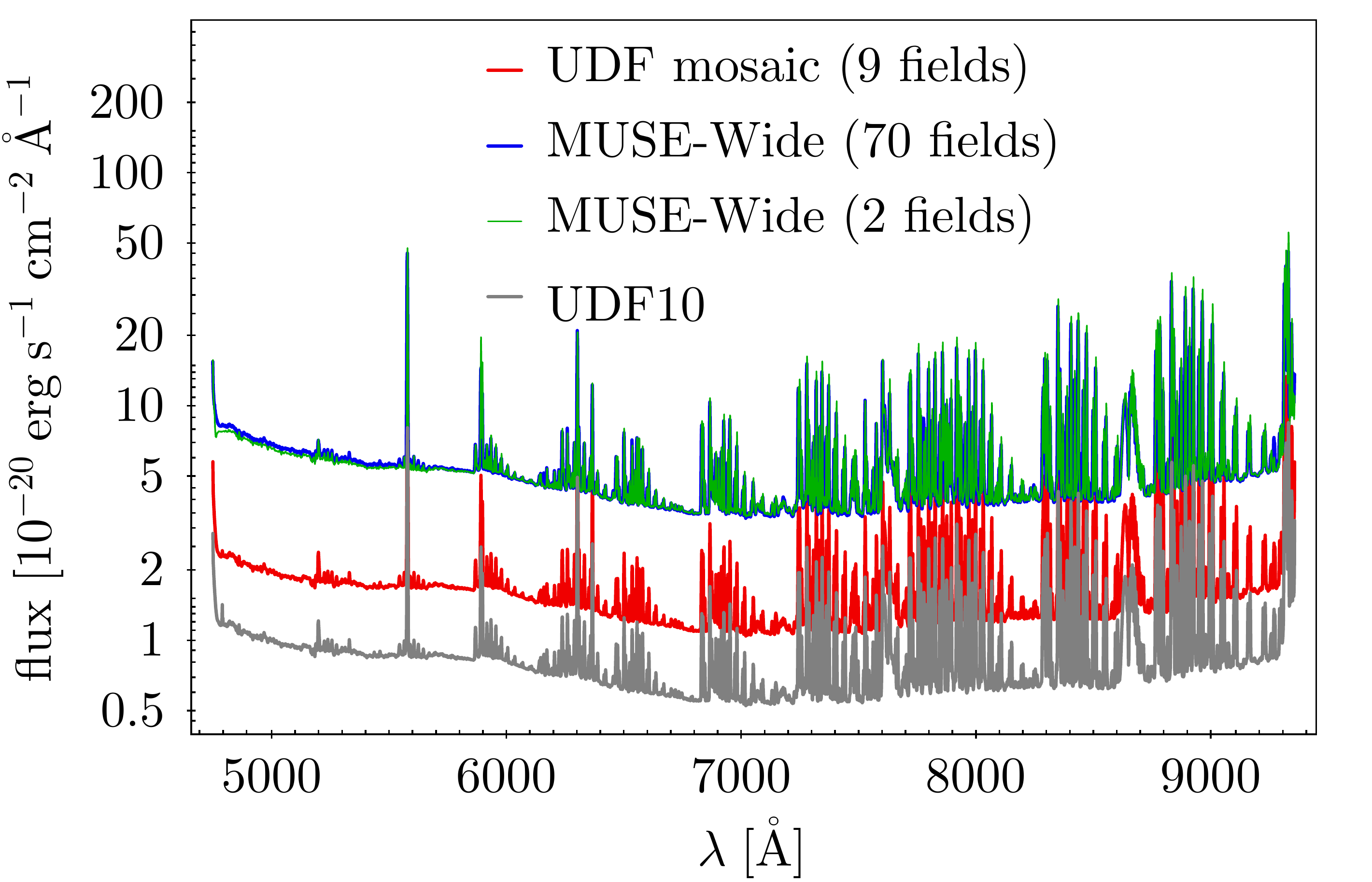}
\caption{Median noise spectrum of two random MUSE-Wide pointings (green) and 70 GOODS-South and COSMOS MUSE-Wide pointings (blue - almost identical to the green curve).
The latter spectrum was used to draw noisy flux values for the MUSE-Wide mock spectra that FELIS was tested on.
For comparison, the median noise spectra of the deeper MUSE UDF mosaic and the UDF10 pointing are shown in red and gray. 
The median noise spectra shown in this panel are displayed by the black filled regions at the bottom of each panel in the spectral overviews in Figure~\ref{fig:ObjSpec} and Figures~\ref{fig:ObjSpec04}--\ref{fig:ObjSpec99}.
}
\label{fig:noisespec}
\end{center}
\end{figure}

We generated a set of noise-free templates as control sample, and a sample of templates with idealized noise added.
Each of the noisy mock spectra was added noise based on the median noise spectrum from MUSE-Wide (shown by the green and blue curves in Figure~\ref{fig:noisespec}) scaled by a factor 5.5. 
The scaling was introduced to resemble noise for a spectrum extracted using a $r=0\farcs6$ aperture, which corresponds to propagating the noise from 30 MUSE spaxels as $\sum_i \sigma_i^2 /\sqrt{N} \approx 5.5 $ for $N=30$.
The flux value in each pixel of the mock spectra was obtained by adding the noise-free template a flux term obtained from drawing a random value from a gaussian centered around 0 with a width corresponding to the value from the MUSE-Wide noise spectrum at each individual wavelength.

\begin{table*}
\small 
\caption{Template parameters of mock MUSE-Wide spectra.}
\centering
\begin{tabular}{lccccccc}
\hline\hline
Line 		&      Line Wavelength    	&    $\sigma_\textrm{Gauss}$	& Line flux scaling					& Flux ratios	& Redshift         	& Gauss skew	 	& N$_\textrm{Spec}$		\\
		&      [\AA]				&    [\AA]					& [$10^{-20}$~erg~s$^{-1}$~cm$^{-2}$~\AA$^{-1}$]				& 			& 		         	& 			 	&					\\
\hline
\ciii		& 1907, 1909			&    0.5, 1.0, 2.0, 4.0			& 50, 100, 300, 600					& 0.3, 1.0, 1.4	&  2.0, 2.7, 3.5		&		0.0		&	144				\\
\civ		& 1548, 1551			&    0.5, 1.0, 2.0, 4.0			& 50, 100, 300, 600					& 0.8, 1.5, 2.0	&  2.5, 3.5, 4.5		&		0.0		&	144				\\
\nv          	& 1239, 1243			&    0.5, 1.0, 2.0, 4.0			& 50, 100, 300, 600					& 0.5, 1.0, 1.5	&  3.0, 4.5, 6.0		&		0.0		&	144				\\
\mgii       	& 2796, 2803			&    0.5, 1.0, 2.0, 4.0			& 50, 100, 300, 600					& 0.5, 1.0, 1.5	&  0.8, 1.5, 2.2		&		0.0		&	144				\\
\oiii 	& 1661, 1666			&    0.5, 1.0, 2.0, 4.0			& 50, 100, 300, 600					& 0.5, 1.0, 1.5	&  2.0, 3.0, 4.0 		&		0.0		&	144				\\
\heii           & 1640           			&    0.5,1.0,2.0,4.0			& 50, 100, 300, 600, 900				& 0.0			&  2.5, 3.5, 4.5 		&		0.0		&	60				\\
\lya            & 1216             		&    3.0,6.0,9.0				& 100, 500, 1000, 3000, 6000, 9000		& 0.0			&  3.0,4.0,5.0,6.0	&	0.0, 5.0, 10.0	&	216				\\
\hline
\end{tabular}
\label{tab:mockspecparam}
\tablefoot{
Each mock spectrum is $\pm$50\AA{} wide around the line wavelength (central wavelength for doublets). The wavelength resolution was set to the average MUSE resolution of 1.25\AA. 
The line width $\sigma_\textrm{Gauss}$ is provided in observed frame and the line seperation of the doublet components is fixed.
}
\end{table*}
\normalsize

To recover the emission lines of the mock spectra we performed a search with FELIS in a $\pm$10~\AA{} window in the rest-frame around the central wavelength of the mock spectra of the targeted emission features.
The mock spectra were positioned at arbitrary observed redshifts to include rest-frame conversions of line widths and interpolation of templates to the mock spectra rest-frame reference grids.
We first matched the templates used in the main text described in Table~\ref{tab:LinesAndTemplates} on the noise-free mock spectra.
In this case the FELIS template matches recovers the input flux values (total flux of emission doublets) to well within 10\% as long as the spectral feature has an integrated S/N above three and the intrinsic rest-frame line width of the feature in the mock spectrum is within the sampled range of the templates listed in Table~\ref{tab:LinesAndTemplates}.
Also the recovered line widths, $\sigma_\textrm{Gauss}$, are accurate to within 10\%. 
Note however, that the template resolution of  $\sigma_\textrm{Gauss}$ is 0.1, so for mock spectra with intrinsic rest-frame line widths below $\approx$0.5~\AA{} the relative inaccuracy grows to roughly 20\% for the noise free mock spectra.
The intrinsic doublet flux ratios of the emission line doublets are recovered correctly to within a few percent (arising from the interpolation of the templates to the ``observed'' mock spectra's rest-frame grid).
In the recovery of the intrinsic parameters of the noise-free mock spectra, there is a correlation between the precision of the recovered total line flux and the intrinsic line widths such that for intrinsic rest-frame line widths comparable to the template resolution (here 0.1) the recovered flux is less precise. 

\begin{figure}
\begin{center}
\includegraphics[width=0.49\textwidth]{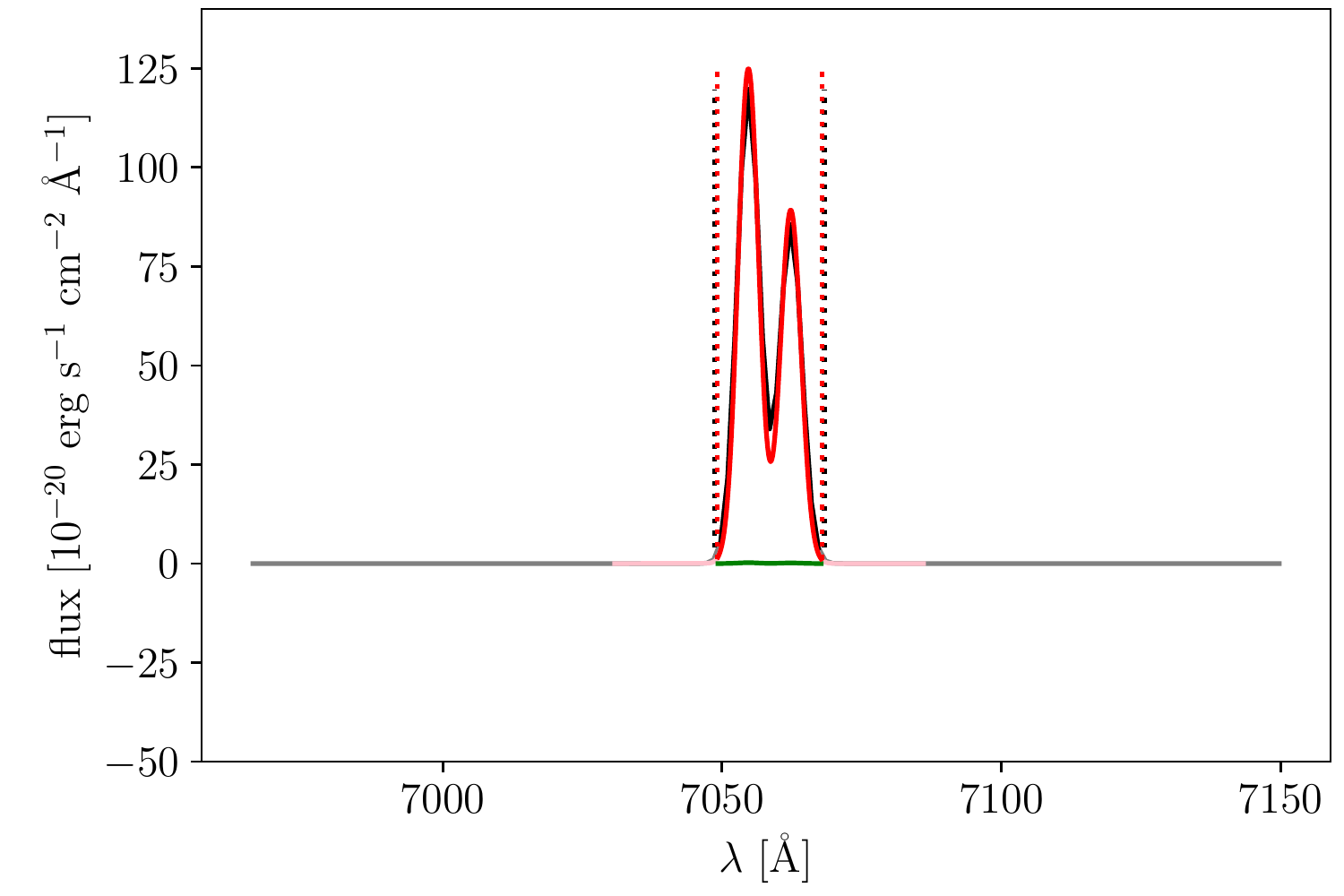}\\
\includegraphics[width=0.49\textwidth]{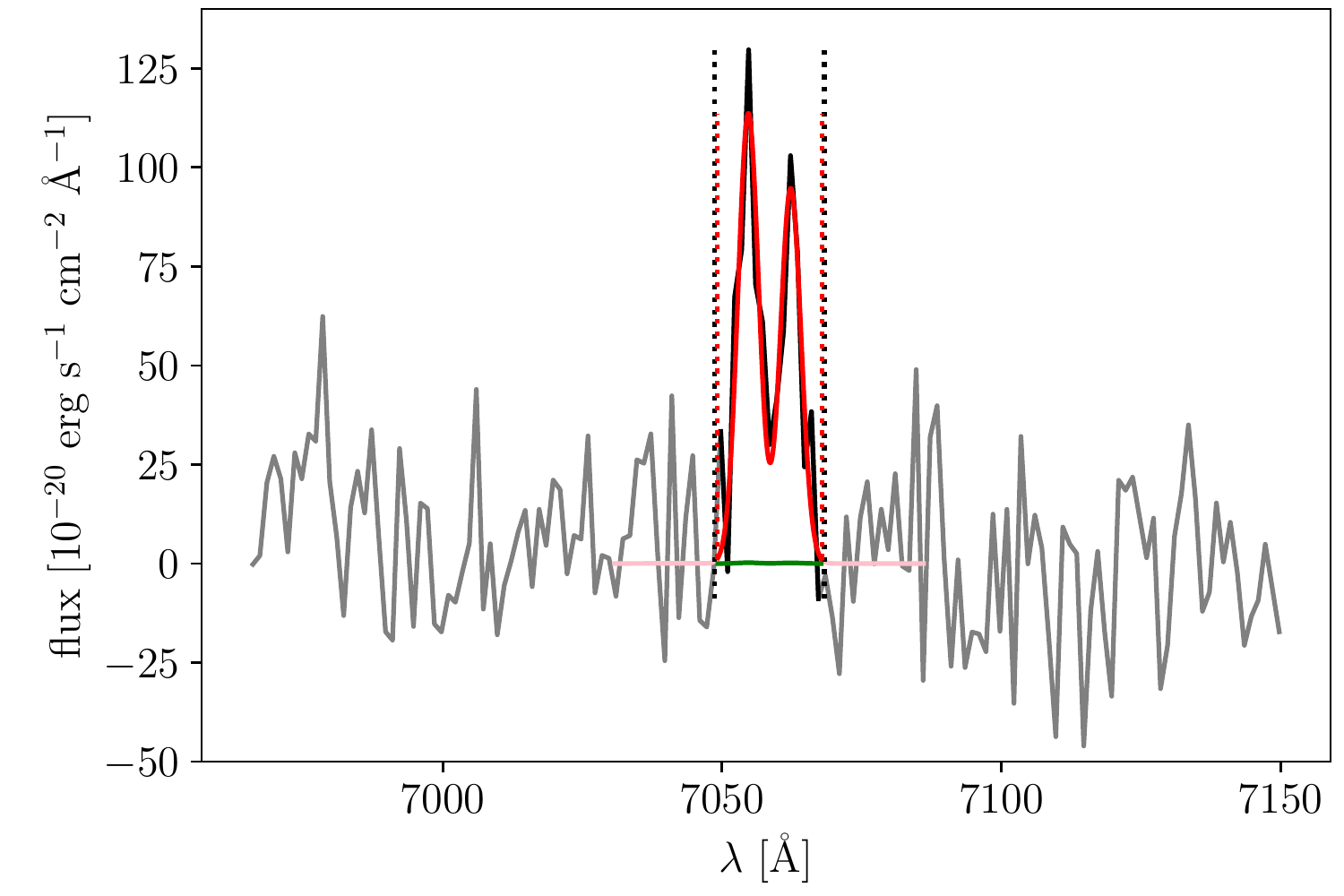}
\caption{Noise-free (top) and noisy (bottom) template matches to the mock spectrum shown in Figure~\ref{fig:FELIStempfit}. The top panel of Figure~\ref{fig:FELIStempfit} shows the same as the bottom panel of this figure. 
The red curve shows the FELIS template match to the part of the spectrum in black (within the vertical dotted lines).
The green curve shows the template match before applying the flux scaling.
The part of the mock spectrum outside the considered wavelength range is marked in gray. 
}
\label{fig:FELIStempfit_both}
\end{center}
\end{figure}

Having fitted the mock spectra without noise we ran FELIS with the same set of templates (Table~\ref{tab:LinesAndTemplates}) on the same set of mock spectra (Table~\ref{tab:mockspecparam}) with noise added as described above.
As expected the ability of the FELIS template match to recover the intrinsic flux values and emission line widths decreases when introducing noise in the mock spectra.
Figure~\ref{fig:FELIStempfit} shows an example of one of the results from a template match to a noisy mock spectrum performed with FELIS. 
The mock spectrum (black curve) shows a \ciii-emitter with a doublet flux ratio of \ciiione/\ciiitwo{} = 1.4 and a total intrinsic (before adding noise) line flux of $1028.57\times10^{-20}$~erg~s$^{-1}$~cm$^{-2}$~\AA$^{-1}$.
The noise in each pixel was drawn from a median effective noise spectrum of the MUSE-Wide fields at an ``observed'' redshift of 2.7 (i.e., at $\lambda_\textrm{obs}\approx7055~\AA$). 
The intrinsic ``observed'' frame Gaussian line width was set to 2.0~\AA{} which corresponds to a rest-frame line width of 0.54~\AA. 
For the noisy mock spectrum the integrated line flux (integrated over [$\lambda_1-3\sigma$,$\lambda_2+3\sigma$] with $\sigma=2$~\AA) is $989.60\times10^{-20}$~erg~s$^{-1}$~cm$^{-2}$~\AA$^{-1}$.
The scaled best-fit template from the set of templates in Table~\ref{tab:LinesAndTemplates} is shown in green.
The best-fit template, returning the maximum cross-correlation (S/N)$_{\mathcal{T}, \textrm{max}}\approx12$ as shown in the third panel of Figure~\ref{fig:FELIStempfit}, has a doublet flux ratio of 1.2, and a rest-frame line width of 0.5~\AA. 
The total flux estimated by the template scaling performed by FELIS is $(966.93\pm80.82)\times10^{-20}$~erg~s$^{-1}$~cm$^{-2}$~\AA$^{-1}$. 
In the case of matching the templates from Table~\ref{tab:LinesAndTemplates} to the noise-free version of the mock spectrum shown in Figure~\ref{fig:FELIStempfit} the estimates of rest-frame line width, flux ratio, and total flux were 0.5~\AA, 1.4, and $993.47\times10^{-20}$~erg~s$^{-1}$~cm$^{-2}$~\AA$^{-1}$, respectively.
The roughly 3\% loss 
in total flux when comparing the noise-free template match to the intrinsic noise-free mock spectrum comes from the mismatch in rest-frame emission line widths, as the line widths of the templates take steps of 0.1~\AA{}. 
Hence, only templates with 0.5~\AA{} and 0.6~\AA{} exist, and the best-fit template thus has a line width of only 0.5~\AA{} compared to the actual rest-frame line width of the mock spectrum of 2.0~\AA/(1+2.7) $\approx$ 0.54~\AA. 
This is illustrated in the panels of Figure~\ref{fig:FELIStempfit_both} showing both the noise-free (top) and noisy (bottom) mock spectra with the best-fit scaled template shown on top in red.

In general, for the template match to the noisy mock spectra we see that the total flux is recovered with an accuracy better than 50\% for the spectral doublets and to within 20\% for single lines for a FELIS match significance S/N~$>3$. 
A FELIS S/N~$>3$ can roughly be translated to a total integrated flux of the spectral feature of $\approx300-500\times10^{-20}$~erg~s$^{-1}$~cm$^{-2}$ with the added noise from the median MUSE-Wide spectrum.
This flux level is of course lowered for deeper data.
The flux ratios are generally robust to within a factor of two of the intrinsic flux ratio, but are in most cases good at the $\approx50\%$ level, especially if a slightly higher S/N than three is required. 
The precision of the recovery of the flux ratio appears to be controlled by the intrinsic line widths and their interpolation  to the rest-frame grid of the ``observed'' mock spectra, as the doublet separations are at least 4.8 times that of the rest-frame pixel resolution given the 1.25~\AA{} spectral resolution of MUSE and the emission line doublets considered here.
The Gaussian line width is recovered within 50\% of the intrinsic value of the mock spectra.
Again this assumes a FELIS S/N~$>3$ of the template match.
Lastly, the redshifts, that is the location in wavelength of the best-fit template, are recovered within $\pm$100~km~s$^{-1}$.
As no velocity offsets were included in the mock spectra, this shows that the expected precision of the recovered velocity offset from FELIS given the spectral sampling of MUSE (1.25~\AA) is at the 100~km~s$^{-1}$ level for the doublets considered here assuming a FELIS S/N~$>3$.  

\section{Rest-frame UV emission line catalog and measurement correlations}\label{sec:mastercat}

Together with the paper we provide a catalog of the full list of MUSE emission line sources at $1.5<z<6.4$ searched for UV emission lines in this paper. 
The content of this catalog is described in Table~\ref{tab:mastercatcol}.
\begin{table*}
\caption{\label{tab:mastercatcol}The columns of the catalog of MUSE emission line sources studied in this work released with the paper.} 
\centering
{
\begin{tabular}{p{2.2cm}p{2.5cm}p{2.5cm}p{10.0cm}}
\hline\hline
 & Unit & Catalog column & Description \\
\hline
ID 								& 						& {\tiny\verb+id+}				& Object ID. Follows the formats described in Section~\ref{sec:objsel}. \\
R.A., Dec. 						& [deg]					& {\tiny\verb+ra+}, {\tiny\verb+dec+}	& Coordinates of the object based on the LSDCat detection  (Section~\ref{sec:objsel})\\
$z_\textrm{lead line}$  				& 						& {\tiny\verb+redshift+}			& Redshift of object based on LSDCat lead line (Section~\ref{sec:objsel})\\
Lead line			  				& 						& {\tiny\verb+lead_line+}			& Lead line from the LSDCat detection (see Section~\ref{sec:objsel})\\
Confidence				  		& 						& {\tiny\verb+confidence+}			& Confidence from QtClassify inspections (see Section~\ref{sec:objsel})\\
Duplication ID				  		& 						& {\tiny\verb+id_duplication+}		& ID of duplicating source in deeper overlapping regions of MUSE fields (see Section~\ref{sec:1Dspec})\\
F$_\textrm{line}$ 					& [10$^{-20}$erg~s$^{-1}$~cm$^{-2}$]	& {\tiny\verb+f_line+}				& Emission line flux where {\tiny\verb+line+} refers to any of the lines \lya{} ({\tiny\verb+lya+}), \nv{} ({\tiny\verb+nv+}), \civ{} ({\tiny\verb+civ+}), \heii{} ({\tiny\verb+heii+}), \oiii{} ({\tiny\verb+oiii+}), \siiii{} ({\tiny\verb+siiii+}), and \ciii{} ({\tiny\verb+ciii+}). For doublets, the values for the individual components are also provided. As an example the columns for \ciii{} are named {\tiny\verb+f_ciii+}, {\tiny\verb+f_ciii1+}, and {\tiny\verb+f_ciii2+}. \\
$\delta$F$_\textrm{line}$   			& [10$^{-20}$erg~s$^{-1}$~cm$^{-2}$]	& {\tiny\verb+ferr_line+}			& Uncertainty on F$_\textrm{line}$. Upper limits have uncertainties set to +99.  \\
(S/N)$_\textrm{line}$	 	 			& 						& {\tiny\verb+s2n_line+}			& S/N of the template match to the emission line performed by FELIS as defined in Equation~\ref{eq:CCs2n}, i.e., the FELIS significance of the detected feature. \\ 
$\sigma_\textrm{line}$ 	 			& [\AA]					& {\tiny\verb+sigma_line+}			& Gaussian $\sigma$ of the best-fit emission line template from the FELIS cross-correlation.   \\
$\Delta v_\textrm{line}$  				& [km~s$^{-1}$]			& {\tiny\verb+vshift_line+}			& Estimated velocity shift of the emission feature with respect to the lead line catalog redshift.  \\
FR$_\textrm{line1line2}$				& 						& {\tiny\verb+fr_line1line2+}		& Flux ratio $f$(line1)/$f$(line2) between the two emission line doublet components of the UV emission line doublets from the FELIS doublet template match (Section~\ref{sec:UVEmissionLineSearch}). \\
$\delta$FR$_\textrm{line1line2}$		& 						& {\tiny\verb+frerr_line1line2+}		& Uncertainty on FR$_\textrm{line1line2}$. Upper and lower limits have uncertainties set to +99 and -99, respectively. \\
FR$_\textrm{line1line2}$ S/N 			& 						& {\tiny\verb+frs2n_line1line2+}		& S/N of the emission line doublet flux ratio.  \\
EW$_{0,\textrm{line}}$ 				& [\AA]					& {\tiny\verb+ew0_line+}			& Rest-frame equivalent width (Section~\ref{sec:EWcoor}) of {\tiny\verb+line+}, where {\tiny\verb+line+} refers to any of the lines \lya, \nv, \civ, \heii, \oiii, \siiii, and \ciii. For doublets, the values for the individual components are also provided. \\
$\delta$EW$_{0,\textrm{line}}$   		& [\AA]					& {\tiny\verb+ew0err_line+}		& Uncertainty on EW$_\textrm{0,line}$. Upper and lower limits have uncertainties set to +99 and -99, respectively. \\
CB						  		& 						& {\tiny\verb+contband_line+}		& HST broad-band used to estimate the continuum level when calculating the EW$_0$ (cf. Equation~\ref{eq:EW0}).\\
CM						  		& [AB mag]				& {\tiny\verb+contmagab_line+}		& AB magnitude of the continuum flux in CB corresponding to $f_\textrm{continuum, observed}$ in Equation~\ref{eq:EW0}.  \\
$\delta$CM 			  			& [AB mag]				& {\tiny\verb+contmagaberr_line+}	& Uncertainty on CM. Lower limits (magnitudes below the detection limit) have their uncertainty set to -99.  \\
photoref	 			  			& 						& {\tiny\verb+photref_line+}		& Reference to the photometric catalog CM was taken from. Here 1 corresponds to photometry from the \cite{Kerutt:2021tr} GALFIT models, 2 refers to magnitudes from the \cite{2014ApJS..214...24S} 3D-HST catalogs, and 3 indicates photometry taken from the \cite{2015AJ....150...31R} catalog. \\
ID$_\textrm{Kerutt}$					& 						& {\tiny\verb+id_kerutt+}			& Object ID used in the \cite{Kerutt:2021tr} catalog. \\
F$_\textrm{\lya}$		 			& [10$^{-20}$erg~s$^{-1}$~cm$^{-2}$]	& {\tiny\verb+f_lya+}				& \lya{} flux from the \cite{Kerutt:2021tr} catalog. \\
$\delta$F$_\textrm{\lya}$		 		& [10$^{-20}$erg~s$^{-1}$~cm$^{-2}$]	& {\tiny\verb+ferr_lya+}			& Uncertainty on F$_\textrm{\lya}$. \\
EW$_{0,\textrm{line}}$ 				& [\AA]					& {\tiny\verb+ew0_lya+}			& Rest-frame equivalent width of \lya{} from the \cite{Kerutt:2021tr} catalog. \\
$\delta$EW$_{0,\textrm{line}}$   		& [\AA]					& {\tiny\verb+ew0err_lya+	}		& Uncertainty on EW$_\textrm{0,\lya}$. \\
M$_\textrm{UV,1500}$		   		& [AB mag]				& {\tiny\verb+magabs_uv+	}		& Estimated UV magnitude at 1500\AA{} from the \cite{Kerutt:2021tr} catalog. \\
$\delta$M$_\textrm{UV,1500}$		   	& [AB mag]				& {\tiny\verb+magabserr_uv+}		& Uncertainty on M$_\textrm{UV,1500}$. \\
FWHM(\lya)				   		& [km~s$^{-1}$]			& {\tiny\verb+fwhm_lya+	}		& Full-width at half maximum of the \lya{} (red component) emission line from the \cite{Kerutt:2021tr} catalog. \\
$\delta$FWHM(\lya)			   		& [km~s$^{-1}$]			& {\tiny\verb+fwhmerr_lya+}		& Uncertainty on FWHM$_\textrm{\lya}$. \\
PeakSep(\lya)						& [km~s$^{-1}$]			& {\tiny\verb+peaksep_lya+}		& Peak separation of LAEs with a two-component \lya{} line from the \cite{Kerutt:2021tr} catalog.  \\
$\delta$PeakSep(\lya)				& [km~s$^{-1}$]			& {\tiny\verb+peakseperr_lya+}		& Uncertainty on the \lya{} peak separation. \\
ID$_\textrm{cat}$ 					& 						& {\tiny\verb+id_cat+	}			& ID of the closest match to a series of catalogs from the literature. Here {\tiny\verb+cat+} is {\tiny\verb+skelton+}, {\tiny\verb+rafelski+}, {\tiny\verb+guo+} or {\tiny\verb+laigle+} and refers to the catalogs presented by \cite{2014ApJS..214...24S}, \cite{2015AJ....150...31R}, \cite{2013ApJS..207...24G} or \cite{2016ApJS..224...24L}, respectively. \\
$\Delta_\textrm{cat}$  				& [arcsec]					& {\tiny\verb+sep_cat+}			& Separation between ID$_\textrm{cat}$ and the catalog object in arc seconds.\\
\hline
\end{tabular}
}
\end{table*}

\begin{figure*}
\begin{center}
\includegraphics[width=0.80\textwidth]{mainlegend.png}\\
\includegraphics[width=0.42\textwidth]{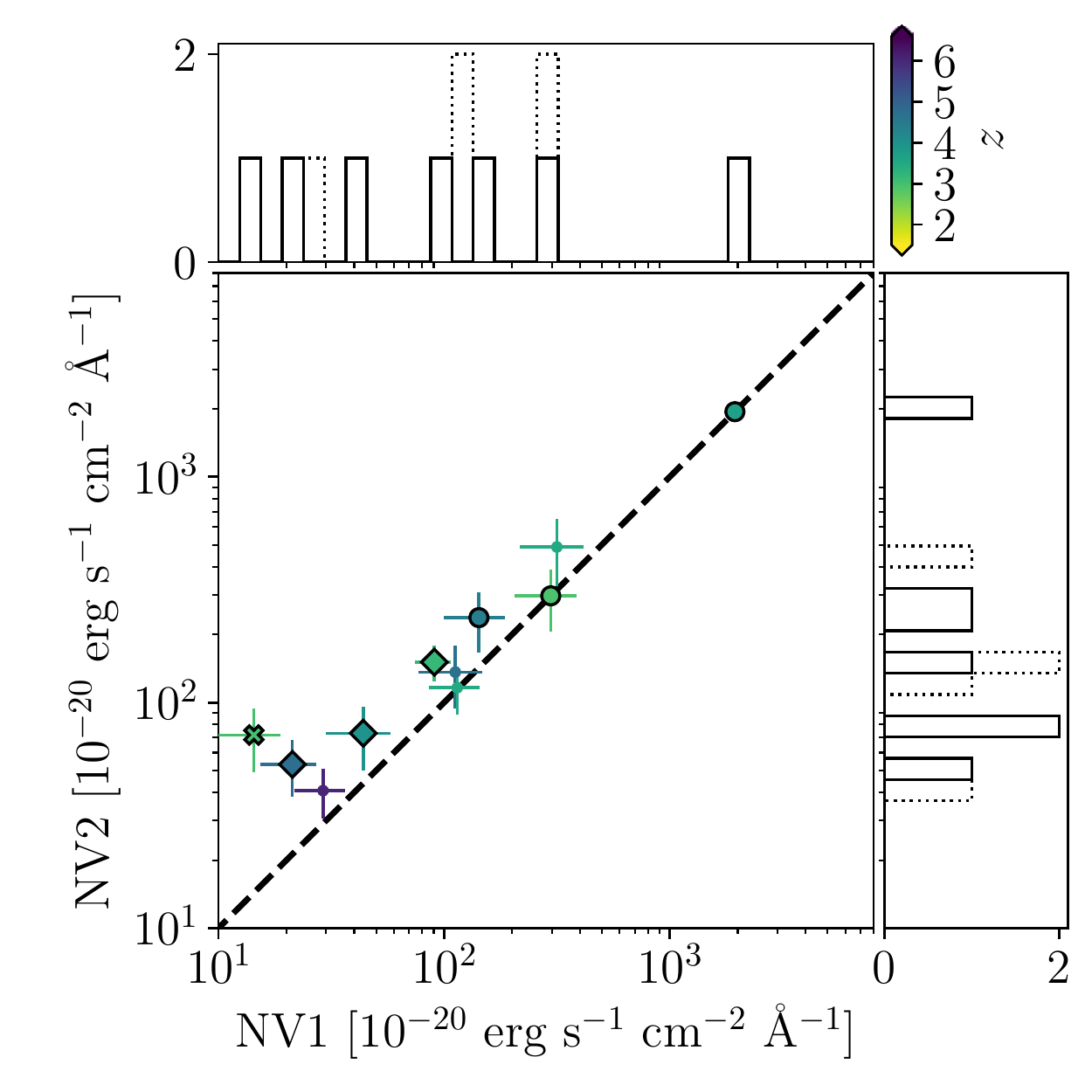}
\includegraphics[width=0.42\textwidth]{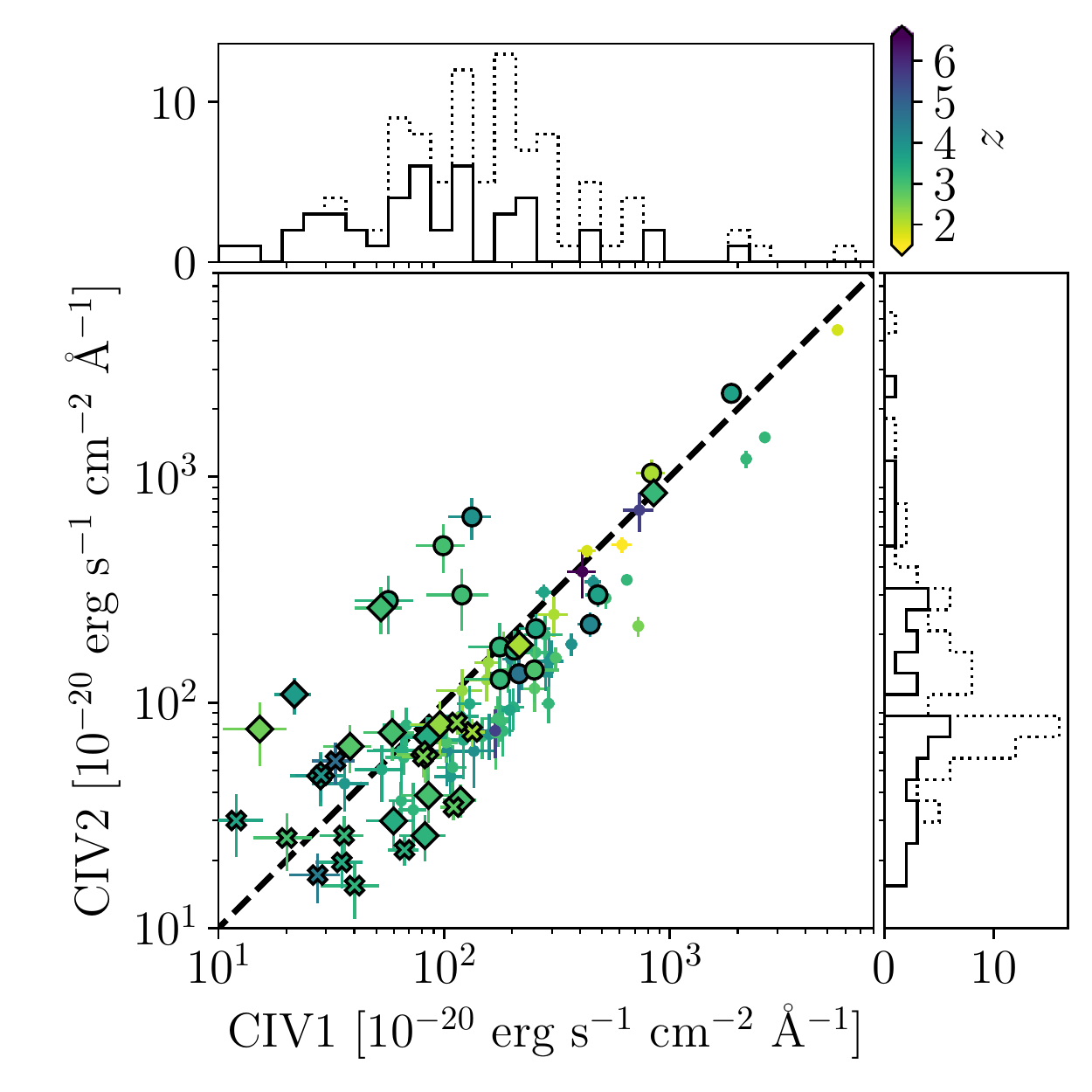}\\
\includegraphics[width=0.42\textwidth]{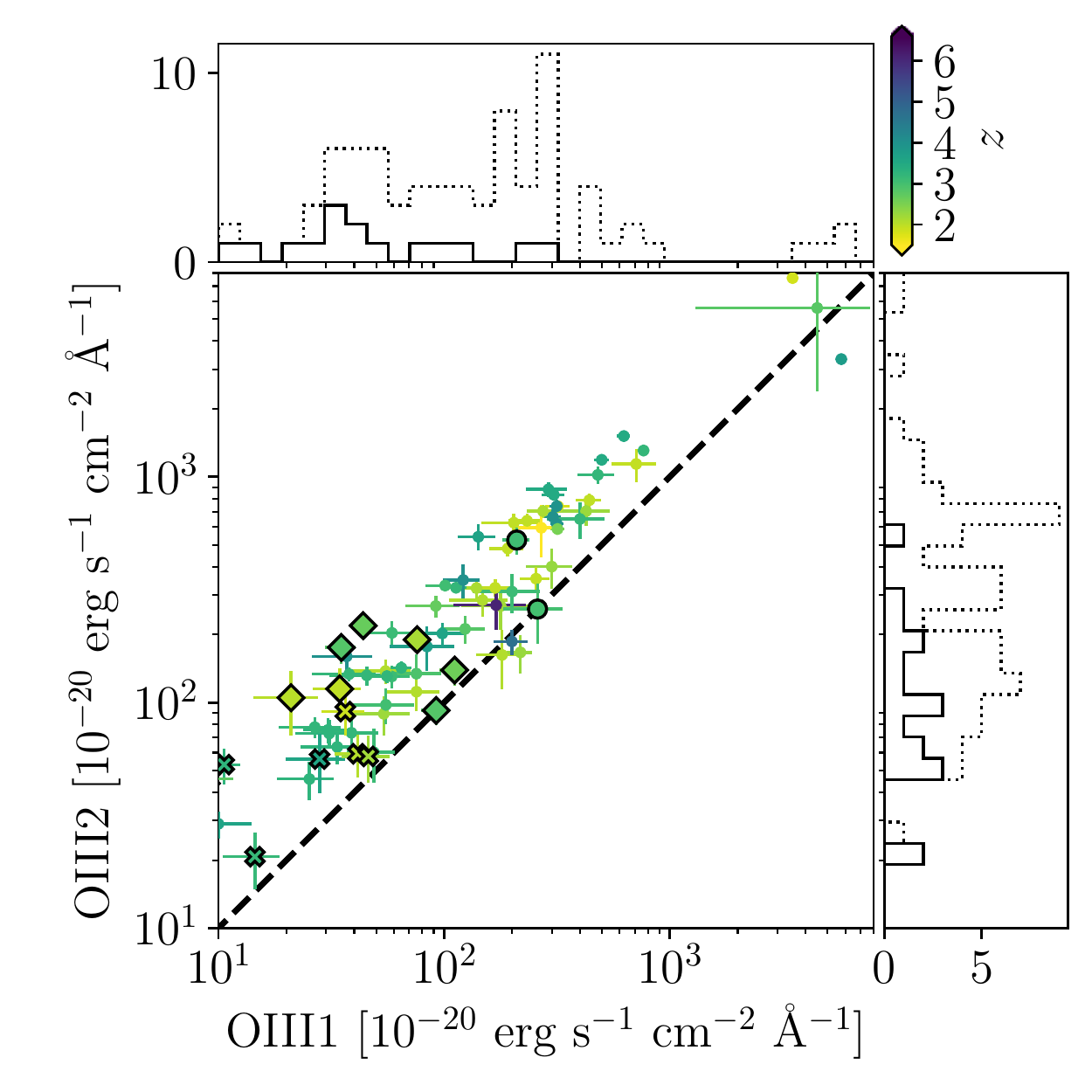}
\includegraphics[width=0.42\textwidth]{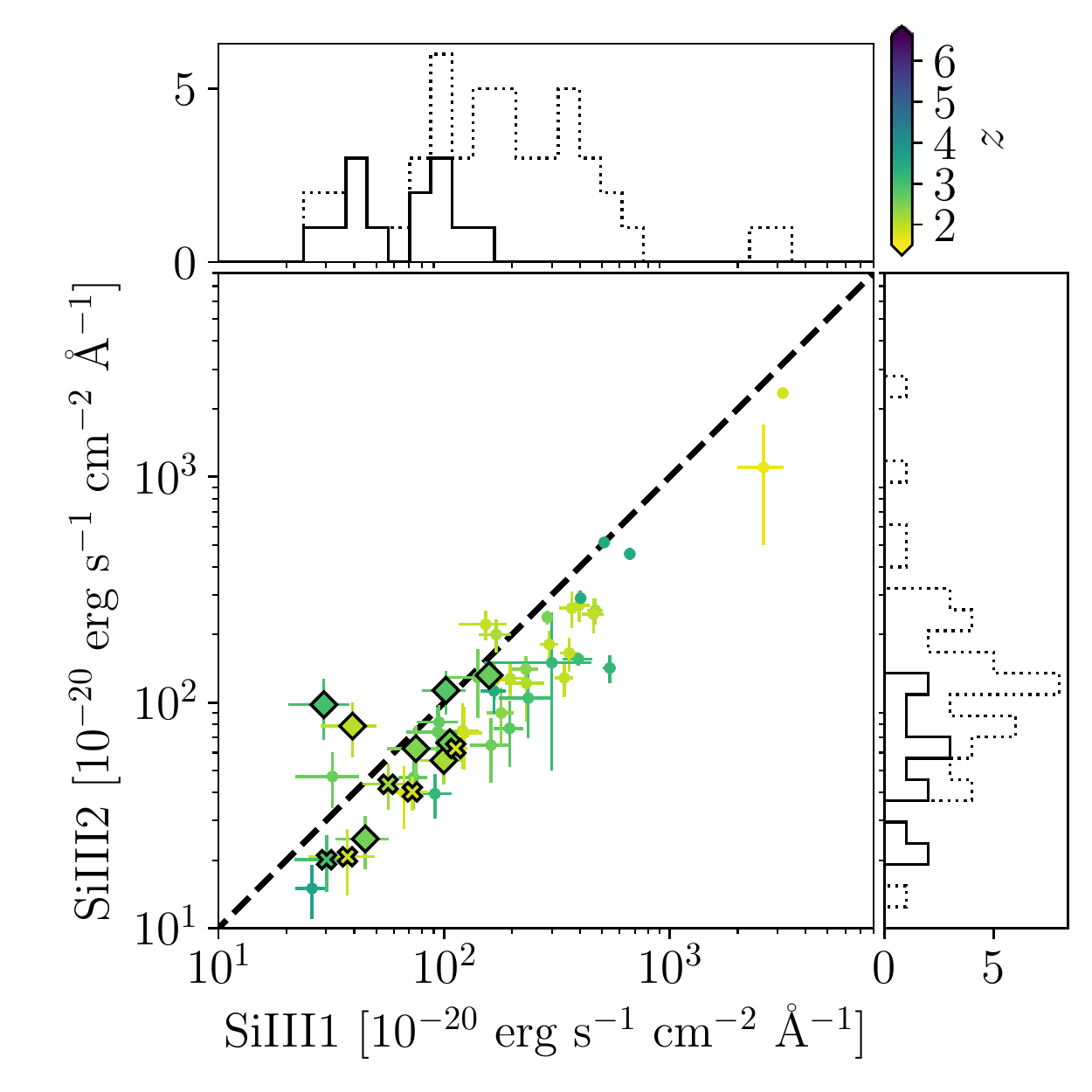}\\
\includegraphics[width=0.42\textwidth]{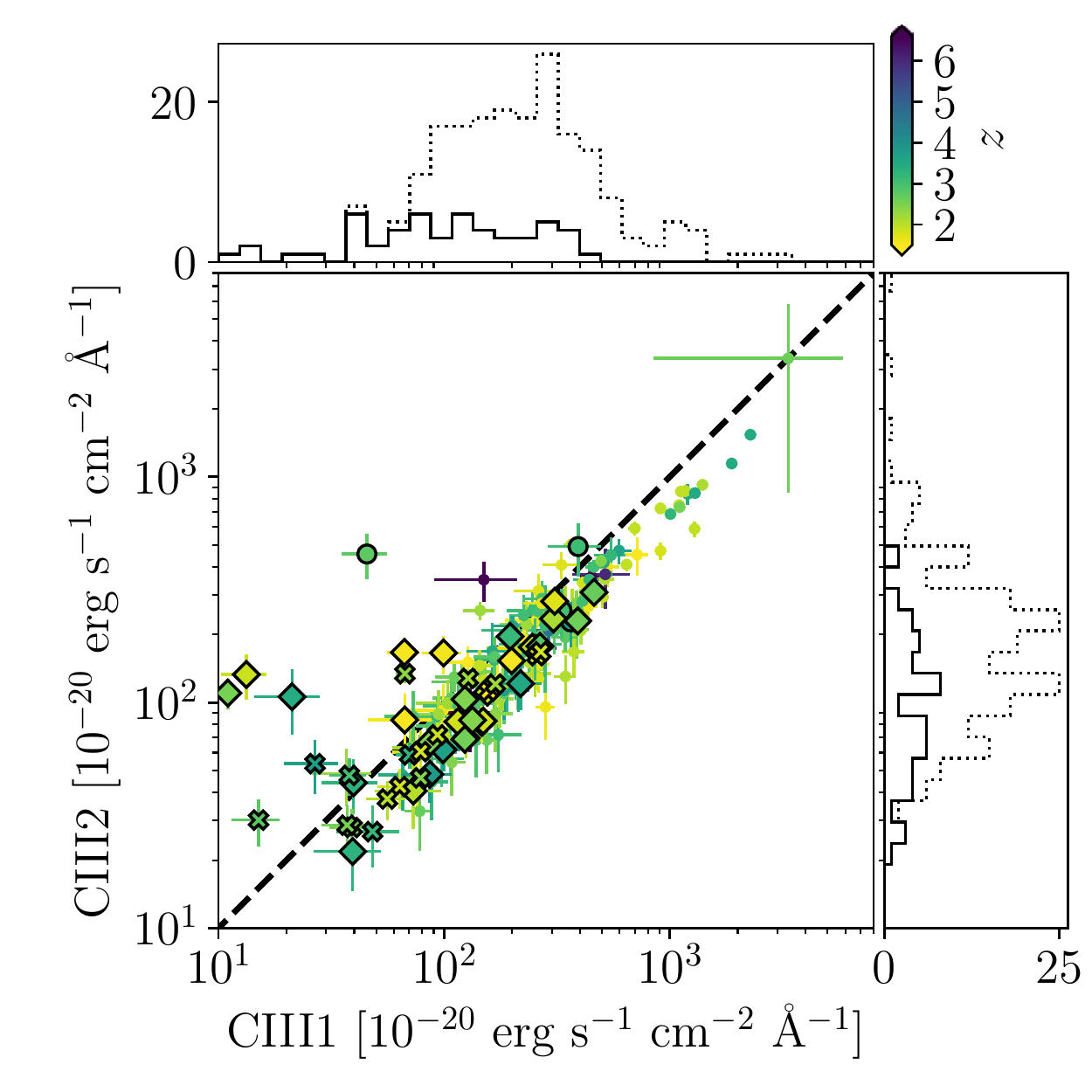}
\caption{Correlations between the individual components of the UV emission line doublets as described in Section~\ref{sec:UVEmissionLineSearch}. The MUSE objects (large symbols) are shown together with the literature comparison sample described in Appendix~\ref{sec:litcol} (small dots) and color coded according to redshift.}
\label{fig:doubletratios}
\end{center}
\end{figure*}

\begin{figure*}
\begin{center}
\includegraphics[width=0.80\textwidth]{mainlegend.png}\\
\includegraphics[width=0.42\textwidth]{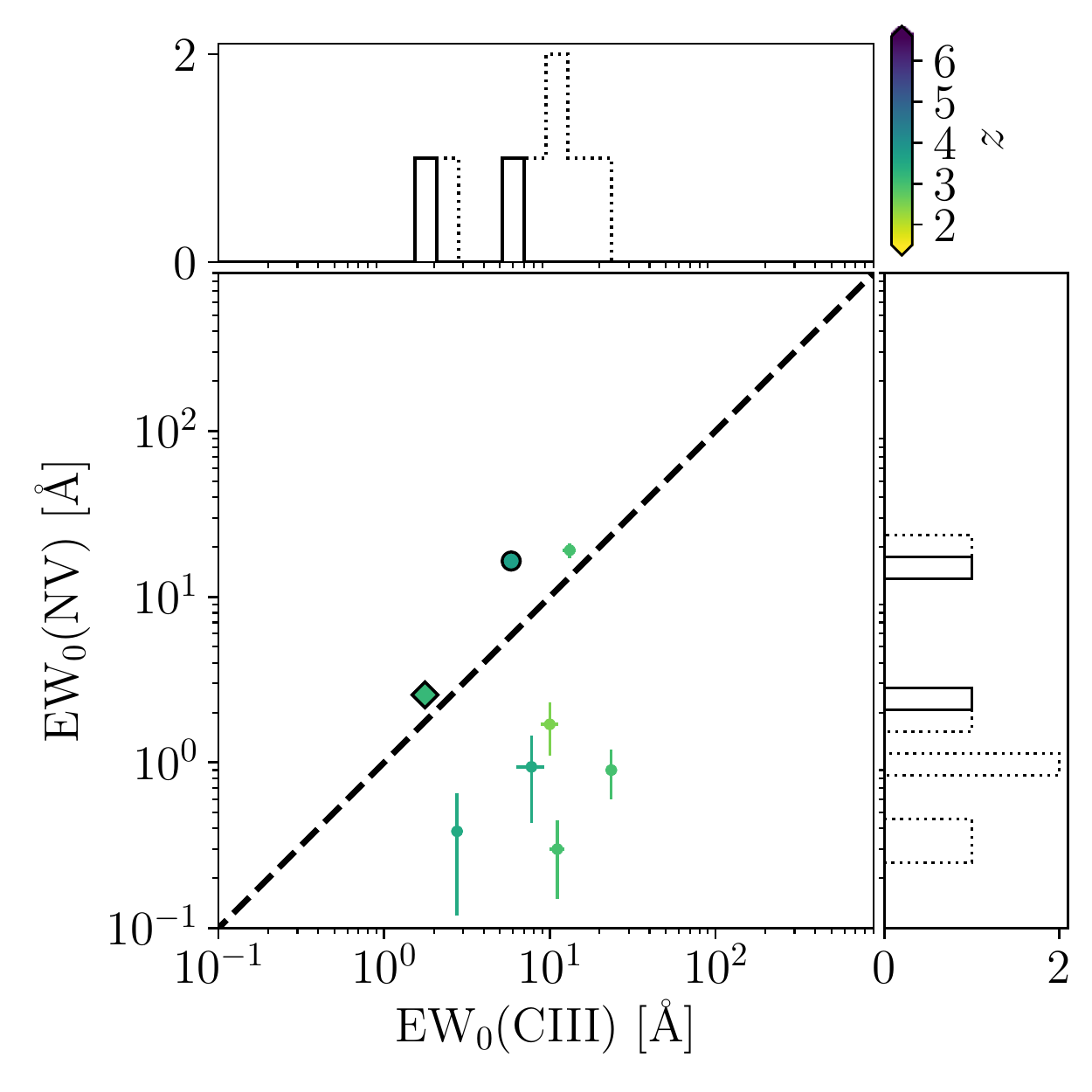}
\includegraphics[width=0.42\textwidth]{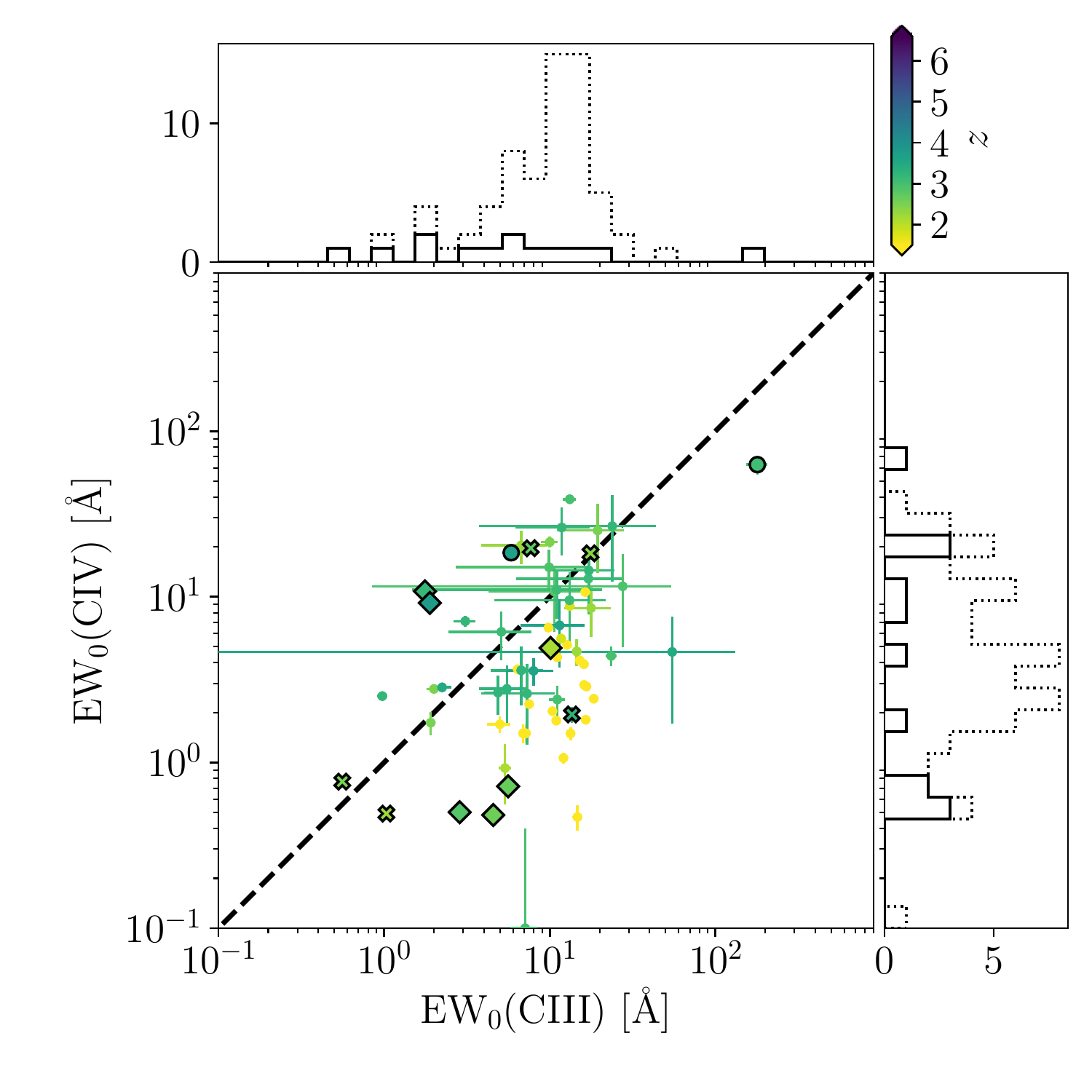}\\
\includegraphics[width=0.42\textwidth]{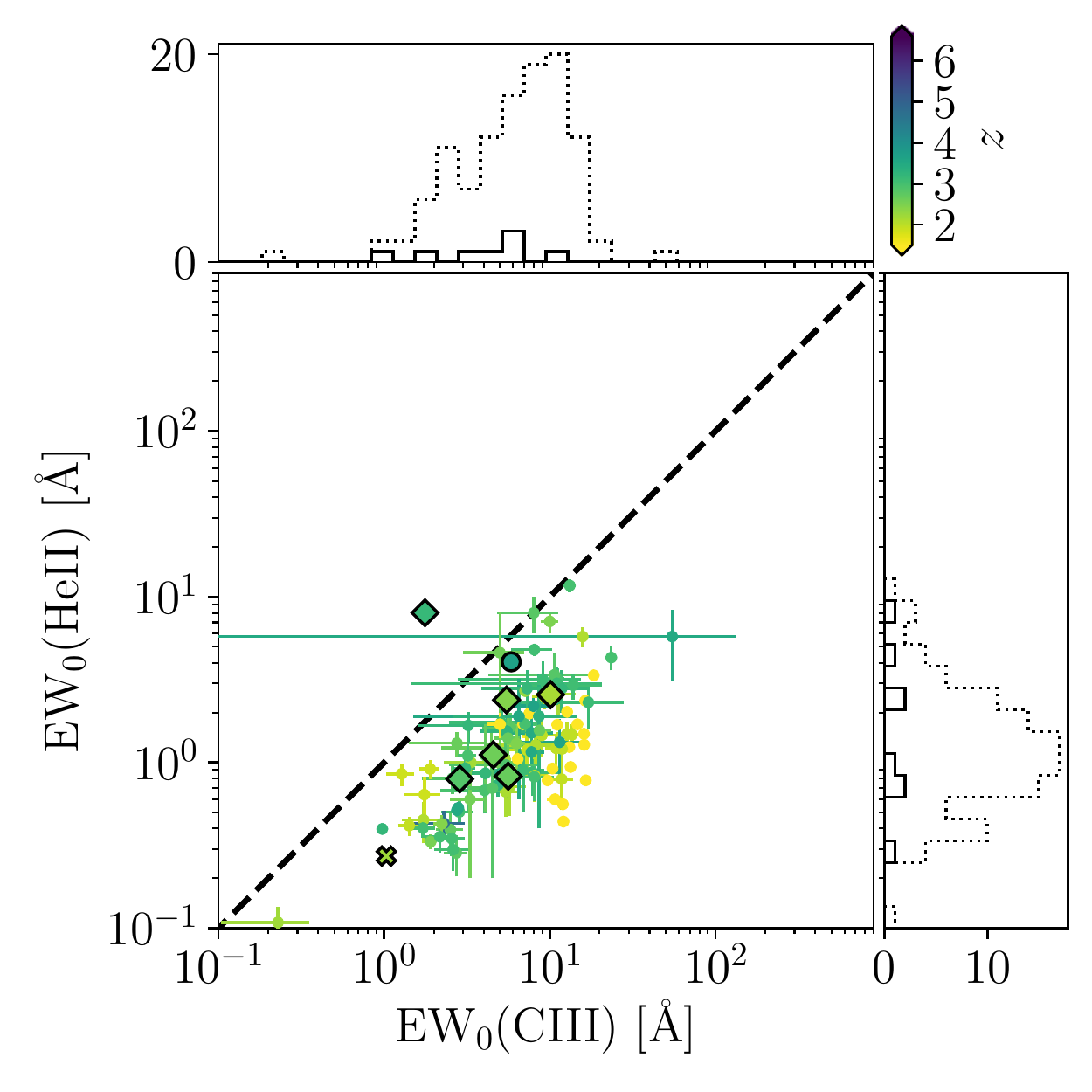}
\includegraphics[width=0.42\textwidth]{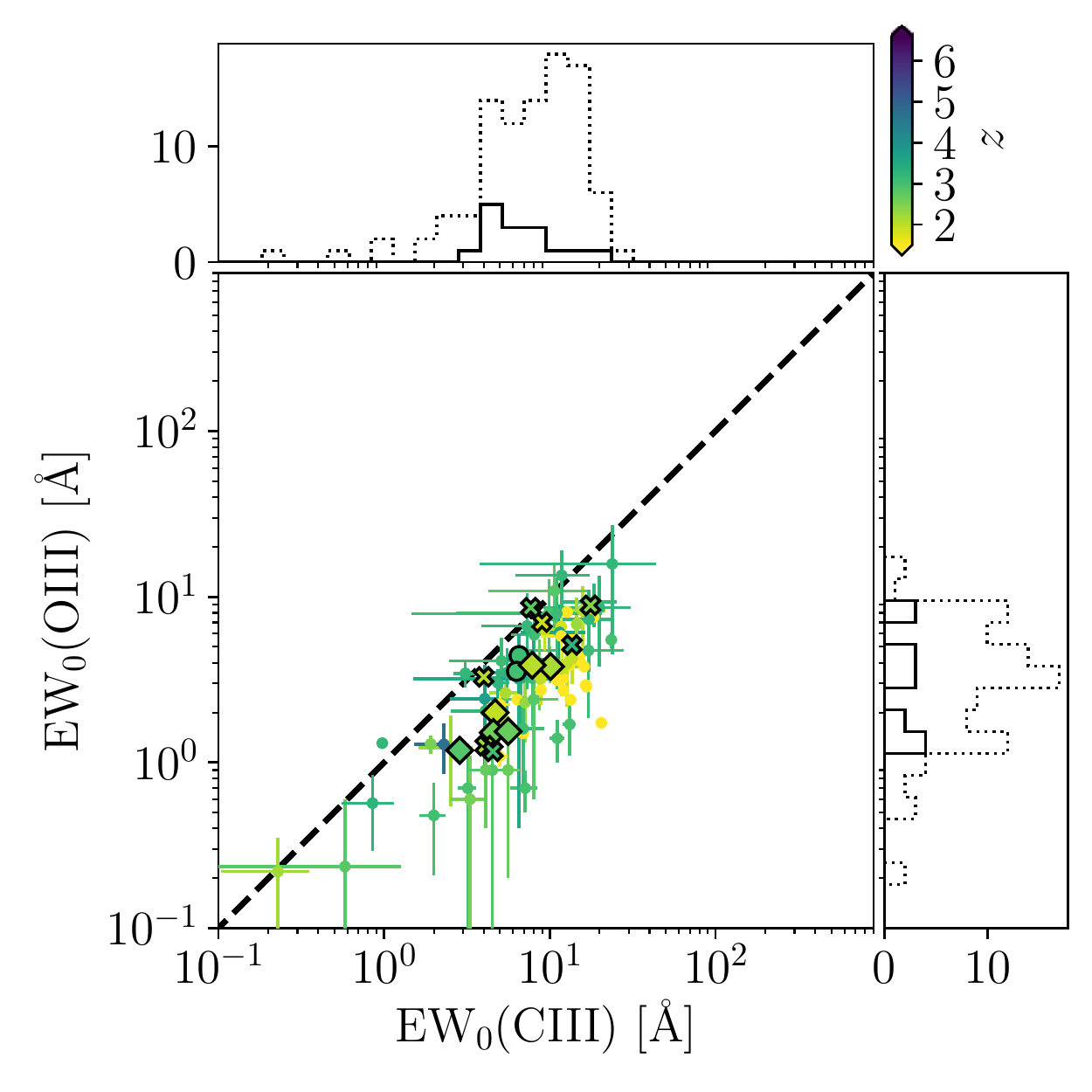}\\
\includegraphics[width=0.42\textwidth]{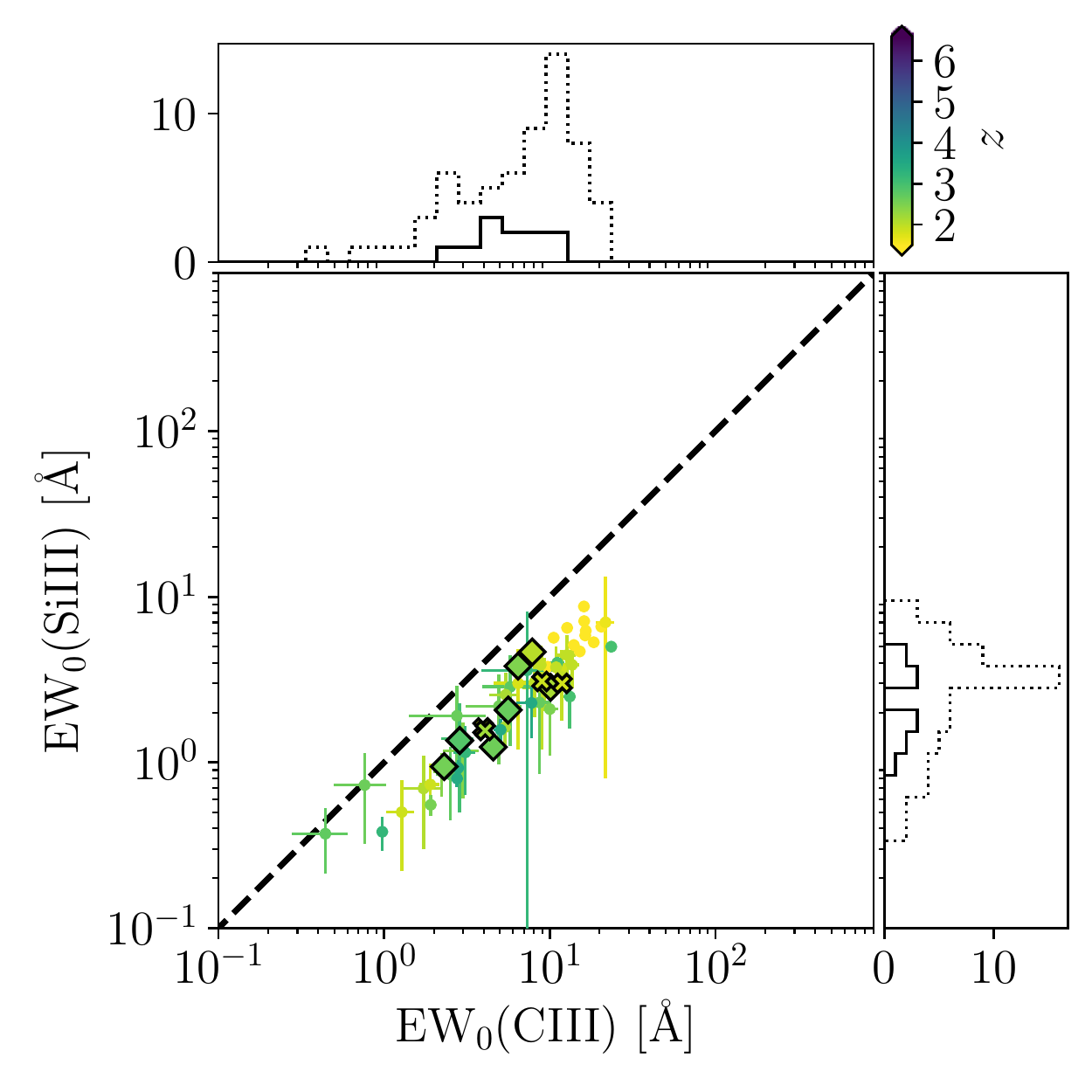}
\caption{Similar to Figure~\ref{fig:fluxesVSciii}, but showing correlations between EW$_0$ estimates as described in Section~\ref{sec:EWcoor} instead of emission line fluxes.
The MUSE objects (large symbols) are shown together with the literature comparison sample described in Section~\ref{sec:litcol} (small dots) and color coded according to redshift.}
\label{fig:EWsciii}
\end{center}
\end{figure*}

\begin{figure*}
\begin{center}
\includegraphics[width=0.60\textwidth]{mainlegend_nolit.png}\\
\includegraphics[width=0.38\textwidth]{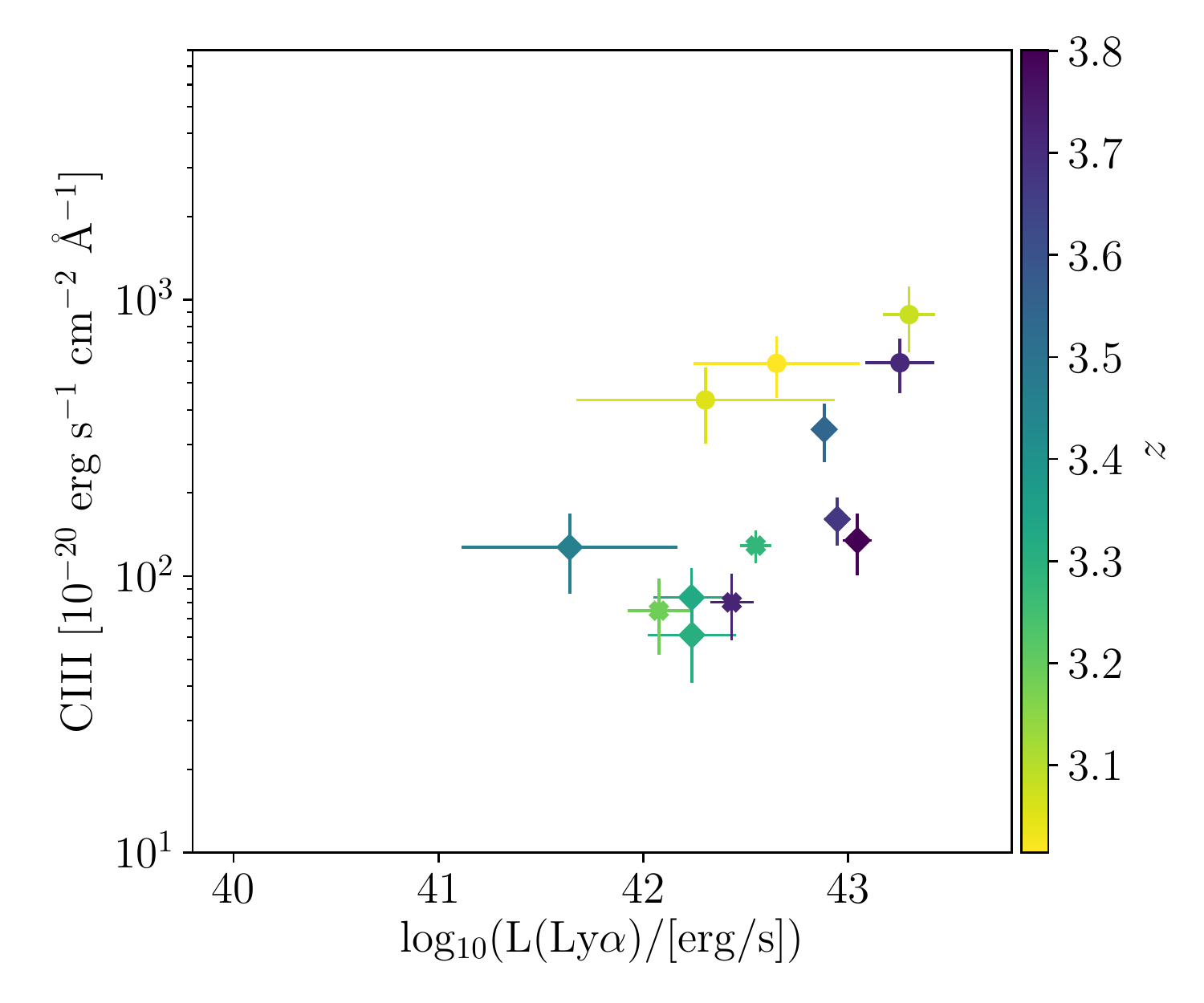}
\includegraphics[width=0.38\textwidth]{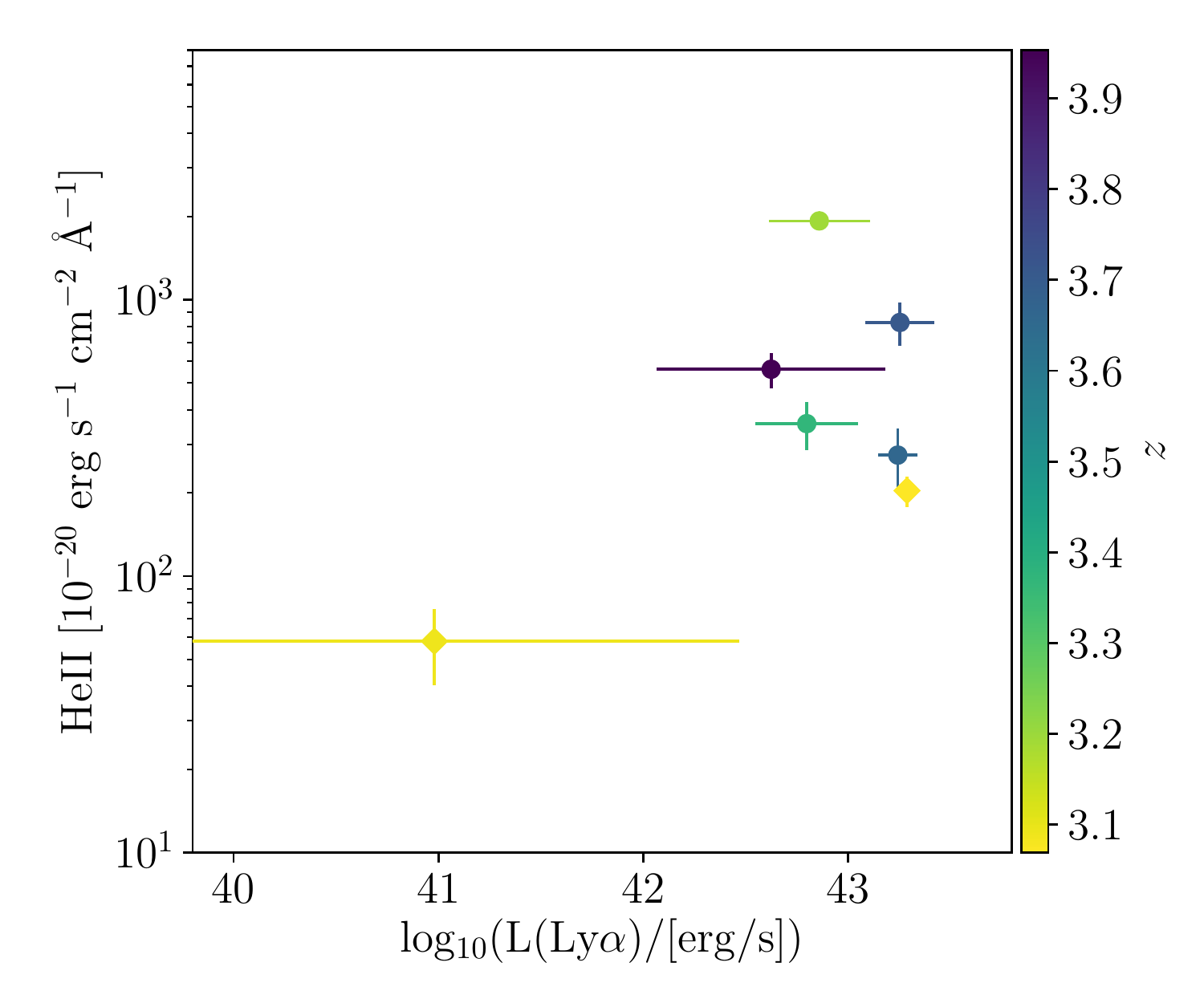}
\includegraphics[width=0.38\textwidth]{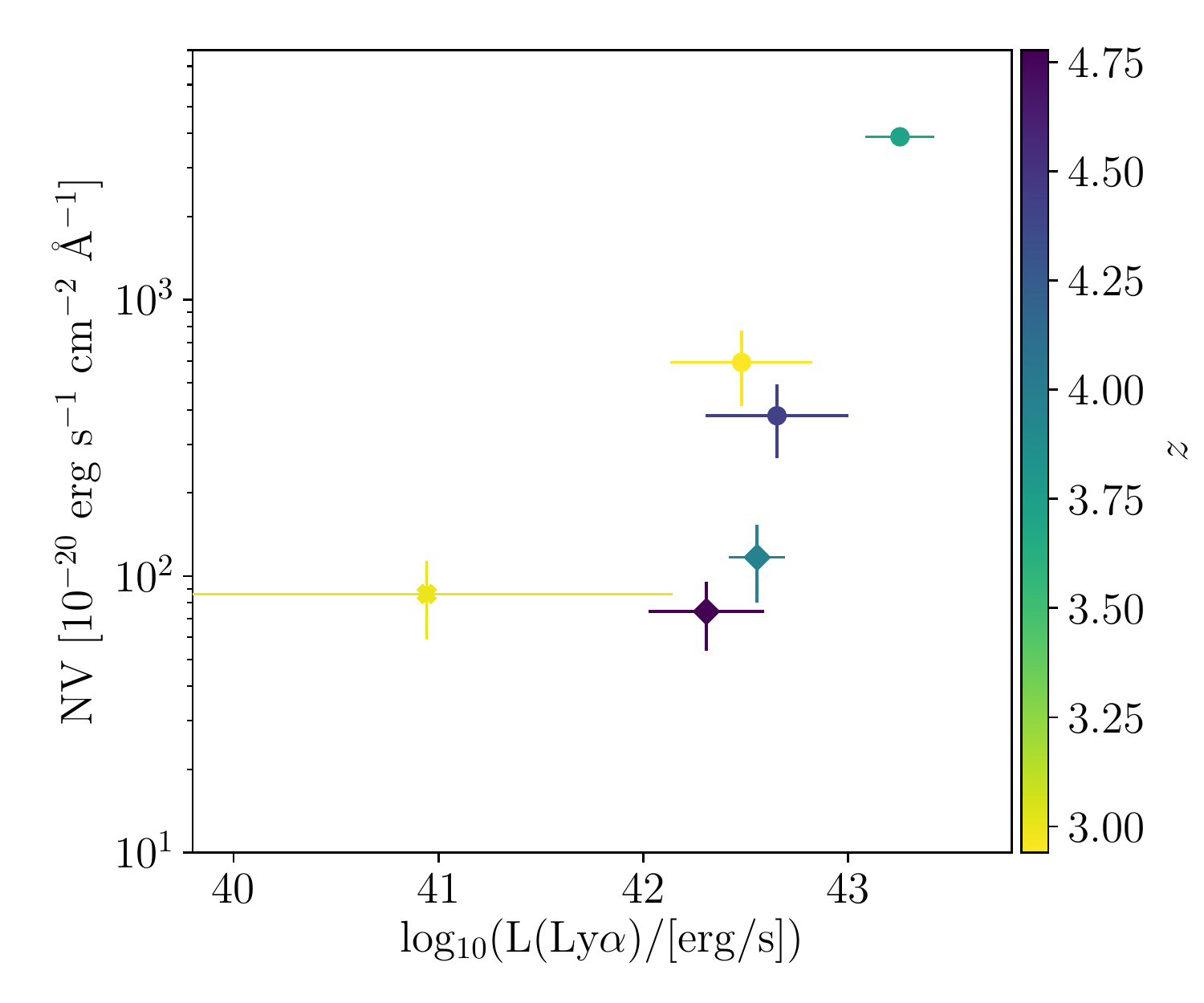}
\includegraphics[width=0.38\textwidth]{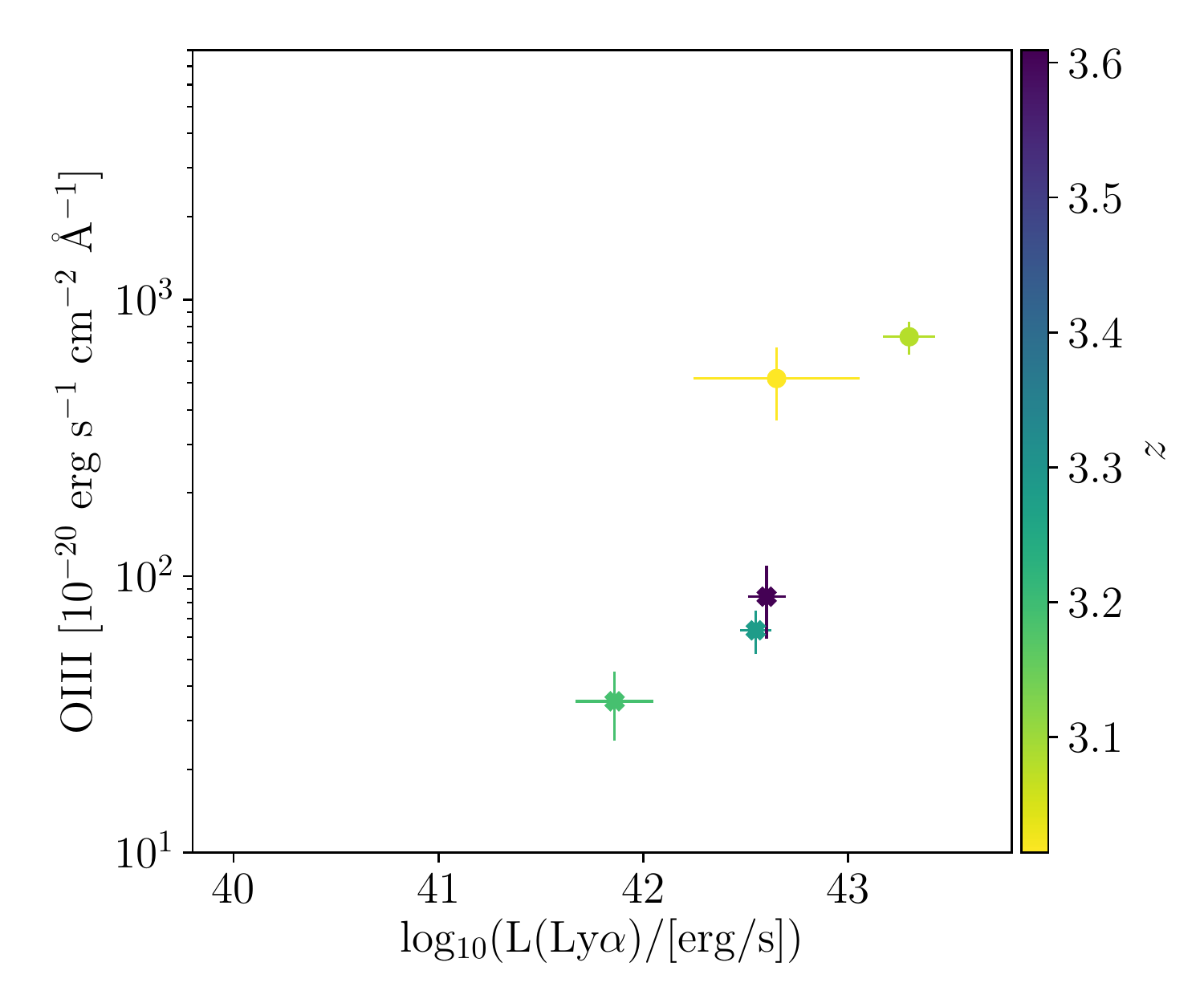}\\
\caption{Potential correlations between UV emission line flux of \ciii, \heii, \nv, and \oiii{} and $\log(L_\textrm{\lya})$ described in Section~\ref{sec:UVandLya}.}
\label{fig:FvsLLya}
\end{center}
\end{figure*}

\begin{figure*}
\begin{center}
\includegraphics[width=0.60\textwidth]{mainlegend_nolit.png}\\
\includegraphics[width=0.38\textwidth]{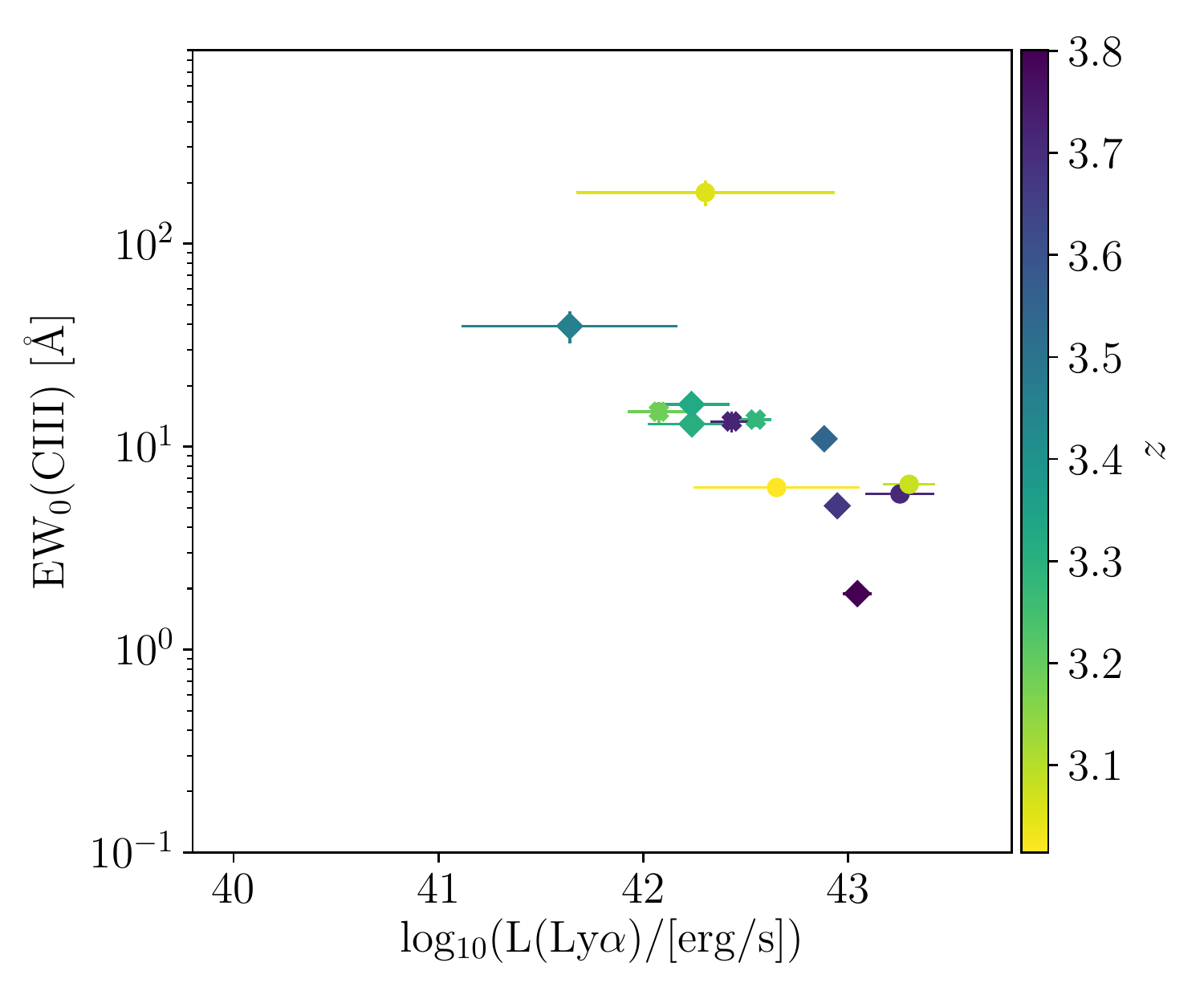}
\includegraphics[width=0.38\textwidth]{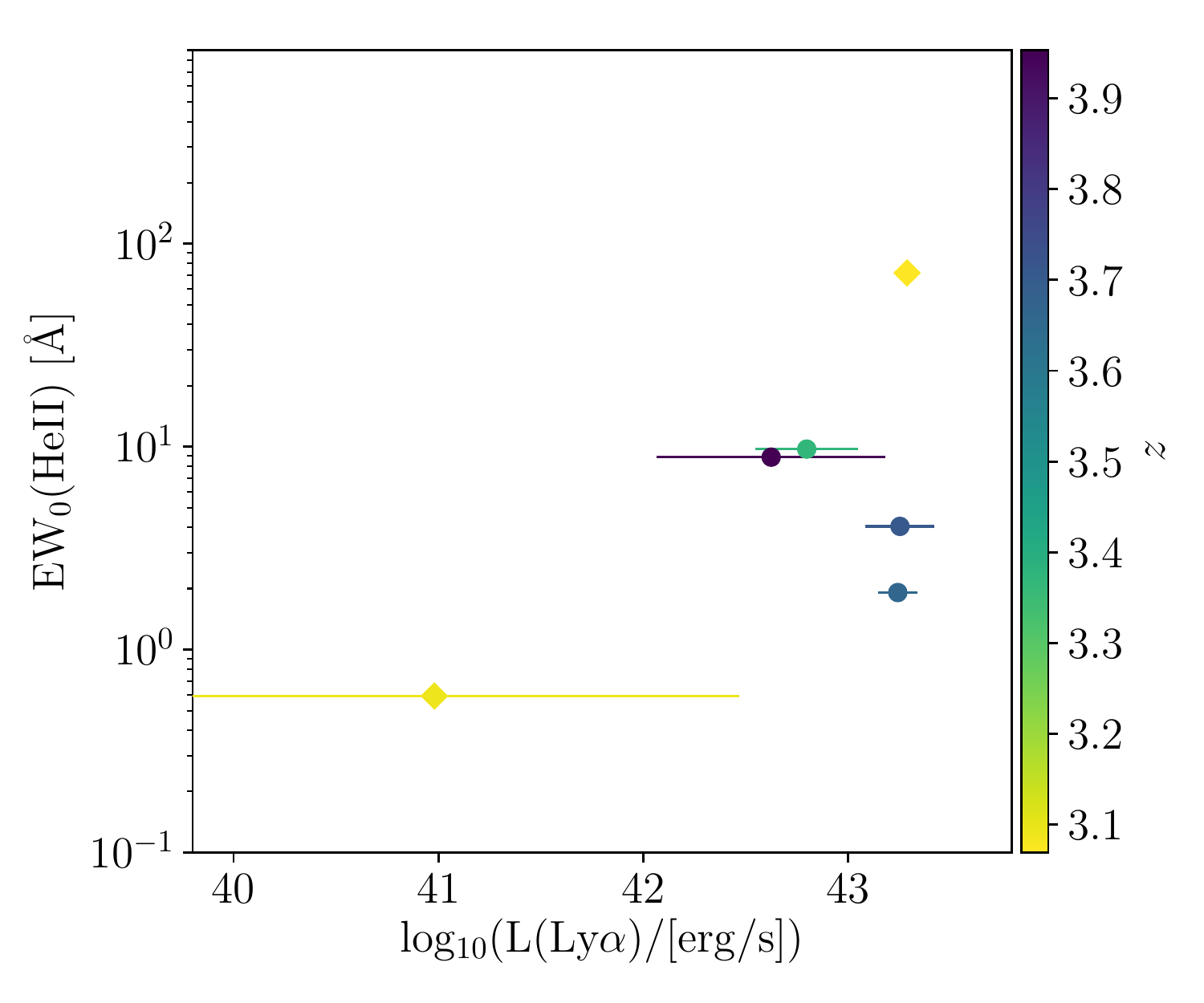}
\includegraphics[width=0.38\textwidth]{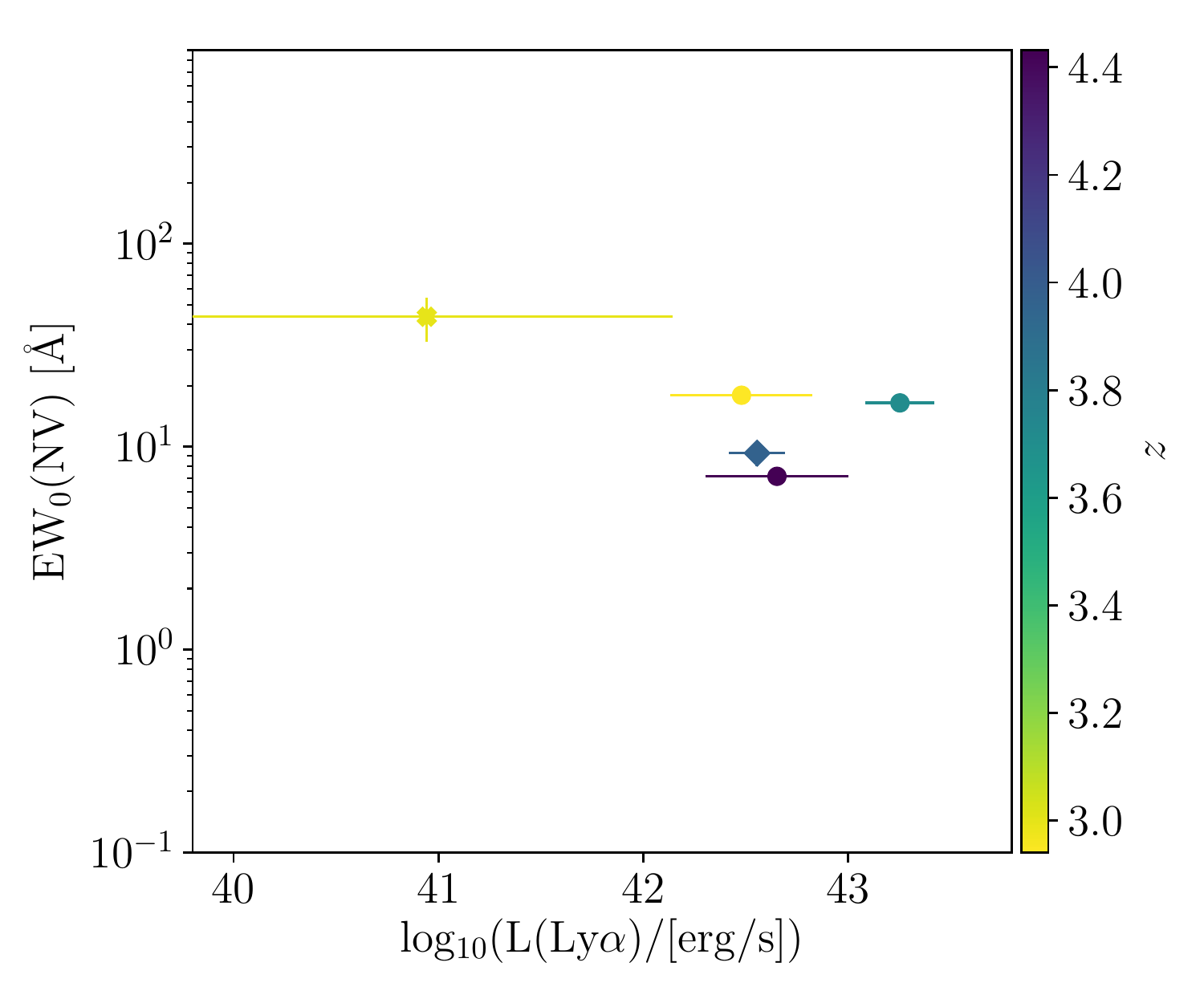}
\includegraphics[width=0.38\textwidth]{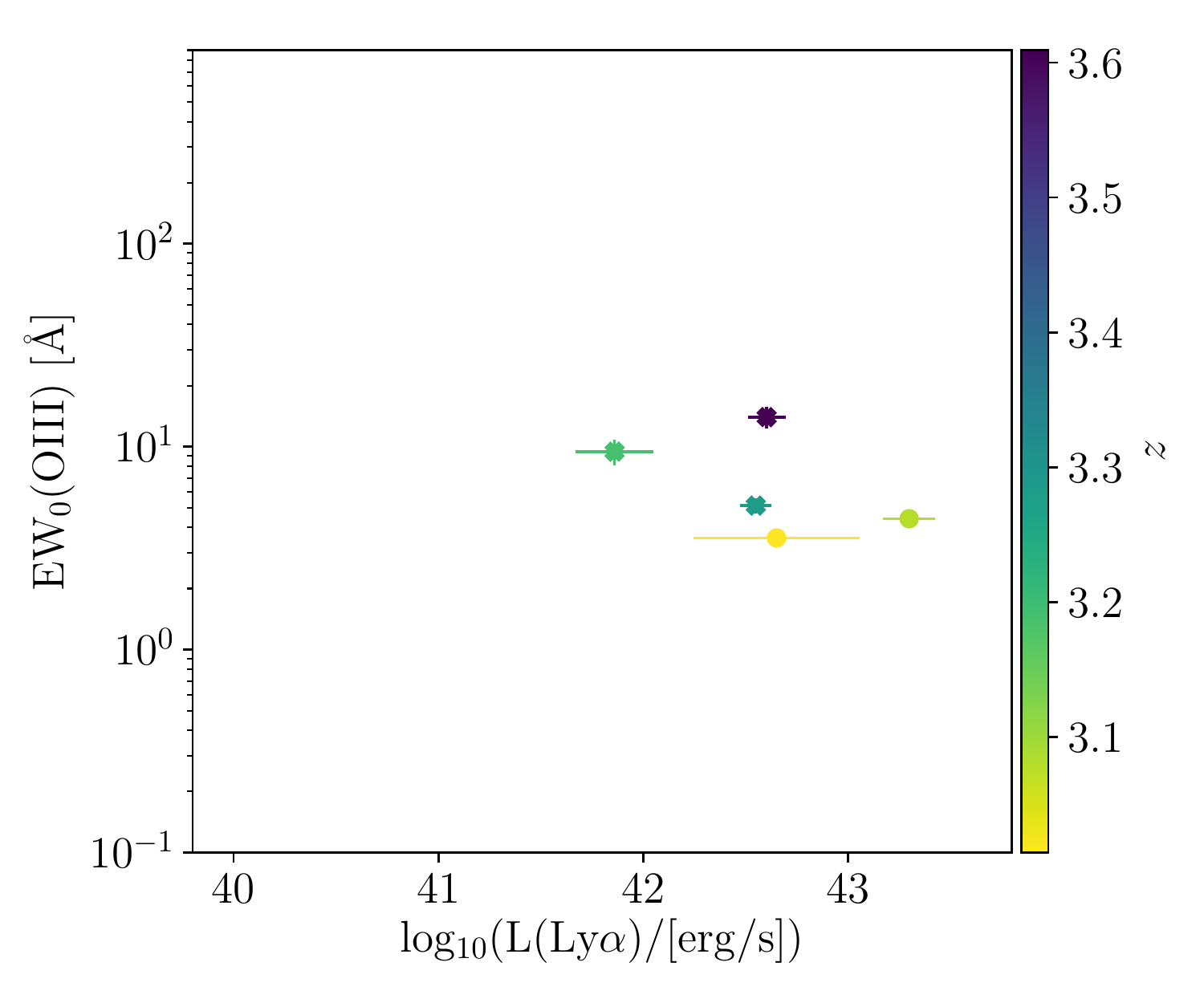}
\caption{Potential correlations between EW$_0$(\ciii), EW$_0$(\heii), EW$_0$(\nv), and EW$_0$(\oiii) and $\log(L_\textrm{\lya})$ described in Section~\ref{sec:UVandLya}.}
\label{fig:EWvsLLya}
\end{center}
\end{figure*}

\begin{figure*}
\begin{center}
\includegraphics[width=0.35\textwidth]{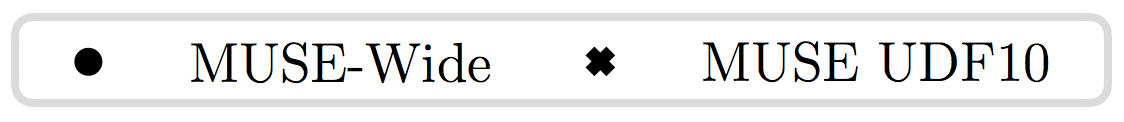}\\
\includegraphics[width=0.4\textwidth]{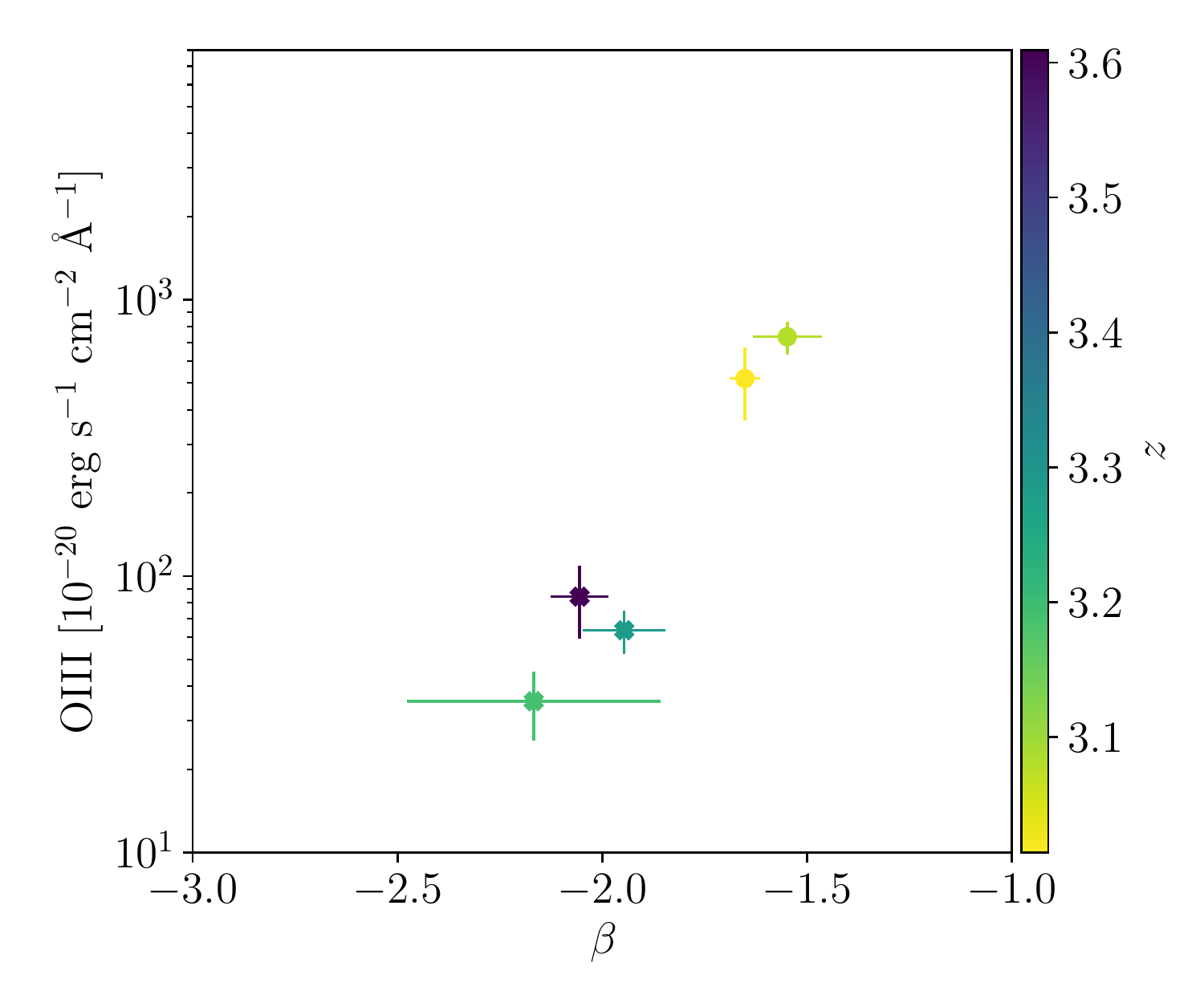}
\includegraphics[width=0.4\textwidth]{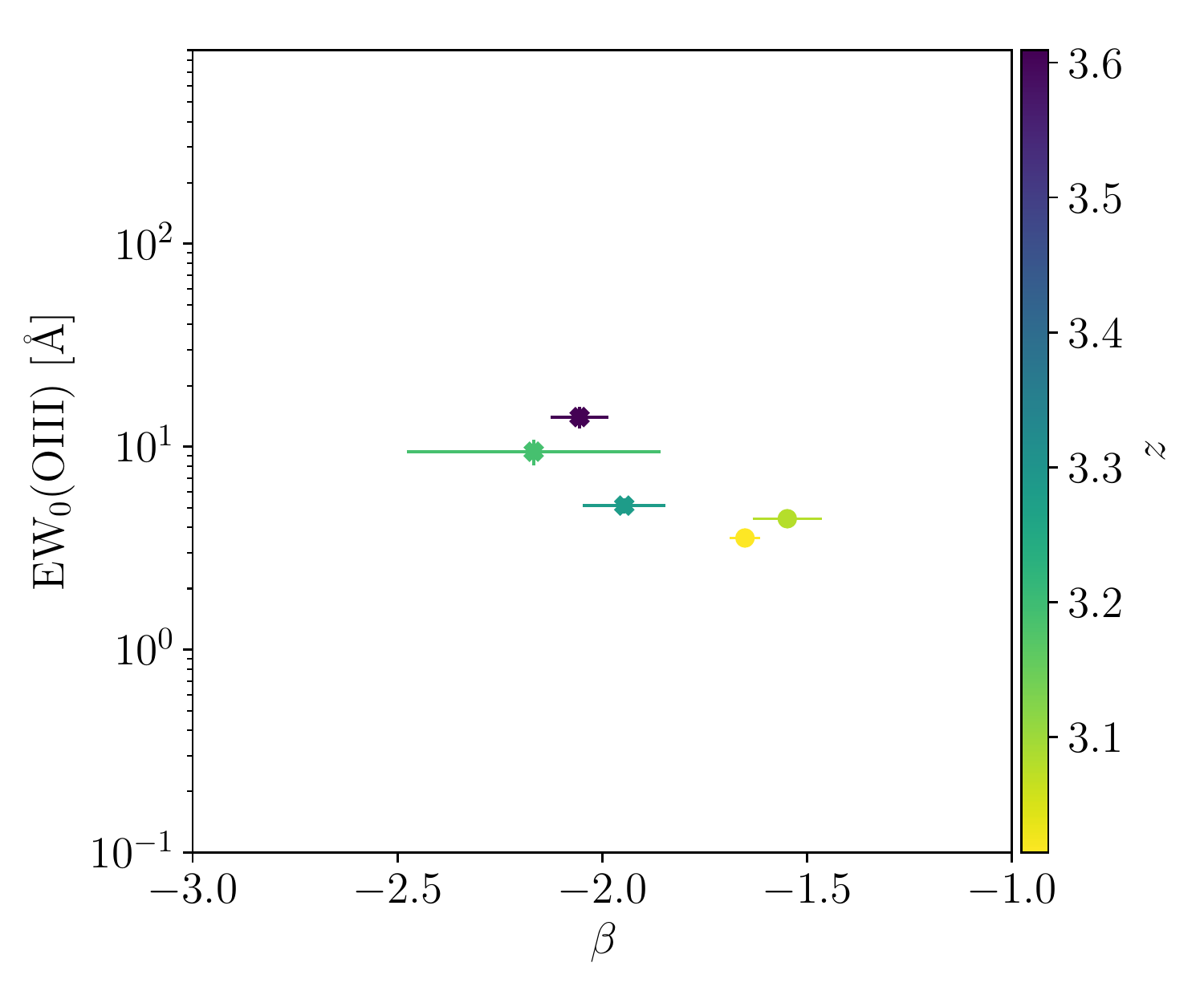}
\caption{Potential correlations described in Section~\ref{sec:UVandLya} between the \oiii{} emission line flux (left) and EW$_0$(\oiii) (right) and the spectral slope $\beta$.}
\label{fig:FandEWvsbeta}
\end{center}
\end{figure*}

\begin{figure}
\begin{center}
\includegraphics[width=0.49\textwidth]{mainlegend_nolit.png}\\
\includegraphics[width=0.49\textwidth]{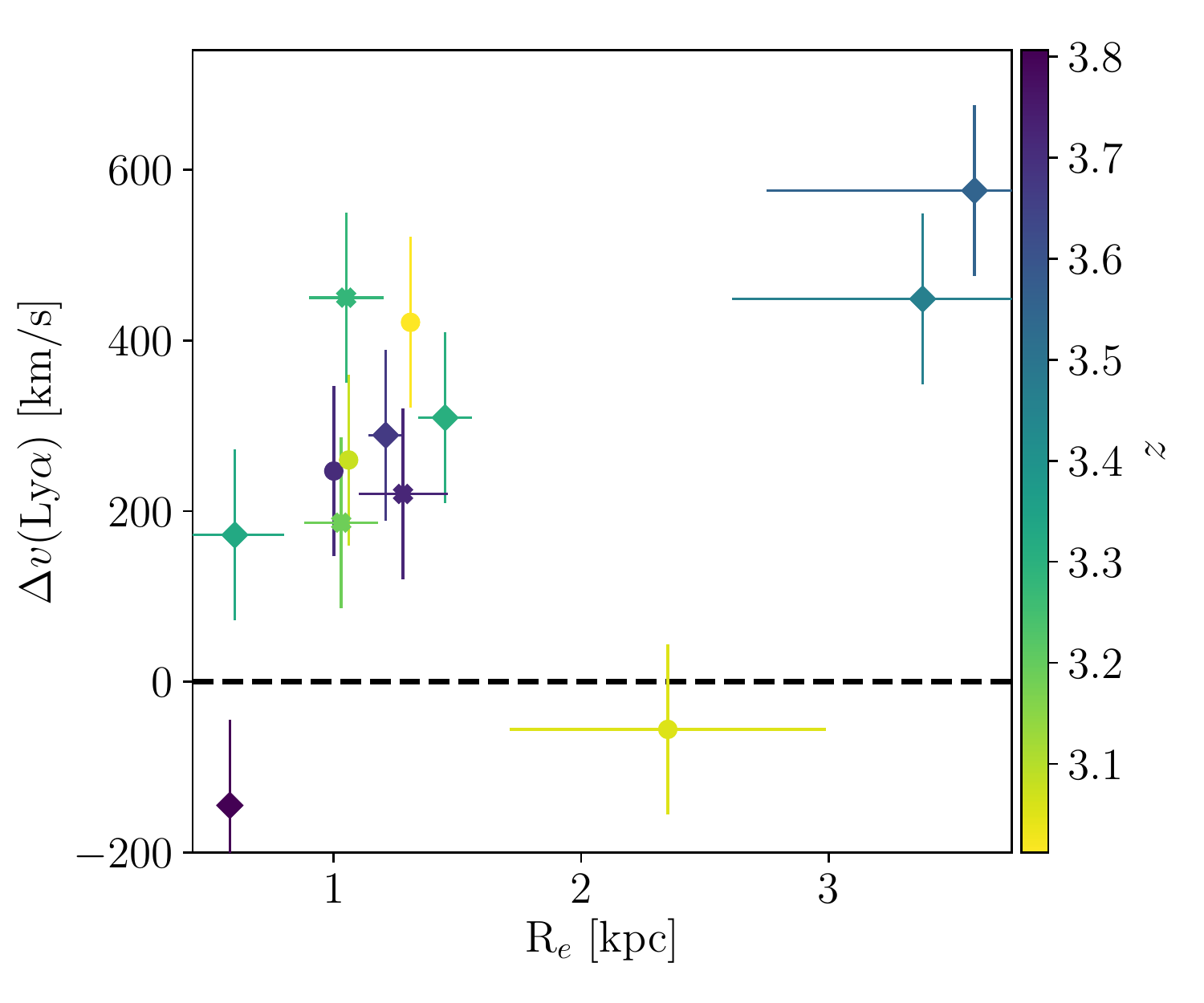}
\caption{Tentative correlation described in Section~\ref{sec:voffset} between the \lya{} velocity offset and the effective radius, R$_{e}$, of the LAEs.}
\label{fig:dvVSRe}
\end{center}
\end{figure}

As presented throughout the paper, several correlations between the UV emission line fluxes and EW$_0$ from this catalog (combined with the literature comparison sample described in Appendix~\ref{sec:litcol}) were found. 
Among those are the correlations between the flux of the individual components of the emission line doublets shown in Figure~\ref{fig:doubletratios} and discussed in Section~\ref{sec:UVEmissionLineSearch}.
The correlations between the UV emission line EW$_0$ estimates described in Section~\ref{sec:EWcoor} are shown in Figure~\ref{fig:EWsciii} .
At the end of Section~\ref{sec:UVandLya} we also present potential correlations between the measurements for a subset of the UV emission lines and the \lya{} luminosity and the spectral slope $\beta$. These correlations are shown in Figures~\ref{fig:FvsLLya}, \ref{fig:EWvsLLya}, and \ref{fig:FandEWvsbeta}.
Lastly, Figure~\ref{fig:dvVSRe} shows the tentative relation between 
R$_{e}$ and $\Delta v_\textrm{\lya}$ described at the end of Section~\ref{sec:voffset}.

\section{Literature collection of UV emission line fluxes}\label{sec:litcol}

\begin{figure*}
\begin{center}
\includegraphics[width=0.99\textwidth]{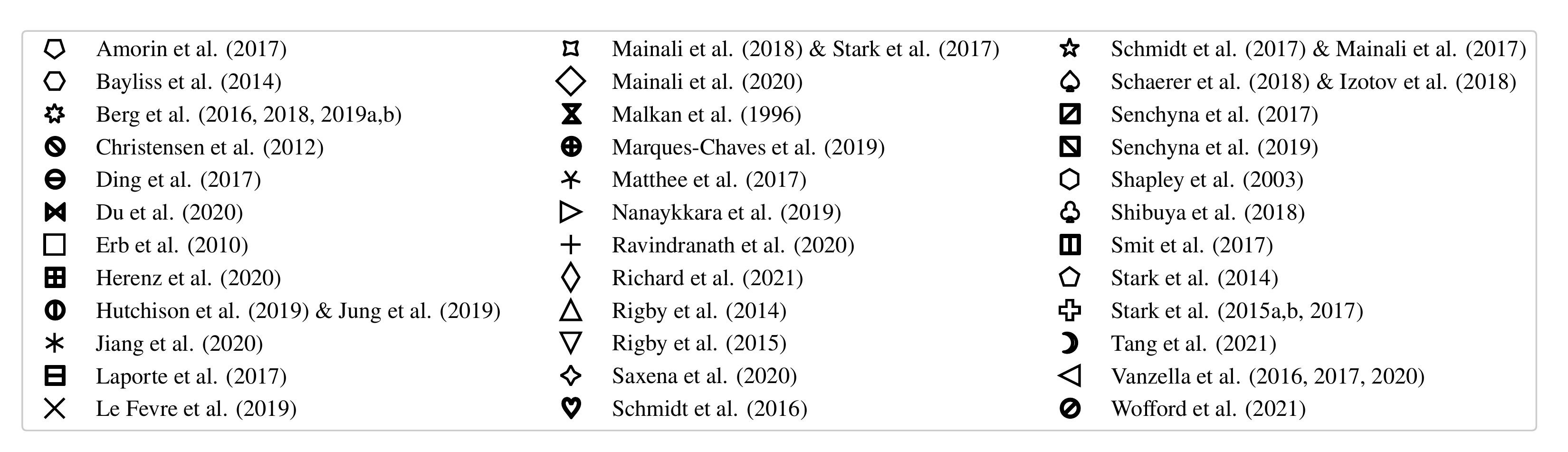}\\
\includegraphics[width=0.49\textwidth]{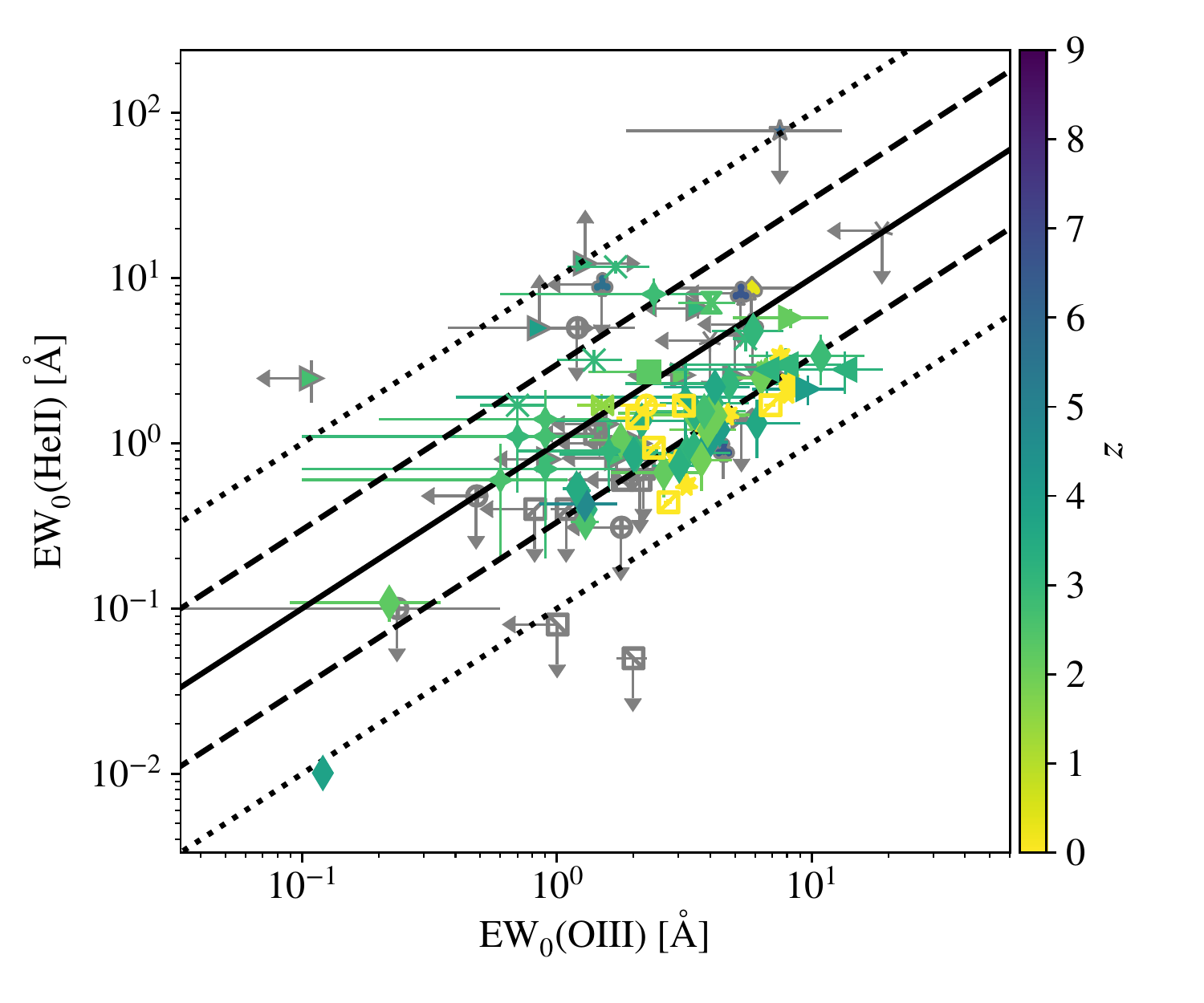}
\includegraphics[width=0.49\textwidth]{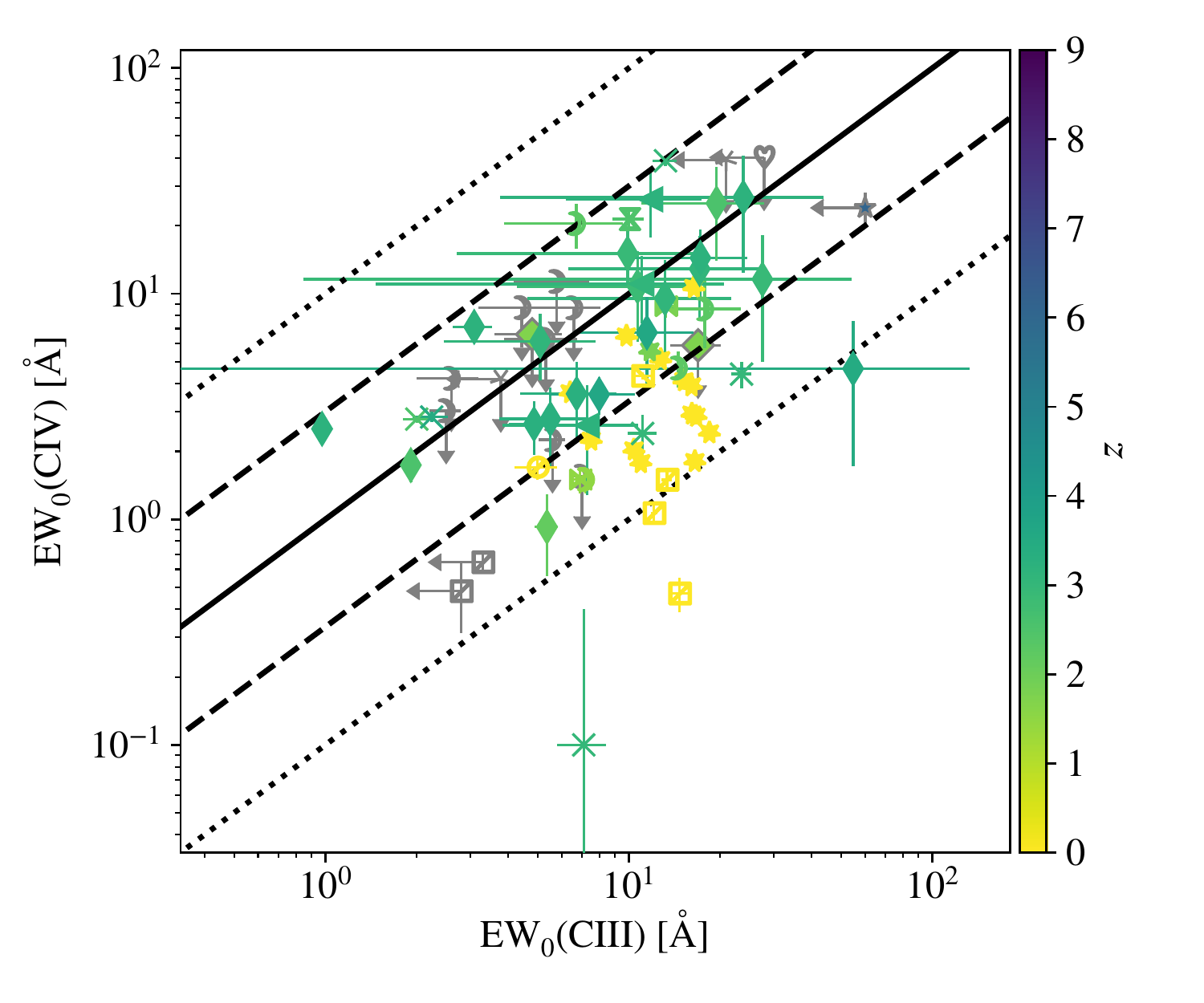}\\
\includegraphics[width=0.49\textwidth]{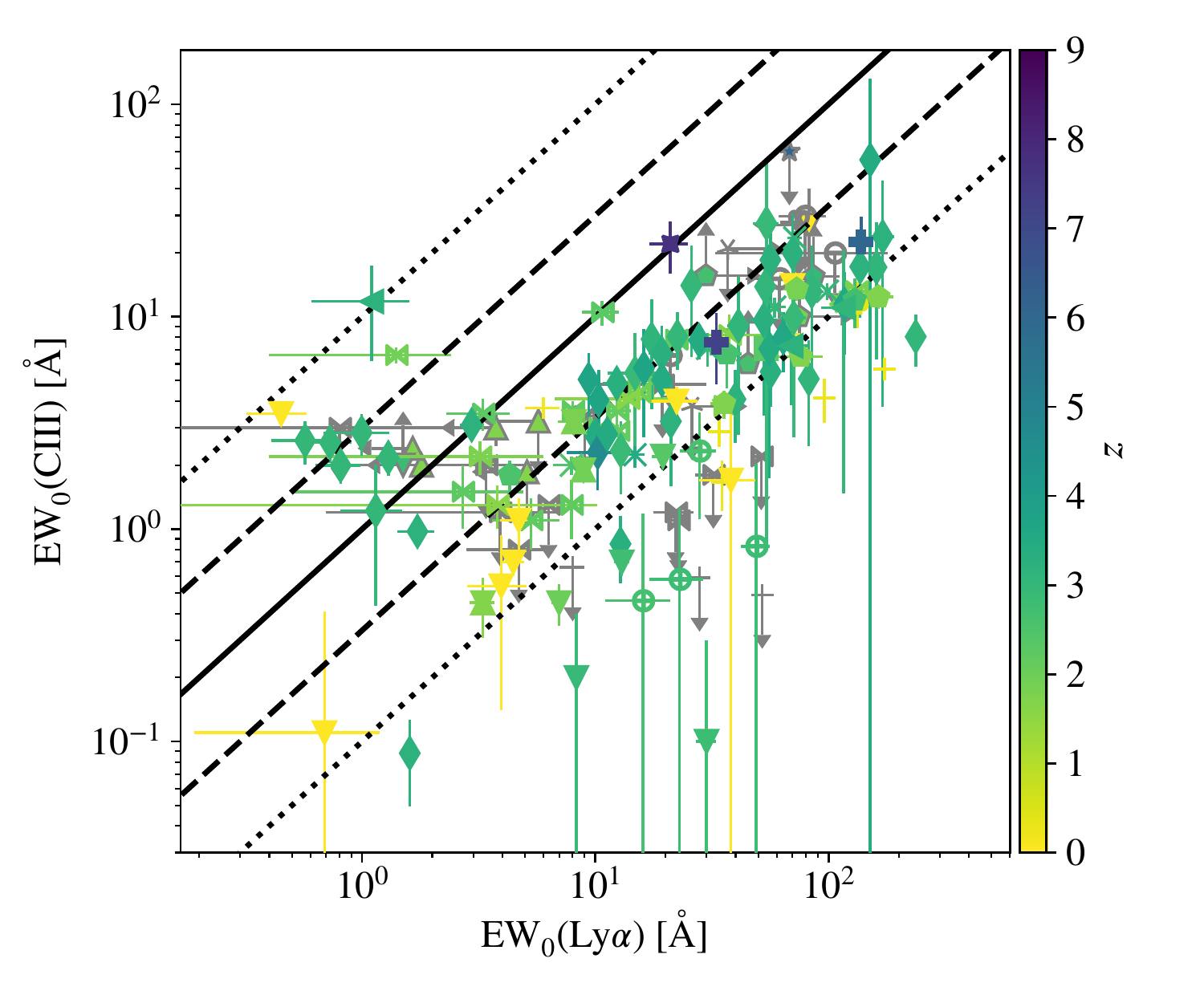}
\includegraphics[width=0.49\textwidth]{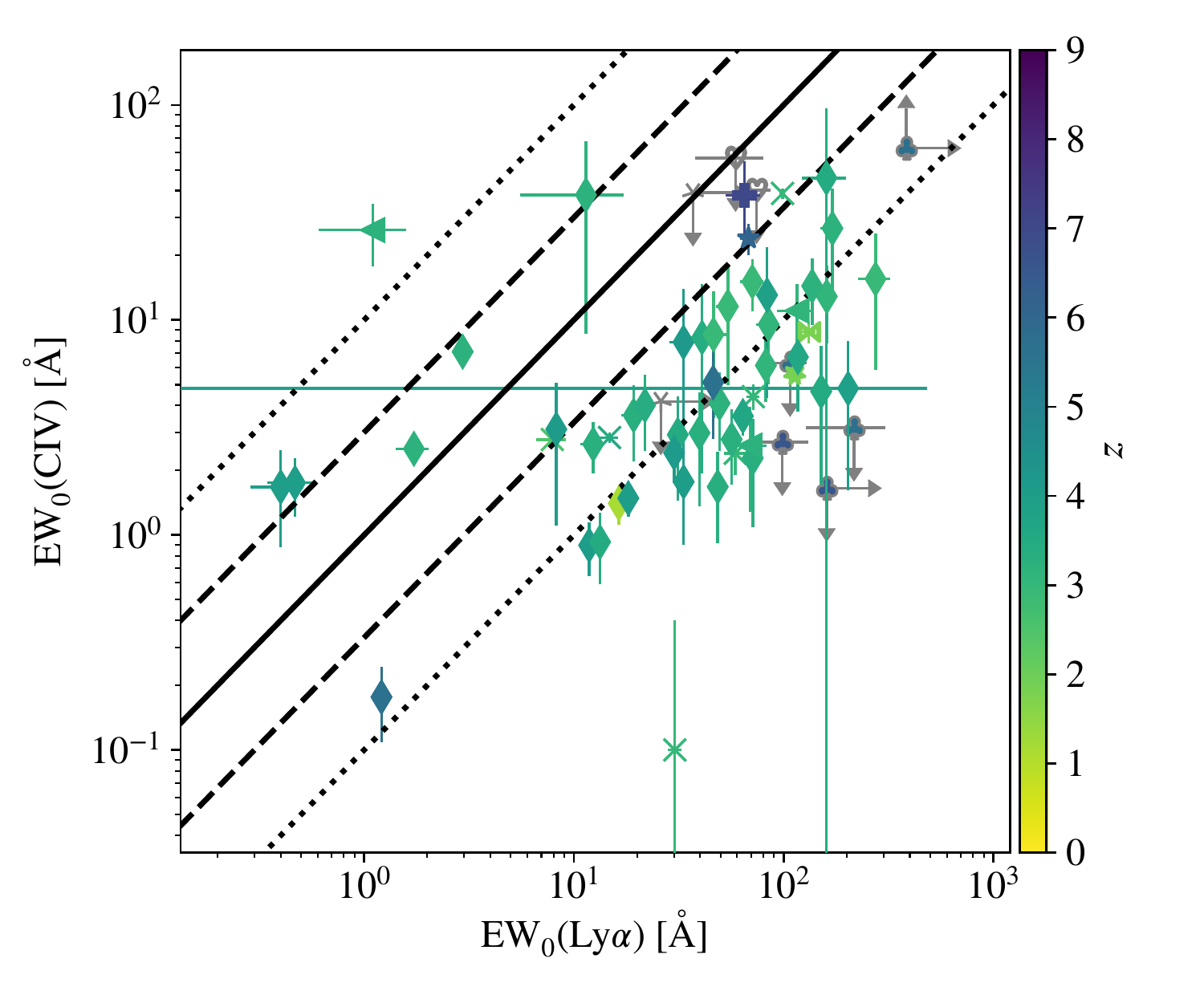}
\caption{Projections of the rest-frame EW space spanned by EW$_0$(\heii), EW$_0$(\oiii), EW$_0$(\ciii), EW$_0$(\civ) and EW$_0$(\lya) from the collection of UV emission line measurements from the literature plotted throughout this paper. 
The bottom panels reproduce the corresponding panels from Figure~\ref{fig:EWs} and the top right panel reproduces the top right panel of Figure~\ref{fig:EWsciii}.
All panels exclude the MUSE measurements presented in this paper, but assign different symbols to the literature data included in Table~\ref{tab:litcatcol} according to the legend at the top.
The points are color coded according to redshift and 3$\sigma$ limits are indicated by the gray arrows.
The diagonal curves correspond to the 10:1, 3:1, 1:1, 1:3, and 1:10 relations.
}
\label{fig:litlegend}
\end{center}
\end{figure*}
In several of the plots in this paper the estimated UV emission line fluxes, EWs, line flux ratios etc. from the FELIS template matches to the MUSE-Wide, MUSE UDF mosaic and MUSE UDF10 spectra are compared to measurements collected from the literature. 
These UV emission line measurements were collected from
\citet[][amo17]{2017NatAs...1E..52A},
\citet[][bay14]{2014ApJ...790..144B},
\citet[][ber19]{2016ApJ...827..126B,2018ApJ...859..164B,2019ApJ...878L...3B,2019ApJ...874...93B},
\citet[][chr12]{2012MNRAS.427.1953C},
\citet[][din17]{2017ApJ...838L..22D},
\citet[][du20]{2020ApJ...890...65D},
\citet[][erb10]{2010ApJ...719.1168E},
\citet[][her20]{2020A&A...642A..55H},
\citet[][hut19]{2019ApJ...879...70H,2019ApJ...877..146J},
\citet[][jia20]{2020NatAs.tmp..246J},
\citet[][lap17]{2017ApJ...851...40L},
\citet[][lef19]{2019A&A...625A..51L},
\citet[][mai18]{2018MNRAS.479.1180M},
\citet[][mai20]{2020MNRAS.494..719M},
\citet[][mal96]{1996ApJ...468L...9M},
\citet[][mar19]{2020MNRAS.492.1257M},
\citet[][mat17]{2017MNRAS.472..772M},
\citet[][nan19]{2019A&A...624A..89N},
\citet[][rav20]{2020ApJ...896..170R},
\citet[][ric21]{2021A&A...646A..83R},
\citet[][rig14]{2014ApJ...790...44R},
\citet[][rig15]{2015ApJ...814L...6R},
\citet[][sax20]{2020A&A...636A..47S},
\citet[][sch16]{2016ApJ...818...38S},
\citet[][sch17]{2017ApJ...839...17S,2017ApJ...836L..14M},
\citet[][sch18]{2018A&A...616L..14S,2018MNRAS.474.4514I},
\citet[][sen17]{2017MNRAS.472.2608S},
\citet[][sen19]{2019MNRAS.488.3492S},
\citet[][sha03]{2003ApJ...588...65S},
\citet[][shi18]{2018PASJ...70S..15S},
\citet[][smi17]{2017MNRAS.467.3306S},
\citet[][sta14]{2014MNRAS.445.3200S},
\citet[][sta15]{2017MNRAS.464..469S,2015MNRAS.454.1393S,2015MNRAS.450.1846S},
\citet[][tan21]{2021MNRAS.501.3238T},
\citet[][van20]{2016ApJ...821L..27V,2017ApJ...842...47V,2020MNRAS.491.1093V}, and
\citet[][wof21]{2021MNRAS.500.2908W}.
The line fluxes and EWs collected for this literature sample are provided with this paper in the catalog described in Table~\ref{tab:litcatcol}. 
The short key in the ``reference'' column of this table is indicated after each of the literature references.
The redshift distribution of this sample of objects is shown as the dotted black histogram in Figure~\ref{fig:zdist}
where it is compared to the MUSE samples studied in this paper.
Figure~\ref{fig:litlegend} shows four projections of the multidimensional rest-frame EW space for the data sets in the literature catalog including plots of EW$_0$(\ciii), EW$_0$(\civ) for LAEs (bottom panels). 
\begin{table*}
\caption{\label{tab:litcatcol}Columns of the catalog of UV emission line sources collected from the literature provided with the paper.}
\centering
{
\begin{tabular}{p{2.2cm}p{2.5cm}p{2.5cm}p{10.0cm}}
\hline\hline
 & Unit & Catalog column & Description \\
\hline
ID 								& 						& {\tiny\verb+id+}				& Object ID. Each literature source was assigned a unique base id on the format $n\times1\textrm{e}9$, where $n$ counts the input references from the reference column. IDs from the literature source were then added to this base ID to form the individual unique object IDs. \\
R.A., Dec. 						& [deg]					& {\tiny\verb+ra+}, {\tiny\verb+dec+}	& Coordinates of the objects in the catalog.\\
Name 							& 						& {\tiny\verb+name+}				& Strings containing names of the objects in the catalog.\\
Reference			  				& 						& {\tiny\verb+reference+}			& Three-letter + publication year of the input reference(s). Appendix~\ref{sec:litcol} lists the full references assigned to each short reference.\\
$z$  								& 						& {\tiny\verb+redshift+}			& Redshift of the object in the catalog.\\
F$_\textrm{line}$ 					& [10$^{-20}$erg~s$^{-1}$~cm$^{-2}$]	& {\tiny\verb+f_line+}				& Emission line flux where {\tiny\verb+line+} refers to any of the lines \lya{} ({\tiny\verb+lya+}), \nv{} ({\tiny\verb+nv+}), \civ{} ({\tiny\verb+civ+}), \heii{} ({\tiny\verb+heii+}), \oiii{} ({\tiny\verb+oiii+}), \siiii{} ({\tiny\verb+siiii+}), and \ciii{} ({\tiny\verb+ciii+}). For doublets, the values for the individual components are also provided. As an example the columns for \ciii{} are named {\tiny\verb+f_ciii+}, {\tiny\verb+f_ciii1+}, and {\tiny\verb+f_ciii2+}. \\
$\delta$F$_\textrm{line}$   			& [10$^{-20}$erg~s$^{-1}$~cm$^{-2}$]	& {\tiny\verb+ferr_line+}			& Uncertainty on F$_\textrm{line}$. Upper limits have uncertainties set to +99. \\
(S/N)$_\textrm{line}$	 	 			& 						& {\tiny\verb+s2n_line+}			& S/N of the catalog emission line. \\
FR$_\textrm{line1line2}$				& 						& {\tiny\verb+fr_line1line2+}		& Emission line flux ratio between emission lines line1 and line2. \\
$\delta$FR$_\textrm{line1line2}$		& 						& {\tiny\verb+frerr_line1line2+}		& Uncertainty on FR$_\textrm{line1line2}$. Upper and lower limits have uncertainties set to +99 and -99, respectively.   \\
FR$_\textrm{line1line2}$ S/N			& 						& {\tiny\verb+frs2n_line1line2+}		& S/N of  FR$_\textrm{line1line2}$. \\
EW$_{0,\textrm{line}}$ 				& [\AA]					& {\tiny\verb+ew0_line+}			& Rest-frame equivalent width of {\tiny\verb+line+}, where line refers to any of the lines \lya, \nv, \civ, \heii, \oiii, \siiii, and \ciii. For doublets, the values for the individual components are also provided. \\
$\delta$EW$_{0,\textrm{line}}$   		& [\AA]					& {\tiny\verb+ew0err_line+}		& Uncertainty on EW$_\textrm{0,line}$. Upper and lower limits have uncertainties set to +99 and -99. \\
\hline
\end{tabular}
}
\end{table*}


\section{The full PIM-PDF of the NEOGAL and BPASS models}\label{sec:pimodels}

\begin{figure*}
\begin{center}
\includegraphics[width=0.95\textwidth]{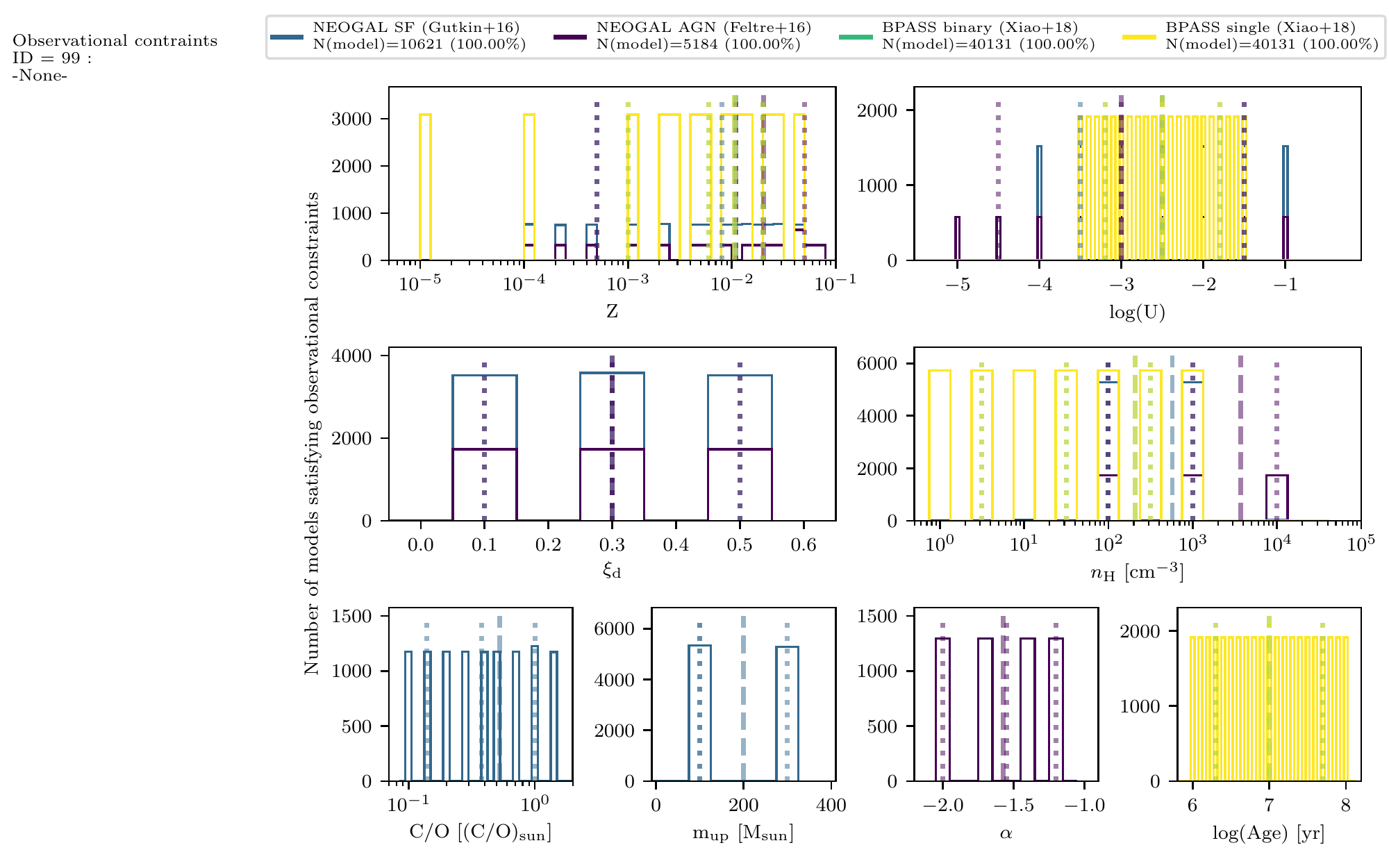}
\caption{Full distribution of the PhotoIonization Model Probability Density Functions (PIM-PDFs) of the NEOGAL star forming (blue), NEOGAL AGN (purple), BPASS single (yellow), and BPASS binary (green) photoionization models (we note that the BPASS models fully overlap in initial parameters sampled).
It is these models that are used for the parameter inference based on the observational constraints from the UV emission line detections presented in Section~\ref{sec:pimodelinference}. 
The actual values of the sampled individual parameters are provided in Table~\ref{tab:photoionnizationparam}.}
\label{fig:pimodelPDFsNoConstraints}
\end{center}
\end{figure*}

Figure~\ref{fig:pimodelPDFsNoConstraints} shows the full prior PIM-PDFs of the physical parameters sampled by the NEOGAL and BPASS-based photoionization models described in Section~\ref{sec:pimodelinference}.
It is this initial photoionization model parameter space (listed in Table~\ref{tab:photoionnizationparam}) that is constrained by the observations and measurements from the TDOSE spectra of the MUSE objects studied in this work resulting in object PIM-PDFs similar to the ones shown in Figure~\ref{fig:pimodelPDFsObj}.

\section{Plots of TDOSE spectra and FELIS UV emission line detections}\label{sec:examplespec}

\begin{figure*}
\begin{center}
\includegraphics[width=0.98\textwidth]{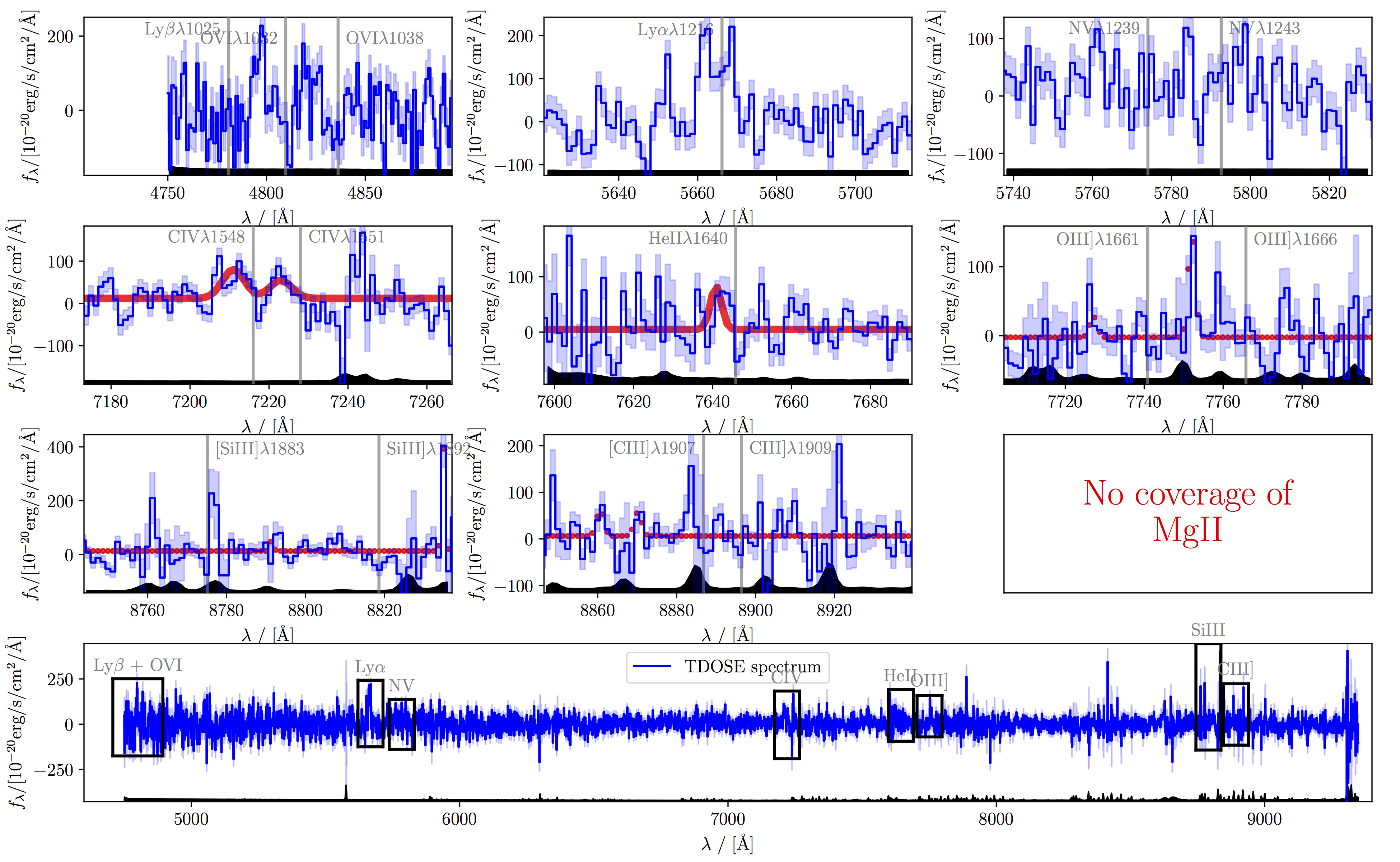}\\
\includegraphics[width=0.98\textwidth]{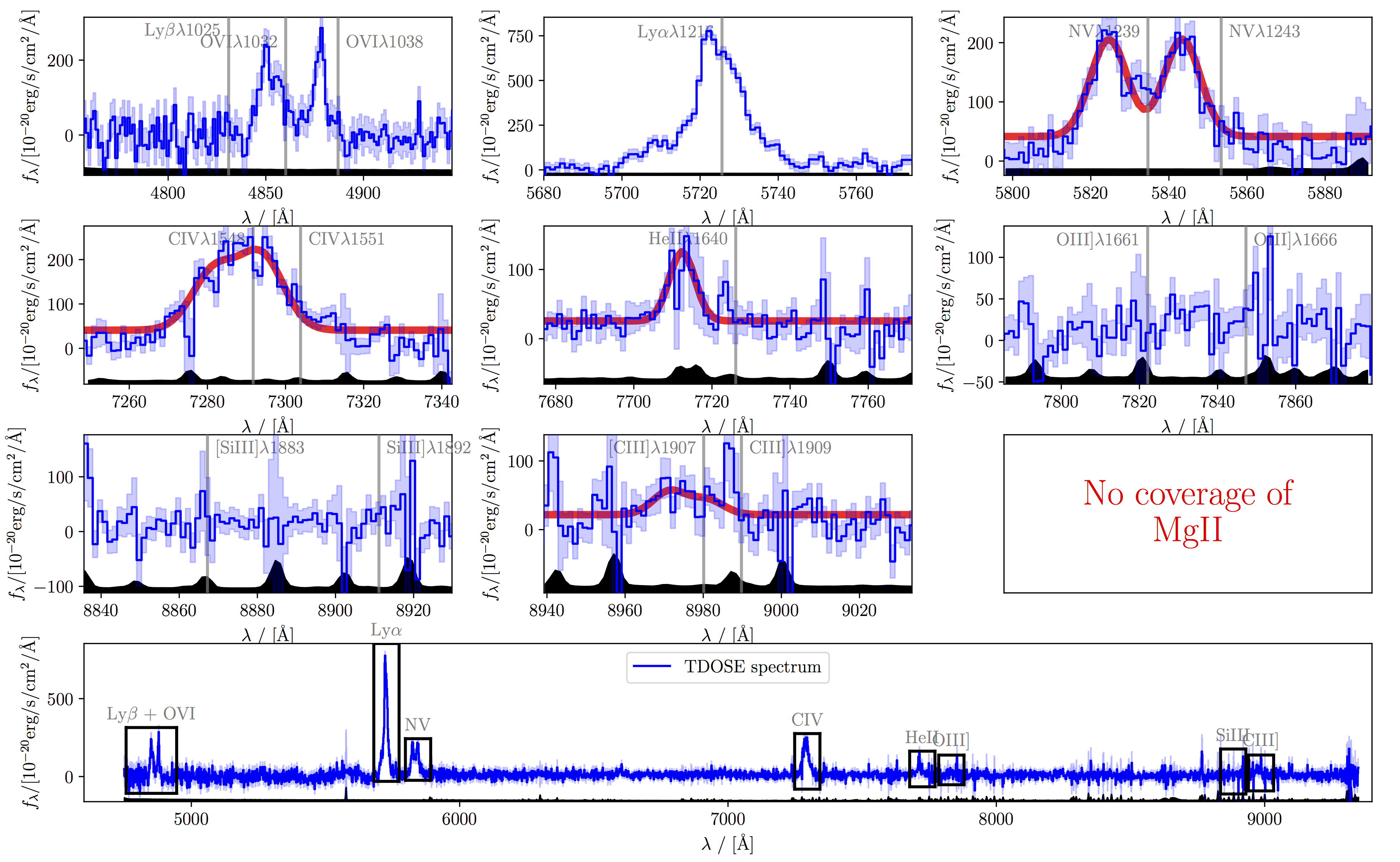}
\caption{Similar to Figure~\ref{fig:ObjSpec}, but showing the MUSE-Wide GOODS-South AGN 104014050 ($z=3.66$) (top) and 115003085 \citep[$z=3.71$, bottom, see also][]{2002ApJ...571..218N}.}
\label{fig:ObjSpec04}
\end{center}
\end{figure*}

\begin{figure*}
\begin{center}
\includegraphics[width=0.98\textwidth]{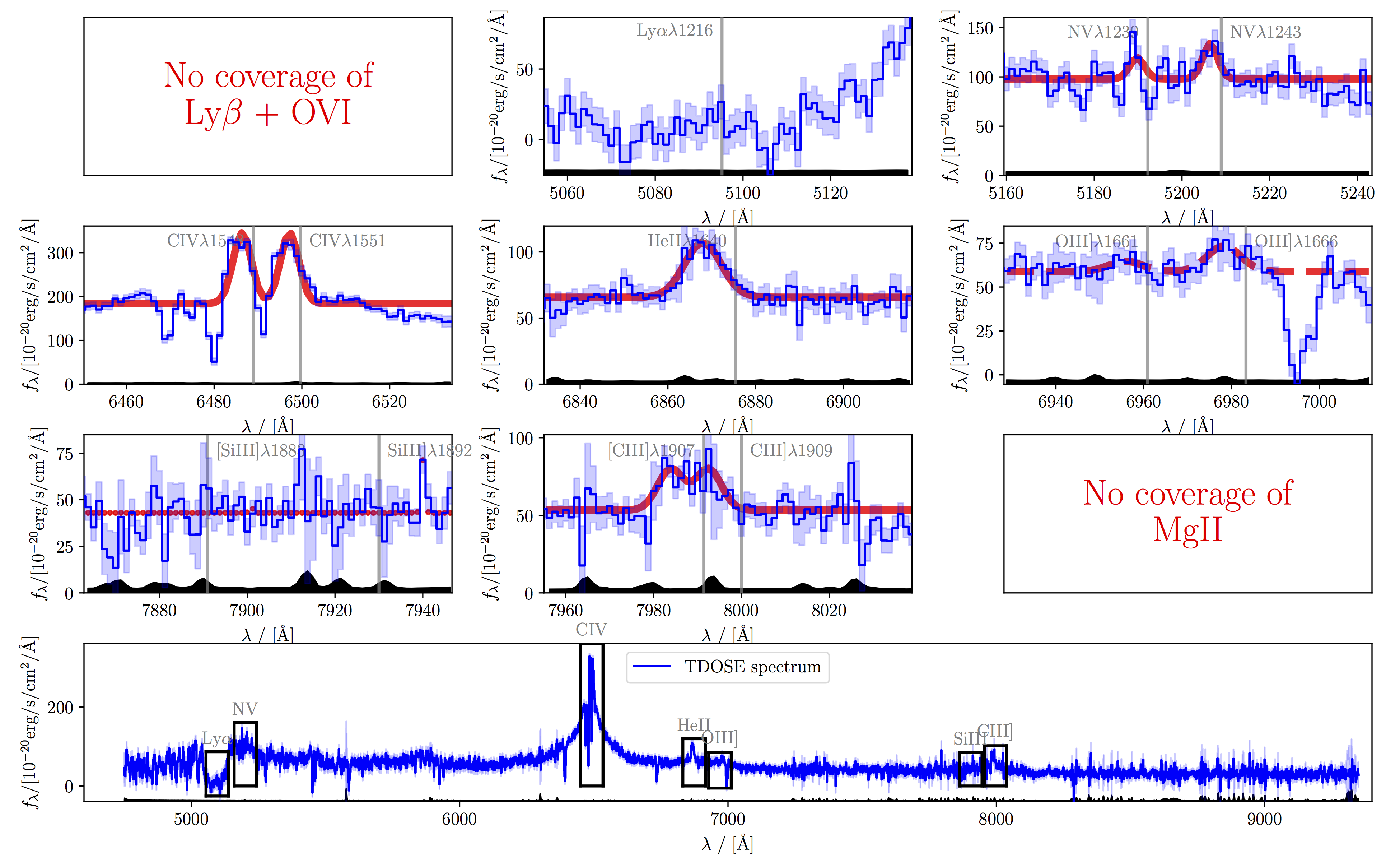}\\
\includegraphics[width=0.98\textwidth]{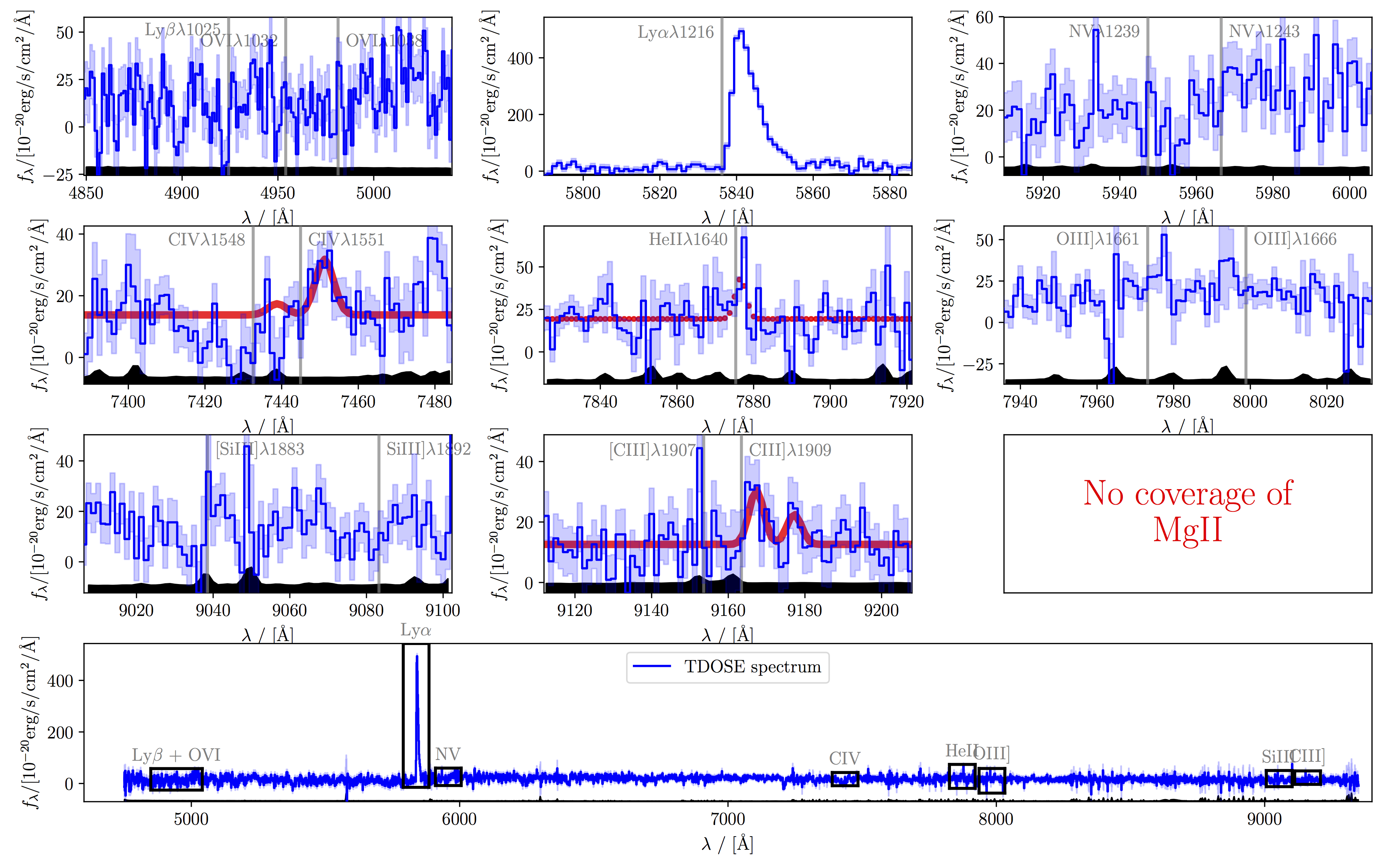}
\caption{Similar to Figure~\ref{fig:ObjSpec}, but showing the UDF mosaic AGN 601381485 ($z=3.19$) (top) and 
object 604992563 ($z=3.80$) (bottom). 
The latter object has a lead line redshift blue-wards of the \lya{} peak as indicated by the vertical gray lines. 
Even when assigning the objects redshift according to a fit of the \lya{} profile, as described in Section~\ref{sec:voffset}, the offset of the \lya{} emission with respect to systemic as traced by the \ciii{} still has an intriguing value of $-145$~km~s$^{-1}$.
}
\label{fig:ObjSpec98}
\end{center}
\end{figure*}

\begin{figure*}
\begin{center}
\includegraphics[width=0.98\textwidth]{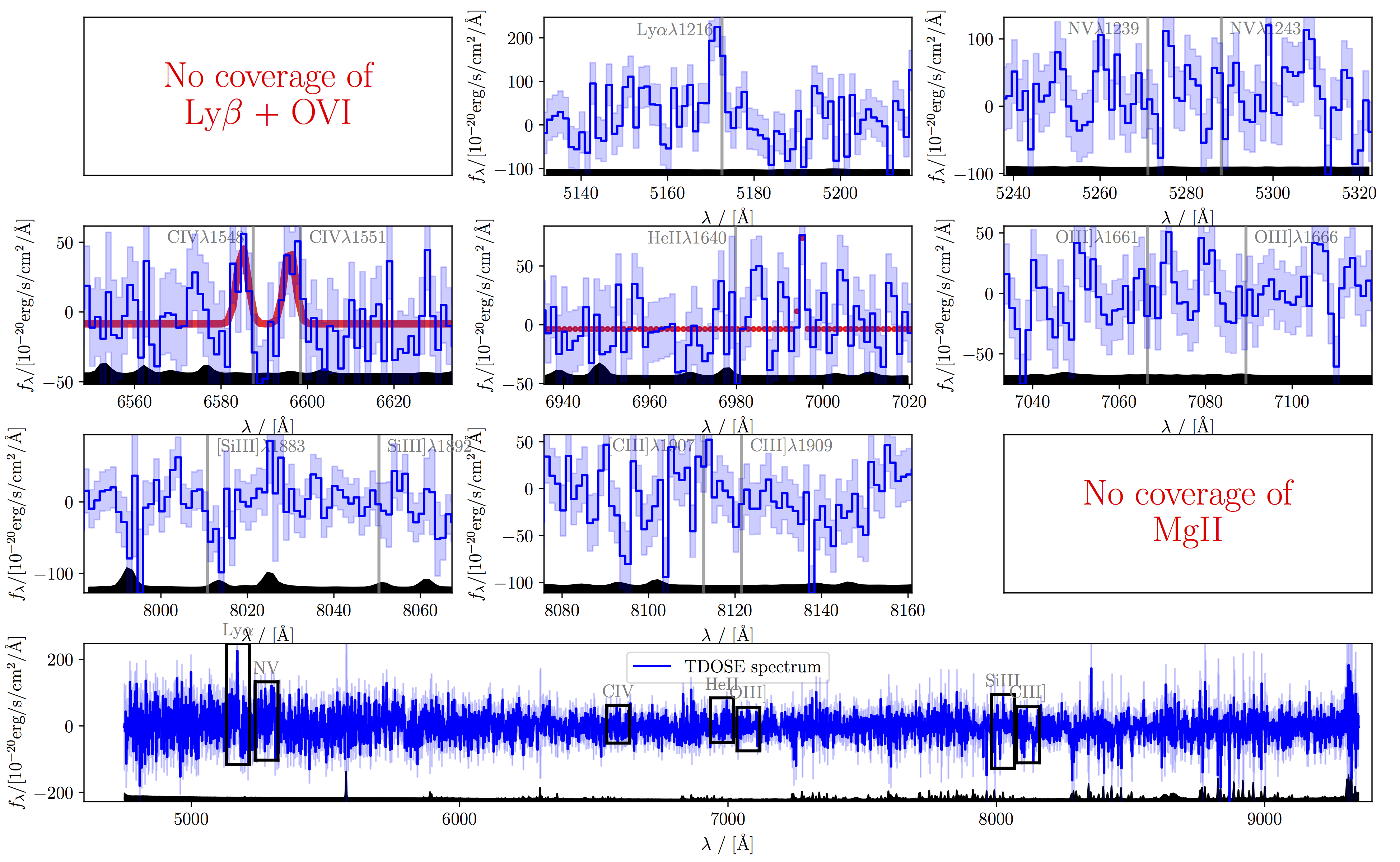}\\
\includegraphics[width=0.98\textwidth]{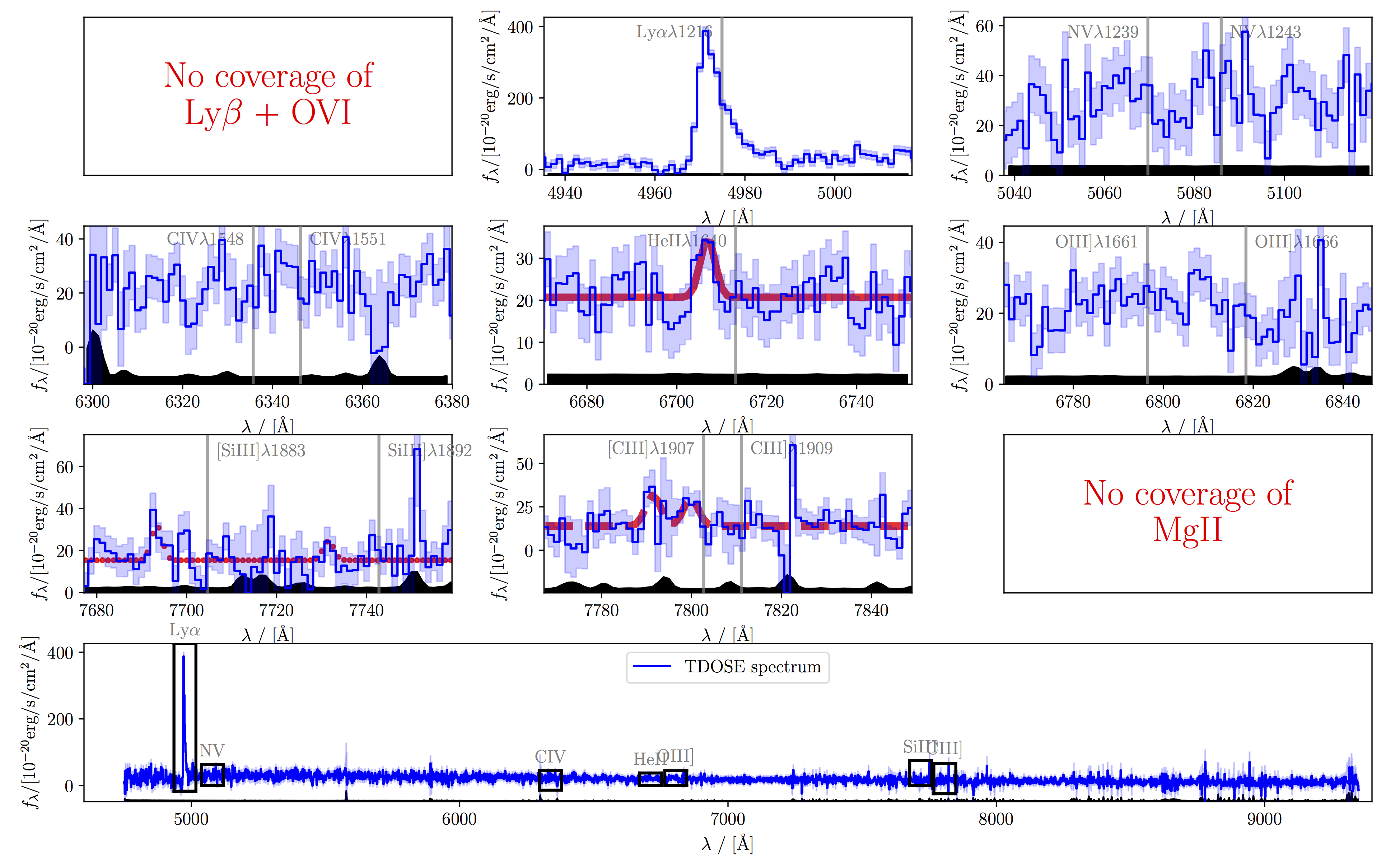}
\caption{Similar to Figure~\ref{fig:ObjSpec}, but showing the objects 102014087 ($z=3.25$) (top) and 601281436 ($z=3.09$) (bottom) for which we show the PIM-PDFs in Figure~\ref{fig:pimodelPDFsObj}.}
\label{fig:ObjSpec02}
\end{center}
\end{figure*}

\begin{figure*}
\begin{center}
\includegraphics[width=0.98\textwidth]{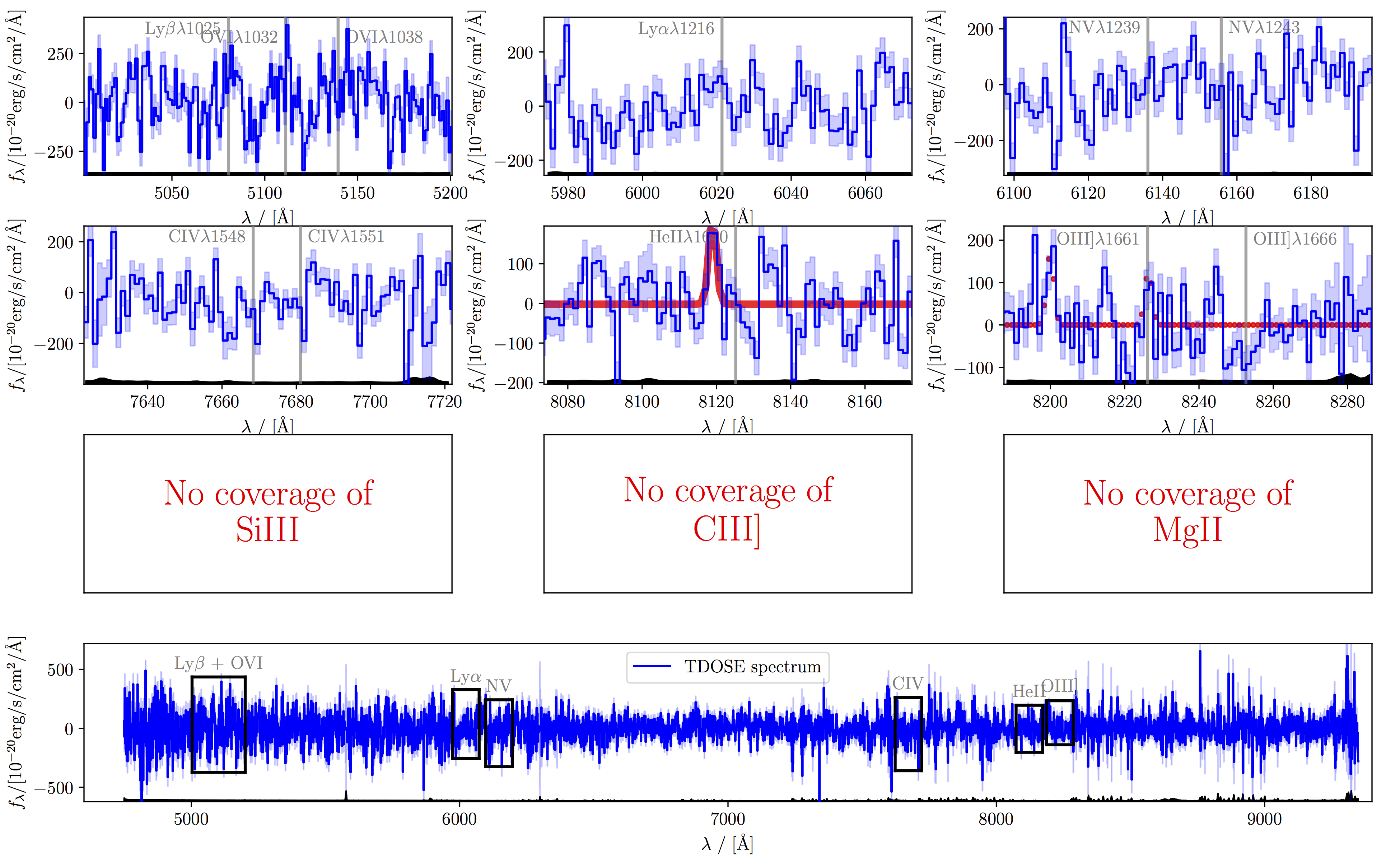}\\
\includegraphics[width=0.98\textwidth]{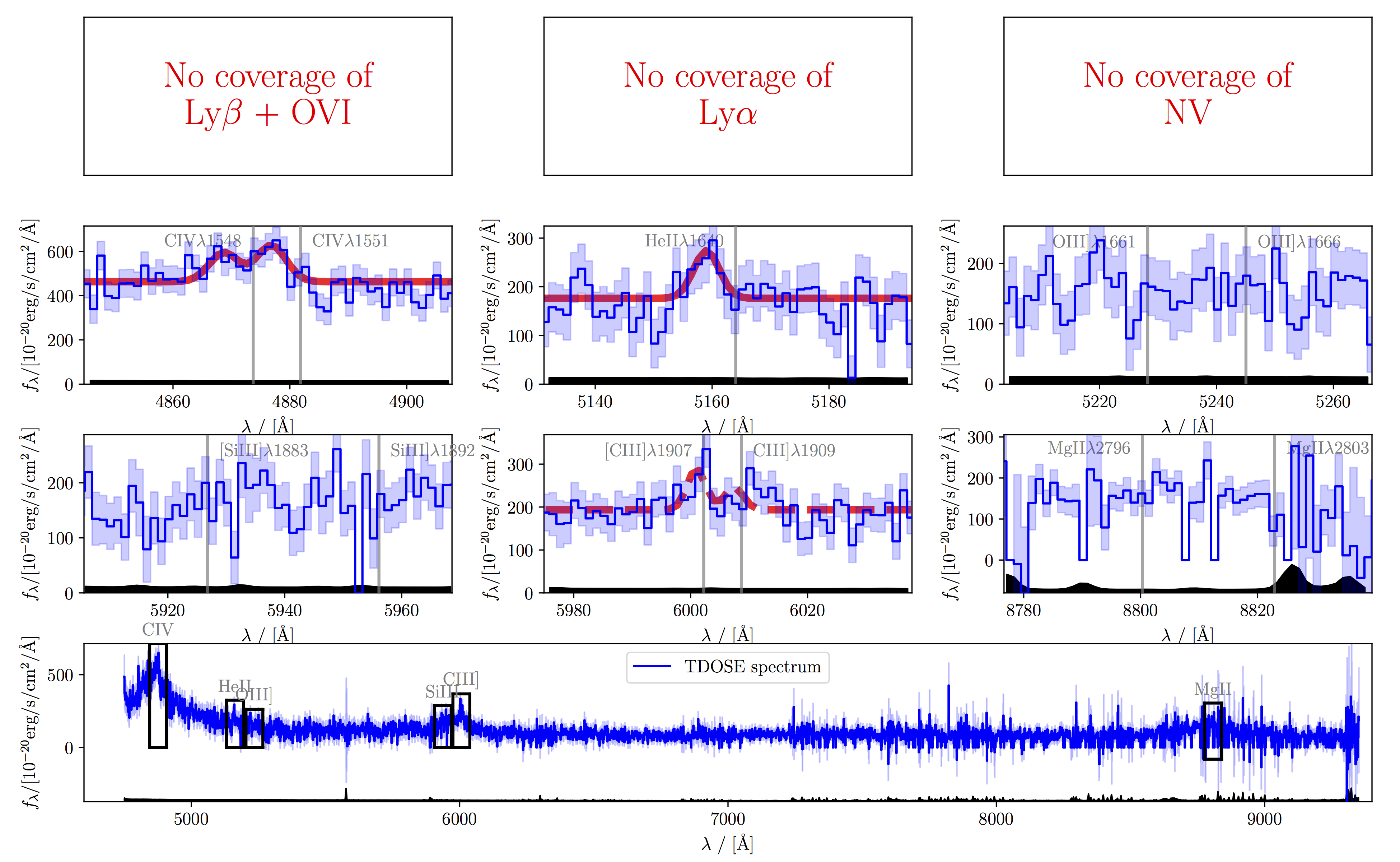}
\caption{Similar to Figure~\ref{fig:ObjSpec}, but showing COSMOS MUSE-Wide objects 202021046 ($z=3.95$) (top) and 221004004 ($z=2.15$) (bottom). The latter showing continuum, \heii{} emission and broad \civ{} emission characteristic of AGN.}
\label{fig:ObjSpec06}
\end{center}
\end{figure*}

\begin{figure*}
\begin{center}
\includegraphics[width=0.98\textwidth]{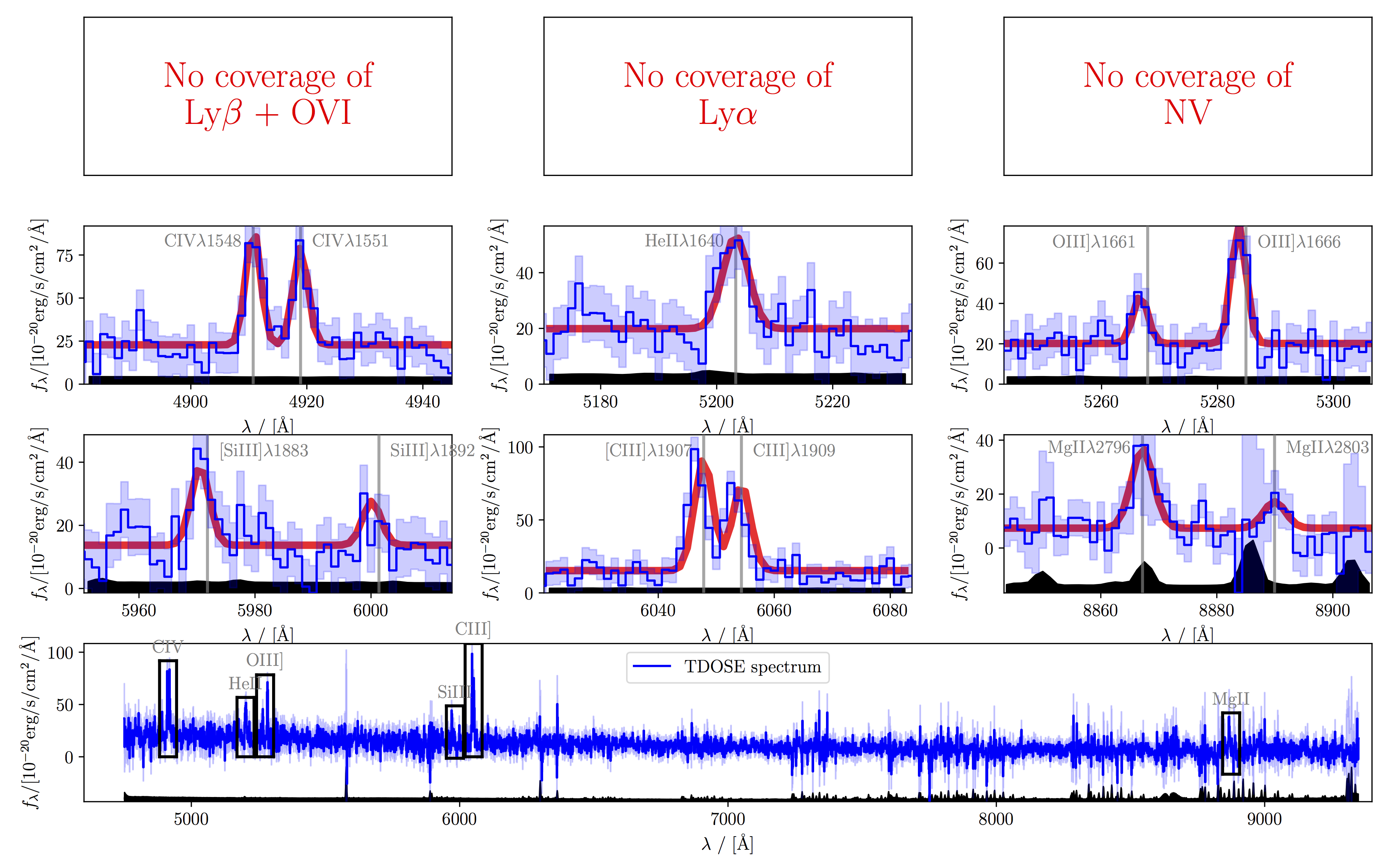}\\
\includegraphics[width=0.98\textwidth]{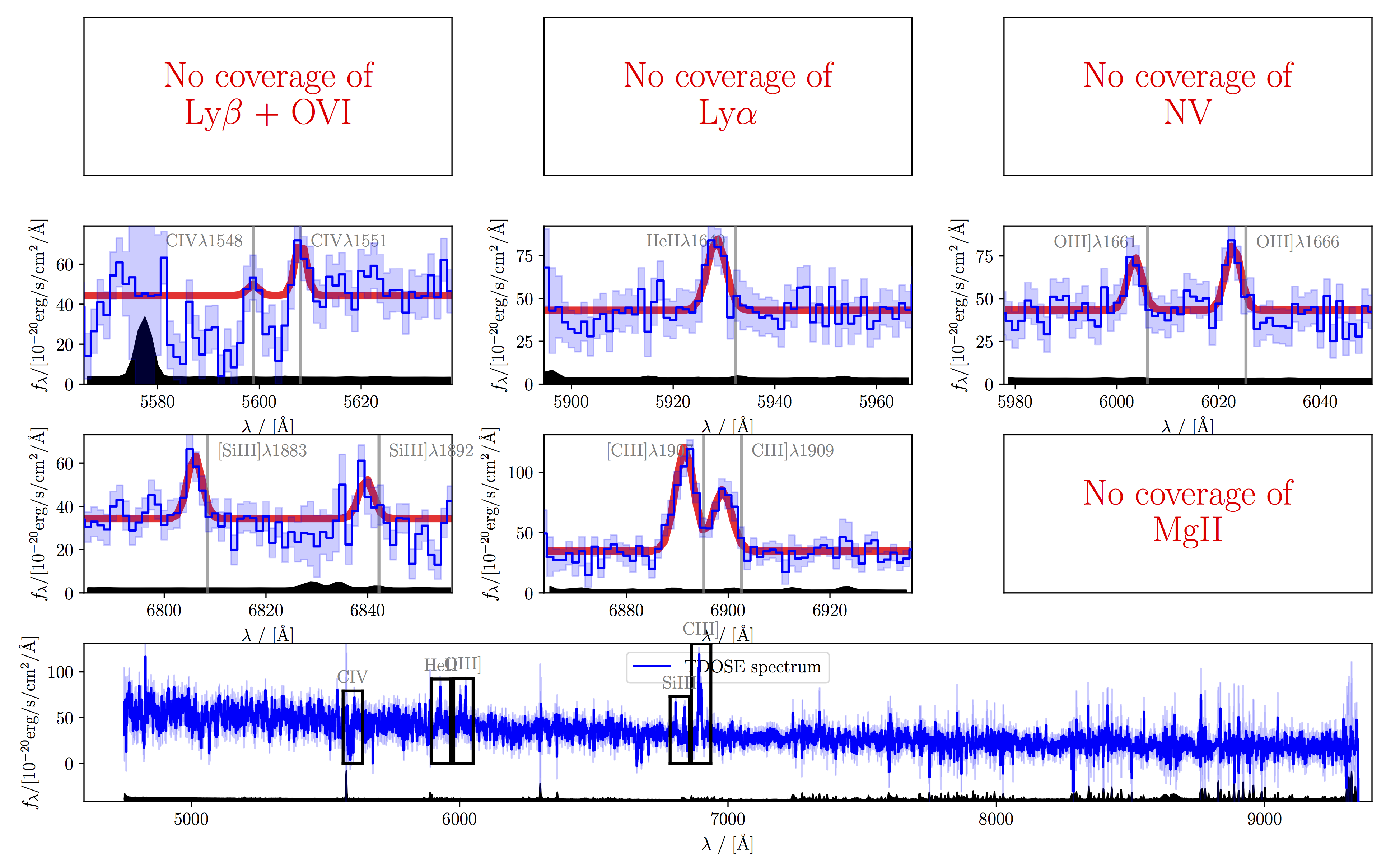}
\caption{Similar to Figure~\ref{fig:ObjSpec}, but showing the two lower-$z$ UDF mosaic objects 600921283 ($z=2.17$) (top) and 605172634 ($z=2.62$) (bottom) where all covered lines were clearly detected.
Object 605172634 is the outlier described in Section~\ref{sec:neTe}.}
\label{fig:ObjSpec03}
\end{center}
\end{figure*}

\begin{figure*}
\begin{center}
\includegraphics[width=0.98\textwidth]{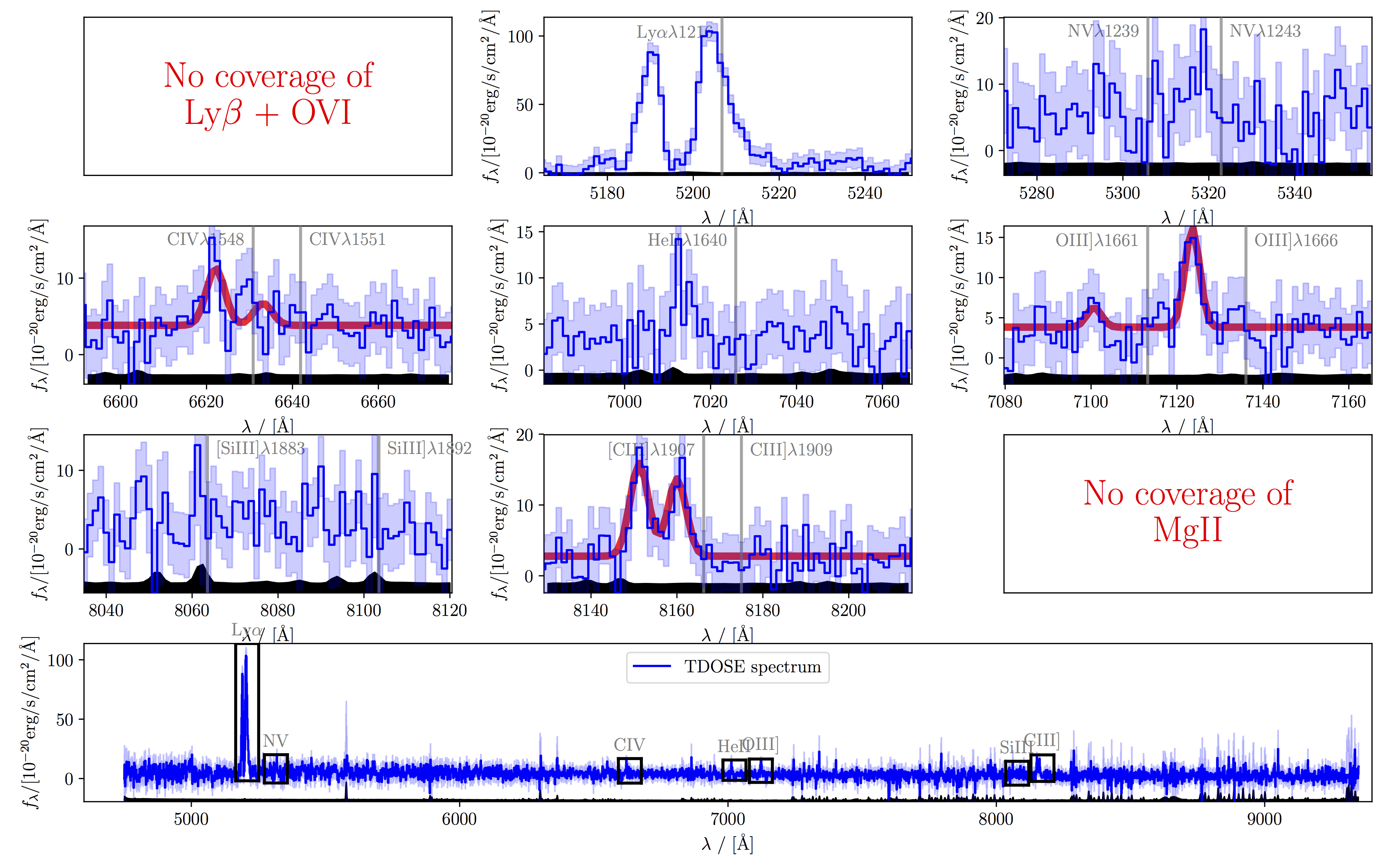}\\
\includegraphics[width=0.98\textwidth]{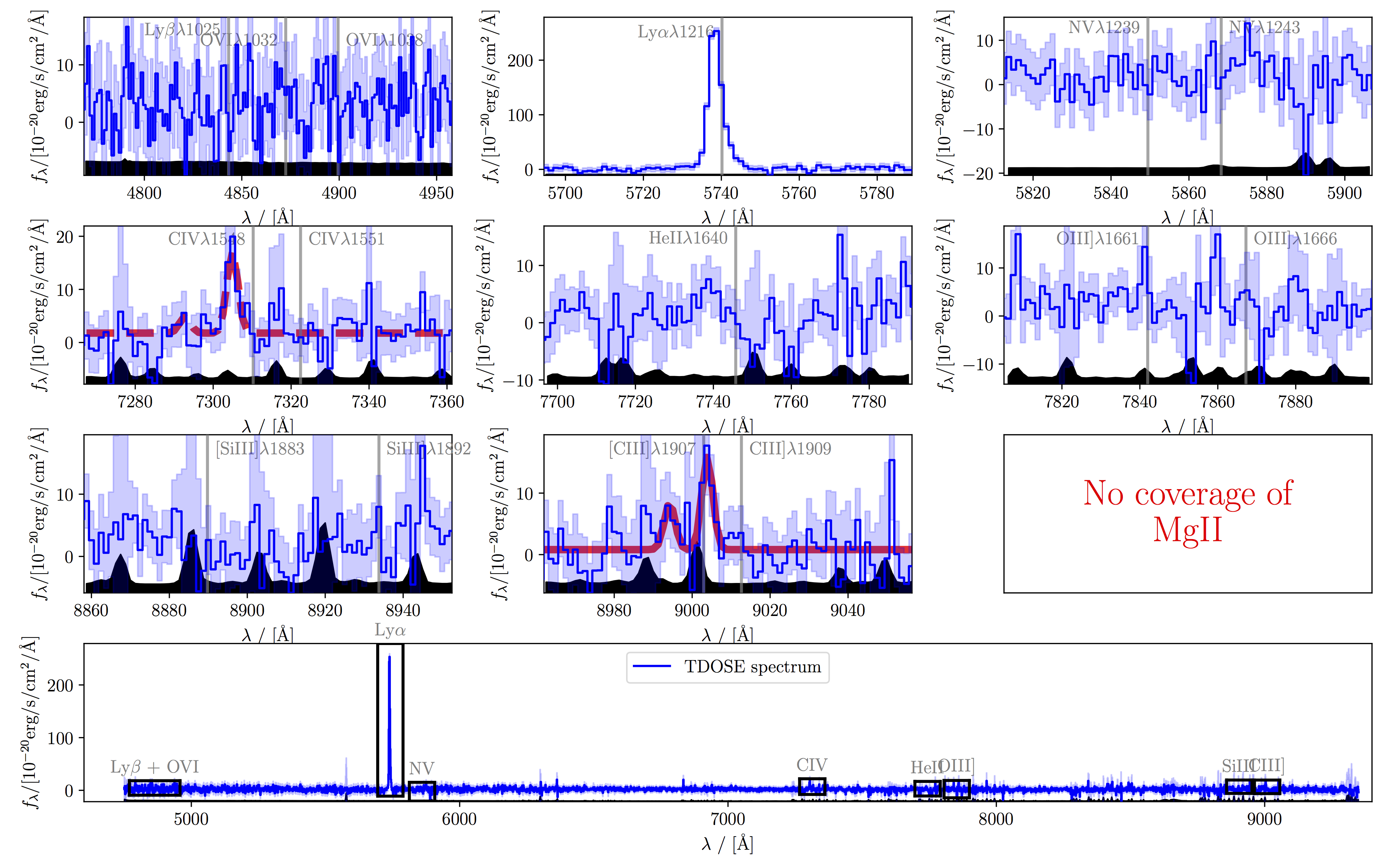}
\caption{Similar to Figure~\ref{fig:ObjSpec}, but showing the two UDF10 LAEs 720500425 ($z=3.28$) (top) and 721250731 ($z=3.72$) (bottom).}
\label{fig:ObjSpec99}
\end{center}
\end{figure*}

Figures~\ref{fig:ObjSpec04}--\ref{fig:ObjSpec99} provide further examples of sources with FELIS UV emission line detections in the TDOSE spectra similar to Figure~\ref{fig:ObjSpec} representing the breadth of sources in the parent sample analyzed in this paper. 
For a detailed caption we refer to Figure~\ref{fig:ObjSpec}.
Several of the panels in these figures support the discussion of individual sources as noted in the respective captions.

\end{appendix}
\end{document}